\documentclass[
  aps,%
  twocolumn,%
  eqsecnum,%
  rmp,%
  letterpaper,%
  amsfonts,%
  floatfix,%
  nofootinbib,%
]{revtex4}

\usepackage{amsmath,amssymb,bm}
\usepackage{graphicx}
\usepackage{subfigure}
\usepackage{hyperref}
\hypersetup{%
  dvips,
  hyperindex = {false},
  citecolor = {blue},
 colorlinks= true,
  linkcolor = red ,
 linktocpage=true,
 pdfauthor = {\textcopyright\ Valeri N. Kotov },
  pdffitwindow = {true},
  pdfpagelayout = {SinglePage},
  pdfstartview = {Fit},
  pdftoolbar = {true},
}

\newcommand{\vF}{{\ensuremath{v}}}

\newcommand{\bsigma}{\ensuremath{\bm\sigma}}

\newcommand{\bnabla}{{\ensuremath{\bm{\nabla}}}}
\newcommand{\bk}{{\ensuremath{\bm{k}}}}
\newcommand{\bK}{{\ensuremath{\bm{K}}}}

\newcommand{\bp}{{\ensuremath{\bm{p}}}}
\newcommand{\bq}{{\ensuremath{\bm{q}}}}
\newcommand{\br}{{\ensuremath{\bm{r}}}}

\newcommand{\e}{{\ensuremath{\varepsilon}}}
\newcommand{\eG}{{\ensuremath{\varepsilon_G}}}

\newcommand{\gt}{\tilde{g}}
\newcommand{\Fref}[1]{Fig.~\ref{#1}}
\newcommand{\Eqref}[1]{Eq.~\eqref{#1}}

\newcommand{\sign}[1]{{\ensuremath{s_{#1}}}}

\begin{document}

\title{Electron-Electron Interactions in Graphene: Current Status and Perspectives}

\author{Valeri N. Kotov}
\affiliation{Department of Physics, University of Vermont, 82 University Place, Burlington, Vermont 05405}

\author{Bruno Uchoa}
\affiliation{Department of Physics,
 University of Illinois at Urbana-Champaign, 1110 West Green Street, Urbana, Illinois 61801}

\author{Vitor M. Pereira}
\affiliation{Graphene Research Centre and Department of Physics, National   
  University of Singapore, 2 Science Drive 3, Singapore 117542}

\author{F. Guinea}
\affiliation{Instituto de Ciencia de Materiales de Madrid, Sor  Juana In\'es de la Cruz 3, E-28049 Madrid,  Spain}

\author{A. H. Castro Neto}
\affiliation{ Department of Physics, Boston University, 590 Commonwealth Avenue, Boston, Massachusetts 02215}
\affiliation{Graphene Research Centre and Department of Physics, National   
  University of Singapore, 2 Science Drive 3, Singapore 117542}

\begin{abstract}
We review the problem of electron-electron interactions in graphene. Starting from the  screening of long range interactions in these systems, we discuss the existence of an emerging Dirac liquid of Lorentz invariant quasi-particles in the weak coupling regime, and strongly correlated electronic states in the strong coupling regime. We also analyze the analogy and connections between the many-body problem and the Coulomb impurity problem. The problem of the magnetic instability and Kondo effect of impurities and/or adatoms in graphene is also discussed in analogy with classical models of many-body effects in ordinary metals. We show that Lorentz invariance plays a fundamental role and leads to effects that span the whole spectrum, from the ultraviolet to the infrared. The effect of an emerging Lorentz invariance is also discussed in the context of finite size and edge effects as well as mesoscopic physics. We also briefly discuss the effects of strong magnetic fields in single layers and review some of the main aspects of the many-body problem in graphene bilayers. In addition to reviewing the fully understood aspects of the many-body problem in graphene, we show that a plethora of interesting issues remain open, both theoretically and experimentally, and that the field of graphene research is still exciting and vibrant.    
\end{abstract}

\date{\today}

\maketitle

\tableofcontents

\section{Introduction}

One of the most important problems in theoretical physics is the understanding of the properties of quantum systems with an infinitely large number of interacting degrees of freedom, the so-called many-body problem. Interactions are present in almost all areas of physics: soft and hard condensed matter, field theory, atomic physics, quantum chemistry, nuclear physics, astrophysics, and so on. Interactions between particles are responsible for a plethora of effects and many-body states, from the band structure of crystals to superconductivity in metals, from the quark-gluon plasma in heavy ion collisions to asymptotic freedom in quantum chromodynamics (QCD). It is the competition between the kinetic energy of the particles, that is, their inertia, and interactions among them that leads to the richness and complexity of these different phases. For these reasons, many-body interactions are very specific, and the hardest to describe theoretically.

 One of the greatest theoretical achievements of the last century, the Landau theory of the Fermi liquid \cite{Baym:1991}, asserts something very simple but, at the same time, very deep: that
the excitations of  a large (indeed, infinite) collection of strongly interacting particles can be described
 as an equally large collection of weakly-interacting quasi-particles that carry the same quantum numbers as the original particles. This statement is far from trivial. Consider, for instance, the behavior of electrons in a metal. The electrons interact among themselves and with the ions in the crystal via strong long-range Coulomb interactions. It is not at all clear what is the outcome of this complex interacting problem. Without having any deep theoretical resources to treat this problem, except an extraordinary intuition, visionaries like Paul Drude \cite{Drude1:1900,Drude2:1900} and Arnold Sommerfeld \cite{Hoddeson:87} settled the foundations for the understanding of this complex problem by postulating, shamelessly, that (1) electrons propagate freely in a non-relativistic (Galilean invariant) way (Drude's contribution), and (2) electrons obey Fermi-Dirac statistics (Sommerfeld's contribution). {\it Galilean invariance} dictates that the electrons have a kinetic energy given by
\begin{eqnarray}
K_0 = \frac{{\bf p}^2}{2 m^*} \, ,
\label{kinetic}
\end{eqnarray}
where ${\bf p}$ is the electron momentum and $m^*$ is a free parameter of the theory that, for 
lack of a better name, is called effective mass. Fermi-Dirac statistics implies that electrons carry spin $1/2$ and that, in the ground state, all states with energy below the so-called Fermi energy, $E_F$, are occupied, and all the states above it are empty. With these two basic assumptions and simple considerations about electron scattering by defects, the Drude-Sommerfeld model was capable
of describing  experimental data of several generations of scientists. 

The understanding of why these two assumptions are valid for a strongly interacting problem, such as electrons in a metal, had to wait for the development of two major concepts: ({\it i}) the {\it band structure} theory that explains that the interaction of the electrons with a periodic lattice of ions produces states that, as the plane waves described by (\ref{kinetic}), are extended over the entire lattice \cite{Bloch:28}; and ({\it ii}) the theory of {\it screening}, that is, that metals are dynamically polarizable materials and that electrons act collectively to screen electric fields in their  interior \cite{Lindhard:54}. Hence,  long range Coulomb interactions become effectively short ranged and weak enough to give substance to Drude's assumptions. In this case the effective mass $m^*$ reflects the change in the inertia of the electron as it moves around in an effective medium. Nevertheless, there are situations when these assumptions fail even in crystalline systems, and that is when interesting things happen, namely, the free electron picture breaks down.

In fact, there are many instances where the Fermi liquid ground state becomes unstable. Electrons not only interact with static ions but also with their vibrations, the phonons. Electron-phonon interactions, in the presence of strong screening, can lead to an effective attractive interaction between electrons producing a catastrophic Fermi surface instability towards a superconducting ground state \cite{tinkham:96}. Fermi surface instabilities also happen in special situations in the presence of Fermi surface nesting which can lead to charge and spin density wave ground states \cite{Gruner:94}. Crystals with inner shell electrons, such as transition metals, can also have many-body instabilities due to the strong {\it local} interactions between the electrons, leading to insulating states with magnetic properties as in the case of Mott insulators \cite{Mott:49}. Another important case of Fermi liquid breakdown is when the electron density is very low and the screening disappears. 

Notice that in quantum mechanics the momentum of the particle relates to its wavelength, $\lambda$, by $p = \hbar/\lambda$ and hence the kinetic energy (\ref{kinetic}) behaves as $K = \hbar^2/(2 m^* \lambda^2)$. If the average distance between electrons is $\ell$ we see that the average kinetic energy per electron has to be of the order $E_K \approx \hbar^2 n_d^{2/d} (2 m^*)$ where $n_d = 1/\ell^d$ is the average electron density in $d$ spatial dimensions. On the other hand, the Coulomb interaction is given by:
\begin{eqnarray}
V(r) = \frac{e^2}{\epsilon_0 r} \, ,
\label{coulomb}
\end{eqnarray}
where $e$ is the electron charge, and $\epsilon_0$ the dielectric constant of the medium. Notice that the  Coulomb energy per electron is of the order $E_C \approx e^2 n_d^{1/d}/\epsilon_0$. Thus, the ratio of   Coulomb 
to kinetic energy is given by $r_s = E_C/E_K \propto (n_0/n_d)^{1/d}$, where $n_0 = [m^* e^2/(\hbar^2 \epsilon_0)]^d$ depends only on material properties. Therefore, at high electron densities, $n_d \gg n_0$, the kinetic energy dominates over the Coulomb energy, which can be disregarded, and the Fermi liquid description is safe. At low densities, $n_d \ll n_0$ the Coulomb energy is dominant and new electronic phases, such as ferromagnetism and Wigner crystallization, can become stable \cite{ceperley:78}. Therefore, the relative strength of the kinetic to Coulomb interactions in Galilean invariant systems is completely controlled by the electron density. Notice that in all the cases discussed above the Galilean invariance was kept intact and the driving force for the many-body instabilities was the enhancement of  the Coulomb relative to the kinetic energy. 

With the advent of graphene \cite{Novoselov:2004}, a two dimensional crystal of pure carbon, this picture has changed and a new example of Fermi liquid breakdown has emerged in a big way. In graphene, due to its peculiar lattice structure, the electrons at the Fermi energy are described in terms of an effective {\it Lorentz} invariant theory where the kinetic energy is given by the Dirac dispersion \cite{CastroNeto:09}
\begin{eqnarray}
K_G = \pm v_F |{\bf p}| \, ,
\label{kgraphene}
\end{eqnarray}
where $v_F$ is the Fermi-Dirac velocity, and the $\pm$ signs refer to two linearly dispersing bands. If we take (\ref{kgraphene}) at face value and reconsider the argument given above on the relevance of the Coulomb interactions we reach very different conclusions. For one, the form of the Coulomb interaction remains the same as in (\ref{coulomb}), since $v_F$ is a material's property and hence much smaller than the speed of light, $c$. This means that the photons which mediate the Coulomb interaction are still much faster than the electrons and, thus, the electron-electron interaction can be considered as instantaneous. Therefore, the Coulomb interaction (\ref{coulomb}) actually breaks the Lorentz invariance of (\ref{kgraphene}). Secondly, because of the linear scaling of the kinetic energy with momentum, we see that the average kinetic energy per electron has to scale like $E_G \approx \hbar v_F n^{1/2}$ and consequently  the ratio of  Coulomb 
to  kinetic energy  is given by
\begin{eqnarray}
\alpha = \frac{E_C}{E_G} = \frac{e^2}{\epsilon_0 \hbar v_F} \, ,
\label{alfa}
\end{eqnarray}
and is {\it independent} of the electronic density $n$, depending only on material properties and environmental conditions, such as $\epsilon_0$.  Here, and from now on, we refer to graphene's electron density
 as $n$. As the electronic properties of graphene are sensitive to environmental conditions, they will be modified by the presence of other layers. In fact, as we are going to show, bilayer graphene has properties which are rather different than its monolayer counterpart. Furthermore, due to the same peculiar dispersion relation, the electronic density of states, $\rho(E)$, vanishes at the Dirac point, $\rho(E) \propto |E|/v_F^2$, and hence graphene is a hybrid between an insulator and a metal: neutral graphene is not a metal because it has vanishing density of states at the Fermi energy, and it is not an insulator because it does not have a gap in the spectrum. This means that  pristine (or lightly doped) 
graphene cannot screen the long range Coulomb interaction in the usual (metallic) way,
 although it is possible to produce electronic excitations at vanishingly small energy. This state of affairs makes of graphene a unique system from the point of view of electron-electron interactions. 
 The long-range interactions lead to non-trivial renormalization of the Dirac quasiparticle
 characteristics near the charge neutrality point, and the resulting electronic state can be called
{\it Dirac liquid}, to be distinguished from the Fermi liquid behavior  at finite chemical
 potential (away from the Dirac point, where conventional screening takes place.)

The unusual relation between kinetic and Coulomb energies not only affects the electron-electron interactions, but also the interactions of the electrons with charged impurities, the so-called Coulomb impurity problem. In a metal described by a Galilean invariant theory of the form (\ref{kinetic}), screening also makes the interaction with the impurity short ranged, and hence the scattering problem  effectively reduces to the one of a short range impurity. In graphene, because of the lack of screening the situation is rather different, and one has to face the problem of the effect of the long range part of the potential. Scattering by long range interactions has a long history in physics and it leads to the issue of logarithmic phase shifts \cite{Baym:69}. In graphene, because of its emergent Lorentz invariance, this issue is magnified. Since Coulomb interactions between electrons and electron scattering by Coulomb impurities are closely related issues, one expects that many of the anomalies of one problem are also 
reflected  in the other. 

Another interesting consequence of the scaling of the kinetic energy with momentum is related to
 the issue of electron confinement. If  electrons are confined to a region of size $L$ the energy of the states is quantized, no matter whether the electrons obey Galilean or Lorentz invariance. However, the quantization of  energy is rather different in these two cases. In a Galilean invariant system, like the one described by (\ref{kinetic}) the energy levels are spaced as $\Delta E_0 \propto 1/L^2$ while in graphene Lorentz invariance, (\ref{kgraphene}), implies $\Delta E_G \propto 1/L$. Hence, the size dependence of the energy levels in sufficiently small samples of graphene is rather different than one would find in normal metals. Moreover, since the Coulomb energy scales like $1/L$ we expect Coulomb effects to be stronger in nanoscopic and mesoscopic graphene samples. 

Furthermore, the fact that graphene is a two dimensional (2D) system has strong consequences for electronic motion in the presence of perpendicular magnetic fields. Since a perpendicular magnetic field $B$ leads to a quantization of the energy in terms of Landau levels, and the electrons cannot propagate along the direction of the field, its effect is singular, in the sense that the problem has a massive degeneracy. So, strong magnetic fields can completely quench the kinetic energy of the electrons that become dispersionless. The electronic orbits are localized in a region of the size of the magnetic length: $\ell_B = \sqrt{\hbar c/(e B)}$. For a Galilean invariant system, such as the one described by (\ref{kinetic}), for $p \approx \hbar/\ell_B$ the  kinetic energy per electron is of order $K \approx \hbar \omega_C \propto B$ where $\omega_C = \hbar/(m^* \ell_B^2)$ is the cyclotron frequency. On the other hand, for graphene, using (\ref{kgraphene}), one has $E_G \approx \hbar \omega_G \propto \sqrt{B}$ where $\omega_G = \sqrt{2} v_F/\ell_B$, which is a consequence of the Lorentz invariance. Notice that in both cases the Coulomb energy per electron scales like 
$E_C \propto e^2/(\epsilon_0 \ell_B) \propto \sqrt{B}$. Hence, in a Galilean invariant system the Coulomb energy is smaller than the kinetic energy at high fields while for Lorentz invariant systems they are always comparable. 
Thus, one expects  Coulomb interactions to be hugely enhanced in the presence of these magnetic fields. In the 2D electron gas (2DEG) this unusual state of affairs is what leads to the fractional quantum Hall effect (FQHE) \cite{Laughlin:83}. 

Given all these unusual circumstances, many questions come to mind: How does screening of the long range Coulomb interaction work in graphene? Can graphene be described in terms a Lorentz invariant theory of quasi-particles? Is the Coulomb impurity problem in graphene the same as in a normal metal? In what circumstances is graphene unstable towards many-body ground states? Are there quantum phase transitions \cite{Sachdev:99} in the phase diagram of graphene? Do magnetic moments form in graphene in the same way as they do in normal metals? What is the ground state of graphene in high magnetic fields?  

The objective of this review is not to cover the basic aspects of graphene physics, since this was already covered in a recent review \cite{CastroNeto:09}, but to try to address some of these questions while keeping others open. The field of many-body physics will always be an open field because a seemingly simple question always leads to another question even more profound and harder to answer in a definitive way. In many ways, what we have done here is to only scratch
the surface of this rich and important field,  and leave open a large number of interesting and unexplored problems.

\section{Charge polarization and linear screening}

\begin{figure}[tb]
  \centering
  \includegraphics[width=0.5\textwidth]{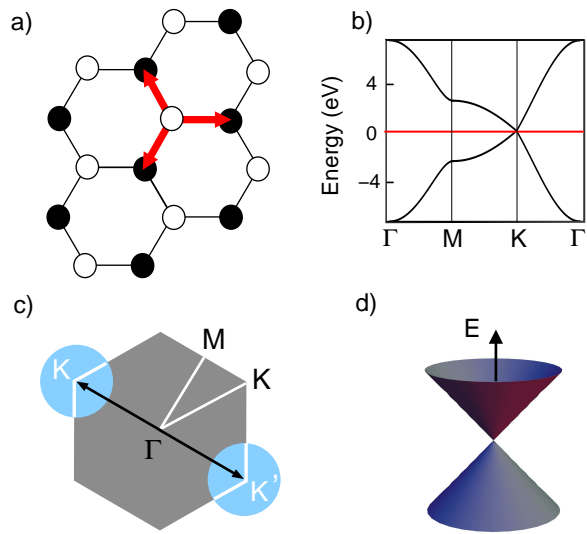}
  \caption{(Color online)
    a) Honeycomb lattice with the two sublattices in graphene.
    The red arrows are nearest neighbor vectors. b) Tight-binding
    spectrum
    for the $\pi-\pi^{*}$ bands. The horizontal line intersecting the
    $K$ point corresponds to the Fermi level at half-filling. c)
    Brillouin
    zone centered around the $\Gamma$ point. d) Dirac cone resulting
    from the linearization of the tight-binding spectrum around the
    $K$ points (blue circles).
  }
  \label{Honeycomb-lattice}
\end{figure}

\subsection{Tight-binding spectrum}

In isolated form, carbon has six electrons in the orbital configuration
$1s^{2}2s^{2}2p^{2}$. When arranged in the honeycomb crystal shown
in Fig.\ref{Honeycomb-lattice}(a), two electrons remain in the core
$1s$ orbital, while the other orbitals hybridize, forming three $sp^{2}$
bonds and one $p_{z}$ orbital. The $sp^{2}$ orbitals form the $\sigma$
band, which contains three localized electrons. The bonding configuration
among the $p_{z}$ orbitals of different lattice sites generates a
valence band, or $\pi$-band, containing one electron, whereas the
antibonding configuration generates the conduction band ($\pi^{*}$),
which is empty. 

From a kinetic energy point of view, the electronic single particle
dispersion in graphene is essentially defined by the hopping of the
electrons between nearest neighbor carbon sites in the honeycomb lattice.
Unlike  square or triangular lattices, the honeycomb lattice
is spanned by two different sets of Bravais lattice generators, forming
a two component basis with one set for each triangular sublattice.
Defining a label for electrons sitting in each of the two sublattices,
say $A$ and $B$, the free hopping Hamiltonian of graphene is \begin{equation}
\mathcal{H}_{0}=-t\sum_{\sigma,\langle ij\rangle}\left[a_{\sigma}^{\dagger}(\mathbf{R}_{i})b_{\sigma}(\mathbf{R}_{j})\right]+{\mbox{h.c.}}-\mu\sum_{\sigma,i}\hat{n}_{\sigma}(\mathbf{R}_{i}),
\label{eq:Ho0}\end{equation}
where $a_{\sigma}(\mathbf{R}_{i}),\, b_{\sigma}(\mathbf{R}_{i})$
are fermionic operators for sublattices $A$ and $B$ respectively,
$\hat{n}_{\sigma}(\mathbf{R}_{i})$ is the number operator, $\sigma=\uparrow,\downarrow$
labels the spin and $\langle ij\rangle$ means summation over nearest
neighbors. The two energy scales in the Hamiltonian are $t\approx2.8$
eV, which is the hopping energy between nearest carbons, and $\mu$, 
 the chemical potential away from half-filling {[}see Fig.\ref{Honeycomb-lattice}(b)].
In a  homogeneous system, deviations from half-filling ($\mu=0$)
are routinely induced either by charge transfer from a substrate \cite{Gio08},
by application of a back gate voltage
\cite{Novoselov:2004,Nov04b,Novoselov:2005}, or else by chemical
doping \cite{Cal07,Gru09b, McC07, Uch08a}. 

In momentum space the free  Hamiltonian of graphene is
\begin{equation}
\mathcal{H}_{0}=\sum_{\mathbf{p},\sigma}\Psi_{\mathbf{p},\sigma}^{\dagger}\left(\begin{array}{cc}
-\mu & -t\phi_{\mathbf{p}}\\
-t\phi_{\mathbf{p}}^{*} & -\mu\end{array}\right)\Psi_{\mathbf{p},\sigma}\,,\label{Ho}\end{equation}
where $\Psi_{\mathbf{p},\sigma}=(a_{\mathbf{p},\sigma},b_{\mathbf{p},\sigma})$
is a two component spinor and \begin{equation}
\phi_{\mathbf{p}}=\sum_{i=1}^{3}\mbox{e}^{i\mathbf{p}\cdot\mathbf{a}_{i}}\label{phi}\end{equation}
is a tight-binding function summed over the nearest neighbor vectors 
\begin{equation}
\mathbf{a}_{1}=a\hat{x},\quad\mathbf{a}_{2}=-\frac{a}{2}\hat{x}+a\frac{\sqrt{3}}{2}\hat{y},\quad\mathbf{a}_{3}=-\frac{a}{2}\hat{x}-a\frac{\sqrt{3}}{2}\hat{y}\,,\label{eq:a_vectors}\end{equation}
where $a\approx1.42\mbox{\AA}$ is the carbon-carbon spacing. The diagonalization
of Hamiltonian (\ref{Ho}) yields the spectrum of the two $\pi$-bands
of graphene in tight-binding approximation \cite{Wal47}, \begin{equation}
E_{\pm}(\mathbf{p})=\pm t|\phi_{\mathbf{p}}|-\mu\,.\label{E_k}\end{equation}
\\
The $+$($-$) sign in the spectrum corresponds to the conduction
(valence) band. 

The hexagonal Brillouin zone (BZ) of graphene shown in Fig.\ref{Honeycomb-lattice}(c)
has three high symmetry points: the $\Gamma$ point, located at the
center of the BZ, the $M$ point, which indicates the position of
the Van Hove singularities of the $\pi$-$\pi^{*}$ bands, where the
density of states (DOS) is logarithmically divergent, and the $K$
points, where the $\pi$-bands touch, 
and the DOS vanishes linearly. An extensive description of the band
structure of graphene and its electronic properties is reviewed in
detail by \onlinecite{CastroNeto:09}.

\subsection{Dirac fermion Hamiltonian}

The topology of the Fermi surface in undoped graphene is defined by
the six $K$ points where the conduction and valence bands touch,
$E_{\pm}(\mathbf{K})=\pm|\phi_{\mathbf{K}}|=0$. These special points
form two sets of nonequivalent points, $K$ and $K^{\prime}$, with
$\mathbf{K}=-\mathbf{K}^{\prime}$ and $|\mathbf{K}|=4\pi/(3\sqrt{3}a)$,
which cannot be connected by the generators of the reciprocal lattice.
The linearization of the spectrum around the valleys centered at  
$\pm\mathbf{K}$  gives rise to an effective low energy description
of the electrons that mimics the spectrum of massless Dirac particles.
In this effective theory, the elementary excitations around the Fermi
surface are described by a Dirac Hamiltonian \cite{Sem84},\\
 \begin{equation}
\mathcal{H}_{0}=\sum_{\sigma\mathbf{k}}\Psi_{\mathbf{k}\sigma}^{\dagger}\left[v\mathbf{k}\cdot
{\bm \gamma}-\mu \tau_0 \otimes \sigma_0 \right]\Psi_{\mathbf{k}\sigma},\label{HoDirac1}\end{equation}
where \begin{equation}
\Psi_{\mathbf{k}\sigma}=\left(a_{\mathbf{K}+\mathbf{k},\sigma},b_{\mathbf{K}+\mathbf{k},\sigma},b_{-\mathbf{K}+\mathbf{k},\sigma},a_{-\mathbf{K}+\mathbf{k},\sigma}\right)\label{eq:Psi}\end{equation}
is a four component spinor for sublattice and valley degrees of freedom.
In this representation, $\gamma_{i}=\tau_{3}\otimes\sigma_{i}$ , where
${\bm \tau}$ and ${\bm \sigma}$ are the usual Pauli matrices, which operate in the
valley and sublattice spaces respectively ($i=1,2,3$ correspond to
$x,y$ and $z$ directions, and $\tau_{0}=1$ and $\sigma_{0}=1$ are identity
matrices). The form of the spectrum mimics the relativistic cone for
massless fermions \cite{Wal47}, \begin{equation}
E_{\pm}(\mathbf{k})=\pm v|\mathbf{k}|-\mu\label{eq: E_k_linear}\end{equation}
where the Fermi velocity $v=(3/2)ta\approx6\mbox{eV}\mbox{\AA}$ is
nearly $300$ times smaller than the speed of light,
 i.e. $v \approx 1\times 10^{6} {\mbox{m/s}}$.  From now on we  set
$\hbar=k_B=1$ everywhere, except where it is needed. For simplicity of notation,
 we call the Fermi velocity $v$  (i.e. $v_F \equiv v$) throughout this review. 

The Hamiltonian (\ref{HoDirac1}) is invariant under a pseudo-time reversal
symmetry operation,
$\mathcal{S}= i (\tau_0 \otimes \sigma_2)\mathcal{C}$,
$\mathcal{S}H\mathcal{S}^{-1}=H$, ($\mathcal{C}$ is the complex conjugation
operator), which is equivalent to a time reversal operation for each
valley separately. It is also invariant under a true time reversal
symmetry (TRS) operation, which involves an additional exchange between
the valleys,
$\mathcal{T}=(\tau_{1}\otimes\sigma_{1})\mathcal{C}$.

In the absence of back scattering connecting the two valleys, the
Hamiltonian can be decomposed in two independent valley species of
Dirac fermions with opposite chiralities:\begin{eqnarray}
\mathcal{H}_{0,+} & = & \sum_{\sigma,\mathbf{k}}\Psi_{+,\mathbf{k}\sigma}^{\dagger}\left[v\mathbf{k}\cdot{\bm \sigma}-\mu\right]\Psi_{+,\mathbf{k}\sigma},\label{eq:H_0+}\\
\mathcal{H}_{0,-} & = & \sum_{\sigma,\mathbf{k}}\Psi_{-,\mathbf{k}\sigma}^{\dagger}\left[-v\mathbf{k}\cdot{\bm \sigma}^{*}-\mu\right]\Psi_{-,\mathbf{k}\sigma},\label{eq:H_0-}\end{eqnarray}
where $\Psi_{\pm,\mathbf{k}\sigma}=\left(a_{\pm\mathbf{K}+\mathbf{k},\sigma},b_{\pm\mathbf{K}+\mathbf{k},\sigma}\right)$
are two component spinors. In this review, unless otherwise specified,
we will arbitrarily choose one of the two cones and assume an additional
valley degeneracy in the Hamiltonian. So valley indexes will be generically
omitted unless explicitly mentioned. A more detailed description of
the symmetry properties of the graphene Hamiltonian can be found in
\cite{Gus07}.

\begin{figure}[bt]
  \centering
  \includegraphics[width=0.15\textwidth]{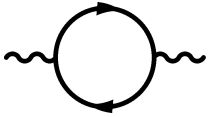}
  \caption{
    Diagram for the polarization bubble corresponding to 
    eq.~\eqref{eq:Pol_E}.
  }
   \label{fig:Diagram-for-the-bubble}
\end{figure}
%

\subsection{Polarization function }

%
\begin{figure}[tb]
  \centering
  \includegraphics[width=\columnwidth]{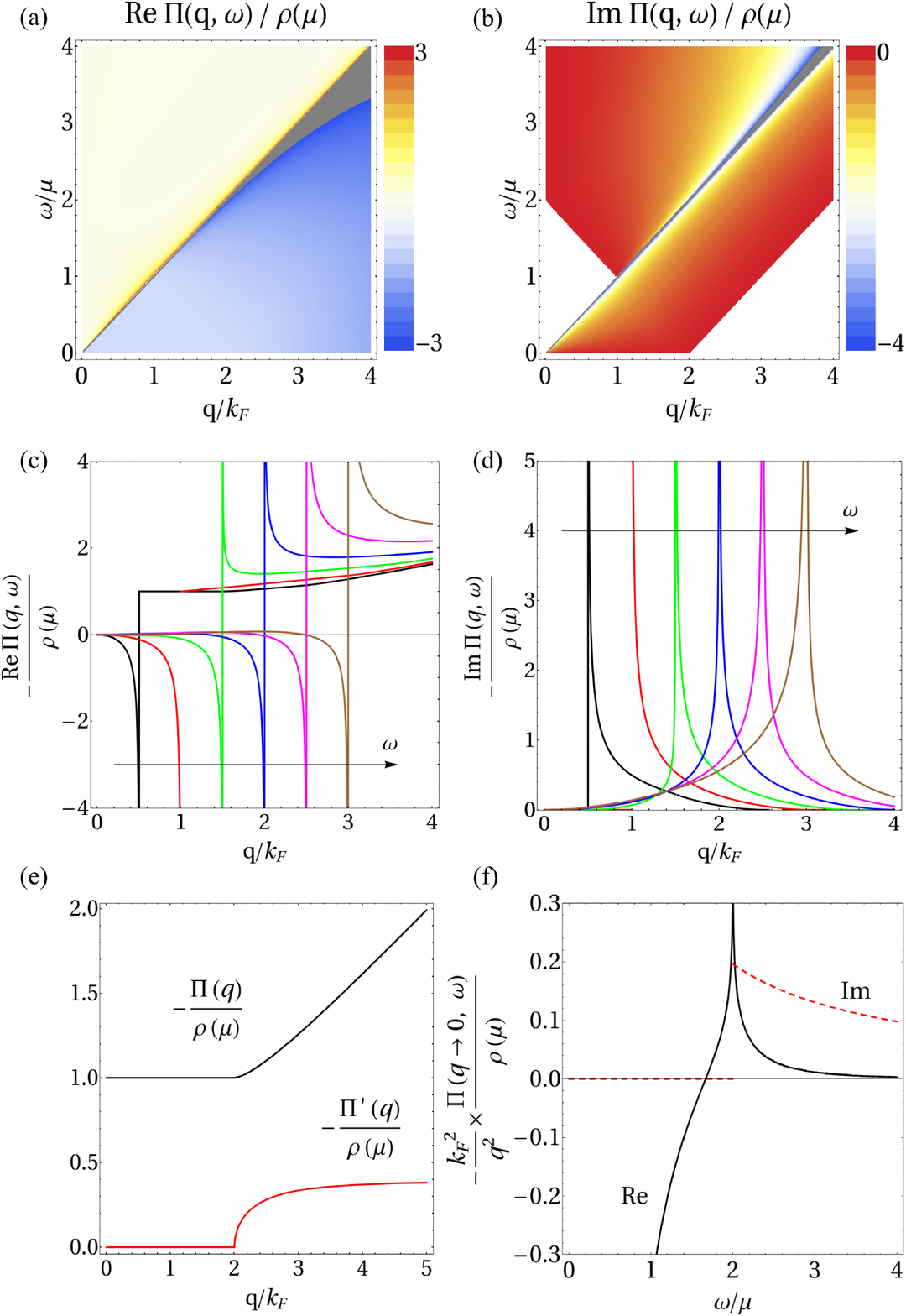}
  \caption{(Color online)
    Polarization bubble $\Pi^{(1)}(q,\omega)$ for graphene, within
    the Dirac approximation.
    Panels (a) and (b) show, respectively, a density plot of the real
    and imaginary parts of the polarization bubble,
    $\Pi^{(1)}(q,\omega)$, defined in eq.~\eqref{eq:Pol_E}, and 
    normalized to the DOS at the Fermi level, $\rho(\mu)$.
    Panels (c) and (d) present constant frequency cuts at 
    $\omega/\mu = 0.5,\,1.0,\,1.5,\,2.0,\,2.5,\,3.0$.
    In panel (e) we show the static limit, $\Pi^{(1)}(q,0)$, whose
    closed form expression is written in eq.~\eqref{eq:Pol1_static}.
    Notice the transition from a constant value ($q<2k_F$) to
    the linear in $q$ dependence at large momenta. The
    derivative of the polarization is shown in the same panel, 
    and can be seen to vary continuously.
    In (f) we plot the real (black/solid) and imaginary
    (red/dashed) parts of the uniform limit \eqref{eq:Pi_omega}.
  }
  \label{fig:Pi-Function}
\end{figure}

The Green's function of graphene is a $2\times2$ matrix represented
in the sublattice basis  by \[
\hat{G}(\mathbf{k},\tau)=\left(\begin{array}{cc}
G_{aa} & G_{ab}\\
G_{ba} & G_{bb}\end{array}\right)\,,\]
where $G_{aa}=-\langle T[a_{\mathbf{k}}(\tau)a_{\mathbf{k}}^{\dagger}(0)]\rangle$
and so on, with $\tau$ as the imaginary time. In the low energy sector
of the spectrum, close to the Dirac points, the non-interacting Green's
function is $\hat{G}^{(0)}(\mathbf{k},i\omega)=\left[i\omega+\mu-v\mathbf{k}\cdot{\bm \sigma}\right]^{-1}$,
or equivalently, in a chiral representation, \begin{equation}
\hat{G}^{(0)}(\mathbf{k},i\omega)=\frac{1}{2}\sum_{s=\pm}\frac{1+s\hat{\sigma}_{\mathbf{k}}}{i\omega+\mu-sv|\mathbf{k}|},\label{eq:G}\end{equation}
where $\hat{\sigma}_{\mathbf{k}}={\bm \sigma}\cdot\mathbf{k}/|\mathbf{k}|$
is twice the quantum mechanical helicity operator for a Dirac fermion
with momentum $\mathbf{k}$, and $s=\pm$ labels the two branches with
positive and negative energy in one cone. It is clear that the positive
and negative branches within the same cone have also opposite helicities. 

The polarization function in one loop is calculated directly from
the bubble diagram shown in Fig.
\ref{fig:Diagram-for-the-bubble},
\begin{align}
\Pi^{(1)}(\mathbf{q},i\omega) & = 
N\sum_{\mathbf{p}}\sum_{s,s^{\prime}}\mathcal{F}_{s,s^{
\prime}}(\mathbf{p},\mathbf{q})\times\nonumber \\
&
\qquad\frac{f[E_{s^{\prime}}(\mathbf{p}+\mathbf{q})]-f[E_{s}(\mathbf{
p})]}{E_{s^{\prime}}(\mathbf{p}+\mathbf{q})-E_{s}(\mathbf{p})-i\omega}\,,\label{eq:Pol_E}
\end{align}
where $f(E)=\left(\mbox{e}^{E/T}+1\right)^{-1}$ is the Dirac-Fermi
distribution, with $T$ as temperature, $N=4$ is the degeneracy for two spins and two valleys,
 and\begin{equation}
\mathcal{F}_{s,s^{\prime}}(\mathbf{p},\mathbf{q})=\frac{1}{4}\textrm{tr}(1+ss^{\prime}\hat{\sigma}_{\mathbf{p}}\hat{\sigma}_{\mathbf{p}+\mathbf{q}})\label{eq:F}\end{equation}
are the matrix elements due to the overlap of wavefunctions for intraband
($s=s^{\prime})$ and interband ($s=-s^{\prime})$ transitions. 'tr'
means trace over the sublattice indexes. In
a more explicit form, $\mathcal{F}_{s,s^{\prime}}(\mathbf{p},\mathbf{q})=\left[1+ss^{\prime}\cos\theta_{\mathbf{p},\mathbf{p}+\mathbf{q}}\right]/2$,
where $\theta$ is the angle between $\mathbf{p}$ and
$\mathbf{p}+\mathbf{q}$.
The full momentum, frequency, and chemical potential dependence of
\eqref{eq:Pol_E} is shown in panels (a-d)
of Fig.~\ref{fig:Pi-Function}.

In metals,  screening is a many-body property directly 
related to the polarizability of the electrons around the Fermi surface.
In graphene, because the density of states (DOS) vanishes linearly
around the Dirac points, $\rho(E)\propto|E-\mu|/v^{2}$, exactly at
the neutrality point ($\mu=0$) the screening of charge is completely
suppressed, and the polarization function describes the susceptibility
of the vacuum to particle-hole pair production, exactly as in the
diagonal time component of the polarization tensor in massless 
Quantum Electrodynamics (QED), QED$_{2+1}$
\cite{App88, Gon94, Pisarski84}, \begin{equation}
\Pi^{(1)}(q,\omega)=-\frac{1}{4}\frac{q^{2}}{\sqrt{v^{2}q^{2}-\omega^{2}}}\,.\label{eq:pol}\end{equation}
Here we have performed a Wick rotation to real frequencies, $i\omega\to\omega+0^{+}$. Since
the Fermi surface in this case is just a point, there is no phase
space for intraband excitations at zero temperature due to the Pauli
principle. The process of creation of particle-hole pairs involves
incoherent excitations of electrons from the lower to the upper band.
The continuum of particle-hole excitations is well defined for all
virtual transitions with $\omega>vq$.

For finite $\mu$ there is a crossover in the behavior of the polarization
function. The DOS around the Fermi level is finite and the intraband
excitations dominate the infrared behavior of the polarization. For
$vq\ll|\mu|$ and $\omega\ll|\mu|$, the leading term in the polarization
function is \cite{Shu86} \begin{equation}
\Pi^{(1)}(q,\omega)\approx-\frac{2|\mu|}{\pi v^{2}}\left(1-\frac{\omega}{\sqrt{\omega^{2}-v^{2}q^{2}}}\right).\label{eq:Pol_mu}\end{equation}
As in a Fermi liquid, there is a particle-hole continuum for $\omega<vq$,
which is due only to intraband transitions. The polarization function
in graphene is a regular function everywhere except at $|\omega|=vq$,
where it has an on-shell singularity delimiting the border of the
particle-hole continuum.

The polarization was derived originally by \onlinecite{Shu86} and later rederived
by a number of authors 
\cite{Ando:2006, Bar07, Hwa07, Wun07}. These results rely on the
cone approximation, which ignores
contributions coming from the non linear part of the spectrum.
In addition, the  band width is assumed to be infinite. Although the charge
polarization for Dirac fermions in 2D is well behaved and does not
require cut-off regularization in the ultraviolet, the physical cut-off
of the band, $D$, generates small corrections that vanish only in
the $D\to\infty$ limit. In this sense, the `exact' expression for
the static polarization function ($\omega=0$) for arbitrary momentum is
\begin{widetext} 
\begin{equation}
\Pi^{(1)}(q,0)=-\frac{2k_{F}}{\pi v}+\theta(q-2k_{F})\frac{q}{2\pi
v}\left[\frac{2k_{F}}{q}\sqrt{1-\left(\frac{2k_{F}}{q}\right)^{2}}
+\sin^{-1}\left(\frac{2k_{F}}{q}\right)-\frac{\pi}{2}\right],
\label{eq:Pol1_static}
\end{equation}
\end{widetext} 
where $k_{F}=|\mu|/v$ is the Fermi momentum, and $\theta(x)$
is a step function. The static polarization is plotted in
Fig.~\ref{fig:Pi-Function}(e).

\begin{figure}[tb]
  \centering
  \includegraphics[width=\columnwidth]{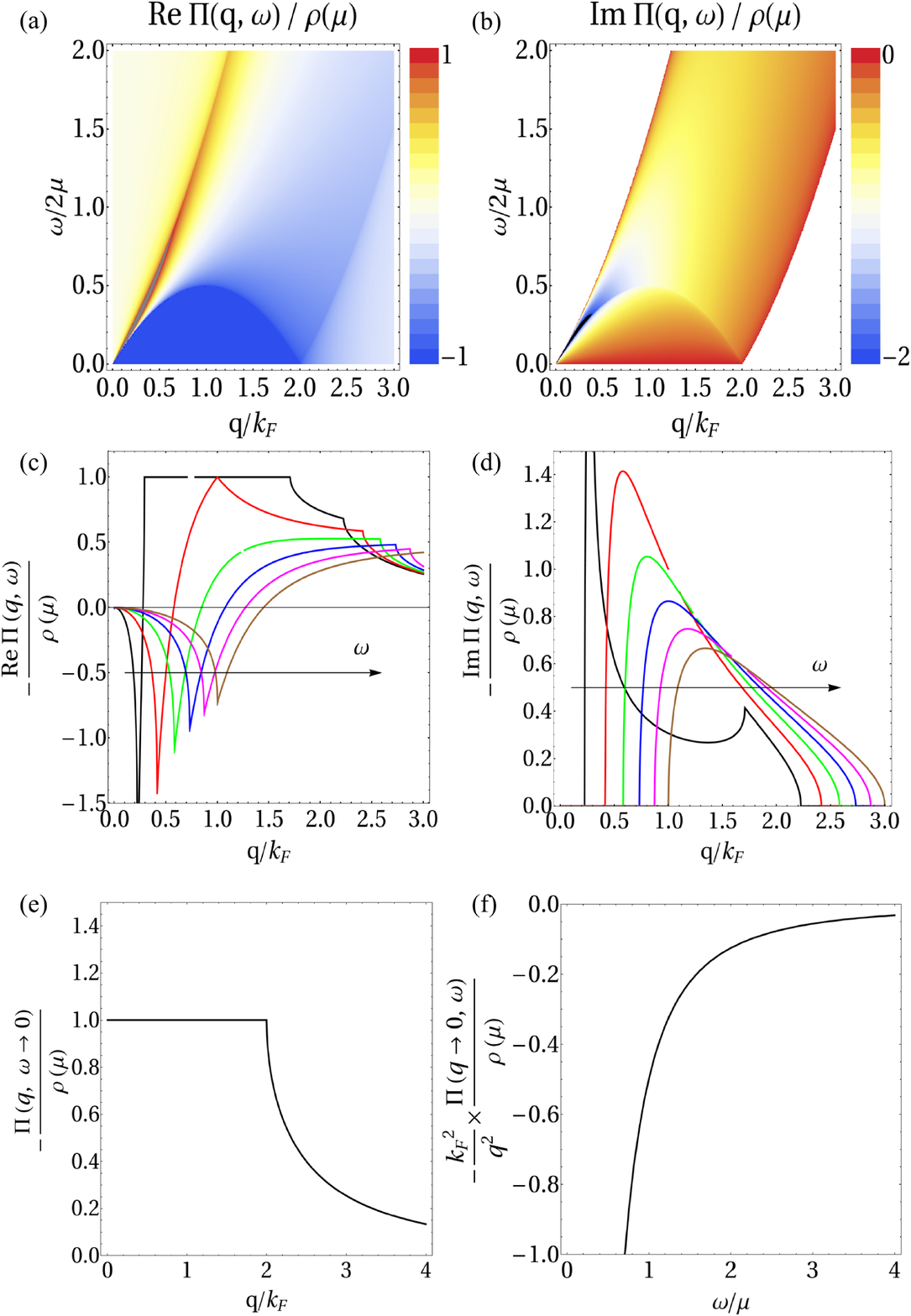}
  \caption{(Color online)
    Polarization bubble $\Pi^{(1)}(q,\omega)$ for the conventional
    2DEG.
    Panels (a) and (b) show, respectively, a density plot of the real
    and imaginary parts of the polarization bubble,
    $\Pi^{(1)}(q,\omega)$, normalized to the DOS at the Fermi
    level.
    Panels (c) and (d) present constant frequency cuts at 
    $\omega/\mu = 0.5,\,1.0,\,1.5,\,2.0,\,2.5,\,3.0$.
    In panel (e) we plot the static limit, $\Pi^{(1)}(q,0)$, and
    in (f) the uniform limit, $\Pi^{(1)}(0,\omega)$.
  }
  \label{fig:Pi-Function-2DEG}
\end{figure}

At $q\approx2k_{F}$  the static polarization exhibits a
crossover from a two dimensional electron gas (2DEG) to  Dirac fermion behavior.
For details of the polarization function in the 2DEG please refer to
 Fig.~\ref{fig:Pi-Function-2DEG}.
As in the  2DEG, the polarization of graphene is constant for $q<2k_{F}$. For $q>2k_{F}$,
it eventually 
becomes linear in $q$ for large momenta. At the crossover, the static polarization
and its first derivative are continuous at $q=2k_{F}$. The discontinuity
only appears in the second derivative. This is distinct
from the 2DEG case, where the first derivative is discontinuous. The
difference will affect the spacial dependence of the Friedel oscillations
in the two systems. 

\begin{figure}[tb]
  \centering
  \includegraphics[width=0.5\textwidth]{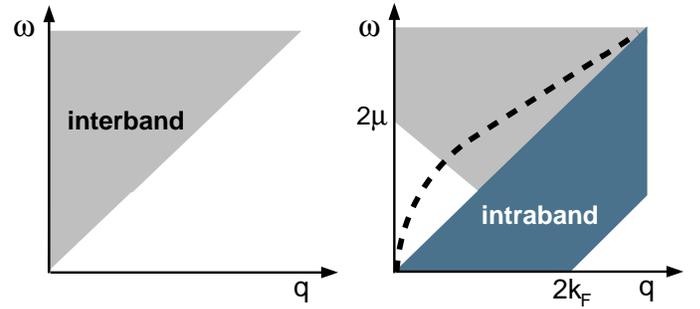}
  \caption{(Color online)
    Colored regions represent the
    particle-hole continuum of graphene
    due to interband (gray area) and intraband (green) transitions. On
    the left: half-filled case; right: finite $\mu$ case, away from
    half filling. Dashed line: acoustic plasmon for the single layer
    ($\omega_{p}\propto\sqrt{\mu q}$).
  }
  \label{fig:particle-hole-continuum}
\end{figure}

In the opposite limit, for arbitrary $\omega$ and $q\to0$, the
polarization function becomes
\begin{equation}
\Pi^{(1)}(q\to0,\omega)=\frac{q^{2}}{2\pi\omega}\left[\frac{2|\mu|}{
\omega}+\frac{1}{2}\ln\!\left(\frac{2|\mu|-\omega}{2|\mu|+\omega}
\right)\right],
\label{eq:Pi_omega}
\end{equation}
which is shown in Fig.~\ref{fig:Pi-Function}(f).
The presence of a pocket of electrons (holes) around the Dirac points
opens a gap in the particle hole continuum for interband excitations
($\omega>vq$). From Eq. (\ref{eq:Pi_omega}), it is clear that the
imaginary part of the polarization function at small momentum is zero
unless $\omega>2|\mu|$ [Fig.~\ref{fig:Pi-Function}(b)]. This is so because the phase space for vertical
interband excitations is Pauli blocked for $\omega<2|\mu|$, generating
a gap for optical absorption in the infrared. At finite $q$, the
threshold for interband transitions is $\omega>2|\mu|-vq$ for $q<2k_{F}$,
as shown schematically in Fig.~\ref{fig:particle-hole-continuum}. 

\subsection{Collective modes and screening}

The Coulomb interaction among the electrons in graphene gives 
rise to collective modes and metallic screening when the Fermi level
is shifted away from the Dirac points. In a 2D system, the bare Coulomb
interaction is given by  \begin{equation}
V(q)=\frac{2\pi e^{2}}{\epsilon_{0}q}\,,\label{eq:V}\end{equation}
where $e$ is the charge of the electron and $\epsilon_{0}$ is the
effective dielectric constant of the medium. For graphene in contact
with air and a substrate with dielectric constant $\kappa$, $\epsilon_{0}=(1+\kappa)/2$.
In most of the experiments, graphene lies on top of some substrate
like SiO$_{2}$ or SiC, where dielectric effects are moderate (for
instance, the dielectric constant of SiO$_{2}$ is $\kappa\approx4$).
The background dielectric constant can be significantly enhanced in
the presence of substrates in contact with strong dielectric liquids such as ethanol (\textbf{$\kappa\approx25$})\textbf{
}or water ($\kappa\approx80$) \cite{Jang08,Ponomarenko:2009}.

As usual, the collective modes follow from the zeros of the dielectric
function \begin{equation}
\epsilon(q,\omega)=\epsilon_0[1-V(q)\Pi^{(1)}(q,\omega)]\,,\label{eq:epsilon}\end{equation}
calculated here in the Random Phase Approximation (RPA). Since graphene
is a 2D system, the collective plasmon mode is gapless. The leading
term in the polarization for small frequency and momenta (compared
to $k_{F}$) is shown in Eq. (\ref{eq:Pol_mu}). From it one can
easily extract the infrared dependence of the plasmon,\begin{equation}
\omega_{p}(q)=\sqrt{(2\mu e^{2}/\epsilon_{0})q}\, ,\label{eq:plasmon}\end{equation}
which follows the same dispersion as the plasmon encountered in the
2DEG. The $\sqrt{q}$  dependence of the plasmon  was
recently confirmed by a high resolution energy loss spectroscopy (EELS)
measurement in graphene \cite{Liu08}. Additional corrections
due to the interband excitations (which are absent in the 2DEG) can
be absorbed into the definition of the background dielectric constant
\cite{Shu86},\begin{equation}
\epsilon_{0}(q)\approx\epsilon_{0}-\frac{qe^{2}}{2\omega_{p}(q)}\ln\!\left(\frac{2|\mu|-\omega_{p}(q)}{2|\mu|+\omega_{p}(q)}\right).\label{eq:epsilon_q}\end{equation}

As in the 2DEG, the screened Coulomb interaction for $q<2k_{F}$ is
\begin{equation}
\frac{V(q)}{\epsilon(q,0)}=\frac{1}{\epsilon_{0}}\frac{2\pi e^{2}}{q+q_{TF}}\label{eq:V/epsilon}\end{equation}
 where $q_{TF}=4\pi e^{2}k_{F}/(v\epsilon_{0})$ is the Thomas-Fermi
momentum ($k_{F}=|\mu|/v$), which sets the size of the screening
cloud. In the presence of an external charged impurity $Ze$, the
induced charge, $\delta Z$, has a non-oscillatory component coming
from the $q\to0$ limit of the polarization that decays as $(k_{F}r^{3})^{-1}$
(as in a 2DEG), and an oscillatory part which corresponds to the Friedel
oscillations at $q=2k_{F}$. The Friedel oscillations in graphene
decay as $\cos(2k_{F}r)/(k_{F}r^{3})$, differently from the
2DEG case, where the decay is  of the form $\cos(2k_{F}r)/r^{2}$. The difference
is caused by the fact that the static polarization function in the
2DEG has a cusp at $q=2k_{F}$, whereas in graphene, the first derivative
is continuous [cfr. Figs.~\ref{fig:Pi-Function}(e) and \ref{fig:Pi-Function-2DEG}(e)]. 

For undoped graphene, $V(q)\Pi^{(1)}=-(\pi/2)[e^{2}/(v\epsilon_{0})]$
{[}see Eq. (\ref{eq:pol})], and the static dielectric function is
a constant. The effective Coulomb interaction in this case is \begin{equation}
\frac{V(q)}{\epsilon(q,0)}=\frac{1}{\epsilon_{RPA}}\frac{2\pi e^{2}}{q}\,,\label{eq:undopedV}\end{equation}
where $\epsilon_{RPA}=\epsilon_{0}+(\pi/2)(e^{2}/v)$ is the
effective background dielectric constant, renormalized by the interband
transitions. Additional many body effects resulting from self-energy
insertions in the bubbles logarithmically renormalize this correction
to zero in the $q\to0$ limit, as  will be clear in Sec.~\ref{sec:QP} of
this review. On the dynamical side, inserting Eq. (\ref{eq:pol})
into Eq. (\ref{eq:epsilon}), one can easily see that no collective
modes are allowed in undoped graphene, at zero temperature, within
the RPA framework. At half-filling, RPA is justified in the limit
of large number of fermionic species, $N$, which favors diagrams
with maximal number of bubbles at each order of perturbation theory.
In graphene, the physical number of species is $N=4$, and additional
corrections beyond RPA coming from the exciton channel near the on-shell
singularity of the bubble, $|\omega|\sim vq$, were shown to generate
a new acoustic plasmon mode \cite{Gan08}. In
the static limit ($\omega\to0$), vertex corrections in the bubble
are perturbatively small and RPA can be justified in the calculation
of the dielectric function even at half-filling \cite{Kot08a}.
 The structure of perturbation theory in graphene will be discussed
in detail in Sec.~\ref{sec:QP}. 

In addition to the low energy acoustic mode due to intraband transitions,
graphene has also two high energy optical plasmons generated by interband
excitations around the Van-Hove singularities of the $\pi-\pi^{*}$
bands, and also by optical transitions between $\sigma-\pi^{*}$ and
$\pi-\sigma^{*}$ bands \cite{Ebe08, Kram08}. The measured optical gaps of the $\pi$ and
$\pi-\sigma$ band plasmons in graphene are 4.5 eV and 15 eV, respectively.
 Similar modes were also observed in graphite, where they appear blue shifted
to 7eV and 24 eV respectively, according to optical data \cite{Taf65}, X-ray measurements 
\cite{Shu88}, and\emph{
ab-initio} calculations \cite{Mar04}.

\subsection{Infinite stack of layers}

In the case of an infinite stack of graphene layers, the Hamiltonian
term for the Coulomb interaction among all the electrons can be written
in real space as 
\begin{equation}
\mathcal{H}_{C}=\frac{e^{2}}{\epsilon_{0}}\int\mbox{d}^{3}\mathbf{r}\mbox{d}^{3}\mathbf{r}^{\prime}\,\hat{n}(\mathbf{r})\frac{1}{|\mathbf{r}-\mathbf{r}^{\prime}|}\hat{n}(\mathbf{r}^{\prime})\,,
\end{equation}
where $\hat{n}(\mathbf{r})$ is the 3D particle density operator. In
the absence of interlayer hopping, as in the case for example of several
graphite intercalated compounds, the electrons remain confined in
each layer, but the unscreened Coulomb lines fill the entire space
in between the layers, coupling all the electrons in the system. In
that case we may constrain the local density operator $\hat{n}$ to
be in the form \cite{Vis70}
\begin{equation}
\hat{n}(\mathbf{r})\to d\sum_{l=-\infty}^{\infty}\hat{n}(\mathbf{r})\delta(z-ld)
\end{equation}
where $l$ is an integer labeling the layers, and $d$ is the distance
between  layers. In momentum space, making a discrete sum over
the layers, the Coulomb interaction between all the electrons is 
\begin{equation}
\mathcal{H}_{C}=\frac{e^{2}}{\epsilon_{0}}\int\mbox{d}^{3}\mathbf{k}\,\hat{n}(-\mathbf{k})V(\mathbf{k})\hat{n}(\mathbf{k})\,,
\end{equation}
where \begin{equation}
V(\mathbf{k})=2\pi\, d\frac{e^{2}}{\epsilon_{0}q}\, S(q,k_{z})\label{eq:V_k}\end{equation}
with $\mathbf{k}=(\mathbf{q},k_{z})$,  $\mathbf{q}$ is an in-plane momentum,
and \cite{Fet74} \begin{equation}
S(q,k_{z})=\frac{\sinh(qd)}{\cosh(qd)-\cos(k_{z}d)}\label{2D}\end{equation}
is the structure factor for a stack with an infinite number of layers.
In the limit when the distance between the layers $d$ is small, Eq.
(\ref{eq:V_k}) recovers the isotropic case $V(k)=4\pi(e^{2}/\epsilon_{0})/(q^{2}+k_{z}^{2})$,
whereas in the opposite limit ($d\to\infty)$ one gets the 2D case, 
$V(k)=2\pi\, d\,(e^{2}/\epsilon_{0})/q$. In any case, the polarization
function must be integrated over a cylindrical Fermi surface of height
$2\pi/d$, and so $\Pi^{(1)}(q,\omega)$ acquires an additional factor
of $1/d$ compared to the single layer case. The extension of this problem 
to include the interlayer hopping dispersion in the polarization 
was considered by \onlinecite{Gui07}. 

\begin{figure}[tb]
  \centering
  \includegraphics[width=0.45\textwidth]{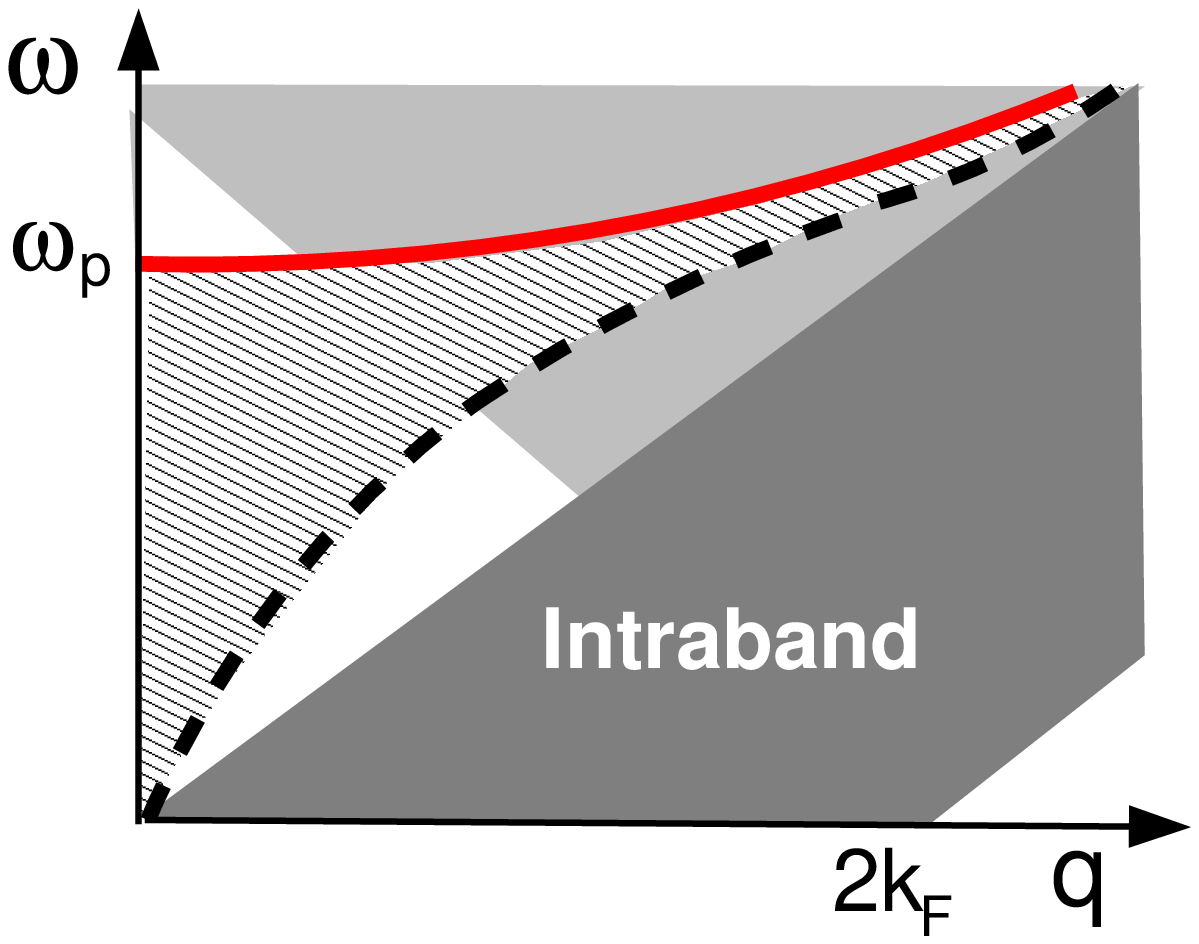}
  \caption{
    (Color online)
    Plasmon band (hatched region) for an infinite
    stack of graphene layers. Red line: optical mode $k_{z}=0$. Dashed
    line: acoustic mode $k_{z}=\pi/d$, $\omega_{p}\propto\sqrt{\mu}q$,
    with linear dispersion, at the edge of the band. All the other
    modes in between are acoustic. Adapted from \onlinecite{Shu86}.
  }
  \label{fig:Infinite-stack-of-layers}
\end{figure}

Away from the neutrality point $(\mu\neq0$), instead of a single
acoustic mode as in the monolayer, the zeroes in the dielectric function of the multilayer  
generate a plasmon band, where the modes are labeled by $k_{z}\in[-\pi/d,\pi/d]$.
For $q\ll1/d$, the plasmon dispersion is \cite{Shu86} \begin{equation}
\omega_{p}^{2}(q,k_{z})=\frac{2\mu e^{2}}{\epsilon_{0}}qS(q,k_{z}).\label{eq:omega_p_layered}\end{equation}
 In
the $k_{z}=0$ mode, the charge fluctuations between different layers
are in-phase, and the resulting plasmon mode is optical, $\omega_{p}^{2}(q,0)\approx(4\mu e^{2}/\epsilon_{0}d)+\frac{3}{4}(vq)^{2}$.
For $\omega_{p}(q)>2\mu$, this mode is damped by the particle-hole
continuum due to interband transitions (see Fig.~\ref{fig:Infinite-stack-of-layers}),
in agreement with energy loss spectroscopy data \cite{Lait96}.
 The out-of-phase modes (for $k_{z}\neq0$) are acoustic.
At the edge of the plasmon band, the mode $k_{z}=\pm\pi/d$ disperses
linearly with the in-plane momentum, $\omega_{p}(q,\pm\pi/d)=\sqrt{\mu e^{2}d/\epsilon_{0}}q$,
in contrast with the 2DEG dispersion ($\omega_{p}\propto\sqrt{q}$)
present in the single layer. Except for the lack of an interband particle-hole
continuum and the associated damping, similar plasmon band features
are also expected in the 2D layered electron gas, for fermions with
quadratic dispersion \cite{Haw87}. 

\subsection{f-sum rule}

The $f$-sum rule is a generic statement about conservation of the
number of particles and results from the analytical properties of
the retarded charge susceptibility. It can be generically defined
as \cite{Noz64} \begin{equation}
\int_{-\infty}^{\infty}\mbox{d}\omega\,\omega\mbox{Im}\chi^R(k,\omega)=\pi\langle\left[\left[\mathcal{H},\hat{n}(-\mathbf{k})\right],\hat{n}(\mathbf{k})\right]\rangle,\label{eq:sum_r}\end{equation}
where $\mathcal{H}$ is the Hamiltonian, $\hat{n}$ is the particle
density operator, $\chi^R$ is a retarded charge susceptibility, $\chi(\mathbf{k},\tau)=\langle T[\hat{n}(\mathbf{k},\tau)\hat{n}(-\mathbf{k},0)]\rangle$,
and $\langle...\rangle$ is an expectation value calculated in some
basis. 

As in any solid, the exact electronic Hamiltonian of graphene can
be decomposed into a Hamiltonian of \emph{free} electrons, plus a periodic
potential due to the lattice, and  interactions. If the interactions
depend only on densities, the commutators in Eq. (\ref{eq:sum_r})
can be calculated exactly, and the only term that survives is the
Kinetic energy due to the free electrons, \begin{equation}
\langle\left[\left[\mathcal{H},\hat{n}(-k)\right],\hat{n}(k\right)]\rangle=N_{e}\frac{k^{2}}{m}\,,\label{eq:comm1}\end{equation}
where $m$ is the bare electron mass and $N_{e}$ is the
number of fermions in the band. Choosing, for example, a basis of non-interacting
fermions, the sum rule in graphene
is \begin{equation}
\int_{-\infty}^{\infty}\mbox{d}\omega\,\omega\mbox{Im}\Pi^{(1)}(k,\omega)=\pi\frac{N_{e}k^{2}}{m},\label{eq:sum_rule2}\end{equation}
as in metals, where $\Pi^{(1)}(k,\omega)$ is the bare polarization
bubble, calculated using the full non-interacting spectrum (dictated
by the lattice symmetry).  The validity of the $f$ sum-rule does not require
Galilean invariance of the quasiparticles, but of the free electrons,
which are not relativistic  and hence obey the Schrodinger equation. 

For low energy effective Hamiltonians, such as the Dirac Hamiltonian
in graphene (which do not include the periodicity of the spectrum
in the Brillouin zone), the $f$-sum rule above is still formally
satisfied when applied for the electrons (holes) in the conduction
(valence) band only, as  can be explicitly checked by direct substitution
of the polarization due to intraband transitions, Eq. (\ref{eq:Pol_mu}),
into Eq. (\ref{eq:sum_rule2}). The number of electrons (holes) in
this band, $N_{e}=k_{F}^{2}A/\pi$, where $A=3\sqrt{3}a^{2}/2$ is
the unit cell area, is set by the size of the Fermi surface, and the
verification of the sum rule follows as in a Fermi liquid.

The Dirac Hamiltonian, however, violates the $f$-sum rule (\ref{eq:sum_rule2})
when interband transitions are taken into account. In that case, the
left hand side of Eq. (\ref{eq:sum_rule2}) becomes independent of
the chemical potential, consistent with the fact that \cite{Sab08} 
\begin{equation}
\langle\left[\left[\mathcal{H},\hat{n}(-\mathbf{k})\right],\hat{n}(\mathbf{k})\right]\rangle=k^{2}\frac{D}{4}\label{eq:comm2}\end{equation}
for a Dirac Hamiltonian, where $D$ is the ultraviolet cut-off. A
similar dependence with the cut-off also occurs in the true 3D relativistic
problem, where the sum rule reflects the number of particles contained
in the vacuum of the theory, which is formally divergent \cite{Cen01,Gol82}. 
In graphene, as in any two band semi-metal or semiconductor, the validity of the $f$-sum rule is physically recovered when 
the  periodicity of the electronic spectrum is restored back into the Hamiltonian.

\section{Quasiparticles in Graphene}
\label{sec:QP}

The quasiparticle properties of graphene
are  modified by the presence of long-range Coulomb
interactions. 
Their effects are especially pronounced when the Fermi
 energy is close to the Dirac point ($\mu \approx 0$), and can result in
 strong renormalization of the Dirac band structure (the Fermi velocity $v$),
 and the quasiparticle residue $(Z)$. Consequently,  many physical
 characteristics, such as the compressibility, spin susceptibility and the specific
 heat can be strongly affected by  interactions.
 Even when the Fermi surface is large 
 and the system is a Fermi liquid, 
 there are strong modifications of the physics near the Dirac point due to the presence of 
 additional peaks in the quasiparticle decay rate, related to plasmon-mediated 
 decay channels. Even reconstruction of the Dirac cone structure near the charge
neutrality point appears possible, as indicated by recent Angle-Resolved Photoemission
Spectroscopy (ARPES) measurements.
 All these effects are sensitive to the value of the Coulomb interaction
 constant in graphene, $\alpha$.

\subsection{Low-energy behavior near the Dirac point}
\label{sec:Low-energy}


\subsubsection{Weak-coupling analysis}
\label{subsubsec:Weak-coupling}

The interaction parameter which characterizes the strength
 of the Coulomb interaction in graphene is 
(Eq.~\eqref{alfa})
\begin{equation}
\alpha = \frac{e^2}{\epsilon_0 v}.
\label{alpha}
\end{equation}
At $k_F  =0$ screening is absent, and the interaction potential
 in momentum space:
\begin{equation}
V(\bp) = \frac{2\pi e^2}{\epsilon_0 p}.
\label{potential}
\end{equation}
The value of $\alpha=2.2/\epsilon_0$ depends on the dielectric environment 
 since, as previously discussed, $\epsilon_0 = (1+\kappa)/2$ for
 graphene in contact with air  and a substrate with dielectric constant $\kappa$.
In vacuum, $\alpha = 2.2$.

In the case of small coupling, $\alpha \ll 1$, we can 
employ standard perturbation theory, involving the
 perturbative computation of the self-energy $\Sigma (\bk,\omega)$,
 which enters in a standard way the Dirac fermion Green's function (GF),
for a given valley:
\begin{equation}
\label{gf}
G({\bk},\omega) = \frac{1}{ \omega \sigma_0 - v {\bm \sigma}\cdot{\bk}
-\Sigma (\bk,\omega) + 
 i \sigma_0 0^{+} {\mbox{sign}} (\omega)}.
\end{equation}
\begin{figure}
\centering
\includegraphics[width=0.7\linewidth]{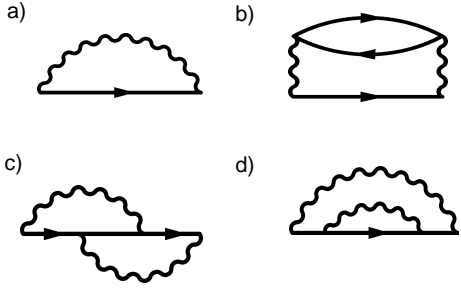}
\caption{Self-energy diagrams: (a) First order Hartree-Fock,
(b) Second order loop diagram (first diagram in the RPA series),
(c) Second order exchange (vertex correction) diagram, (d) Rainbow
diagram.
}
\label{figdiag1}
\end{figure}

\noindent
It is convenient to decompose the self-energy into
 two pieces with different pseudo-spin structure
\begin{equation}
\Sigma (\bk,\omega) = \Sigma_0 (\bk,\omega)+\Sigma_v (\bk,\omega), 
\ \Sigma_0 \propto \sigma_0, \ \Sigma_v \propto{\bm \sigma}\cdot{\bf k},
\label{sedecomposition}
\end{equation}
where $ \sigma_0 =1$ is the unit matrix, which from now on will not be written 
 explicitly. Then we have
\begin{equation}
\label{gf2}
G({\bk},\omega) =
 \frac{Z}{\omega  - Z(v{\bm \sigma}\cdot\bk +\Sigma_{v})} \, ,
\end{equation}
where $Z$ is the quasiparticle residue
\begin{equation}
\label{Z}
Z^{-1} = 1 - \partial\Sigma_{0}/\partial \omega \, ,
\end{equation}
and $\Sigma_{v}$ is responsible solely for the velocity renormalization.

 The first order diagram shown in Fig.~\ref{figdiag1}(a) is the
 Hartree-Fock exchange contribution, and can be readily evaluated
(we denote by $G^{(0)}$ the non-interacting GF):
\begin{equation}
\label{hf}
\Sigma^{(1)}(\bk,\omega) = i \int \frac{d^{2}p d \varepsilon }{(2\pi)^{3}} 
G^{(0)}(\bk +\bp,\omega +\varepsilon )
V(\bp) \, ,
\end{equation}
which  at low external momenta exhibits a logarithmic singularity
\begin{equation}
\label{hf2}
\Sigma^{(1)}(\bk,\omega) = \Sigma^{(1)}_{v}({\bk}) = \frac{\alpha}{4} v {\bm \sigma}\cdot{\bk}  \ln(\Lambda/k),
 \  \   \Lambda/k \gg 1.
\end{equation}
At this order we have $\Sigma_{0}=0$, i.e. $Z=1$ due to the frequency independence
 of the interaction potential, and the quasiparticle velocity increases:
\begin{equation}
\label{hf3}
v(k) = v\left(1 +\frac{\alpha}{4}  \ln(\Lambda/k)\right),
 \  \   \Lambda/k \gg 1.
\end{equation}
The ultraviolet cutoff $\Lambda \sim 1/a$ represents the 
 momentum scale up to which the spectrum is Dirac-like.

 While the linearity of the spectrum in graphene was realized a long time
 ago \cite{Wal47}, in the context of studying graphite formed by layers
 of graphene,  the  self-energy correction Eq.~\eqref{hf2} due to interactions
was first investigated perturbatively much later  by \onlinecite{Gon94}.
The non-trivial velocity renormalization is due to the unscreened,
 long-range Coulomb interactions. Similar  logarithmic divergencies were also found in  
 gapless 3D semiconductors, where the Dirac spectrum originated  from special symmetries
 \cite{Abr71}. 

The above calculation forms the basis of the Renormalization Group (RG) analysis.
 In the RG spirit one integrates out the high momentum degrees of 
freedom, i.e. regions of momenta  $\Lambda > |\bp| > \Lambda_1 $, 
and the results vary with the quantity $\ln(\Lambda/\Lambda_1) \equiv l$.
 Here we denote by $l$ the RG parameter, so that the infrared limit
 corresponds to $l \rightarrow \infty$ (i.e. one  integrates down to
 the infrared scale $k \rightarrow 0$, $l = \ln(\Lambda/k))$.
 From Eq.~\eqref{hf3} we obtain
\begin{equation}
\label{vrun}
\frac{dv}{dl}= \frac{\alpha}{4} v = \frac{e^{2}}{4\epsilon_0} \, .
\end{equation}
This equation has to be supplemented with  an additional
 equation reflecting the absence of charge ($e^2$) renormalization:
\begin{equation}
\label{erun}
\frac{de^2}{dl}= 0\  .
\end{equation}
There are several ways to understand this. It was argued early on that the vertex function 
 does not acquire any divergent contributions, which is related to the expected regular behavior
 of the polarization operator to all orders in graphene \cite{Gon94}. 
 More recently, explicit calculations up to two loop order were performed \cite{Kot08a,Vozm10};
 it was confirmed that the vertex function is finite in the low-energy limit.
 In addition, direct examination of the polarization function  at two loop level \cite{Kot08a}
 found that the self-energy correction, Fig.~\ref{figdiag2}(a),  acquires a logarithmic divergence
 which can be absorbed into the renormalized velocity $v(k)$ (Eq.~\eqref{hf3}),  while
 the vertex correction of  Fig.~\ref{figdiag2}(b) is finite:
\begin{figure}
\centering
\includegraphics[width=0.65\linewidth]{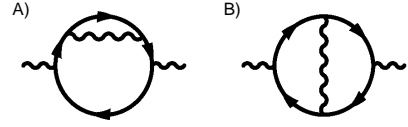}
\caption{(a) Self-energy  and (b) Vertex corrections to the polarization bubble}
\label{figdiag2}
\end{figure}
\begin{equation}
\frac{2\pi e^{2}}{\epsilon_0q}\Pi^{(2b)}(q,0) =\mbox{finite}= - 0.53 \alpha^{2} \ .
\end{equation}
Incidentally, this  contribution leads to enhancement of the dielectric static screening 
(i.e. the dielectric constant beyond linear (RPA) order becomes $\epsilon
 = 1 + \frac{\pi}{2} \alpha + 0.53 \ \alpha^{2}$.)

Alternatively,  one can argue that in two-dimensional field theories with Coulomb interactions
the charge $e^{2}$ does not flow because it appears as a coefficient in a nonanalytic term
 in the action \cite{Ye98,Her06}. 
The conclusion then is that only the quasiparticle velocity and residue (see below) are
 renormalized. In particular, at first order we can combine  Eqs.~\eqref{vrun},\eqref{erun} into
 a single one reflecting the renormalization (running) of the coupling $\alpha$:
\begin{equation}
\label{alpharun}
\frac{d\alpha}{dl}= -\frac{\alpha^2}{4}\  .
\end{equation}
Therefore we have an infrared stable fixed point at $\alpha =0$,
 and the flow towards it is logarithmic: 
\begin{equation}
\label{alpharun2}
\alpha(k) \sim \frac{4}{\ln(\Lambda/k)}, \  \  k \rightarrow 0\  .
\end{equation}
Thus the Coulomb interactions are marginally irrelevant. 
This is equivalent to a logarithmically divergent velocity:
 $v(k) \sim (e^2/4) \ln(\Lambda/k), \ k \rightarrow 0$.

\paragraph{Two-loop results.}
It is instructive to examine corrections beyond first order
\cite{Mi07,Vafek:2008}, since additional
 effects appear, such as renormalization of $Z$. For example the
 first diagram in the RPA series shown in Fig.~\ref{figdiag1}(b) is 
\begin{equation}
\label{serpa1}
\Sigma^{(2b)}(\bk,\omega)\! = \! i \int \frac{d^{2}p d \varepsilon }{(2\pi)^{3}} 
G^{(0)}(\bk +\bp,\omega +\varepsilon )
(V(\bp))^{2} \Pi^{(1)}(\bp, \varepsilon)\, .
\end{equation}
An explicit evaluation at low energies and momenta gives a single logarithmic
 divergence
\begin{equation}
\label{serpa2}
\Sigma^{(2b)}(\bk,\omega) = -\frac{N\alpha^2}{24}(\omega +v{\bm \sigma}\cdot\bk )\ln(\Lambda/k), \
 k/\Lambda \rightarrow 0 ,
\end{equation}
i.e. $\Sigma^{(2b)}_0= -(N\alpha^2/24)(\omega)\ln(\Lambda/k)$, and
$ \Sigma^{(2b)}_v =-(N\alpha^2/24)v{\bm \sigma}\cdot{\bk}\ln(\Lambda/k)$. Because the polarization
 bubble is proportional to the number of fermion flavors $N=4$ (valley+spin),
 we have explicitly written the $N$ dependence.
By comparing with Eq.~\eqref{gf2}, we find that the velocity
 is changed by an amount $(-N/24 -N/24)\alpha^2 v\ln{(\Lambda/k)}$.

In addition,  other diagrams at second order have to be added, such as
 the vertex correction of Fig.~\ref{figdiag1}(c). Most importantly, this diagram
 is also proportional to $\ln{\Lambda}$. Collecting all contributions 
 one finds the RG equation for the velocity flow \cite{Vafek:2008}
\begin{equation}
\label{vrun2}
\frac{dv}{dl}= \frac{\alpha}{4} v - \left(\frac{N}{12} - \delta \right) \alpha^2 v \, ,
\end{equation}
with $\delta \approx 0.03$. 
One observes that the contribution of the ``RPA" diagram is numerically dominant
 at second order (it is larger than the rest  by a factor of $10$ for $N=4$.)
In addition, the second order tendency is a decrease of the velocity. 
Consequently a  finite coupling fixed point is  possible at $\alpha_c\approx0.8$. This
 fixed point  is infrared unstable since near $\alpha_c$, $\frac{dv}{dl}=-C(\alpha -\alpha_c)v, \ C>0$,
 i.e. for $\alpha > \alpha_c$, $v$ flows towards zero ($\alpha$ flows to $\infty$)
 while for $\alpha < \alpha_c$, $v$ flows towards $\infty$ ($\alpha$ flows to zero.)
 Of course it is not clear that this estimate is reliable since the fixed point value
$\alpha_c$ is not small, and we  used perturbation theory ($\alpha \ll 1$) to
 derive this result. On the other hand,  a flow towards
 strong coupling for $\alpha > \alpha_c$ is consistent with the formation
 of an excitonic insulator (mass generation), for which strong evidence has
 accumulated by now, as we discuss in Section \ref{sec:Mass}. Recent numerical
simulations give the value $\alpha_c \sim 1$ (see Section \ref{sec:Mass}).

 Finally, we also find that $Z$ is renormalized at second order, since
 the self-energy is frequency dependent. From Eq.~\eqref{Z}  we can expand to
 second order of bare perturbation theory 
$Z \approx 1-\frac{N\alpha^2}{24}\ln(\Lambda/k)$, which would lead
 us to an RG equation for $Z$: $\frac{dZ}{dl} = -\frac{N\alpha^2}{24}Z$,
  to be solved together with Eq.~\eqref{alpharun}, or  Eq.~\eqref{vrun2},
 depending on the desired level of approximation.
 Alternatively, Eq.~\eqref{Z} is already written in a ``nonperturbative" way.
 Ignoring for the moment the running
 of $\alpha$, we have at low energies
\begin{equation}
\label{ZRPA}
Z = \frac{1}{1+\frac{N\alpha^2}{24}\ln(\Lambda/k)} \rightarrow \frac{24}{N\alpha^2 \ln(\Lambda/k)},\
\ k/\Lambda \rightarrow 0. \
\end{equation}
This result, along with the previous one for $\Sigma^{(2b)}_0$,
 brings us to the infrared behavior (we use $\omega$ and $k$ interchangeably in the infrared
 limit):
\begin{equation}
\label{marginalFL}
 Z \sim \frac{1}{\alpha^{2}|\ln(\omega)|}, \  \  \Sigma_0 \sim \alpha^{2}\omega|\ln(\omega)|, \ \  \omega \rightarrow 0.
\end{equation}
 This is  characteristic of a marginal Fermi liquid \cite{DSar07,Gon94}.
 However, this regime is never achieved if the running of $\alpha$ is taken into account,
 as is intuitively clear from the above equations. As we will see later from the solution
 of the RG equations for $Z$ and $\alpha$, in fact $Z$ tends to level
 off in the infrared, and the system has well-defined quasiparticles.

 It is interesting to note that
 trigonal distortions, which change the band structure away from the
 Dirac equation,  are modified by the electron-electron interaction, and their irrelevance
 at low energies is enhanced \cite{FA08}. As a result, the linear
 dispersion becomes an even more robust feature of
 graphene \cite{RLG07}.

\begin{figure}
\centering
\includegraphics[width=0.65\linewidth]{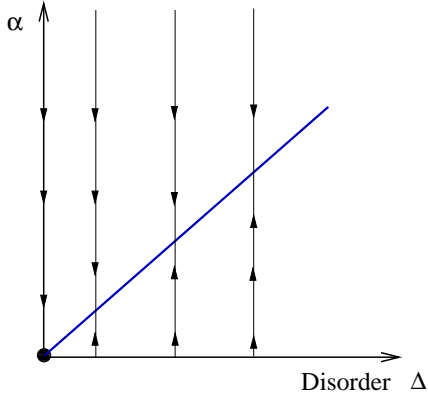}
\caption{An attractive line of fixed pints for interactions and gauge field disorder.}
\label{figdisorder}
\end{figure}

 \paragraph{Influence of disorder.}
Before we proceed,  let us briefly address the effect of disorder.
Two major sources of disorder are scalar potential random fluctuations
(e.g. formation of electron-hole puddles), and vector gauge field randomness,
 related to formation of ripples. Starting with the latter, i.e. a gauge field
coupled to the Dirac fermion pseudospin ${\bm \sigma}\cdot{\bf A}$, and characterized by
 variance $\Delta$,  
$\langle A_{\mu}({\bf r_1}) A_{\nu}({\bf r_2})\rangle = \Delta \delta_{\mu \nu} \delta({\bf r_1}-{\bf r_2})$,
 one can readily derive the corresponding RG equations in the weak disorder and
 interactions limit \cite{Stauber05,HJV08}
\begin{equation}
\label{rgdisorder}
\frac{d\Delta}{dl}=0,  \   \  \frac{d\alpha}{dl}= -\frac{\alpha^2}{4} + \frac{\Delta}{\pi} \alpha \  .
\end{equation}
Gauge field disorder itself is not renormalized, while the interplay of disorder
 and interactions leads to a line of attractive fixed points located at: $\alpha^{*} = \frac{4}{\pi}\Delta$,
 as shown in Fig.~\ref{figdisorder}. 
Physically the variance is related to the characteristic height $h$, and length $L$ of the corrugations
of the surface, $\Delta \sim h^{4}/(L^2a^2)$. Thus weak disorder generically shifts the fixed point 
away from $\alpha =0$, while strong disorder can have an even more profound effect (Section 
\ref{sec:PacoDefects}).

In addition,  for weak interactions,  the inclusion of scalar (density fluctuations) disorder turns out to be a relevant perturbation
 which grows under renormalization,  and thus
away from the perturbative regime  \cite{Aleiner06}.  Moreover,  gauge field disorder, when combined
 with strong-enough interactions, can  cause the interactions to grow
\cite{Vafek:2008}.
 It has been argued that the strong-coupling regime  for disorder and interactions
generically occurs when
 all types of disorder  consistent with graphene's symmetries are
included \cite{FA08}.

 A detailed  analysis of this complex situation is beyond the scope of this work, and
 from now on we continue our discussion of  clean graphene. 

\subsubsection{Strong-coupling/RPA analysis}
\begin{figure}
\centering
\includegraphics[width=0.9\linewidth]{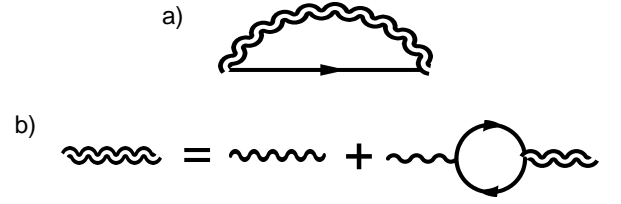}
\caption{RPA self-energy, which includes an infinite resummation of polarization bubbles.}
\label{figdiagramrpa}
\end{figure}

The  full RPA  treatment was  performed by many authors
\cite{Gon99,Son07,FA08,DSar07, Pol07,Kot1N}.
Here we mostly follow \onlinecite{Son07}. The RPA self-energy is shown diagrammatically
in Fig.~\ref{figdiagramrpa}, and corresponds to the equation
\begin{equation}
\label{serpa}
\Sigma^{(RPA)}(\bk,\omega) = i \int \frac{d^{2}p d \varepsilon }{(2\pi)^{3}} 
G^{(0)}(\bk +\bp,\omega +\varepsilon )
V^{RPA}(\bp,\varepsilon) \, .
\end{equation}
The RPA potential is given by
\begin{equation}
V^{RPA}(\bp,\varepsilon) = \frac{2\pi e^2}{\epsilon_0 p-2\pi e^2 \Pi^{(1)}(p, \varepsilon)}.
\label{rpapotential}
\end{equation}
Quite remarkably, at low momenta one can evaluate the singular contribution to
the self-energy analytically
\begin{equation}
\label{StrongRPA}
\Sigma^{(RPA)}(\bk,\omega) = \frac{8}{N\pi^2}
[- F_0(\lambda) \omega + F_1(\lambda)v{\bm \sigma}\cdot\bk ]
\ln(\Lambda/k)\,,
\end{equation}
where we have defined
\begin{equation}
 \lambda = \frac{ \pi}{8}N\alpha .
\end{equation}
This parameter is measuring the importance of polarization loop contributions
relative to the bare Coulomb term (i.e. the ratio of the second term to the first in the denominator
 of Eq.~\eqref{rpapotential}).  
The RPA is generally  expected to be valid when the loops dominate over other diagrams,
 i.e. $N\gg1$. Provided this condition is satisfied, we can also analyze the strong-coupling
 regime $\lambda \gg 1$, and the crossover toward the weak-coupling one ($\lambda \ll 1$),
 i.e. we can hope to cover a wide range of $\alpha$ values.

The calculated functions $F_0$ and $F_1$ in  Eq.~\eqref{StrongRPA} are
\begin{equation}\
\label{F1}
  F_1(\lambda) = \left\{ \begin{array}{ll}
    -\displaystyle{\frac{\sqrt{1-\lambda^2}}\lambda}\arccos\lambda 
       -1+ \frac\pi{2\lambda}\,,  \  \  \  \  \   \  \   \  \    \  \  \lambda<1, \\
    \displaystyle{\frac{\sqrt{\lambda^2-1}}\lambda} 
    \ln\left(\lambda+\sqrt{\lambda^2-1}\right)
       -1+ \frac\pi{2\lambda}\,,    \lambda>1,
  \end{array}\right.
\end{equation}
\begin{equation}
\label{F0}
  F_0(\lambda) = \left\{ \begin{array}{ll}
    -\displaystyle{\frac{2-\lambda^2}
      {\lambda\sqrt{1-\lambda^2}}}\arccos\lambda 
       -2+\frac\pi\lambda\,, \  \  \  \  \  \  \   \  \  \  \   \lambda<1,\\
    \displaystyle{\frac{\lambda^2-2}{\lambda\sqrt{\lambda^2-1}}} 
    \ln\left(\lambda+\sqrt{\lambda^2-1}\right)
        -2+\frac\pi\lambda\,,  \lambda>1.
  \end{array}\right.
\end{equation}
This leads to the system of RG equations for $v$ and $Z$, to leading order in $1/N$
\begin{eqnarray}
\frac{dv}{dl}= \frac{8}{N\pi^2}\left(F_1(\lambda) -F_0(\lambda) \right)v \,
\label{rgrpa1} ,
\end{eqnarray}
\begin{eqnarray}
\frac{dZ}{dl}=- \frac{8}{N\pi^2}F_0(\lambda)Z \, .
\label{rgrpa2}
\end{eqnarray}

At strong-coupling, $\lambda \gg 1$, one finds
\begin{eqnarray}
\frac{dv}{dl}= \frac{8}{N\pi^2}v \,  ,
\label{rgstrong1}
\end{eqnarray}
\begin{eqnarray}
\frac{dZ}{dl}=- \frac{8}{N\pi^2}\ln{(2\lambda)}Z \, .
\label{rgstrong2}
\end{eqnarray}
The first equation, after integration, leads to the low-energy result  ($k \rightarrow 0$)
\begin{equation}
v(k)/v =\left (\frac{\Lambda}{k} \right)^{\eta}, \  \eta = \frac{8}{N\pi^2} ,
\label{vanomalous}
\end{equation} 
which implies that the quasiparticle dispersion is of the form
\begin{equation}
\omega(k) \sim k^{z}, \   \   \  z=1 - \frac{8}{N\pi^2}  .
\label{disanomalous}
\end{equation}
The existence of the anomalous velocity  dimension, $\eta$, and consequently
 $z\neq1$, is characteristic of the strong-coupling regime $N\alpha \rightarrow \infty$ \cite{Son07}.
 However this strongly-coupled   fixed point is infrared unstable,
 since, due to the velocity increase, the RG for $\alpha$ flows towards weak coupling.
 (One also  expects that for certain $N<N_c$ and $\alpha \gg 1$
an excitonic gap can appear, which will be discussed in  Section \ref{sec:Mass}.)
In this regime $Z$ can be approximated perturbatively (in $1/N$) as
\begin{equation}
Z  \approx 1 - \frac{8}{\pi^2}\frac{1}{N}\ln(N\alpha\pi/4) \ln(\Lambda/k) , \  \  \ N\alpha \gg 1 ,
\label{Zstrong}
\end{equation}
which can be obtained from Eq.~\eqref{rgstrong2} by ignoring the scale
 dependence of $\lambda$.

In the weak-coupling limit $\lambda \ll 1$, it is easy to verify that we recover
 the previous result  \eqref{vrun} for the velocity $v$ (leading to
 a flow for $\alpha$ towards zero), and the
previously encountered perturbative
equation for $Z$ 
\begin{equation}
\frac{dZ}{dl}=-\frac{8}{N\pi^2}\frac{\lambda^{2}}{3}Z \ , \mbox{i.e.} \  \  \ Z \approx 1 - \frac{N}{24}\alpha^{2} \ln(\Lambda/k) .
\label{Zweak}
\end{equation}
The last formula is written to first order in $N\alpha$.
\begin{figure}
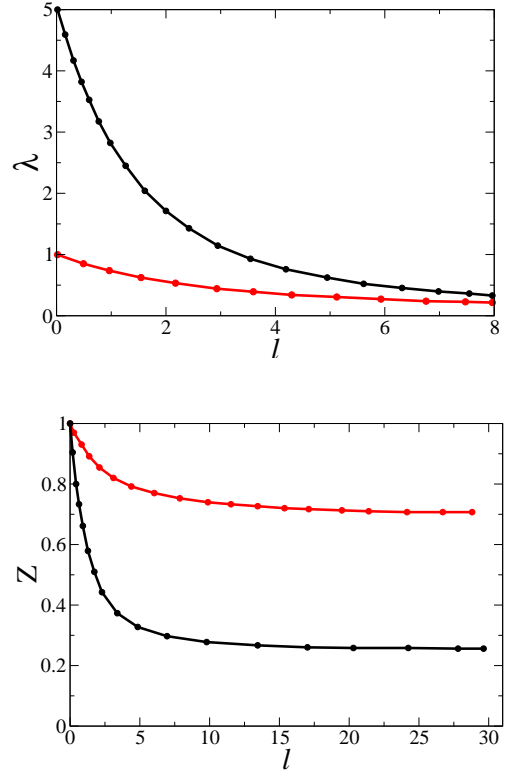

\centering
\includegraphics[width=0.75\linewidth]{Figs/Fig-10-1}
\vspace{0.7cm}

\includegraphics[width=0.75\linewidth]{Figs/Fig-10-2}
\caption{RG flow of the coupling $\lambda$ and the quasiparticle residue $Z$
 as a function of the RG scale $l$; the infrared limit
 is at $l \rightarrow \infty$. From \onlinecite{Gon99}.}
\label{gonstrong}
\end{figure}

Eqs.~\eqref{Zstrong},\eqref{Zweak} allow us to have a qualitative understanding
 of the  behavior of  $Z$  as a function of the RG scale $l$. If the initial value of $\alpha$ is
 large, at the initial RG steps $Z$ decreases  logarithmically fairly fast (due to the weak 
 $\ln(\alpha)$ dependence  in Eq.~\eqref{Zstrong}, even though $\alpha$ itself decreases).
Eventually, when $\alpha$ has decreased substantially 
($\alpha \sim (\ln(\Lambda/k))^{-1}$),  $Z$ is governed by Eq.~\eqref{Zweak}, 
meaning that $Z$ will stop decreasing, and will level off for $l= \ln(\Lambda/k) \rightarrow \infty$.
 
A numerical evaluation of the system of equations  \eqref{rgrpa1},\eqref{rgrpa2}
confirms the anticipated behavior and is shown in Fig.~\ref{gonstrong}, \cite{Gon99}.
(The equation for the coupling $\lambda =\frac{ \pi}{8}Ne^2/(\epsilon_0v)$ is obtained
 by observing that $(d\lambda/dl) = (-1/v^2)  \frac{ \pi}{8}N(e^2/\epsilon_0) (dv/dl)$, due to the lack
 of charge renormalization.)
\begin{figure}
\centering
\includegraphics[width=0.9\linewidth]{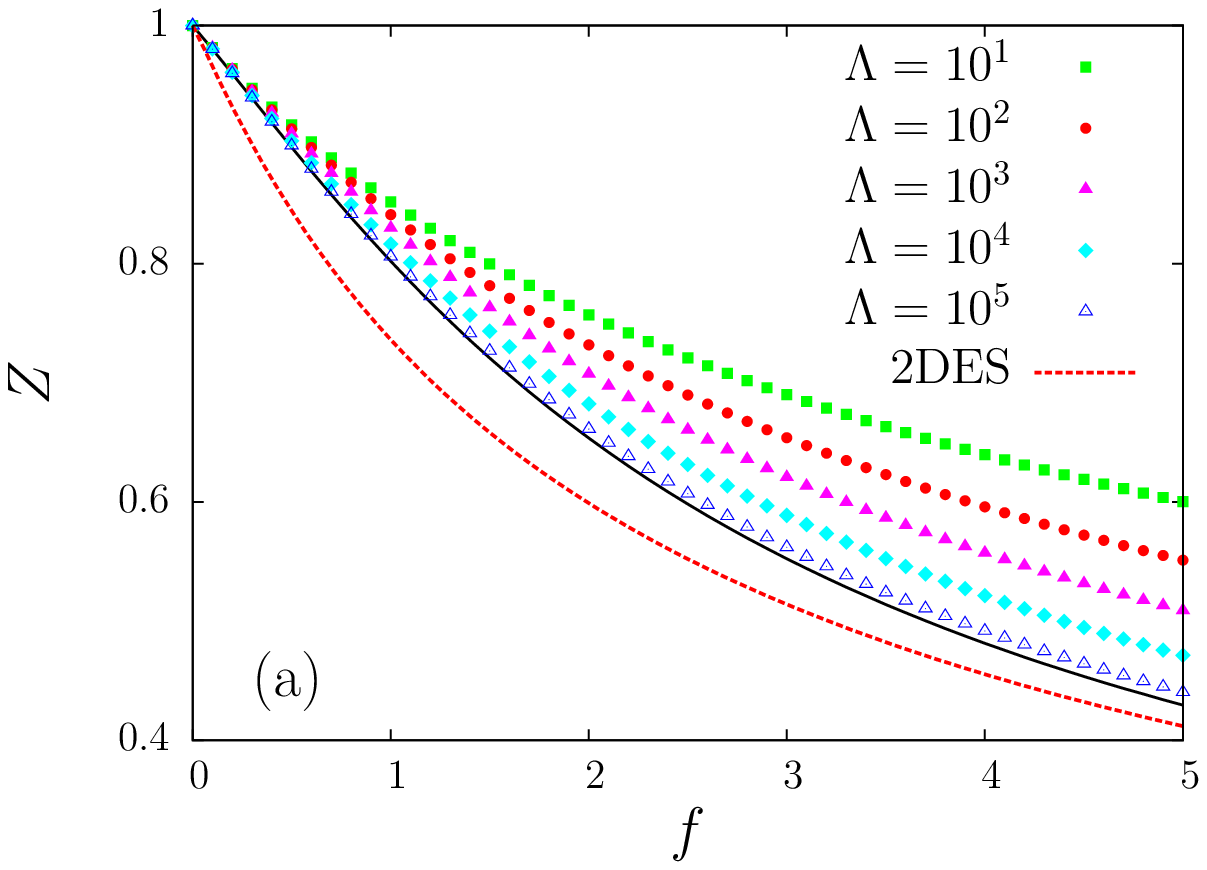}
\includegraphics[width=0.9\linewidth]{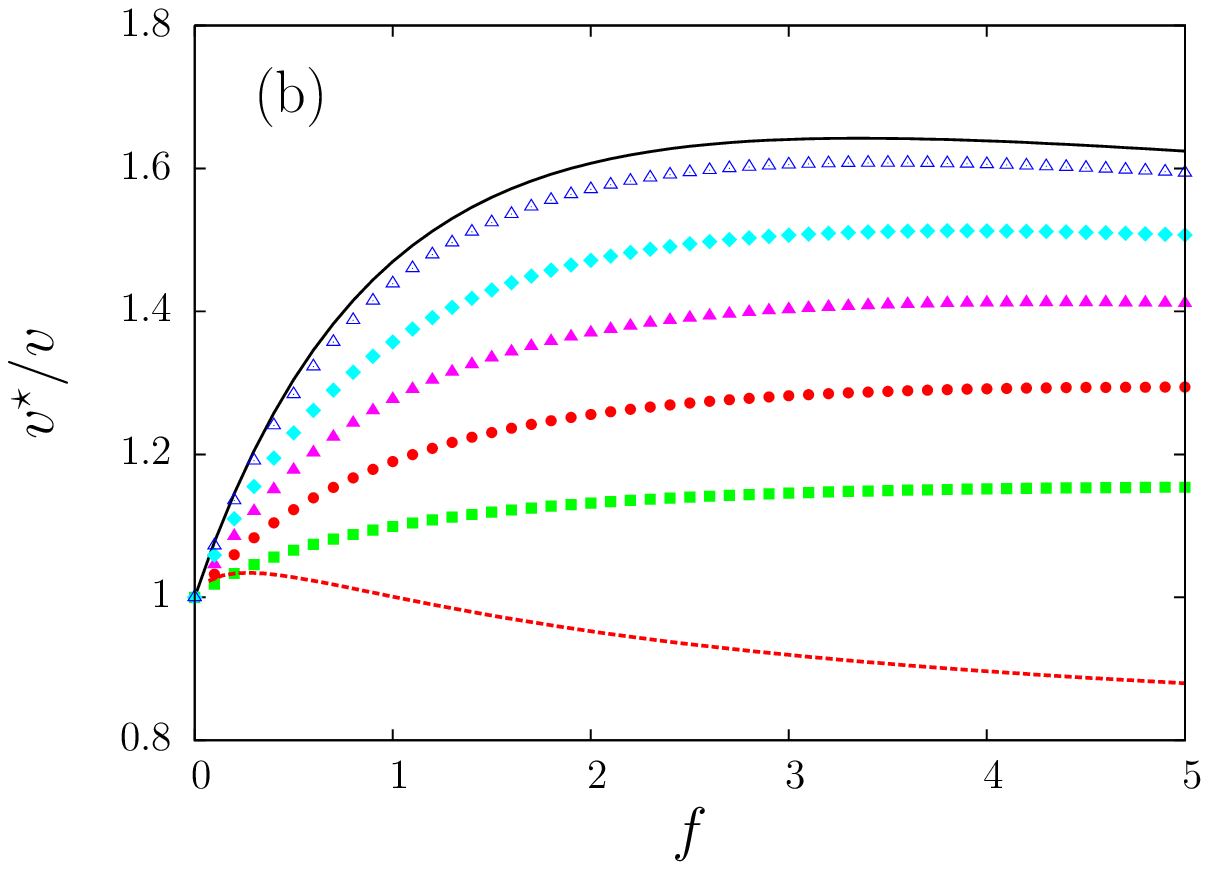}
\caption{(Color online) Exact evaluation of the RPA equations for (a) the quasiparticle residue,
 and (b) the Fermi velocity. On the horizontal axis $f$ is defined as $f \equiv N\alpha$.
$\Lambda$ is in units of $k_F$ . Values of $\Lambda$ from $\sim 10^2$ to $10^1$
correspond to density $n$ from $n \sim 10^{11}{\mbox{cm}}^{-2}$ to 
 $n \sim 10^{13}{\mbox{cm}}^{-2}$
(while $\Lambda \sim 10^3$ is ultra low density $n \sim 10^{9}{\mbox{cm}}^{-2}$).
The values of $\Lambda$ (in units of $k_F$) 
 can be converted into density $n$ via: $\Lambda/k_F \approx 220/\sqrt{\tilde{n}}$,
$\tilde{n}=n/(10^{10} {\mbox{cm}}^{-2})$.
The curves labeled 2DES refer to  the case of 2DEG with parabolic bands, where $f=\sqrt{2} r_s$,
 and $r_s \sim 1/\sqrt{n}$.
 From   \cite{Pol07}.}
\label{polini}
\end{figure}
We conclude that the flow of $\lambda$ is towards weak coupling, no matter how large its initial
 value is. $Z$ does not renormalize to zero at low energy due to the RG decrease of $\lambda$.
 Thus, near the weak-coupling infrared fixed point, the marginal Fermi liquid
(Eq.~\eqref{marginalFL}) is ultimately
 not reached, and the system behaves as a Fermi liquid (although the quasiparticle
 decay rate is non-Fermi liquid like, see below.) At higher energies  however (away
 from the fixed point but still  much lower than the bandwidth $v\Lambda$), the system
 exhibits marginal Fermi liquid behavior. 

At finite (but still small) density away from the Dirac point, i.e. $k \neq 0$,
the logarithmic behavior in the infrared is  cut-off by the Fermi momentum,
 i.e. $\ln(\Lambda/k_F), \ k_F/\Lambda \rightarrow 0$,
 and  the RG stops away from the fixed point.
 For comparison with experiments, the flow toward this stable fixed point should
be stopped at a scale set by the (small) density, temperature, or frequency,
 whichever is higher.

One can also perform a numerical evaluation of the main RPA equation
Eq.~\eqref{serpa} \cite{Pol07}. For small density, and with logarithmic
 accuracy ($\ln(\Lambda/k_F)$), this is equivalent to evaluating,
by using the notation of Eq.~\eqref{StrongRPA}, and taking into account
Eqs.~\eqref{sedecomposition},\eqref{gf2},\eqref{Z}
\begin{equation}
\label{rpapolini1}
Z=(1-\partial\Sigma^{(RPA)}/\partial\omega)^{-1}
= \frac{1}{1+\frac{8}{N\pi^2}F_0(\lambda)\ln(\Lambda/k_F)}\ ,
\end{equation}
\begin{equation}
\label{polini2}
v^{*}/v = Z\left(1+\frac{8}{N\pi^2}F_1(\lambda)\ln(\Lambda/k_F)\right) \ .
\end{equation}
Here $v^{*}$ is the renormalized velocity.
At any finite density the numerical evaluation of  $\Sigma^{(RPA)}$
also picks up finite (subleading) contributions, while
it can be shown \cite{Pol07} that the leading perturbative results such as 
 Eqs.~\eqref{Zstrong},\eqref{Zweak} are readily reproduced.
 The RPA results  are shown in Fig.~\ref{polini}, and exhibit the natural density
dependence tendency, i.e. the strongest renormalization occurs at the lowest
 densities. Similar RPA results have been obtained by \onlinecite{DSar07}. 

A  significant
 velocity enhancement was observed in the infrared conductivity 
 \cite{Li08a}, which reported around $15\%$ increase of
 the Fermi velocity,  having value as high as $v^{*} \approx 1.25\times 10^{6} {\mbox{m/s}}$  at
the lowest densities (compared to $v \approx 1.1\times 10^{6} {\mbox{m/s}}$ at higher density). 
The  system is at  a finite Fermi energy
 $\mu \approx 0.2 {\mbox{eV}}$. 
However the velocity renormalization  is not logarithmic, and 
it is not clear  what is the origin of this effect.

\begin{figure}
\centering
\includegraphics[width=1.2\linewidth]{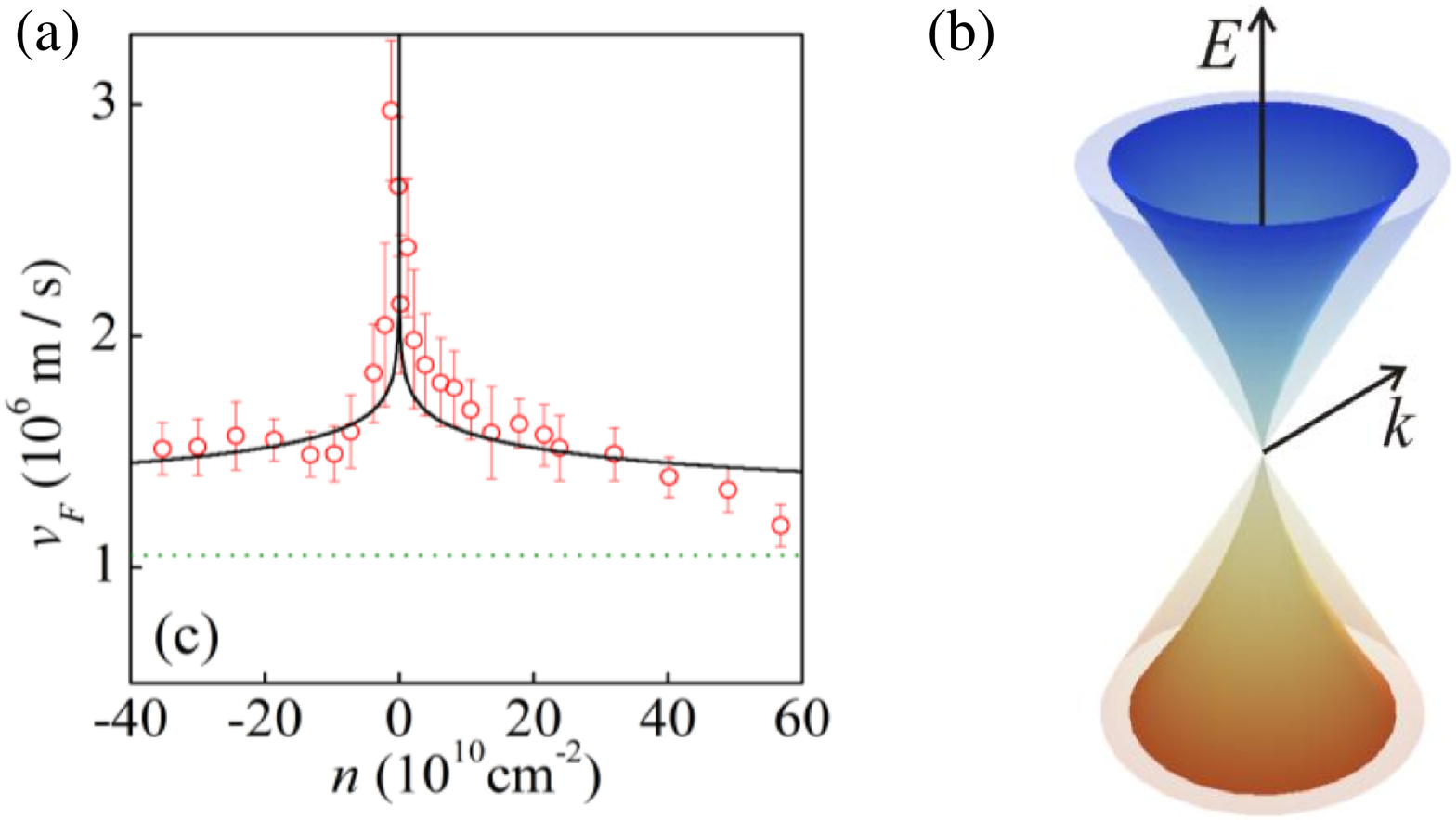}
\caption{(Color online) (a) Density dependence of the velocity for suspended graphene,  from \cite{Elias11}. The solid
line is the result of RG treatment within RPA (Eq.~\eqref{rgrpa1}).
(b) Reshaping of the Dirac cone due to the interaction-driven
 renormalization (increase) of the Fermi velocity at low 
momenta.
 The outer cone represents the linear Dirac spectrum without many-body effects.}
\label{reshapedcones}
\end{figure}

A recent study  of suspended graphene which measures the cyclotron mass  \cite{Elias11}
 has detected significant logarithmic renormalization 
of the Fermi velocity, having the high value 
 $v^{*} \approx 3\times 10^{6} {\mbox{m/s}}$ at the lowest densities
$n < 10^{10}{\mbox{cm}}^{-2}$,  almost three times the value at high density 
($n > 4 \times 10^{11}{\mbox{cm}}^{-2}$), Fig.~\ref{reshapedcones}(a). The logarithmic renormalization of the velocity
predicted  by theory fits the data fairly  well, 
and thus offers a direct proof that the Dirac cones
 can be reshaped by long-range electron-electron interactions near the Dirac point, as schematically 
 shown in Fig.~\ref{reshapedcones}(b).  Finally, ARPES measurements of quasi-freestanding graphene grown on the carbon face of SiC 
 have also detected logarithmic velocity renormalization \cite{LanzaraMB}.

\subsubsection{Quasiparticle lifetime}
The inverse quasiparticle lifetime (decay rate) due to electron-electron interactions, $1/\tau_{ee}$, 
is an important quantity  which is relevant to many properties of   graphene  (and Fermi systems in 
general). In particular
the dependence of $1/\tau_{ee}$ on energy (or temperature)  determines  the importance
 of the electron-electron interaction contribution, relative to other processes, 
 to transport, and  interpretation of spectroscopic features, such as ARPES.

The decay rate is determined by the imaginary part of the self-energy,
 $\mbox{Im}\Sigma (\bk,\omega)$. The first diagram which has energy dependence, 
and thus a non-zero imaginary part, is the one bubble  diagram of Fig.~\ref{figdiag1}(b),
 whose real part is given by Eq.~\eqref{serpa2}, i.e. behaves as in  Eq.~\eqref{marginalFL}
 at low energies. We can therefore deduce, for energies and momenta close
 to the mass shell \cite{Gon96},
\begin{equation}
\label{lf}
\mbox{Im}\Sigma^{(2b)}(\bk,\omega)  
\sim \alpha^{2} \theta(\omega- vk) \ \omega, \ \ \omega \approx vk \ ,
\end{equation}
i.e. the decay rate is linear in energy. In addition, there is an on-shell (``light cone", $\omega = vk$)  discontinuity, where the rate experiences a jump.
This on-shell behavior is due to the fact that, for $\omega < vk$, there is no phase
 space available for virtual interband particle-hole excitations 
(see Fig.~\ref{fig:particle-hole-continuum}),
 whereas such excitations are possible for $\omega > vk$. 

The above behavior is valid at the Dirac point and   $T=\mu=0$, while for
 small  $T,\mu$, it is valid for  energies of order ${\mbox{max}}(T,\mu)$.
 Notice also that  the linear energy behavior of Eq.~\eqref{lf} is very different from
 the conventional Fermi liquid result $\mbox{Im}\Sigma  \sim \omega^2$ \cite{DSar07},
 which would occur for a finite Fermi surface ($\mu \neq 0$) and is due to intra-band
 particle-hole excitations. 

The on-shell discontinuity present at the one-loop level Eq.~\eqref{lf} disappears
 when the full RPA self-energy is evaluated (Fig.~\ref {figdiagramrpa}).
In this case one obtains \cite{Khve06b}
\begin{equation}
\label{lfrpa}
\mbox{Im}\Sigma^{(RPA)}(\bk,\omega) \sim
 \ln{(\pi \alpha)} \theta(\omega-vk) (\omega-vk), \ \ \omega \approx vk \  .
\end{equation}
Away from the mass shell, the energy dependence is naturally  linear: 
\begin{equation}
\label{lfrpa1}
\mbox{Im}\Sigma^{(RPA)}(\bk,\omega) \sim
 \ln{(\pi \alpha)} \ \omega, \ \ \omega \gg vk \  .
\end{equation}
The full dependence $\mbox{Im}\Sigma^{(RPA)}(\bk,\omega)$ has to be evaluated
numerically \cite{DSar07}, and the results confirm the smooth rise of $\mbox{Im}\Sigma^{(RPA)}$
from the point $\omega = vk$.

In the limit of  zero doping $\mu \rightarrow 0$, when the system approaches the fixed point
 $\alpha=0$, we argued previously that the residue $Z$ does not approach zero (i.e. the marginal
 Fermi liquid behavior ultimately does not manifest itself.)
 On the other hand the marginal Fermi-liquid behavior is expected to be much more robust 
as far as  the inverse lifetime, $\mbox{Im}\Sigma \sim \omega$,
 is concerned, because the running of the coupling $\alpha(\omega)$
 only introduces logarithmic variation on top of a much stronger linear energy dependence.

The linear decay rate  discussed above is  consistent with ARPES experiments  \cite{Bos07,Zhou08},
 and STM measurements of graphene on graphite \cite{Li08b} (see also  the discussion
in \onlinecite{Grushin09}).

\subsection{Spontaneous mass generation}
\label{sec:Mass}
It is an intriguing possibility that graphene can undergo a metal-insulator
 transition for strong  enough Coulomb interaction $\alpha$, due to an excitonic
pairing mechanism. We  restrict ourselves to the charge neutrality point $\mu =0$
 since the excitonic pairing tendency decreases quickly beyond that.

\subsubsection{Finite explicit mass}
Before we outline
 the main results, let us mention that
 an explicit gap can also  open  in graphene under certain conditions
 that depend on graphene's environment.
 For example  there are suggestions of a detectable gap in  
situations when   graphene is  on  a substrate with specific symmetry, creating sublattice 
asymmetry in the graphene plane, and thus making the graphene electrons 
massive (gapped)  \cite{Zhou:2007}.  Gaps can also be produced by confining the electrons
 into finite-size configurations, such as quantum dots  \cite{Petal08}.
 In these cases the gap generation mechanism is not intrinsic to graphene,
 and the value of the gap depends strongly on the external conditions.
 However even in such situations  interactions can play an important role
 by increasing the gap. 

Consider a gap arising from an external potential that alternates between the two
 sublattices
\begin{equation}
\mathcal{H}_{mass} =  \Delta_0 \sum_{\sigma,i \in A}n_{\sigma}(\mathbf{R}_{i})
- \Delta_0 \sum_{\sigma,i \in B}n_{\sigma}(\mathbf{R}_{i})\ .
\label{eq:H-Mass}
\end{equation}
 Consequently an additional pseudospinor structure related
 to $\sigma_3$ is generated, and the new Green's function
 has the form
\begin{equation}
\label{gfmass}
G({\bk},\omega) = \frac{1}{ \omega  - v {\bm \sigma}\cdot{\bk}
-\Delta_0 \sigma_3 -\Sigma (\bk,\omega)}.
\end{equation}
Here $\Delta_0$ is the explicit ``mass" of the graphene electrons (while $\Sigma (\bk,\omega)$
 contains the information 
about interactions, assumed to perturbatively renormalize all the other terms.)
 The new spectrum is then $E(k) =\pm \sqrt{v^2k^2 + \Delta^2_0}$, with a gap of $2\Delta_0$.
Computing the  Hartree-Fock interaction correction to $\Delta_0$
 leads to a renormalized mass $\tilde{\Delta}_0$  \cite{Kan05,Kotov:2008} 
\begin{equation}
\label{renormmass}
\tilde{\Delta}_0/ \Delta_0 \approx 1 + \frac{\alpha}{2} \ln{(D/\Delta_0)} .
\end{equation}

The above enhancement can be substantial. For example
   for a  bare gap due to spin-orbit coupling  $\Delta_0 \sim 10^{-3}{\mbox{meV}}$
\cite{Min06,Yao07},
and taking into account the bandwidth $D=v\Lambda \approx 7  {\mbox{eV}}$, the
 logarithmic factor is around $15$. In fact one should integrate the RG equation
 for the renormalized mass $\tilde{\Delta}_0$ as a function of $\ln(\Lambda)$
simultaneously with the equation for the running coupling $\alpha(\ln(\Lambda))$, 
Eq.~\eqref{alpharun},
 down to the  lowest infrared scale $\sim \Delta_0$ (bare gap). This leads to
the stronger dependence $\tilde{\Delta}_0/ \Delta_0= (1 + \frac{\alpha}{4} \ln{(D/\Delta_0)})^2$
\cite{Kan05}, and the perturbative expansion of this result is Eq.~\eqref{renormmass}.
It is interesting to note that the logarithmic mass renormalization formula in graphene
Eq.~\eqref{renormmass} is similar to the well-known expression for the 
electromagnetic mass of the electron (accounting for 
radiative corrections) in 3D relativistic QED \cite{Weiss}.

\begin{figure}
\centering
\includegraphics[width=0.7\linewidth]{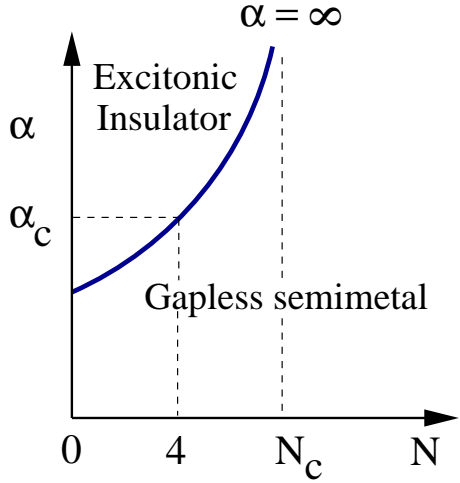}
\caption{Schematic phase diagram in the  $\alpha-N$ plane.}
\label{phasegraphenegap}
\end{figure}

\subsubsection{Excitonic mass generation}
We now turn to the possibility of  spontaneous gap generation due to long-range Coulomb interactions
(we set the explicit gap $\Delta_0=0$ in Eq.~\eqref{gfmass}).
In relativistic QED in two space (plus one time) 
dimensions, QED$_{2+1}$, the study  of this phenomenon, called chiral symmetry
 breaking, started quite a while 
ago \cite{Pisarski84,Appel86}, and is still going strong today. Graphene is actually 
different from QED$_{2+1}$ because only the fermions are confined to a 2D 
plane, while the field lines extend through the whole 3D space. In addition, 
the Coulomb interaction in graphene can be considered instantaneous since
the speed of light $c$ is much larger than the Fermi velocity 
($v \approx c/300$). Hence, Lorenz invariance is not respected, 
which reflects the non-relativistic, purely band origin of the Dirac 
quasiparticles.  The analysis in relativistic QED reveals that dynamical
mass can be generated  below a critical
 number of fermion flavors $N_c$, with the mass scale
 set by the coupling itself, which has dimension of energy in pure QED$_{2+1}$.
A transition is also found in non-relativistic graphene, where the generated mass scale
 is related to the ultraviolet energy cutoff (bandwidth $D=v\Lambda$)
 since the coupling $\alpha$ is dimensionless in this case.

The gap equation can be obtained  as  a self-consistent
 solution for the self-energy within RPA (i.e. vertex corrections are neglected),
 and is referred to as the Schwinger-Dyson equation.
It has the form 
\begin{equation}
\Delta({\bp},\varepsilon) = i  \int \frac{d^{2}k  d\omega}{(2\pi)^{3}}
\frac{V^{RPA}({\bp} -{\bk}, \varepsilon-\omega)
\Delta({\bk},\omega)}{\omega^2 - v^2k^2 - \Delta^{2}(k) + i0^{+}}.
\label{gapeq}
\end{equation}
The structure of the solution has been  analyzed extensively  
\cite{Khve08,Khv04,Gorbar02,Gorbar10,Liu09} at different
 levels of approximation. The equation is simplified significantly
 if the static RPA potential is used $V^{RPA}({\bf p},0)$
\cite{Khv04},
 while the dynamical equation has also been studied on-shell ($\Delta({\bf p},\varepsilon=vp)$)
  \cite{Khve08}, 
 as well as numerically \cite{Liu09}. 

 The mass gap $\Delta(p)$ has  strong momentum dependence, due to the
 long-range nature of the Coulomb interaction. $\Delta(p)$ decreases at large momenta
 and reaches maximum value at small momenta where it levels off.
 For   fixed physical
 value of $N=4$,  a transition to a gapped state is found above a critical coupling $\alpha_c$.
 Some of the calculated values are: $\alpha_c=0.92$ \cite{Gorbar10},   $\alpha_c=1.13$ \cite{Khve08}.
 At strong coupling $\alpha \rightarrow \infty$  the gap is non-zero only below a critical
 number of fermion flavors (since the effective interaction scales as $1/N$ in this limit);
for example $N_c \approx 7.2$ \cite{Khve08}, $N_c \approx 7$ \cite{Liu09}.

Near the critical coupling the low-momentum gap scales as
\begin{equation}
\Delta(0) \propto D \exp{\left(-\frac{C}{\sqrt{\alpha_{eff} -
 \alpha_{eff,c}}}\right)} \ ,
\label{gapcrit}
\end{equation} 
where $C$ is a constant, the critical  $\alpha_{eff,c}=1/2$, and the form of  the effective coupling
$\alpha_{eff}$ depends on the level
 of approximation used --- for example an improvement over the static RPA potential
 leads to: $\alpha_{eff} = \alpha/(1+N\pi\alpha/8\sqrt{2})$
(which gives $N_c \approx 7.2, \ \alpha \gg 1$,  and $\alpha_c=1.13, \ N=4$  \cite{Khve08}).
The form of Eq.~\eqref{gapcrit} suggests that the transition is of infinite order 
(Berezinskii-Kosterlitz-Thouless type).
Even though Eq.~\eqref{gapcrit} is only valid near the critical coupling,
 numerical results find that the gap in units of the bandwidth, $\Delta(0)/D$, is
 exponentially small in a wide range of couplings \cite{Khv04}. 
 Since $D \approx 7{\mbox{ eV}}$, this implies $\Delta(0) \sim {\mbox{meV}}$,
i.e. a rather small gap value.
Finally, recent work  that takes into account the renormalization
 of the coupling constant  and the quasiparticle residue  suggests
that $\alpha_c$ could be much larger than previously found \cite{Sabio10,Gonzalez10}.

The above results are  based on various  approximation schemes and
 it is therefore important to compare them with direct numerical simulations
of the lattice field theory model. Recent Monte Carlo calculations
\cite{Drut09b,Hands08,Drut09a,Drut09c} provide strong evidence
that spontaneous mass generation does occur, and give comparable
values for the critical couplings:   $N_c \approx 9.6, \ \alpha \gg 1$ \cite{Hands08},
$\alpha_c=1.1, \ N=4$  \cite{Drut09a}. 
Unfortunately the Monte Carlo simulations do not allow for an exact determination of the gap size,
 and for that we can only rely on the previously described Schwinger-Dyson equation
 (leading to small gaps). For graphene deposited on SiO the 
value of $\alpha_{SiO_2} \approx 0.79$
  and is therefore not enough to generate a gap; only experiments on ultrahigh mobility suspended
 samples can potentially reveal the  insulating state.

The overall phase diagram of graphene in the $\alpha-N$ plane is expected
 to look as shown in Fig.~\ref{phasegraphenegap}, with $\alpha_c \approx 1$ and
$N_c \approx 7-9$.  At finite temperature one expects the existence of a critical
 temperature $T_c \sim \Delta(0)$,
 while finite doping $\mu$ very quickly destroys the gap \cite{Liu09}.
Application of magnetic field perpendicular to the graphene layer leads to enhancement of the
 excitonic  instability due to the formation of Landau levels  
\cite{Khv01a,GMSS06,Gorbar02}.
In addition, it has been suggested
 that an in-plane magnetic field  favors a gapped excitonic state \cite{Kharzeev07},
due to the instability of a system of electrons and holes polarized in opposite directions.


The physical structure of the gapped   state 
depends on the nature of pairing between the valleys --- 
for example one can  have charge density wave states \cite{Khv01b} with modulation of the electronic
 density around the two sublattices (which corresponds to intravalley paring),
 or Kekule dimerization \cite{Hou07} which corresponds to tripling of the unit cell (intervalley pairing).
 One generally expects that interactions beyond the long-range Coulomb potential,
such as short-range repulsion, would favor particular states, including time-reversal
symmetry broken (spin) states. Further discussion appears in Section \ref{sec:gapsBruno}.

\subsection{Finite density Fermi-liquid regime}
\label{sec:FL}
\begin{figure}
\centering
\includegraphics[width=0.87\linewidth]{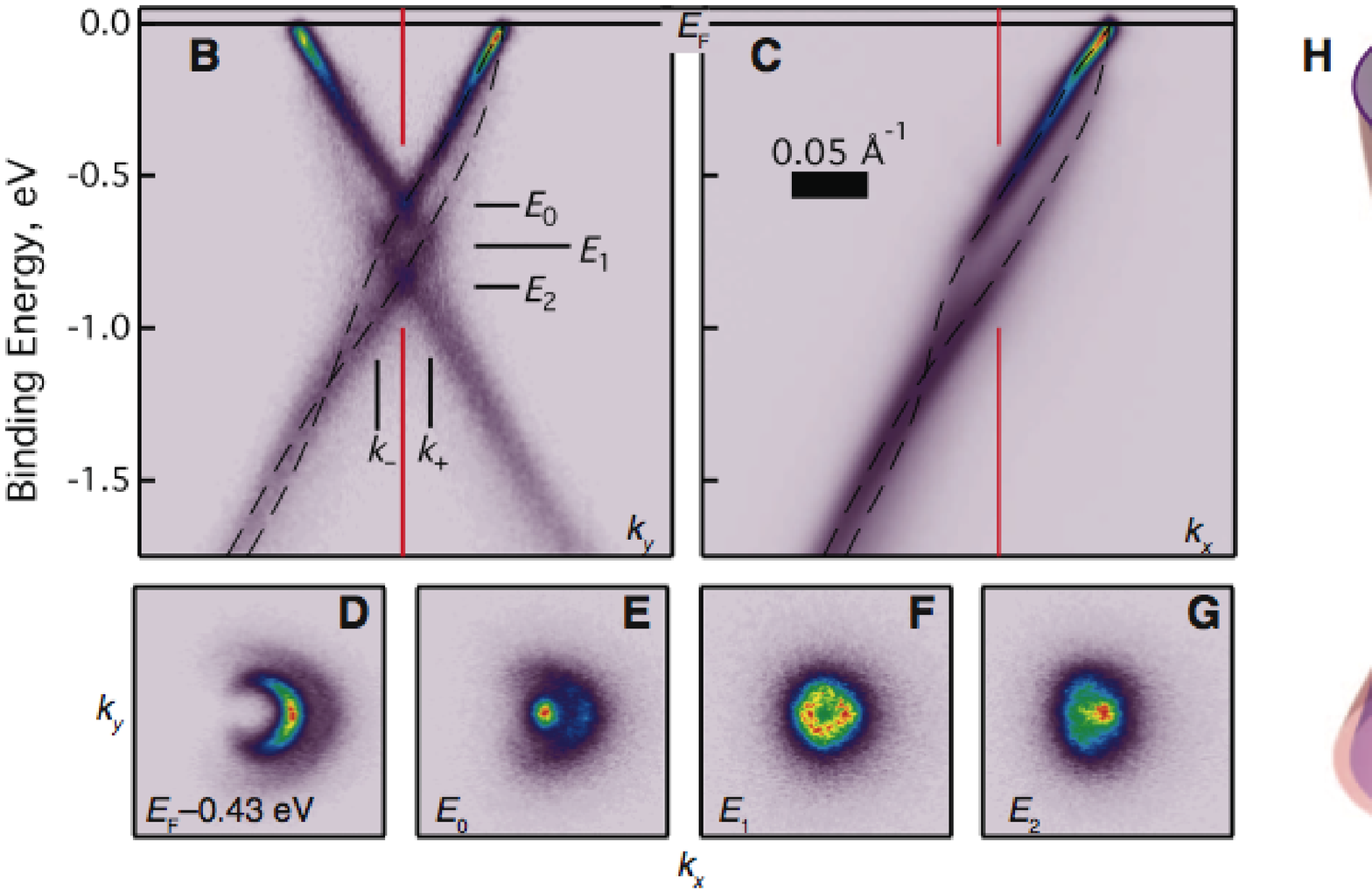}
\vspace{0.4cm}

\includegraphics[width=0.95\linewidth]{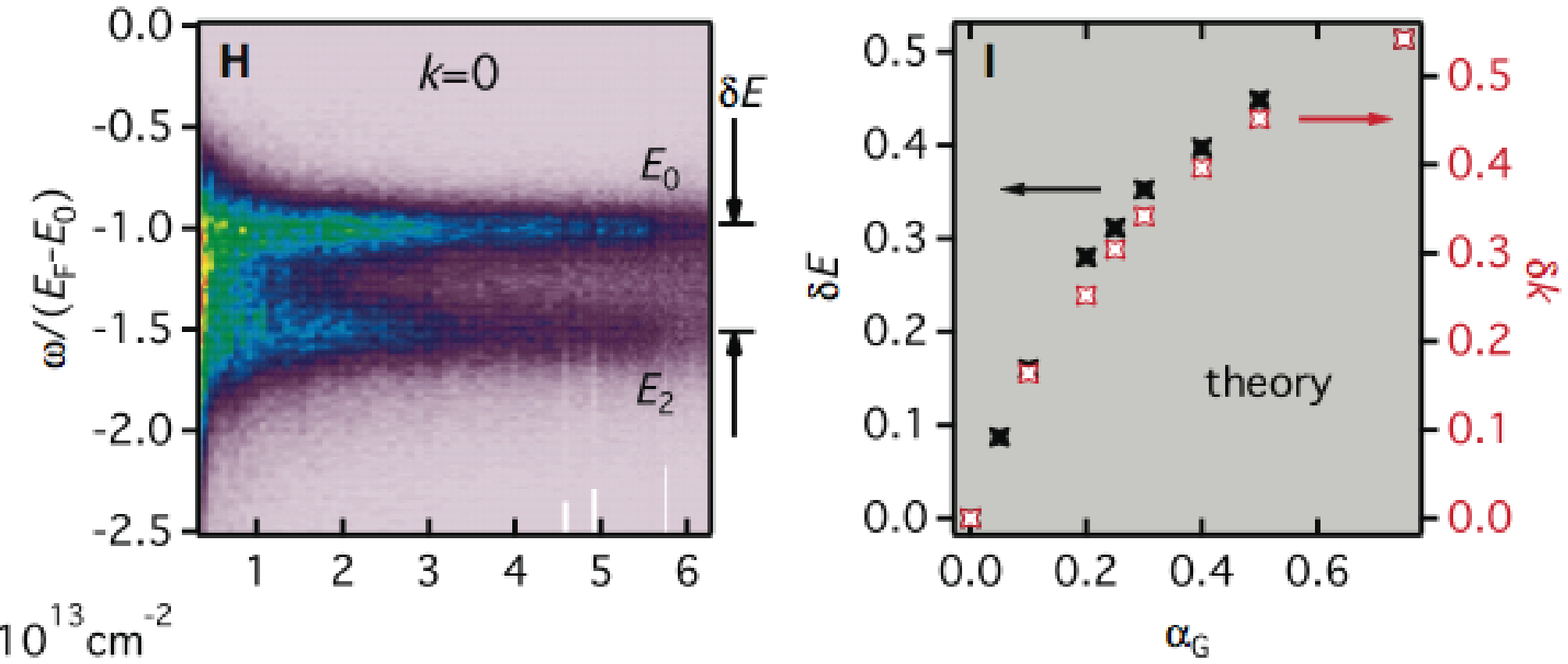}
\caption{(Color online) ARPES data  from \cite{Bos10}, showing strong features at the Dirac point,
 which is below the Fermi energy (at $0$). The splitting shown in (H) is attributed to the presence
 of ``plasmarons" --- quasiparticles strongly bound to plasmons --- and depends on the value of $\alpha$
($\alpha \approx 0.5$ fits the data.) }
\label{ARPES}
\end{figure}

 As the density increases above half-filling, i.e. graphene is  at a finite,
   not necessarily small,  chemical
 potential $\mu$, with a finite Fermi surface, a crossover towards a Fermi liquid regime takes place.
 In this case the lower (hole) band becomes irrelevant and the
 physics  near the Fermi surface is dominated by intra-band transitions
 in the conduction (upper) band (assuming $\mu >0$).
However the physics near the Dirac point can still  be very strongly
 affected due to the presence of plasmon and ``plasmaron" features in
 the quasiparticle spectral function.

 The quasiparticle width near $k_F$ is quite similar to the case of an ordinary 2D electron
 gas \cite{Pol08a,DSar07,Hwa08}, and is proportional to the second power of
 energy (or temperature),  as in a Fermi liquid, while the quasiparticle
 residue is finite at the Fermi surface. 

The existence of a plasmon-related peak  in the quasiparticle decay rate, 
which originates  from intraband transitions
 in which an electron can decay into a plasmon, was pointed out in the context
 of intercalated graphite, where the physics is dominated by graphene layers
\cite{ShungLi,Lin96}.
 For n-doped graphene ($\mu > 0$), which is relevant to ARPES
 experiments,   a double-feature is   found in the decay rate ${\mbox{Im}}\Sigma$:
  a peak at positive energies, signaling an onset of plasmon emission, and
a sharp spectral feature at negative energies, below the Dirac point, and 
 separated from it by  an amount  proportional to the plasmon frequency 
\cite{Pol08a,Hwa08}. This is the so-called ``plasmaron" --- a resonance 
which consists of a quasiparticle strongly coupled to plasmons \cite{Lundqvist}. 
Plasmaron features have been previously detected for example  in optical measurements of
 Bismuth \cite{Armitage}.

The above calculations were done within RPA theory. 
 Line widths    have also been analyzed via ab-initio many-body methods
\cite{Park09,Trevis08}.
 Experiments  generally show  a well-pronounced linear quasiparticle
spectrum 
\cite{Bos07,Zhou:2007,Zhou08,Sprinkle09}, with additional features
near the Dirac point 
 which seem to depend on the way graphene is prepared, and its
 purity.  For example, gap-like features have been observed near the
Dirac point \cite{Zhou:2007},
 and attributed to external, substrate-related factors. Bending of the Dirac spectrum
(kink-like feature) 
 was attributed to  plasmons \cite{Bos07}.  Most recently  manifestations of
 the sharp plasmaron spectral intensities have been observed in quasi-freestanding graphene \cite{Bos10},
 where a reconstruction of the Dirac point crossing seems to take place, as shown in Fig.~\ref{ARPES}.
A diamond-like shape appears due to crossing  of charge and plasmaron bands.
 Comparison of the RPA calculation for the energy splitting with experiment leads to
 the value of $\alpha \approx 0.5$ (Fig.~\ref{ARPES}.)  \onlinecite{Bos10} also suggest
 that the plasmaron features were obscured in earlier measurements on non free-standing
 graphene \cite{Bos07},  due to the   several times stronger screening (and consequently
 smaller $\alpha$.)
 Perhaps most importantly, all the current activity in ARPES on different graphene
 samples reveals that the electron-electron interactions can affect strongly the physics
 around the Dirac point, even for relatively large density (Fermi energy).

Tunneling spectroscopy measurements, combined with ab-initio
calculations, have also found evidence for  density-dependent interactions
 effects  in the tunneling current  \cite{Brar10}  which arise from the sharp
 spectral features in the quasiparticle decay rate below the Dirac point,
 as discussed above.


\begin{figure}
\centering
\includegraphics[width=0.86\linewidth]{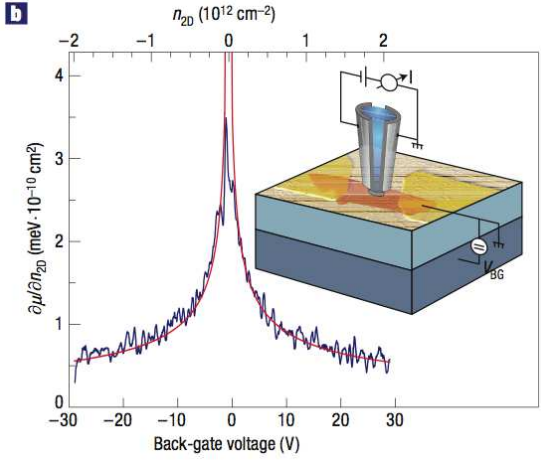}
\caption{(Color online) Inverse compressibility, measured by \onlinecite{Mar08}. 
The red line is  the compressibility of non-interacting Dirac fermions.}
\label{figcomp}
\end{figure}

\subsection{Physical Observables}
\label{sec:PR}
The interaction-driven singular logarithmic structure near the Dirac point (for $\mu \approx 0$) 
encountered in the fermion self-energy,  and in  particular the renormalization
 of the Fermi velocity, can manifest itself in numerous physical observables,
 such as the  charge compressibility and the  spin susceptibility,  which
 exhibit  non Fermi-liquid behavior. Interactions can also affect the conductivity 
near the Dirac point, leading to deviations from the celebrated
 quantized value $\sigma_0=e^2/4\hbar$ expected for free Dirac fermions \cite{CastroNeto:09}.

\subsubsection{Charge and spin response}

\paragraph{Compressibility.}
First we discuss the compressibility $\kappa$, which  was recently measured \cite{Mar08},
Fig.~\ref{figcomp}, 
 and it was concluded that no interaction effects were clearly visible in those samples.
 Theory predicts significant ($\alpha$  dependent) deviations from the free electron
 behavior \cite{Bar07,Pol08b,Hwang07,She07}.

The computation of the compressibility requires knowledge of the ground state
 energy, which contains the first order Hartree-Fock exchange contribution $E_{ex}$,
 and the correlation energy $E_{Corr}$, describing all the higher order effects.
 Keeping in mind applications of the theory for fairly strong coupling ($\alpha \sim 1$),
 the contribution of $E_{Corr}$ can be substantial. The correlation energy can
 be readily  calculated within the RPA approximation, i.e. we take $E_{Corr}=E_{RPA}$.
The total ground state energy $E$, per unit area,  is  the sum
$E = E_{kin} + E_{ex} + E_{RPA}$.
 The  kinetic energy  $E_{kin} = (2/3) vk_{F} n$, and  $n=(k_F)^2/\pi$ is the particle
 density. 
The inverse compressibility is then calculated as $1/\kappa  = \partial^2E/\partial n^2$,
 which is equivalent to the usual definition involving the variation of the chemical
 potential with density, $(1/\kappa)  = \partial \mu/\partial n$.
 For free Dirac particles this gives $(1/\kappa_0) =v\sqrt{\pi/(4n)}$ --- 
 behavior which can be clearly seen in experiment Fig.~\ref{figcomp}.

The interaction effects in the ground state energy acquire divergent contributions
 in the limit of small density $k_F/\Lambda \approx 0$, similarly to the previously discussed
 self-energy (velocity) renormalization. Ignoring any finite (non-diverging) terms, one finds \cite{Bar07}
\begin{equation}
 E_{ex}/n = \frac{\alpha}{6}  (vk_{F}) \ln(\Lambda/k_F), \  (k_F/\Lambda) \rightarrow 0,
\label{gsex}
\end{equation}
\begin{equation}
 E_{RPA}/n = -\frac{N \alpha^{2}}{6}  G(\alpha) (vk_{F}) \ln(\Lambda/k_F),
\label{gsrpa}
\end{equation}
where the function $G(\alpha)$ is defined as 
$G(\alpha) = (1/2) \int_{0}^{\infty}dx(1+x^{2})^{-2}(\sqrt{x^2 +1} + N\pi\alpha/8)^{-1}$,
 and, in particular, at zero coupling $G(0) = 1/3$.
The above results exactly follow the velocity renormalization, i.e. are
 equivalent to the substitution $v \rightarrow v(k_F)$ in the free
 compressibility  $(1/\kappa_0) =v\sqrt{\pi/(4n)}$, where  $v(k_F)$ is the running
velocity  calculated within RPA at the infrared scale $k_F$.
The result is particularly simple at the Hartree-Fock (exchange) level
(when the velocity follows Eq.~\eqref{hf3}):
\begin{equation}
\frac{1}{\kappa} = v\sqrt{\frac{\pi}{4n}}
 \left ( 1 + \frac{\alpha}{4}  \ln(\Lambda/k_F) + O(\alpha^2)\right ) \ ,
\end{equation} 
and was obtained by a number of authors \cite{Bar07,Hwang07,She07}.

The above results are valid at zero temperature.
We also point out that exactly at zero density $k_F=0$,  but $T\neq0$,
the compressibility behaves as: $\kappa^{-1} \sim (v^2/T)
(1 + (\alpha/4)  \ln(T_0/T))^2$, where $T_0$ is the temperature related to the ultraviolet
 cutoff; since $\Lambda v \approx 7 {\mbox{ eV}}$, then 
$T_0 \approx 8 \times 10^4 {\mbox{K}}$. This is easily understood since in
 the infrared limit near the ``critical point" $n=T=0$ it's the larger scale,
 either $vk_F$, or the temperature $T$, which enters the physical
 observables \cite{She07}.

Of course Eqs.~\eqref{gsex},\eqref{gsrpa} are valid only asymptotically ($k_F \rightarrow 0$), 
 and at any finite density the compressibility should be calculated numerically.
 This was achieved by expressing the ground state energy via the charge response
function \cite{Bar07}.

\begin{figure}
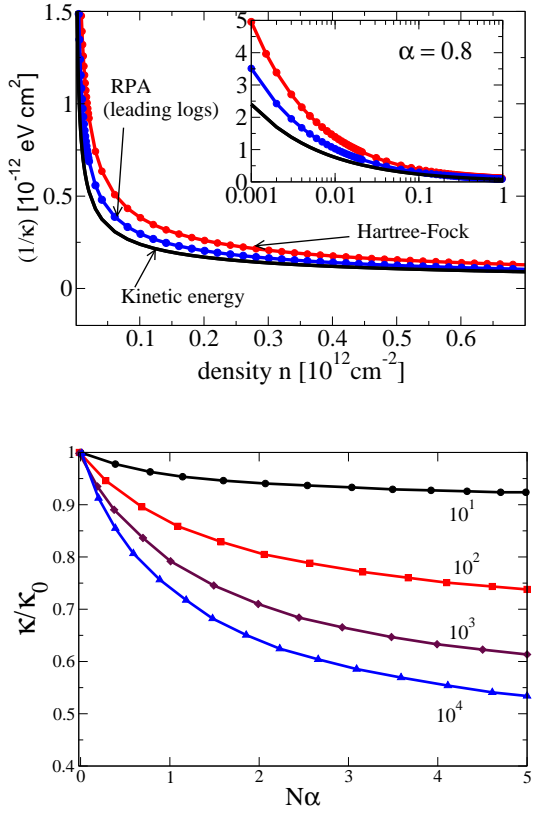

\centering
\includegraphics[width=0.8\linewidth]{Figs/Fig-15-1}
\vspace{0.75cm}

\includegraphics[width=0.8\linewidth]{Figs/Fig-15-2}
\caption{Upper panel: Inverse compressibility calculated at different levels
 of approximation as a function of density. The inset enlarges the low-density region. Lower panel
 (adapted from \onlinecite{Bar07}):  
 Compressibility calculated within RPA,
 relative to the free level for different couplings and densities. Here $N=4$ is
 the Dirac fermion degeneracy. The numbers refer to the values of $\Lambda/k_F$,
 which can be converted into density $n$ via: $\Lambda/k_F \approx 220/\sqrt{\tilde{n}}$,
$\tilde{n}=n/(10^{10} {\mbox{cm}}^{-2})$. This implies $(\Lambda/k_F) \sim 10^2$
 for $n \sim  10^{11} {\mbox{cm}}^{-2}$, and $(\Lambda/k_F) \sim 10$ for
$n \sim  10^{13} {\mbox{cm}}^{-2}$.}
\label{theorycomp}
\end{figure}

Fig.~\ref{theorycomp}, upper panel, illustrates the variation
 of $1/\kappa$ with density for fixed interaction. Most notably, $1/\kappa$ is larger
 than the free value $1/\kappa_0$. Also, the  full RPA implementation weakens the first order
 Hartree-Fock (exchange) result, due to the different signs in Eqs.~\eqref{gsex},\eqref{gsrpa}.
 For example, at $\alpha=0.8$ the RPA term is approximately $1/2$ of the exchange,
 and thus has to be taken into account 
(although the RPA effects become weaker for $\alpha \rightarrow 0$).
 Asymptotically, 
 $(\kappa^{-1}/\kappa_0^{-1}) \sim \ln(\Lambda/k_F)$, as $k_F/\Lambda \rightarrow 0$.
The lower panel gives the variation $\kappa/\kappa_0$ as a function of the interaction
 for different densities; naturally the deviation from the free limit increases with
 increasing interaction and decreasing density.

 The increase of the inverse compressibility,   $\kappa_0/\kappa$, as a function of the interaction
$\alpha$ (at fixed
density),
 and with decreasing density  (for fixed interaction), 
 represents  non-Fermi liquid behavior,   and reflects
 the lack of screening.
 By contrast, in a 3D (and 2D) Fermi liquid with a screened potential  $\kappa_0/\kappa$  decreases;
 for example within Hartree-Fock, $\kappa_0/\kappa \approx 1 - r_s/6 <1$,
 and eventually goes through zero, signaling an instability \cite{Mahan}
(although the critical value of $r_s$  depends strongly on the level of approximation.)
  Such an instability does not occur in graphene, which is 
 related to the impossibility of Wigner crystallization \cite{Dahal06}.
It should be noted that for larger densities (larger than the density range  shown
 in Fig.~\ref{theorycomp}) the logarithmic corrections become unimportant
 and the system recovers the Fermi liquid behavior,
 i.e. eventually $\kappa/\kappa_0$ becomes larger than 1.

Fits of the experimental data  for $\kappa$ with adjusted (slightly larger) velocity
 $v=1.1 \times  10^6 {\mbox{m/s}}$
show that $\alpha \approx 0$ (Fig.~\ref{figcomp}), while the use of  
$v=10^6 {\mbox{m/s}}$  by \onlinecite{She07} at the Hartree-Fock level
 produced $\alpha \approx 0.4$. On the other hand, the application
 of the full RPA analysis led us to conclude that $\alpha < 0.1$.
It has also been argued that exchange and correlation effects vanish and 
 do not manifest themselves at all in the compressibility \cite{Abergel09}.
 These discrepancies indicate that the issue is still unsettled,
 while it's also possible (indeed, quite probable)  that interaction effects are obscured by charge
 inhomogeneities (electron-hole puddles) in these samples.
Nevertheless theory predicts strong systemic (albeit logarithmic) deviations from
 Fermi-liquid theory, and it would be important to  test these predictions  in cleaner,
 more uniform,  high-mobility, low-density
 samples.

\paragraph{Spin susceptibility.}
 The paramagnetic spin susceptibility, $\chi_s$,  
shows behavior very similar to the charge compressibility, i.e. $ (\chi_{s}/\chi_{s,0})$
decreases  as the interaction increases \cite{Bar07}. This is again related to the fact that $\chi_s^{-1}$
 is calculated via the ground state energy, and is proportional to the Fermi velocity $v$.
It was also pointed out that the same effect, i.e. the logarithmic growth of the  exchange energy, 
Eq.~\eqref{gsex}, 
 can lead to suppression of  ferromagnetism 
 in graphene at low densities  \cite{Peres05}. The full calculation of $\chi_s$ within  RPA
 was carried out by \onlinecite{Bar07}.

On the other hand the orbital diamagnetic susceptibility, $\chi_{dia}$, is proportional
 to $v^2$, because the quasiparticle current that couples to the vector
 potential  contains $v$ (the magnetic field is perpendicular to  the graphene plane).
  Therefore interaction corrections lead to an increase of  $\chi_{dia}$  \cite{She07} and, consequently,
 orbital effects are expected to dominate in the susceptibility.
At the Dirac point, $k_F=0$,  one finds at finite temperature
\begin{equation}
\chi_{dia}/\chi_{dia,0} = \left ( 1 + \frac{\alpha}{4}  \ln(T_0/T) \right)^{2} \ ,
\end{equation}
where  the non-interacting $\chi_{dia,0}=-e^2v^2/(6\pi c^2 T)$ \cite{Sudip07}.
Here $c$ is the speed of light. 
At $T=0$, $n \neq 0$,  we have  $\chi_{dia,0} \sim - e^2v/(c^2 \sqrt{n})$,
 and interaction corrections readily follow from the $v$ dependence. 
This result is, strictly speaking, valid for $T \ll B \ll \mu=v\sqrt{\pi n}$, whereas for $B=0$
 the orbital susceptibility is zero for $\mu \neq 0$ as $T \rightarrow 0$, and is finite only when
the Fermi energy is at the
Dirac point.  It has been  suggested that an interaction driven positive (paramagnetic) 
 contribution to the orbital susceptibility can therefore become  dominant
 in doped graphene, $\chi_{orb} \sim  [e^2v^2/(\mu c^2)] \alpha |\ln{\alpha}|, \ \alpha \ll 1$
\cite{Principi10}.

\paragraph{Specific heat.}
The specific heat is logarithmically suppressed due to the suppression
 of the DOS $\sim v^{-2}$. Consequently $C_V \sim C_{V,0}/(\ln(T_0/T))^2, \ T/T_0 \ll 1$,
 where $C_{V,0} \sim T^2/v^2$ is the free Dirac fermion specific heat.
 The full RPA calculation, valid also for large coupling, was carried out by \onlinecite{Vaf07}.

\paragraph{Graphene as a quantum critical system.}
A unified view of the above behavior is presented in \citet{She07}, where it was stressed
that the logarithmic corrections are manifestations of scaling behavior
 around the quantum critical point at $n=0, T=0$. As discussed previously,
 at finite chemical potential, $T=0, n\neq0$, graphene behaves as a  Fermi liquid,
 whereas at $T\neq0$, a quantum critical region fans out of the point $n=0, T=0$.
In  the critical region it is natural to call graphene a Dirac liquid,
 where the proximity to the Dirac point is important for physical phenomena at finite $T$.
This puts graphene's behavior into the general framework of quantum critical 
 phenomena \cite{Sachdev:99}. In practical terms, it implies that the logarithmically
 divergent velocity contributions are cut-off by the largest scale: temperature $T$,
 $k_F\sim\sqrt{n}$, or magnetic field. Computing physical quantities
 in perturbation theory (Hartree-Fock or RPA) naturally involves these infrared scales.
 The separation between the Dirac liquid and
 the Fermi liquid regimes in the $n-T$ plane
 is defined by the crossover temperature $T^{*}(n) = vk_F(1+(\alpha/4) \ln(\Lambda/k_F))$,
$k_F=\sqrt{\pi n}$, and thus the temperature dependencies quoted previously,
 are valid for $T_0>T>T^{*}(n)$. The ultraviolet temperature scale
 $T_0 \approx 8 \times 10^4 {\mbox{K}}$, while for typical graphene densities
$n \lesssim  10^{12} {\mbox{cm}}^{-2}$,  $T^{*}(n) \sim  10^2 {\mbox{K}}$.

\subsubsection{Conductivity}
The  behavior of the electrical conductivity  in graphene has been extensively reviewed
\cite{DasSarmaRMP, NunoRMP}.
 It is believed that charged impurities
 and resonant scatterers  are the main sources of scattering away from the Dirac point,
 and to extent the long- or short-range part of the Coulomb potential contributes to 
 scattering is a matter of ongoing debate \cite{Ni:2010, Monteverde,
Ponomarenko:2009,Chen:2008, Bruno}.

 Here we will only mention effects related to long-range electron-electron
 interactions near the Dirac point. 
Interaction corrections to the minimum metallic conductivity
 of free Dirac fermions, $\sigma_0=e^2/(4\hbar)= \frac{\pi}{2}e^2/h$ \cite{Fradkin86,Lee93}, 
are more involved, because this expression does not
contain the quasiparticle velocity, while the electric charge is not renormalized.
The debate was fueled in part by  electrical
measurements of the minimum conductivity (at the Dirac point)
 which turned out to be somewhat larger than $\sigma_0$ \cite{Geim07,Tan:2007}.
Theoretically,  at $T=0$ (or $T \ll \omega$ where $\omega$ is the external frequency),
 it is expected that any interaction effect should have sub-leading
 character, and the frequency can enter only through the running of
 the coupling $\alpha(\omega)$.
Even though some debate still exists \cite{Mi08,HJV08,She09,Juricic11} as to the
implementation
 of the cut-off regularization procedure, the conductivity should have the form
\begin{equation}
\sigma(\omega)/\sigma_0 = 1 + \frac{\tilde{C} \alpha}{1+ \frac{\alpha}{4} \ln{(\Lambda  v/\omega)}} \ ,
\end{equation}
where the constant $\tilde{C}\approx 0.01$, as argued by \onlinecite{Mi08,She09}.
The smallness of $\tilde{C}$ reflects the near cancellation of self-energy and vertex corrections, 
 and thus the effect of interactions is  small. 
 This value is also consistent with optical measurements on suspended samples \cite{Nair08}, 
 as well as graphene on a substrate \cite{Li08a}, which find $\sigma(\omega)$
 to be very close to $\sigma_0$, and frequency independent in a wide range of energies.

In the strict DC limit $\omega=0$, the presence of disorder, in combination with interactions,
 can alter the conductivity. For example, for weak gauge field disorder ($\Delta$) where an attractive
line of fixed points exists (Fig.~\ref{figdisorder}) with $\alpha^{*} = \frac{4}{\pi}\Delta$,
calculations show that the conductivity (on the fixed line) increases relatively to the free limit
\cite{HJV08}: 
 $\sigma = [\pi/2 + (4-\pi)\Delta]e^2/h$.
 For stronger scalar and vector disorder/interactions where the couplings run away to infinity
 the problem is  non-perturbative, and a complex variety of behavior is
expected \cite{FA08}. 
 
For clean graphene at $\mu=vk_F=0$ it was pointed out \cite{Fri08,Mu08,Kashuba08} that at high temperature (compared to the frequency),  the conductivity is expected to have  the form:
\begin{equation}
\sigma =\frac{0.76}{\alpha^2} \frac{e^2}{h},  \  \   T \alpha^2 \gg \omega,
\end{equation}
where $\alpha(T) = 4/\ln(\Lambda v/T)$
 is the running Coulomb coupling.
 This form reflects electron-electron  inelastic collisions 
 with scattering rate $1/\tau_{ee} \sim \alpha^{2} T$. The linear temperature dependence
 is characteristic for Dirac particles. The above formula is valid as long as $1/\tau_{ee}$ is
 the dominant scattering mechanism (collision-dominated transport), and implies that
 clean graphene at the neutrality point should exhibit a universal, interaction-limited
 conductivity, reflecting essentially the quantum critical behavior of graphene
 in this regime ($T \gg \mu$). 
 With increased doping ($\mu/T$), a crossover  takes place to a Fermi liquid
 regime with screened interactions, where $\tau_{ee}^{-1} \sim \alpha^{2} T^2/\mu$, 
\cite{Mu08} and the conductivity is dominated by charged impurity scattering.

It has also been pointed out that for $\mu=0$ graphene behaves as an almost ``perfect"
 fluid, in a sense that its shear viscosity, $\eta$, relative to the entropy density $s$
 is anomalously small: $\eta/s = (0.13/\alpha^{2}(T)) (\hbar/k_B)$  \cite{Viscosity09}.
This ratio measures how strongly the excitations in a fluid interact.
 At room  temperature  $\eta/s$ of graphene
 is smaller than $\eta/s$ of any known 
correlated quantum fluid, and is close to the lower bound
 of $\frac{1}{4\pi} \frac{\hbar}{k_B}$  proposed to exist for a large class of strongly interacting quantum
 field theories \cite{Kovtun05}. Therefore, due to its quantum critical nature near the Dirac point,
 graphene  is suggested to behave as a strongly correlated quantum liquid and should exhibit 
 signatures of electronic turbulence \cite{Viscosity09}.

\subsection{Overview of main results}
 Before we proceed with further topics related to interactions in graphene,
 let us broadly  summarize the main findings  and questions raised so far:
\begin{enumerate}
\item{For clean graphene at the neutrality point $\mu=0$,
 interactions  are not screened  and are marginally irrelevant; the fixed point $\alpha^{*}=0$
is approached logarithmically (or, equivalently, the quasiparticle velocity increases logarithmically).
 From a theory standpoint, the approach towards this
 fixed point is well understood both from weak and strong-coupling (RPA) perspectives.
 Since in graphene one can have $\alpha \sim 1$ under rather conventional
experimental conditions, our understanding of RPA calculations
 is important. RPA is justified only in the limit of large number of fermion species ($N \gg 1$),
 while for $N=4$ it should work for weak to moderate coupling; however there are indications,
 coming mostly from two-loop calculations, that vertex corrections are numerically small,
and thus RPA should work well. 
 Disorder generally drives the system away from the clean fixed point,  towards finite
 or even strong coupling, depending on disorder type.
}
\item{The resulting  behavior near the Dirac point is that of a non-Fermi-liquid with
a quasiparticle decay rate which is linear in energy, and decreasing  quasiparticle residue.
 All physical characteristics related to the quasiparticle velocity (which increases logarithmically)
are affected, and predicted to exhibit systemic, interaction dependent, deviations from their non-interacting
 values as the Dirac point is approached, either as a function of density or temperature.
}
\item{Can graphene be driven into an excitonic insulating state? At the Dirac point
 the long-range Coulomb interactions can lead to bound electron-hole pairs,
 creating a gap. There has been intense debate whether this can happen under
 realistic conditions --- since the critical interaction strength  appears to be $\alpha_c \sim 1$,
 it seems possible to occur in suspended samples ($\alpha=2.2$). So far no experimental
 indications have been observed. }
\item{What is the value of the interaction $\alpha$? Clearly, since $\alpha=2.2/\epsilon_0$ is
 dielectric constant  dependent, working with  different substrates could
 lead to changes in interaction-dependent effects  \cite{Jang08}.
 There are also suggestions that graphene has an ``intrinsic" value of $\alpha$
\cite{Bruno}, arising from dynamical  dielectric screening. 
 The polarizability of the Dirac
fermions was found to be  amplified by excitonic effects, improving screening of interactions between
quasiparticles. This analysis leads to values of $\alpha$ ranging from $\alpha \approx 1/7$ in the static
 limit  to  $\alpha \approx 2$ at high frequencies. 
Very recent measurements  of the cyclotron mass in suspended graphene \cite{Elias11} have found  logarithmic
 velocity renormalization and extract, within the RPA scheme,  an effective
 value of graphene's dielectric constant 
$\epsilon_{G} \approx 3.5$.
One can also expect that near the Dirac point,  where interactions
 lead to singular effects, additional factors  can be important such as disorder, inhomogeneities,
rippling, etc., and thus obscure the  clean behavior.
}
\item{In the Fermi-liquid regime, where  interactions are screened,  the physics
 near the Dirac point can still be strongly affected --- this is due to  resonant
 features in the quasiparticle self-energy, reflecting interactions of quasiparticles
 with plasmons.}
\end{enumerate}


%




\section{The Coulomb Problem and Charged Impurities}
\label{sec:Coulomb}

%
%

The consideration of non-interacting Dirac electrons in 2D under a
Coulomb field is of paramount relevance for graphene, and for several
reasons. First of all, the Coulomb problem for relativistic fermions
has many features that are unfamiliar in condensed matter systems,
and which resemble long standing predictions made in the context of
QED in strong fields. As such, and given that having $\alpha \sim 1$
makes graphene intrinsically strongly coupled, it can provide the
first
experimental ground for testing many elusive predictions from
strong-coupling QED.

On the other hand, the single particle Coulomb problem constitutes
the first step in addressing nontrivial features of the full,
many-body interacting problem. Characteristics like non-linear
screening, or the supercritical instabilities, provide valuable
insight in grasping some proposed many-body effects, like exciton
condensation, or spontaneous mass generation in graphene.

Historically, however, the motivation for studying the Coulomb
problem comes from the seminal experimental observations
\cite{Novoselov:2004} that the field effect
in graphene prepared on SiO$_2$ is characterized by carrier mobilities
that do not depend on the Fermi energy or carrier density (the DC
conductivity, $\sigma = m e |n|$, with $m\simeq \text{const.}$),
and that carriers are chiral Dirac fermions in 2D
\cite{Zhang:2005,Novoselov:2005}. Early semiclassical investigations
\cite{Ando:2006,NM06,Nomura:2007,Adam:2007} showed that such
linear-in-density conductivity could be explained by
scattering of unscreened Coulomb impurities, which are
typically seen in silica in concentrations of
$\sim 10^{10} \text{cm}^{-2}$ \cite{Ando:1982}. As a result,
transport in the presence of charged impurities rapidly became one of
the most studied topics in the quest for the ultimate mobility in
graphene. Since, as we saw before, Coulomb's law is exactly preserved
in undoped graphene, and approximately preserved for small and
moderate doping, the scattering processes are essentially governed by
the bare Coulomb problem, unlike conventional metals, where screening
is perfect. A thorough understanding of this problem is
therefore important not only for its theoretical relevance and its
import on electron-electron interactions, but also for its
experimental implications, and our understanding of transport in
graphene. 

Finally, it is highly significant that this is an exactly solvable
problem. This means that most quantities can be obtained exactly,
allowing us to unveil many interacting and non-interacting effects
that are not within reach of the perturbative approaches already
discussed. We proceed to show several such features. On account of
the long range nature of the Coulomb field, inter-valley processes
are not relevant, and hence we will solve the problem within each
(independent) valley in the Dirac description of fermions in
graphene.

%
%
\subsection{Exact Solution of the Coulomb Problem}
\label{subsec:ExactSolutionOfTheCoulombProblem}
%

%
\subsubsection{Wave Equations and Spectrum}
\label{subsubsec:WaveEquationsAndSpectrum}

A Coulomb center of charge $Z|e|$ generates the potential
$U(r)=Ze^2/(\epsilon_0 r)$ for the electrons. Without any loss of generality let us
consider $Z>0$. The electronic dynamics is governed by the wave
equation
\begin{equation}
  \vF \biggl( 
    -i\bsigma\cdot\bnabla - \frac{g}{r} + \sigma_3 Mv
  \biggr)
  \Psi(\br) = E \Psi(\br)
  \label{eq:WaveEquation}
  .
\end{equation}
Here we use $g=Z\alpha = Ze^2/(\epsilon_0\vF)$, with $\epsilon_0$
reflecting the effective dielectric constant of the embedding medium,
and the mass $M$ accounts for the more general possibility of a
symmetry breaking gap. Throughout this chapter we shall use the scaled
energy and mass $\e=E/v$, $m=M v$, and $k = \sqrt{\e^2
- m^2}$.
Even though $m=0$ for ideal graphene without interactions, nonzero
$m$ can be induced in many ways.
One of them is through interaction with suitable substrates, of which
some experimental hints have been reported 
\cite{Zhou:2007,Grneis:2008,Martinazzo:2009,Li08b}. In terms of the
original tight-binding Hamiltonian, the mass $M$ arising from a
sublattice symmetry is related to the parameter $\Delta_0$ introduced
in eq.~\eqref{eq:H-Mass} via $Mv^2=\Delta_0$.
The axial symmetry of the potential allows us to use the
eigenstates of the total pseudo angular momentum, $J_z = L_z +
\sigma_z/2$, which is conserved \cite{DiVincenzo:1984}. We write 
$\Psi_j^\dagger = r^{-1/2}
[F_j(r)\Phi_{j-1/2}(\phi),\,iG_j(r)\Phi_{j+1/2}(\phi)]$, where
$j=\pm1/2,\pm3/2,\dots$ are the eigenvalues of $J_z$, and
the cylindrical harmonics read $\Phi_p(\phi) =
e^{ip\phi}/\sqrt{2\pi}$. 
A detailed derivation of the 2D Dirac equation for general radial
potentials is given by \onlinecite{Novikov:2007}. In our case,
\Eqref{eq:WaveEquation} reduces to the following radial equations
\cite{Khalilov:1998,Novikov:2007}
\begin{subequations}\label{eq:RadialEqs}
\begin{align}
  \bigl[ m - \e -g/r \bigr] &F_j(r)   +
  [ \partial_r + j/r ] G_j(r) =0
  \\
  \bigl[ \partial_r - j/r \bigr] &F_j(r)   +
  \bigl[ m + \e +  g/r \bigr] G_j(r) =0
  .
\end{align}
\end{subequations}
This coupled pair of first order equations can be straightforwardly
reduced to two decoupled second order equations. Free solutions
($g=0$) of \eqref{eq:WaveEquation} exist when $|\e|>|m|$, and are
simple spherical waves whose $k$-normalized version reads
\begin{equation}
  \Psi_j = \sqrt{\frac{k}{2|\e|}}
    \begin{bmatrix}
      \sqrt{|\e+m|} \, J_{j-1/2}(kr)\, \Phi_{j-1/2} \\
      i \sign{\e} \sqrt{|\e-m|} \, J_{j+1/2}(kr)\, \Phi_{j+1/2}
    \end{bmatrix}
  \label{eq:FreeSolution}
\end{equation}
($s_x\equiv\text{sgn}(x)$). For nonzero $g$, one readily sees from
\eqref{eq:RadialEqs} that the solutions at $r\sim0$ behave as
\begin{equation}
  F(r), G(r) \sim r^{\pm\gamma}, \qquad \gamma = \sqrt{j^2 - g^2}
  .
  \label{eq:ShortDist}
\end{equation}
The general exact solution is given in terms of
confluent hypergeometric, or Whittaker's functions, both in the
massive \cite{Khalilov:1998,Novikov:2007,Gupta:2008,Pereira:2008,
Gamayun:2009, Gupta:2010}, and massless cases
\cite{Pereira:2007,Shytov:2007-1,Gupta:2009}. In the massless case,
one can map \eqref{eq:RadialEqs} into the familiar Coulomb
radial Schrodinger equation in 3D \cite{Pereira:2007}:
\begin{equation}
  \partial_r^2 f_\pm +
  \left[
    \e^2 + 2g\e/r - \gamma(\gamma\mp1)/r^2
  \right]
  f_\pm(r) = 0
  ,
  \label{eq:3DCoulomb}
\end{equation}
where the $f_\pm$ are linear combinations of $F$ and $G$, $\e^2$ takes
the place of the Schrodinger energy, and $\gamma$ plays the 
role of angular momentum. Since the solution is formally the same,
the appearance of $\e^2$ instead of $\e$ means that the massless case
admits no bound solutions, as we expect on account of the absence of
a spectral (mass) gap. The massive case, however, has a well defined
infinite spectrum of bound solutions when $|\e|<|m|$, given by
\cite{Khalilov:1998}
\begin{equation}
   \e_{n,j} = \sign{g} m 
    \frac{n+\sqrt{j^2-g^2}}{\sqrt{g^2 +
    \bigl[n+\sqrt{j^2-g^2}\bigr]^2}}
    \label{eq:FineStructure}
    ,
\end{equation}
%
lowest level is given by $\eG \equiv \e_{0,1/2} =
\sign{g}m\sqrt{1-(2g)^2}$.

%
\subsubsection{Supercritical Instabilities}
\label{subsubsec:Supercritical Instabilities}

Consideration of eq.~\eqref{eq:ShortDist} immediately reveals a
complication if $g>g_c=1/2$, because $\gamma$ becomes
imaginary for the lowest angular momentum channels ($j=\pm1/2$). The
solution \eqref{eq:ShortDist} is neither regular nor divergent, but
rather oscillates endlessly towards $r=0$. This
is pathological because the space of solutions is of dimension 2, and
we can no longer discard an irregular contribution since both
linearly independent solutions are square integrable. In other words,
there is no boundary condition at the origin to univocally select the
solution. Secondly, in the massive case the level $\eG$ becomes
imaginary, signaling a loss of self-adjointness of the Dirac
Hamiltonian for $g>1/2$.

Physically, both effects are a symptom that the potential has such a 
strong divergence that particles are inexorably attracted and ``fall''
into the origin, leading to a collapse of the
system (for example, the endless oscillations can be read as an
infinite phase shift). This ``fall to the center'' is a general
characteristic of diverging potentials in any dimension of space. For
power law potentials, one particular power signals the threshold of
criticality. The Coulomb potential is the marginal case for the Dirac
equation (both in 2D and 3D), just like the potential $1/r^2$ is the
marginal case of the 3D Schrodinger equation \cite{Landau-QM:1981}.
This, of course, begs the question of regularization.
Regularizing the potential introduces an additional boundary condition
at some short distance $R$, which allows a formal solution, and cures
the total collapse of the system \cite{Case:1960,Perelomov:1970}. 
In graphene the lattice is the natural regulator and there are no
ultraviolet issues. But the physics in the supercritical regime
depends explicitly on the short range details. 

This supercritical collapse has a long history in the context of QED,
where the Dirac equation stands as the basis for understanding the
stability of matter. In QED the collapse would occur for
$Z\alpha_{QED} > 1$, which lead to extensive investigations regarding
the stability of heavy nuclei having $Z>Z_c=137$
\cite{Case:1960,Popov:1971,Popov:1971-1,Zeldovich:1972,Greiner:1985}.
After regularization
$Z_c\to 170$, which makes the problem highly academic,
and QED's predictions untestable. In graphene, on the contrary,
$Z_c \sim 1$, which opens the real possibility of testing the
supercritical instability in a condensed matter setting.

\begin{figure}[tb]
  \centering
  \includegraphics[width=0.4\textwidth]{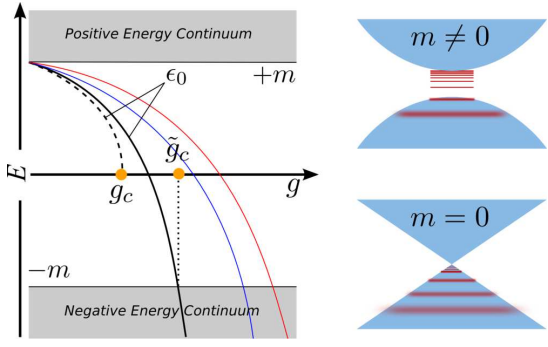}
  \caption{(Color online)
    Schematic drawing of the level diving process in the supercritical
    regime, and of the resulting quasi-spectrum of levels for massive
    and massless fermions.
  }
  \label{fig:Diving}
\end{figure}

\paragraph{Massive Electrons.}
\label{par:Massive Electrons}

To understand the physics in the supercritical regime we can follow
the level $\eG$ as the coupling increases (\Fref{fig:Diving})
\cite{Zeldovich:1972,Greiner:1985,Pereira:2008}.
For the pure Coulomb case, $\eG(g)$ decreases towards zero in a
singular way at $g=g_c$. In a regularized potential, $\eG$
depends also on the cutoff radius $R$, and is allowed to monotonically
penetrate the negative energy region, until eventually touching the
lower continuum at $\e=-m$. If $g$ is further increased,
$\eG$ dives into the hole (positron) continuum and becomes a
resonance. Other levels will sequentially follow at higher $g$. The
diving point for $\eG(g)$ defines a renormalized critical
coupling, $\gt_c>g_c$ that is characterized by a log singularity at
$m R\sim0$: $\gt_c \simeq g_c + \pi^2/\log^2(m R)$
\cite{Khalilov:1998,Pereira:2008,Zhu:2009,Gamayun:2009}, strongly
depending on the regularization. 

This diving of bound levels entails a complete restructuring of
the vacuum. If the level was empty, an electron-hole pair will be
immediately created: the electron remains tightly bound and shielding
the center, while the hole is ejected to infinity
\cite{Zeldovich:1972,Greiner:1985}. The supercritical regime is thus
characterized by spontaneous pair creation, or a spontaneous Schwinger
mechanism \cite{Schwinger:1951}. One expected consequence is a strong
signature of these resonances in the hole sector of the
scattering and transport cross sections.

An essential detail is that these resonances are not
usual bound levels diluted inside a continuum, where their lifetime
essentially disappears. One consequence of the chiral nature of
Dirac fermions, combined with the long range tail of the Coulomb
potential, is that the supercritical levels in the
relativistic Coulomb remain sharply defined, with diverging
lifetime. For example, for $S$ states ($j=1/2$), one shows that 
these resonances follow \cite{Gamayun:2009}
\begin{equation}
  \e_n \approx -m 
    \Bigl( 1 + \xi + i \frac{3\pi}{8} e^{-\pi/\sqrt{2\xi}} \Bigr)
  ,
  \quad \xi = \frac{3\pi(\beta-\beta_c)}{8\beta\beta_c}
  \label{eq:ResonantSpectrum-Massive}
  ,
\end{equation}
when $g\gtrsim\gt_c$, and where $\beta=i\gamma$,
$\beta_c=\sqrt{\gt_c^2-1/4}$. In real space the localization of the
supercritical levels is controlled by the reduced Compton
wavelength: $\lambda_C=1/(mv)$. The modulus squared of their
wavefunction decays as 
$\Psi^\dagger\Psi\propto\exp(-\sqrt{8gr/\lambda_C})$ and,
consequently, even inside the continuum, such levels retain a highly
localized nature, which is why they are so relevant, in particular in
their potential for screening \cite{Pereira:2008}.

\paragraph{Massless Electrons.}
\label{par:Massless Electrons}

The spectrum in this case is continuous everywhere, and thus
there is no sequential diving and restructuring of
the hole continuum as described above. But the pathology associated
with \Eqref{eq:ShortDist} still exists. Physically, the massless
situation is rather more catastrophic since the solution in a
regularized potential reveals an infinite number of quasi-localized
resonances in the hole sector
\cite{Pereira:2007,Shytov:2007-1,Gamayun:2009}. This is a highly
non-trivial effect for several reasons: 
(i) in the massless case there is no natural length scale in the
problem to characterize such localized states; 
(ii) the system abruptly develops an infinite quasi-bound spectrum at
$g>g_c$, when its spectral fingerprint is rather featureless for
$g<g_c$;
(iii) the infinite spectrum has the potential to over-screen the
Coulomb center.
In addition, unlike the massive case, here the critical coupling
remains unchanged at $g_c=1/2$, and no qualitative features
(like how many, if any, states have dived) depend on the
magnitude of the regularization distance. The spectrum of
supercritical resonances behaves as
\cite{Shytov:2007-1,Gamayun:2009,Gupta:2009}
\begin{equation}
  \e_n \approx - \frac{a + i b}{R} 
    e^{-\pi n / \sqrt{g^2-g_c^2}}
  ,
  \quad (a,b) \sim \mathcal{O}(g)
  \label{eq:ResonantSpectrum-Massless}
  ,
\end{equation}
which has an essential singularity at $g_c$, an energy scale/lower
bound set explicitly by the regularization distance, $R$, and
diverging lifetimes close to the critical point. Since the width of
these states vanishes linearly, they are practically bound states
(hence the designation quasi-bound states). In real space, the
localization scale is determined by the regularization distance $R$
itself.

Since meso and nanoscopic devices are of high interest, it is
pertinent pointing out that \emph{massless} Dirac fermions in a
finite-sized system mimic in all aspects the physics of
\emph{massive} electrons, as a result of the linearly vanishing DOS
and the effective gap coming from finite-size quantization
\cite{Pereira:2008}.

\begin{figure}[tb]
  \centering
  \includegraphics[width=\columnwidth]{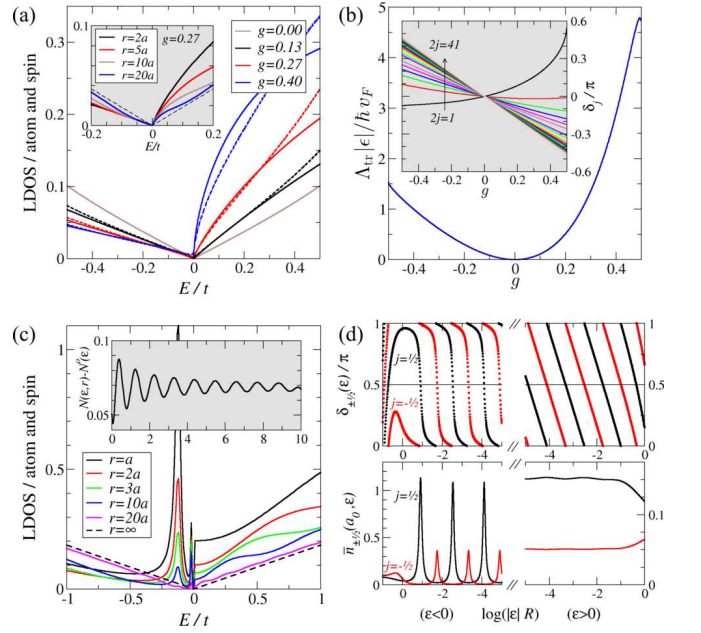}
  \caption{(Color online)
  (a)
  LDOS, $N(\e,r)$ at $r=a$ for several couplings $g<g_c$. 
  The inset shows $N(\e,r)$ for $g=0.27$ and different $r$.
  For comparison, the exact LDOS calculated in the 
  full tight-binding lattice for the same parameters is shown as
  dashed lines. In the horizontal axis the energy is in units of the
  hopping  $t$.
  (b)
  The weak coupling transport cross section as a function of $g$.
  The inset shows the phase-shifts for different $j$.
  (c)
  LDOS, $N(\e,r)$ at several distances $r$, for $g=1>g_c$.
  The inset shows the oscillating LDOS correction for $\e>0$.
  (d)
  Energy dependence of the phase shifts (top) and the
  supercritical contribution $\bar{n}_j(\e,r)$ to the LDOS (bottom)
  for $g=1.0$.
  }
  \label{fig:UnderSuper}
\end{figure}

%
\subsubsection{DOS, Scattering and Transport Cross Sections}
\label{subsubsec:DOSScatteringTransport}

Here and in the coming sections we shall be concerned mostly with
massless Dirac fermions, except when explicitly stated otherwise.
The local density of states (LDOS) and cross sections are useful quantities insofar as they are
directly accessible in local probe and transport experiments. 
The LDOS per unit area and spin is isotropic, and can be written in
closed form in terms of partial waves as $N(\e,r) = \sum_j
n_j(\e,r)$, \cite{Pereira:2007} with
\begin{equation}
  n_j(\e,r)\!=\!\frac{j^2}{2\pi^2\gamma^2r}
  \biggl[
    F_{\gamma-1}^2 + F_{\gamma}^2 + \frac{2g\sign{\e}}{|j|}
    F_{\gamma}F_{\gamma-1}
  \biggr]
  \label{eq:LDOS-Weak}
\end{equation}
for $g<g_c$, and $F_l$ represents the Coulomb function
$F_l(-g\sign{\e},|\e|r)$ \cite{Abramowitz:1964}. The function
$N(\e,r)$ is plotted in \Fref{fig:UnderSuper}(a) for different
couplings and distances. Apart from the
evident particle-hole asymmetry, the LDOS remains rather featureless,
even at the shortest distances.
If $g>g_c$ the corresponding analytical expression obtained in the
regularized potential is more complex, but still has a closed form
\cite{Pereira:2007}. In this case, supercritical channels ($|j|<1/2$)
need to be isolated from undercritical ones ($|j|>1/2$), yielding two
contributions to the LDOS: 
\begin{equation}
  N(\e,r) = \sum_{|j|<|g|} \bar{n}_j(\e,r) + \sum_{|j|>|g|} n_j(\e,r) 
  .
  \label{eq:LDOS-def}
\end{equation}
The total LDOS for this case is shown in \Fref{fig:UnderSuper}(c) for
$g=1.0$, and at different distances to the impurity. It is now clear
that strong resonances, decaying rapidly with distance, appear in
the  vicinity of the Dirac point, signaling the presence of the
quasi-bound levels \cite{Pereira:2007,Shytov:2007-1}.
Their exponential accumulation at $\e=0$ is confirmed in
\Fref{fig:UnderSuper}(d) where we show the supercritical contribution
$\bar{n}_j(\e,r)$ as a function of $\log(|\e|)$. At positive energies
the LDOS exhibits periodically decaying oscillations in $\e r$ [inset
of \Fref{fig:UnderSuper}(c)], with extrema separated by 
$\approx n\pi$, within logarithmic accuracy
\cite{Shytov:2007}. When directly measured
in STM such oscillations can be used to extract the electronic
dispersion, as done by \onlinecite{Ouyang:2002}. 

We point out that, since the solution of the supercritical problem
involves a nontrivial \emph{ad-hoc} regularization, these results have
been checked numerically against exact solution of the full
tight-binding problem in the honeycomb lattice, being found that the
analytical Dirac results reproduce the full lattice problem down to
distances as small as the lattice scale \cite{Pereira:2007}.

The striking differences between the two regimes and the violent
modification of the ground state at strong coupling are likewise
evident in the behavior of the scattering phase-shits, $\delta_j(\e)$.
They admit closed formed expressions at both $g<g_c$
\cite{Novikov:2007,Pereira:2007,Shytov:2007}
and $g>g_c$ \cite{Shytov:2007-1,CastroNeto:2009}. For example, the
undercritical $S$-matrix reads \cite{Novikov:2007}
\begin{equation}
 S_j(\e) = e^{2i\delta_j(\e)} =
  \frac{je^{i\pi(j-\gamma)}}{\gamma-ig\sign{\e}}
  \frac{\Gamma(1+\gamma-ig\sign{\e})}{\Gamma(1+\gamma+ig\sign{\e})}
  \label{eq:Smatrix-Weak}
  ,
\end{equation}
which is energy independent, but considerably asymmetric with respect
to the sign of $g$. The corresponding $\delta_j$ are shown in
\Fref{fig:UnderSuper}(b)(inset) as a function of coupling strength.
Note how $\delta_{1/2}$ (the most important partial wave) behaves
rather differently from the others: only $\delta_{1/2}$ shows the
expected sign for the attractive/repulsive situations. On the other
hand, in the supercritical regime there is a strong $\e$-dependence of
$\delta_j$. In the top row of \Fref{fig:UnderSuper}(d) we present
$(\delta_j\!\!\mod\pi)$ as a function of $\log(\e)$. In the attractive
sector ($\e<0$ if $g>0$) the abrupt steps centered around $\pi/2$ mark
the position of the infinite quasi-bound spectrum (which,
as per \eqref{eq:ResonantSpectrum-Massless}, accumulates exponentially
at $\e=0$), whereas in the attractive sector $\delta_j(\e)$ is
smooth.

Knowledge of the phase-shifts allows direct calculation of the full
transport cross-sections for our 2D Dirac fermions:
\begin{equation}
  \Lambda_\text{tr}(\e) = \frac{2}{\e} 
    \sum_j \sin^2\bigl( \delta_{j+1/2}(\e) - \delta_{j-1/2}(\e) \bigr)
  \label{eq:TrCrossSection}
\end{equation}
\cite{Novikov:2007,Katsnelson:2006}. The profile of
$\Lambda_\text{tr}\times\e$ at weak coupling is shown in
\Fref{fig:UnderSuper}(b). When scattering is due only to unscreened
charges, the marked asymmetry between $g>0$ and $g<0$ can be used to
extract the density of positively and negatively charged impurities
($n_i^\pm$) from a single measurement of the electrical conductivity,
$\sigma$, as a function of carrier density \cite{Novikov:2007-1}. This
technique has been used in some experiments
\cite{Chen:2008,Chen:2009-1,Chen:2009-2}, but the asymmetry effect can
be easily masked by other spurious influences
\cite{Huard:2008,BarrazaLopez:2010,Nouchi:2010}. 
Moreover, on account of the $\e$-independence of $\delta_j$ in
\eqref{eq:Smatrix-Weak}, the corresponding Drude conductivity, 
$
\sigma = 4\pi e^2 \mu / (\vF n_i \Lambda_\text{tr} h^2 )
$,
is immediately seen to scale linearly with density: $\sigma\propto
\mu^2\propto n$. Therefore, the linear-in-density conductivity,
which appears already in the first Born approximation, remains when
the cross section is calculated exactly.

For supercritical potentials, and similarly to the LDOS, there will be
undercritical and supercritical partial waves contributing to
$\Lambda_\text{tr}(\e)$ [cfr. eq.~\eqref{eq:LDOS-def}]. The latter
give rise to strong peaks in the transport cross-section at
densities for which the Fermi energy matches the levels
$\e_n$ \cite{Shytov:2007-1}, tallying with the behavior of the
DOS.

%
%
\subsection{Induced Charge and Screening}
\label{subsec:InducedChargeScreening}

First attempts at understanding screening in graphene date back to
\onlinecite{DiVincenzo:1984}, where it was recognized that
conventional procedures of the theory of metals, like self-consistent
screening, linear response or Friedel sum rules, are not
straightforward in this system. For example, within the
Dirac (effective mass) approximation, the ultraviolet cutoff scale
enters explicitly in Friedel's sum rule, and
Levinson's theorem is modified \cite{Lin:2006}
(Levinson's theorem is one of the fundamental results in quantum
scattering theory, asserting that in the 
Schr\"odinger's equation with a non-singular spherically symmetric
potential the zero energy scattering phase-shift exactly counts the
number of bound states: $\delta_l(0) = N_l \pi$).
One consequence is
that a na\"ive application of Friedel's sum rule can yield divergent
displaced charges \cite{DiVincenzo:1984}. Even though these
divergences are artificial in the target lattice problem, they
point, already at a single particle level, to the anomalous
screening properties of graphene.

\begin{figure}[tb]
  \centering
  \includegraphics[width=0.75\columnwidth]{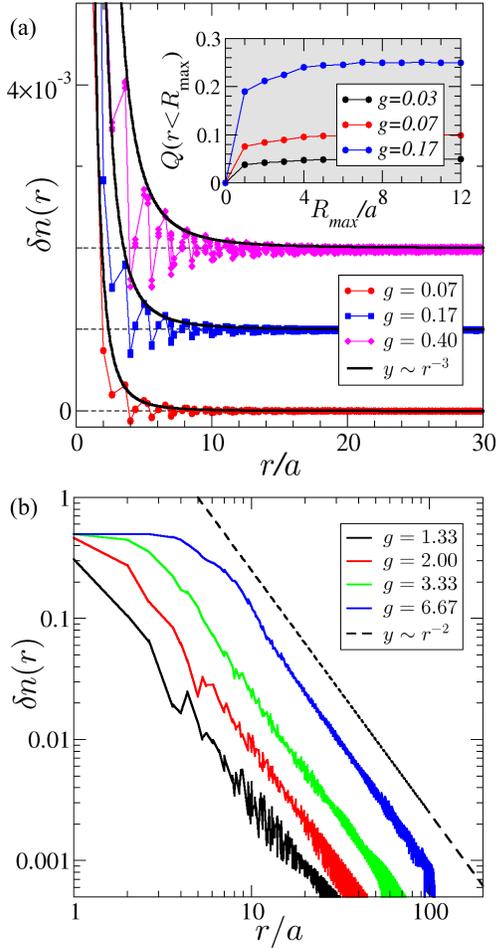}
  \caption{(Color online)
  (a) 
  Induced electron density, $\delta n(\br)$, plotted as a function
  of distance to the Coulomb center, for different
  impurity strengths, $g<g_c$. 
  Data obtained from full diagonalization of 
  the tight-binding Hamiltonian in a lattice with $124^2$ atoms.
  Black lines are $\propto 1/r^3$, and guides for the eye.
  Inset shows the saturation of the integrated charge accumulated
  inside $r<R_\text{max}$, as a function of $R_\text{max}$.
  (b) 
  Same as (a), but for the supercritical case, $g>g_c$, and the dashed
  line is now $\propto 1/r^2$.
  }
  \label{fig:IndCharge-1}
\end{figure}

\begin{figure}[tb]
  \centering
 
\includegraphics*[width=\columnwidth]{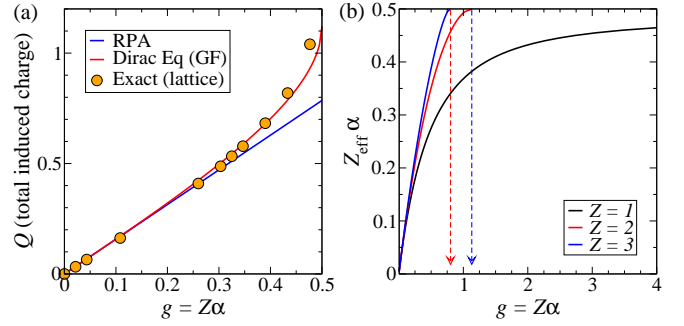}
  \caption{(Color online)
  (a)
  Total integrated charge in the vicinity of the impurity, $Q$, 
  obtained: from exact diagonalization in the lattice (dots), from RPA
  \eqref{eq:InducedCharge-RPA} (blue), and from the exact Green's
  function in the Coulomb field \eqref{eq:InducedCharge-Valeri} (red).
  (b)
  The self-consistent $Z_\text{eff}$, obtained
  from \Eqref{eq:Zeff-Valeri} \cite{Terekhov:2008}. Numerical
  data (dots) is plotted after accounting for finite-size 
  renormalization of $g_c$ \cite{Pereira:2008}.
  }
  \label{fig:IndCharge-2}
\end{figure}

%
\subsubsection{Weak Coupling ($g<g_c$)}
\label{subsubsec:WeakCoupling}

\paragraph{Non-interacting Induced Charge.}
\label{par:Non-interactingInducedCharge}

Knowledge of the exact LDOS within the Dirac approximation
(Sec.~\ref{subsubsec:DOSScatteringTransport}) allows the
straightforward calculation of the perturbation to the electronic
density induced by the Coulomb center. The induced density is defined
as $\delta n(\br) = n(\br) - n^0(\br)$, and is related to the LDOS via
(for undoped graphene at zero temperature)
$n(\br)=\sum_j n_j(r) = \sum_j \int^0_{-D} n_j(\e,r)d\e$, where $D$
is the cutoff scale for the linearly dispersing band. The induced
charge density is just $\delta\rho(\br)=-|e|\delta n(\br)$.
Closed form expressions for $n_j(r)$ are provided in
\eqref{eq:LDOS-Weak}. One difficulty with this approach is that
the resulting density per partial wave behaves asymptotically as
\begin{equation}
  \delta n_j(r\to\infty) \!\sim\! \frac{1}{r}
    \left[
      D \!-\! \frac{g}{r} \!-\! D^0 \!+\!
      \mathcal{O}(r^{-2})
    \right]
  ,
  \label{eq:InducedCharge-Weak}
\end{equation}
which diverges upon summation over $j$ (a reminiscence of the
problems associated with the ultraviolet scale alluded to above).
In the above expression $D$ and $D^0$ represent the cutoff in the
presence and in the absence of the coulomb center, respectively.
Since the subleading terms in \eqref{eq:InducedCharge-Weak} are
convergent in $j$, we regularize it by taking a position dependent
cutoff: $D\to D^0+\frac{g}{r}$. As a result, the total induced density
acquires the form $\delta n(r) \sim H(D^0 r)/r^3$, where $H(x)$ is
a constant-amplitude oscillating function \cite{Pereira:2007}. 
Since it is desirable to have control over the validity of the
regularization procedure outlined above, we have calculated the
total induced density $\delta n(\br)$ in the full
tight-binding problem, via exact diagonalization. The result is
plotted in \Fref{fig:IndCharge-1}(a), and unequivocally shows the
predicted $1/r^3$ decay, with oscillations on the scale of the
lattice.
Such fast decay implies that the induced charge concentrates within
a small vicinity of the impurity. Moreover, the numerical results in
the lattice further suggest that such distance is of the order of the
lattice parameter $a$: the inset in \Fref{fig:IndCharge-1}(a)
reveals that the total charge pulled inside a region $r<R_\text{max}$
saturates within very few lattice spacings.
In fact, since $D^0\propto 1/a$, in the limit $a\to 0$ (where the
effective mass description is meaningful) the analytical expression 
$\delta n(r) \sim H(D^0 r)/r^3$ can be seen as a representation of the
2D Dirac-delta function. In other words, we expect the
induced charge density to behave as 
\begin{equation}
  \delta \rho(\br) = -|e|\,\delta n(\br)
    \stackrel{a\to0}{\longrightarrow} -Q|e|\,\delta(\br)
  \label{eq:InducedCharge-Weak-Delta}
  .
\end{equation}
The same conclusion follows from a modified Friedel argument
\cite{Shytov:2007}, and from the exact calculation of the
non-interacting Green's function in the Coulomb field (see below)
\cite{Terekhov:2008}. The induced charge has a screening sign, as
expected, but the strongly localized distribution of the induced
charge \eqref{eq:InducedCharge-Weak-Delta} 
implies that undoped graphene cannot screen in the usual sense,
because it merely renormalizes the strength of the impurity: $Z\to
Z_\text{eff} = Z - Q$. This leaves Coulomb's law unaltered,
except for the substitution $Z\to Z_\text{eff}$.

\paragraph{Linear (RPA) Screening.}
\label{par:LineaRPAScreening}

Single particle results, like the one above, are not
generally sufficient to draw conclusions about screening. 
Consider now the same problem in linear response, at the RPA
level, which is justified for small, undercritical couplings. Within
the RPA, the Fourier transform of the statically screened potential is
given by $U_s(q)=U_0(q)/[1-\Pi^{(1)}(q)V(q)]$ \cite{Fetter:1971},
where $V(q)=2\pi e^2/(\epsilon_0 q)$ is the electron-electron
interaction, and $U_0(q)=ZV(q)$ the external impurity potential. From
\eqref{eq:pol} we know that
$\Pi^{(1)}(q\to 0)\approx - q / (4 v)$, and hence 
\begin{equation}
  U_s(q) \approx U_0(q)\Bigl(1+\frac{\pi}{2}\alpha\Bigr)^{-1}
        = \frac{ U_0(q) }{ \epsilon_\text{RPA} }
  .
  \label{eq:ScreenedPotential-RPA}
\end{equation}
Therefore linear response confirms the absence of screening, except
for the trivial renormalization of the static dielectric constant: 
$\epsilon_0\to\epsilon_\text{RPA} = \epsilon_0(1 + \pi\alpha/2)$
\cite{Ando:2006}. Likewise, the induced density can be computed in
linear response from $\delta n(\bq)=-ZV(\bq)\Pi(\bq)$ or, in the
RPA:
\begin{equation}
  \delta n(\br) = -Z\int d\bq
    \frac{\Pi^{(1)}(\bq)V(\bq)}{1-\Pi^{(1)}(\bq)V(\bq)}
    e^{i\bq.\br}
  \label{eq:InducedCharge-RPA}
  ,
\end{equation}
yielding 
$\delta \rho(r\gg a) \sim -\delta(\br) Z|e|\pi\alpha/2$ to linear
order in $\alpha$ \cite{Kolezhuk:2006}. This is exactly what was
obtained in \eqref{eq:InducedCharge-Weak-Delta} from a single
particle, wavefunction, perspective. In addition, the argument that
the Fourier transform of
$\delta n(\br)$ is dimensionless can be used to show that
it should be a pure constant in undoped graphene, for
which there is no natural length scale. As a result, that $\delta
\rho(\br)\propto\delta(\br)$ remains true in all orders of
perturbation theory \cite{Biswas:2007}. 
For consistency, the total induced charge,
$Q$, introduced in \eqref{eq:InducedCharge-Weak-Delta} is then given
by
\begin{equation}
  Q = \frac{\pi}{2} Z\alpha + (\text{higher orders in } Z\alpha)
  \label{eq:Q}
  .
\end{equation}
To verify this correspondence we can compare \eqref{eq:Q} with the
value of $Q$ extracted from the non-interacting exact diagonalization
in the honeycomb lattice. As shown in \Fref{fig:IndCharge-2}(a), the
numerical $Q$ for different values of $Z$ follows the relation
\eqref{eq:Q} for most of the range $0<g<g_c$, thereby confirming the
correspondence, and showing how weakly undoped graphene
screens \cite{Pereira:2007,Shytov:2007}.
Given that only the global dielectric constant is affected, one can
say that undoped graphene screens like an insulator.

At finite densities, however, the system screens like a conventional
metal. This derives at once from the fact that, at finite
Fermi momentum, $\Pi^{(1)}(q\approx0)\approx -2k_F/(\pi v)$, no
longer vanishing, and leading to the screened potential
\begin{equation}
  U_s(q) = \frac{U_0(q)}{\epsilon_\text{RPA}(q)}
  ,\qquad
  \epsilon_\text{RPA}(q) = 1 + \frac{q_s}{q}
  ,
  \label{eq:V-RPA}
\end{equation}
$q_s=4\alpha k_F$ playing here the role of inverse screening length
\cite{Ando:2006,NM06}. Contributions from interband transitions
can be simply incorporated by renormalizing the background dielectric
constant by the factor $(1+\pi\alpha/2)$, as in
eq.~\eqref{eq:ScreenedPotential-RPA}.
Using \eqref{eq:InducedCharge-RPA}, the
total integrated charge is now seen to be $\int \delta \rho(\br) d\br
= -Z|e|$. This means that, unlike the undoped situation, at finite
electron densities the system completely screens the Coulomb center,
just as expected in a metallic system \cite{CastroNeto:2009}.
 
For transport considerations it is important to underline that, even
though at finite densities charged impurities have a finite range
determined by $q_s$, the Boltzmann conductivity remains linear in
density. This happens because the screened potential \eqref{eq:V-RPA}
entering in the relaxation time calculation, maintains the same
dependence with $k_F$. From this perspective, the mobility remains
constant in density for both screened and unscreened charges,
differing only by an overall constant related to
$\epsilon_\text{RPA}(k_F)$
\cite{NM06}.

\paragraph{Nonlinear Screening.}
\label{par:NonlinearScreening}

As \Fref{fig:IndCharge-2}(a) documents, even as linear response is
acceptable at small values of $g=Z\alpha$, the approximation becomes
increasingly unwarranted as $g$ nears the critical threshold,
$g_c=1/2$, which is non-perturbative.
Rather than analyze this limit
on the basis of exact wavefunctions in the Coulomb field, as was
done in Sec.~\ref{par:Non-interactingInducedCharge}, we now describe
the solution obtained by \onlinecite{Terekhov:2008}. These authors
bypass the solution of the Dirac equation, obtaining instead an exact
integral expression for the Green's function in a Coulomb field,
using a proper-time approach common in QED \cite{Milshtein:1982}.
The main result is that
\begin{equation}
  \delta \rho(\br) = -Q\,\delta(\br) + \delta \rho_\text{dist}
  \label{eq:InducedCharge-Valeri}
\end{equation}
where $\delta\rho_\text{dist}(\br)$ represents a positive charge
distributed at $r=\infty$ (needed to satisfy the constraint of total
zero induced charge). It is significant that this approach affords an
exact expression for the dependence of $Q$ upon $g=Z\alpha$, which is
shown in panel (a) of \Fref{fig:IndCharge-2}. A series expansion
of this dependence yields the following:
\begin{equation}
  Q(g)\approx \frac{\pi}{2}\, g + 0.783\, g^3 + 1.398\, g^5+\cdots
  \label{eq:Terekhov}
,
\end{equation}
with each term corresponding to successive orders in perturbation
theory. The linear term is the one that appeared already in
\eqref{eq:Q}, at the RPA level. The next term in the expansion
was also calculated perturbatively by \onlinecite{Biswas:2007}.
Interestingly, even though this problem is analogous to conventional
QED vacuum polarization of a point charge, the perturbative
coefficients in $Q(g)$ are not small, and increase with order, in
stark opposition with the behavior known in 3D QED \cite{Brown:1975}.
This offers another perspective upon the uniqueness of
electron-electron interactions in graphene, for, even though the
problem is on the surface analogous to the QED situation, the
physics can be qualitatively different. In this particular case, the
difference seems to arise from the 2D dimensionality
of the problem and the absence of Lorentz invariance in graphene,
which renders the Coulomb interactions instantaneous.

Inspection of the curve $Q(g)$ in \Fref{fig:IndCharge-2}(a)
reveals that it reaches 1 at $g=0.49$, slightly before $g_c$.
This implies that, for a monovalent impurity ($Z=1$), the
non-interacting result predicts complete shielding before
$g_c$, insofar as $Z_\text{eff}(Z,\alpha) = Z - Q(g) \to 0$. Such
strong renormalization of the potential source immediately begs the
consideration of interaction and correlation effects. They can be
incorporated at the Hartree level by solving the self-consistent
equation
\begin{equation}
  Z_\text{eff} \alpha = Z \alpha - \alpha Q(Z_\text{eff}\alpha)
  \label{eq:Zeff-Valeri}
  ,
\end{equation}
which encodes an infinite summation of a selected set
of bubble diagrams \cite{Terekhov:2008}.
Since $Q(g)$ is obtained exactly, one obtains the renormalized
effective potential strength, $Z_\text{eff} \alpha$, with an accuracy
much beyond the RPA. In addition, the reduction of $Z_\text{eff}$ with
respect to the bare $Z$ means that $g_c$ is also self-consistently
renormalized to $\tilde{g}_c = Z_\text{eff} \alpha$. The effect is
shown in \Fref{fig:IndCharge-2}(b), which reveals that,
as $\tilde{g}_c>g_c$, self-consistent screening delays the
supercritical threshold because the condition $Z_\text{eff}
\alpha=0.5$ requires a higher bare $Z$.
This phenomenon is most striking for $Z=1$, in which
case the supercritical point disappears altogether ($\tilde{g}_c<1/2$
even as $Z\to\infty$), whereas $g_c^{Z=2}=1.136$ and 
$g_c^{Z=3}=0.798$. The prediction of this self-consistent Hartree
renormalization of $Z_\text{eff}$ would then be that impurities with
$Z=1$ can never become supercritical.
In addition, Hartree screening is sufficient
to suppress the tendency for over-shielding of the Coulomb
center: as seen in the inset of \Fref{fig:IndCharge-2}(b),
$Z_\text{eff}$ remains always positive.

An alternative approach to the Hartree screening consists in treating
the induced charge in linear response, 
$\delta\rho(\bq)=ZV(\bq)\Pi(\bq)$, but taking into account
electron-electron interactions perturbatively, via the renormalization
of the coupling constant \cite{Biswas:2007}. This is valid for small
$\alpha$ (weak interaction), and leads to a result formally equivalent
to \eqref{eq:InducedCharge-Valeri}, but where $\delta\rho_\text{dist}$
now arises from the electronic correlations. The distributed charge in
the interacting case also has an anti-screening sign, but decays as
$1/r^2$, while the non-interacting $\delta\rho_\text{dist}(\br)$ is
zero everywhere, except at infinity.

Even though the above considerations pertain to undoped graphene,
since all screening charge accumulates completely within a narrow
distance, finite densities are not expected to alter the picture for
as long as $q_s=4\alpha k_F$ remains large compared to the lattice
scale $a$.

%
\subsubsection{Strong Coupling  ($g>g_c$)}
\label{subsubsec:StrongCoupling}

In Sec.~\ref{par:NonlinearScreening} Hartree screening was shown to
renormalize $g_c$ and delay the critical threshold. Two
important questions naturally arise: (i) since the self-consistent
solution of \eqref{eq:Zeff-Valeri} is uncontrolled, how certain can
one be that the critical regime is reachable at all?  (ii) So far we
looked only at screening from the undercritical side (i.e. as long as
$Z_\text{eff}\alpha<1/2$). How can one address screening from the
supercritical side, given that this regime cannot be reached
perturbatively?

The answer to these questions is far from trivial. In QED it is
related to the ground state and stability of super-heavy nuclei
($Z\gtrsim170$), when the bound spectrum dives into the positron
continuum (\Fref{fig:Diving}). Despite having received considerable
attention throughout the 1970-80's \cite{Greiner:1985}, the fact
that these systems require such high $Z$'s, has turned it
largely into an academic problem. The exciting prospect about graphene
is that impurities with $Z=1,\,2$ might already display supercritical
physics, in which case it would afford a bench-top test of some yet
untested QED predictions.

The essence of the difficulties in treating the supercritical regime
clearly lies in its non-perturbative nature. Graphene, being gapless,
is even more pathological because of the infinite quasi-spectrum that
appears in the hole channel [\Fref{fig:Diving}]. This quasi-spectrum
is akin to an atom filled with infinitely many electrons and, as
known from studies of heavy atoms \cite{Landau-QM:1981}, it requires
full consideration of correlations and interactions, and
self-consistent techniques like Thomas-Fermi
\cite{Thomas:1927,Fermi:1927}.

\paragraph{Non-interacting Induced Charge.}
\label{par:NonInteractingInducedCharge}

In Sec.~\ref{subsubsec:DOSScatteringTransport} we saw some
unusual consequences for the DOS and cross-sections extracted from the
exact solution of the Dirac equation for $g>g_c$. Now we address
the corresponding induced charge obtained using the same procedure as
in Sec.~\ref{par:Non-interactingInducedCharge}. 
Consideration of the exact wavefunctions \cite{Pereira:2007} or the
exact phase-shifts \cite{Shytov:2007} leads to the conclusion that the
supercritical partial waves contribute with an induced charge
$\propto 1/r^2$. This could be expected on dimensional grounds: 
 $\delta(\br)$ and $1/r^2$ are the only dimensionally consistent
possibilities in the absence of any intrinsic length scale in massless
graphene. The exact induced density per partial wave reads
\cite{Shytov:2007}
\begin{equation}
  \delta \bar{n}_j(\br) = 
    \frac{2\sign{g}}{\pi^2r^2}\sqrt{g^2-j^2}
  ,
  \label{eq:InducedCharge-Shytov}
\end{equation}
and, like the undercritical contributions, has a screening sign. The
full induced charge is obtained from 
$\delta\rho(\br)=-|e|\delta n(\br),\, 
n(\br)=\sum_{|j|<g_c}\delta \bar{n}_j + \sum_{|j|>g_c}\delta n_j
$, and has the general form 
\begin{equation}
  \delta n(\br) = 
    \sign{g}A \frac{1}{r^2} + B\sign{g}\delta(\br)
  .
  \label{eq:InducedCharge-Supercritical}
\end{equation}
If $1/2<g<3/2$ eq.~\eqref{eq:InducedCharge-Supercritical} reduces to 
$\delta n(\br) = (\pi g/2)\delta(\br)
+ 2\sign{g}\sqrt{g^2-g_c^2}/(\pi^2r^2)$.
The general behavior \eqref{eq:InducedCharge-Supercritical} is also
confirmed numerically by exact diagonalization of the tight-binding
Hamiltonian in the honeycomb lattice, whose results are plotted in
\Fref{fig:IndCharge-1}(b).

\paragraph{Supercritical Protection.}
\label{par:UndercriticalProtection}

Unlike the undercritical regime, the additional power law decay in
\eqref{eq:InducedCharge-Supercritical} causes a modification of
Coulomb's law at large distances. But since we have a quasi-atom with
all levels \eqref{eq:ResonantSpectrum-Massless} filled, the
non-interacting result in \Eqref{eq:InducedCharge-Supercritical}
cannot be the final answer. Each level is quasi-localized on the
lattice scale, and should contribute significantly to shield the
Coulomb center. For $g$ not too much above $g_c$ we can follow an
argument advanced by \cite{Shytov:2007} that assumes electrons at
some distance $r$ feel the effect of a point charge consisting of the
impurity subtracted from all the accumulated screening charge up to
$r$. In other words, we introduce a distance dependent impurity
strength, $Z_\text{eff}(r) = Z - \int_R^r \delta n(\br) d\br$, and
substitute \eqref{eq:InducedCharge-Supercritical} for $\delta n(\br)$:
\begin{equation}
  Z_\text{eff}(r) = Z - \frac{\pi}{2}g -
\frac{4\sqrt{g^2-g_c^2}}{\pi} 
  \log\frac{r}{R}
\end{equation}
Since the log term represents the renormalization coming from
screening at distances away from the center, we should replace 
$(g\equiv Z\alpha)\to (Z_\text{eff}\alpha\equiv g_\text{eff})$. This
leads to a self-consistent renormalization of the coupling that can
be written in an appealing RG fashion as 
$dg_\text{eff}/d\log(r)=-4\alpha\sqrt{g^2_\text{eff}-g^2_c}$. In
this way, it can be immediately seen that the coupling $g_\text{eff}$
will ``flow'' to the constant value $g_c$ within a finite distance 
[see also \cite{Gupta:2009} for a related renormalization
procedure]. As such, irrespective of the bare $Z$, the system
self-consistently rearranges itself so that electrons at large
distances never feel a supercritical effective coupling. The
undercritical (stable) situation is therefore protected. This
reasoning agrees with expectations for the corresponding problem in
QED, where it was shown that, within Thomas-Fermi, the vacuum
polarization charge in super-heavy nuclei behaves in such a way as to
reduce $Z$ to the threshold value \cite{Muller:1975}. 

This is quite different from a metal, to the extent that graphene
always leaves an universal amount of charge ($Z_c=g_c/\alpha$)
unscreened at large distances. Such behavior derives from the sharp
transition between the under and supercritical regimes. On the one
hand, the system wishes to screen as much charge as it possibly can.
But, on the other, it cannot screen if $g<g_c$, therein lying the
compromise that makes screening stop when $Z$ reaches $Z_c$.

\paragraph{Nonlinear Thomas-Fermi and Beyond.}
\label{par:NonlinearTFBeyond}

While the above approach is valid in principle only for $g\gtrsim
g_c$, the fact that qualitatively supercritical graphene resembles
a super-heavy atom suggests the use of TF theory, which is
exact for atoms with $Z\to\infty$ \cite{Lieb:1981}, and affords an
approximation from the opposite limit $g\gg g_c$. If we wish to
calculate how Coulomb's law is modified in this regime we can
calculate the total potential 
$V_\text{eff}(\br)=V(\br)+\delta V(\br)$, where 
$\delta V(\br)= \frac{e^2}{\epsilon_0} \int \frac{\delta
n(\br')}{|\br-\br'|}d\br' $ 
is the potential induced by the screening charge. Within TF we
replace $\delta n(\br')=n[\mu-V(\br)] - n(\mu)$, and the homogeneous
density depends on $\mu$ via $n=\sign{E}\mu^2/(\pi v^2)$.
Solution of the resulting integral equation leads to the correction
to Coulomb's law, which asymptotically reads \cite{Katsnelson:2006}
\begin{subequations}\label{eq:TF-Screening}%
\begin{align}
  V_\text{eff}(\br) &\approx \frac{e^2}{\epsilon_0 r}
    \left[ \frac{Z}{1+2Z\alpha^2\log(r/R)} \right]
    ,\label{eq:TF-Screening-1} \\
  V_\text{eff}(\br) &\approx \frac{e^2}{\epsilon_0 r (q_s r)^2}
    \left[ \frac{Z}{1-2Z\alpha^2\log(q_s R)} \right]
    \label{eq:TF-Screening-2}
  ,
\end{align}
\end{subequations}
valid for $\mu=0,\,r\gg R$ and $\mu\ne 0,\,rq_s\gg 1$ respectively,
where $q_s= 4\alpha \mu/v$ is the screening length
\eqref{eq:V-RPA}. One notes that the overall space dependence is
formally the same as the one obtained within RPA, both at zero and
finite density. Hence the bracketed coefficients in
\eqref{eq:TF-Screening} can be interpreted as a renormalization of the
valence. The important difference is that, in the limit $Z\to\infty$
of interest in the context of TF, the nominal valence $Z$ disappears
from $V_\text{eff}(\br)$, which thus becomes universal (and
undercritical). Hence, even for strong impurities one can formally use
perturbative expressions for the screened potential, corrected for
this renormalization of $Z$.

It is important to emphasize that, since at this stage we are
concerned with screening and corrections to the induced charge coming
from electron-electron interactions, $g=Z\alpha$ is no longer
the relevant parameter alone, but both $Z$ and $\alpha$ (that
controls the interaction) independently. For this
reason, \onlinecite{Fogler:2007} have argued that the result 
\eqref{eq:TF-Screening} is valid only for small $\alpha$. More
precisely, it applies for $1/Z\ll \alpha \ll 1/\sqrt{Z}$, and
provided that $\log(r/R)<1/\alpha$. Otherwise, for intermediate
electron-electron coupling ($\alpha\sim 1$), the asymptotic screened
potential should follow $V_\text{eff}\approx Z_c e^2 /(\epsilon_0 r)$,
with $Z_c=g_c/\alpha=1/(2\alpha)$. This result embodies the
undercritical protection discussed above in
Sec.~\ref{par:UndercriticalProtection}, insofar as the supercritical
core is always self-consistently screened so that $Z_\text{eff}\to
Z_c$. Moreover, within the supercritical core region, $r<2Z\alpha^2
R$, the effective potential decays as $\propto 1/r^{3/2}$. This
obtains treating graphene as an ideal classical metal, under the
assumption of quasi-complete screening in the core region
\cite{Fogler:2007}.

%
\subsubsection{Finite Mass}
\label{subsubsec:Finite Mass}

We now briefly address the differences expected in the screening
properties of charged impurities in massive graphene. We shall
consider only the undoped situation, and assume $\mu = -m$, such that
none of the bound levels \eqref{eq:FineStructure} are occupied.

\paragraph{Weak Coupling ($g<g_c$).}
\label{par:WeakCoupling}

It is clear that at weak coupling one can directly rely on
perturbative results (Sec.~\ref{subsubsec:WeakCoupling}), and obtain
the induced density from $\delta n(\bq) = - ZV_0(\bq)\Pi(\bq)$.
$\Pi^{(1)}(\bq)$ has been calculated in
\eqref{eq:pol}, and simple substitution yields
the following asymptotics:
\begin{equation}
  \delta n(\br) \sim Z\alpha
    \begin{cases}
      \frac{\pi}{2} \delta(\br)  & r\simeq a\to0 \\
      - \lambda_C^{-2} \log\frac{\lambda_C}{r} 
          & a \ll r \ll \lambda_C \\ 
      -\lambda_C r^{-3} & r \gg \lambda_C
    \end{cases}
  ,
  \label{eq:InducedCharge-Massive-Weak}
\end{equation}
where $\lambda_C=1/(mv)$ is the Compton wavelength, and $a$ the
lattice parameter of graphene.
The short distance term is the same as found in the massless case
(\ref{eq:InducedCharge-Weak-Delta},\ref{eq:Q}), which makes sense
given that when $r\ll\lambda_C$ the system does not  ``feel'' the mass
yet. It has a screening sign. However as the distance increases
screening is increasingly suppressed, first weakly up to $\lambda_C$,
and then strongly, beyond $\lambda_C$. In fact, since here
$\delta n(\bq=0)=0$, we have exactly $\int \delta n(\br) d\br = 0$.
The meaning of this is simple: the total induced charge is zero. The
system cannot screen beyond $r\gtrsim \lambda_C$ because it is
essentially an insulator (or a semiconductor with $\mu$ in the middle
of the gap). Notwithstanding, unlike a conventional insulator, gapped
graphene shows a novel screening behavior at short distances,
reflected in the live dependence of $\delta n(\br)$ on the distance up
to $\lambda_C$.

\paragraph{Strong Coupling ($g>g_c$).}
\label{par:StrongCoupling}

In gapped graphene, screening in the supercritical regime is 
qualitatively easier to understand, at least when
$g\gtrsim\tilde{g}_c$. If the first level has just merged inside to
hole continuum, its effective probability density, $|\Psi_c(\br)|^2$,
remains exponentially localized, as described in \ref{par:Massive
Electrons}. Invoking completeness of the set of single-particle
states, one can easily show that the non-interacting induced charge
follows \cite{Pereira:2008}
\begin{equation}
  \delta n(\br) \approx |\Psi_c(\br)|^2 + \delta n_\text{pol}(\br)
  ,
  \label{eq:InducedCharge-Massive-Super}
\end{equation}
where 
$\delta n_\text{pol}(\br)\approx \sum_{E<-m} 
|\chi_E(\br)|^2 - |\chi_E^0(\br)|$
represents the vacuum polarization
(i.e.: the induced charge coming from the full set of plane wave
states), and is the same quantity that obtains in RPA
\eqref{eq:InducedCharge-Massive-Weak}. Clearly, the contribution from
the supercritical state alone makes $\delta n(\br)$ in
\eqref{eq:InducedCharge-Massive-Super} highly localized within the
Compton wavelength, $\lambda_C$. For all purposes, this state screens
like a bound state would, and consequently one expects the impurity
valence to be reduced by one unity times the degeneracy, $N$, of the
level. But since $N=4$, this would imply, for the
experimentally significant cases of $Z\sim 1$, a tendency to
over-screen the Coulomb center. This bring us again to the role of
interactions. The above would be true in the limit of weak
interaction $\alpha\ll 1$. But, in that case, the supercritical
regime would require $Z\gg1$, which is not feasible. In the end,
if supercritical systems are to be produced, electron-electron
interactions should be strong which, besides requiring the
computation of the vacuum polarization in strong-coupling, brings the
question of the renormalization of the bound levels themselves (Lamb
shift). This situation, however, is completely
analogous to the problem of super-heavy nuclei in QED, and
an extensive account of its particular features and difficulties can
be found in \onlinecite{Greiner:1985}.

%
%
\subsection{From Single to Many Particle Interactions}
\label{subsec:FromSingleManyParticleInteractions}

Coupling to an external Coulomb field can be seen as the zero-th
order approach to the full many body electron interactions in
graphene. The decisive difference that leaves  graphene apart from
standard electronic systems is the existence of the supercritical
region, which, for the Coulomb field, has the peculiarities discussed
so far. Since the coupling constant in vacuum is $\alpha\approx 2$,
one can justifiably ask whether supercritical effects carry to
electrons interacting among themselves. After all, even if a
simplification, from a reference frame moving with an electron the
problem becomes an impurity one again.

%
\subsubsection{Interacting Two Body problem}
\label{subsubsec:InteractingTwoBodyProblem}

The two particle problem has traditionally provided valuable insights
into the full many-body phenomena in condensed matter [e.g. the
Cooper pairing \cite{Cooper:1956}]. The chiral nature of the
electronic
states, however, precludes the usual decoupling between center-of-mass
and
relative coordinates, except for $s$-states in a quiescent
center-of-mass \cite{Sabio:2009}. Even so, these authors show that the
supercritical
collapse is a general effect present in the two body problem. In this
case the critical coupling occurs at $\alpha_c=1$ and $\alpha_c=2.24$
for $s$ and $p$ channels, respectively. 
The interacting two-body problem usually encodes much of the physics
that the many-body system displays. One example is the study
 of pairing, pair condensation,  and other processes which are dominated by two
particle channel events. This has a clear relation with the issue of spontaneous
 gap generation, discussed in Sec.~\ref{sec:Mass}. The prospect of 
exact solution of  the two particle problem would afford more controllable means
 to explore this instability in graphene. 

%
\subsubsection{Excitons and Spontaneous Mass Generation}
\label{subsubsec:ExcitonsSpontaneousMassGeneration}

It is noteworthy that the value $\alpha_c=1$ quoted above is
tantalizingly close to recent calculations of the
critical coupling which precipitates a spontaneous mass
generation and metal-insulator transition in undoped graphene. Those
values range from
$\alpha_c=0.8$ \cite{Vafek:2008}, to $\alpha_c=1.1$
obtained within Monte Carlo \cite{Drut09a} or
by using the Schwinger-Dyson equation \cite{Khve08}. 
As described at length in
Sec.~\ref{sec:Mass}, this metal-insulator transition in
graphene has been ascribed to the emergence of an excitonic
instability beyond $\alpha_c$. 

Recently the excitonic problem has
been considered vis-a-vis the supercritical instability of the
Coulomb center.
Instabilities in the particle-hole channel appear at critical
couplings consistent with the above \cite{Gamayun:2009,Wang:2009}.
For example, \onlinecite{Gamayun:2009} show that solving the
Bethe-Salpeter equation in graphene leads to instability-prone
tachyonic states  ($E^2<0$) at $\alpha_c=1.6$. Such states are the
analogue in the two channel many-body language of the quasi-bound
resonances for supercritical impurities, and a glimmer of
supercritical effects in the fully interacting problem.

%
%
\subsection{Supercritical Physics in Experiments}
\label{subsec:ExperimentalDevelopments}

The non-perturbative nature of supercritical Coulomb impurities, and
the associated analytical difficulties, preclude unequivocal
predictions regarding the possibility of crossing the supercritical
threshold.
Experimental investigation of this problem requires the
ability to vary the strength of the Coulomb impurity and/or the
electron-electron interactions. 
Control over the dielectric environment provides a handle to tune
interactions and impurity strength at the same time, via selection of
$\epsilon_0$. Experiments in this vein have been performed by
\onlinecite{Jang08} and \onlinecite{Ponomarenko:2009}, showing
that it is possible to controllably tune the value of $\epsilon$ by
exploring substrates with different dielectric properties. 
Variation of $Z$ is a more delicate issue. 
\onlinecite{Chen:2008} have devised a way to add monovalent ions to
graphene via K-irradiation, in quantities that can be controlled with
some precision. But exploration of the supercritical regime might
require higher valences. For real impurities, the
valence is determined by the nature of the impurity atom and the host
system, and cannot be changed. One can, in principle, use ions of
different valence, but here the difficulty lies in the fact that
valences higher than $Z=2$ are very unlikely. 
One possible alternative to this constraint imposed by nature, would
be to resort to sharp STM tips, whose strong local field could
mimic a strong local charge. As mentioned in the beginning of this
chapter, the experimental exploration/confirmation of the
supercritical state would be rather important a milestone. Not only in
understanding the physics of graphene, but because it would
afford a glimpse to what might happen in the more fundamental QED
situation.

\section{Strong correlations in graphene}

\subsection{Mass gaps in the honeycomb lattice}
\label{sec:gapsBruno}
Graphene is a semi-metal (SM) with gapless quasiparticles. The Dirac
points in graphene are protected by the combination of  sublattice
and translational symmetries  of the honeycomb lattice. The point group
symmetry of the honeycomb lattice, $C_{6v}$, can be decomposed into
the point group of the triangular sublattice and the $\mathbb{Z}_{2}$
sublattice symmetry group, $C_{3v}\otimes\mathbb{Z}_{2}$. Violation
of sublattice symmetry leads to the opening of a mass gap in the Dirac
Hamiltonian. This broken symmetry can be physically implemented either
by the Semenoff gap \cite{Sem84}, which is induced by a staggered
scalar potential that breaks the sublattice inversion symmetry,
as previously discussed in Eq.~\eqref{eq:H-Mass}, or
by the Haldane gap \cite{Hal88}, where there is an additional broken
time reversal symmetry (TRS) induced by the inclusion of circulating
current loops with zero magnetic flux per unit cell, corresponding
to a staggered magnetic field. In particular, a system that breaks
inversion and TRS is susceptible to a ``parity'' anomaly,  where the application of an electric field
generates a net axial current flowing between the two valleys in graphene
\cite{Jac84}. 

In the presence of mirror symmetry along the $z$-axis, the spin-orbit
interaction in graphene has the form \cite{Kan05}
 \begin{equation}
\mathcal{H}_{SO}=\Delta_{SO}\sum_{\mathbf{k},\sigma}\Psi_{\mathbf{k},\sigma}^{\dagger}\tau_{0}\otimes\sigma_{3}\otimes s_{3}\Psi_{\mathbf{k},\sigma},\label{eq:SO}\end{equation}
where $\Delta_{SO}$ is the spin orbit coupling gap, and $s_{3}$
is the diagonal Pauli matrix in  spin space. The other matrices follow
the convention in the Dirac Hamiltonian (\ref{HoDirac1}). The spin-orbit
interaction in graphene breaks the spin degeneracy in the valleys,
giving rise to spin polarized currents that flow along the edge states
of the system --- a quantum spin Hall state \cite{Kan05}.
Although the spin orbit coupling gap in graphene is rather small,
$\Delta_{SO}\approx10^{-3}$meV, \cite{Hue06, Min06, Yao07}, it can be
drastically enhanced either by curvature effects \cite{Hue06}, or by impurities 
\cite{Cas09}.
The spin-orbit coupling is also logarithmically enhanced by Coulomb
interactions \cite{Kan05}, as discussed in Sec. \ref{sec:Mass}. When the mirror symmetry is
broken either by a substrate or external electric field, 
 an additional Rashba term is allowed  \begin{equation}
\mathcal{H}_{R}=\lambda_{R}\sum_{\mathbf{k},\sigma}\Psi_{\mathbf{k},\sigma}^{\dagger}\tau_{3}\otimes(\sigma_{1}\otimes s_{2}-\sigma_{2}\otimes s_{1})\Psi_{\mathbf{k},\sigma}\,,\label{eq:rashba}\end{equation}
where $\lambda_{R}>0$ is the Rashba coupling. The induced gap is $2 (\Delta_{SO}-\lambda_R$) for $\lambda_R< \Delta_{SO}$, closing to zero when $\lambda_R > \lambda_{SO}$ \cite{Kan05}.

Kekule lattice distortions \cite{Hou07}, which break
the translational symmetry of the lattice, also lead to the opening
of gaps in graphene, whereas lowering the rotational symmetry of the
$C_{3v}$ group, by stretching the honeycomb lattice in one direction,
does not. In the presence of topological defects in the order parameter,
such as vortices, the midgap states which are bounded to them allow
the emergence of excitations with fractional statistics under vortex
exchange \cite{Cha08a, Cha08b, Hou07,Ser08a}. In the superconducting case, the vortex core may
sustain a quantum Hall state in the presence of a strong Zeeman coupling of the electrons
with the magnetic field, which lifts the spin degeneracy \cite{Her10}.
 In the most general case, where any spin, valley and pairing
symmetries are allowed, 36 different types of instabilities that generate
mass gaps in graphene have been classified \cite{Ryu09}.

\subsection{Charge and magnetic instabilities}

Although no evidence of mass gaps has been found in graphene, numerical
results have predicted a semi-metal-insulator (SM-I) transition in
the presence of strong correlations. Quantum Monte Carlo (QMC) calculations
on the Hubbard model for the honeycomb lattice at half filling predicted
the opening of a Mott gap above the critical ratio $U/t\gtrsim5$
\cite{Mar97, Pai05, Sor92}, where $t\approx2.8$ eV is the hopping energy and $U$
is the on-site electronic repulsion. A more recent QMC calculation  
has found a gapped AF state at half filling
for $U/t > 4.3$, preceded by an intermediate coupling insulating phase for $3.5 < U/t < 4.3$,
which has been attributed to a gapped spin liquid state formed by 
short-range resonating valence bonds \cite{Meng}. An insulating antiferromagnetic
(AF) ground state has been also predicted above $U/t\gtrsim4$ \cite{Fur01, Mar97}.
 Variational \cite{Han95} and mean field calculations 
\cite{Per04} predicted 
the possibility of Nagaoka ferromagnetism (where the polarization
is maximal) above a critical coupling both in the half filled and
in the doped regimes.  
Although the validity of the  Hubbard model in graphene may be questioned 
since it does not include long range Coulomb interactions, it could be in principle
justified if one accounts for strong screening effect from a substrate  which can 
deplete the long range part of the interactions  (or also, perhaps, by accounting for 
dynamical screening effects from graphene itself  \cite{Bruno}), leaving only the 
short-range part of the electron-electron interactions. The extent of validity of the 
Hubbard model in graphene is a subject of ongoing debate.

The bare spin polarization in graphene is a 2$\times$2 tensor \cite{Per04}, 
\begin{equation}
\Pi_{x,y}^{+-}(\mathbf{q},\tau)=\langle S_{x}^{+}(\mathbf{q},\tau)S_{y}^{-}(-\mathbf{q},0)\rangle,
\label{eq:PiSpin}
\end{equation}
 where $S_{x}^{+}$ and $S_{x}^{-}$  are the spin raising and lowering
operators in the two sublattices, $x=a,b$. Written in terms of the
Green's function (\ref{eq:G}) with additional spin labels, \begin{eqnarray}
\Pi_{x,y}^{(1)+-}(\mathbf{q},i\omega) & = & -\frac{1}{4}\sum_{\mathbf{k},s,s^{\prime}=\pm}\mathcal{A}_{x,y}^{s}(\mathbf{k})\mathcal{A}_{y,x}^{s^{\prime}}(\mathbf{k}+\mathbf{p})\times\,\nonumber \\
 &  & \negmedspace\negmedspace\negmedspace\frac{f\left[E_{s,\uparrow}(\mathbf{k})\right]-f\left[E_{s^{\prime},\downarrow}(\mathbf{k}+\mathbf{q})\right]}{i\omega+E_{s,\uparrow}(\mathbf{k})-E_{s^{\prime},\downarrow}(\mathbf{k}+\mathbf{q})}\,,\label{eq:Pi_spin}\end{eqnarray}
where $\hat{\mathcal{A}^{s}}\equiv1+s\mathbf{k}\cdot{\bm \sigma}/k$,
and $E_{s,\sigma}(\mathbf{k})=sv|\mathbf{k}|-\mu$ describes the two branches
of the spectrum near the Dirac points. Since $\Pi_{a,a}=\Pi_{b,b}$
and $\Pi_{a,b}=\Pi_{b,a}^{*}$ by the honeycomb
lattice symmetry, the eigenvalues of the spin polarization are
$\Pi_{F/AF}=\Pi_{a,a}^{+-}\pm|\Pi_{a,b}^{+-}|\,,$
which correspond to ferromagnetic ($+$) and AF $(-)$ states. In
RPA, the spin susceptibility is $\hat{\chi}=[\hat{\mathbf{1}}-U\hat{\Pi}^{(1)}]^{-1}\hat{\Pi}^{(1)}$,
and the critical Hubbard coupling required for a divergence in the
spin susceptibility in graphene is \cite{Per04} \begin{equation}
U_{c}^{F/AF}=\frac{1}{\Pi_{F/AF}^{(1)}(0)}\,.\label{eq:U_c}\end{equation}
The ferromagnetic transition translates in the condition $U_{c}^{F}=2/\rho(\mu)\approx D^{2}/|\mu|$,
which is the Stoner criterion, where $\rho(E)$ is the DOS and $D$ the band width. The AF
transition occurs at $U_{c}^{AF}\approx D^{2}/(D-|\mu|)$. 

The application of an in plane magnetic field, $B$, splits the spin
degeneracy at the Dirac points, creating two  Fermi surface (FS) pockets with opposite
spins. Including the Zeeman coupling, $H_{B}=\sum_{\sigma}\sigma B\hat{n}_{\mathbf{k},\sigma}$
into the Hamiltonian, the spin polarized energy spectrum is $E_{s,\sigma}(\mathbf{k})=sv|\mathbf{k}|+\sigma B-\mu$.
The nesting between the two Fermi surface sheets can produce a logarithmic divergence
in the spin polarization in the limit $|B|\gg {\mbox{max}}(T,|\mu|)$ \cite{Berc09}, 
\begin{equation}
\Pi_{AF}^{(1)}(0)\sim\rho(B)\,\ln\!\left(\frac{|B|}{{\mbox{max}}(T,|\mu|)}\right)\,.\label{eq:xhi}
\end{equation}
 This instability brings the possibility of a canted
AF state in graphene. In the presence of Landau level quantization
due to the application of an out of plane magnetic field, electronic interactions 
may lead to the formation of quantum Hall ferromagnetic states
at integer values of the filling factor \cite{NM06}.
The magnetic field has been also proposed as a source of a charge density wave (CDW) Peierls
distortion in the zero Landau level in graphene, breaking the parity
symmetry between the valleys \cite{FL07}. For a discussion of interaction effects at strong magnetic fields, see sec. VIII. 

For Dirac fermions in 2+1 dimensions, a CDW 
instability translates into the phenomenon of chiral symmetry breaking
(CSB), with spontaneous generation of a mass term that breaks the
sublattice symmetry. The AF state is favored by strong on site repulsion
and competes with the long range part of the Coulomb field, which
can favor either strong coupling ferromagnetism \cite{Per04} 
or else excitonic CDW instabilities at strong coupling \cite{Drut09a,
Drut09c, Liu09, Khv01a,
Khv01b, Khv04}. 

\begin{figure}[tb]
\centering
\includegraphics[width=0.45\textwidth]{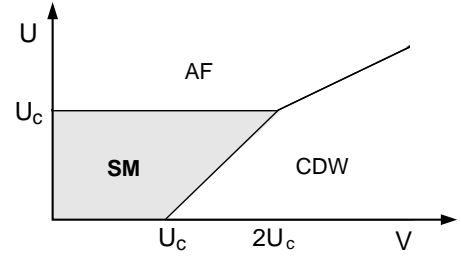}
\caption{{\small Semi-metal (SM) insulator transition predicted by the renormalization
group analysis of the extended Hubbard model, in large $N$ expansion.
$U$ is the on-site Hubbard coupling and $V$ is the nearest neighbor
site repulsion. $U_{c}$ stands for the critical coupling. AF: antiferromagnetic
phase; CDW: charge density wave state \cite{Her06}. }}
 \label{fig:Semi-metal-(SM)-insulator}
\end{figure}

At large $N$, with $N$ the number of fermionic flavors, the continuum
limit of the Hubbard model in the honeycomb lattice falls in the universality
class of the Gross-Neveu model \cite{Gro74} for massless
Dirac fermions in 2+1 dimensions, with four-fermion contact interactions.
The extended  version of this model accommodates the short
range piece of the Coulomb interaction involving the repulsion between nearest
neighbor sites, $V$ \cite{Her06}. In addition to the Gaussian fixed
point, which controls the semi-metal (SM) phase, the RG flow of the extended model
was shown to be controlled by two other fixed points at large $N$:
an AF fixed point, and a CDW fixed point,  both unstable
towards the Gaussian fixed point at weak coupling, and having a runaway
direction to strong coupling when $U$ or $V$ are sufficiently large.
The two fixed points compete, resulting in the phase diagram shown
in Fig.~\ref{fig:Semi-metal-(SM)-insulator}. The fact that the
AF fixed point has only one unstable direction to leading order in
$1/N$ motivated the conjecture that the SM-I transition to the AF
state is continuous and of the Gross Neveu type \cite{Her06}. The
symmetry analysis of the possible quartic terms has been discussed
by \onlinecite{Her09}.

The $1/N$ results were confirmed qualitatively by numerical renormalization
group (NRG) calculations for the extended Hubbard model in the honeycomb
lattice \cite{Rag08}. In the presence of next-nearest
neighbors repulsion, the NRG calculations suggested the possibility
of competition between the CDW and spin density wave (SDW) phases with non-trivial topological
insulating states, such as the quantum spin Hall (QSH) state, where
TRS is spontaneously broken \cite{Rag08}. Functional
renormalization group (FRG) calculations for the $t-J$ model on the
honeycomb lattice with on site and nearest neighbor repulsion also
suggested the possibility of strong coupling CDW and SDW instabilities
in graphene at half filling \cite{Hon08}. In the doped regime,
the $t-J$ model can favor the formation of superconducting states
for $J>2t$, either in the triplet or in the $d$-wave singlet channels
\cite{Hon08}. 

In the high doping regime, the proximity of the Fermi level to the
Van-Hove singularities, where the graphene DOS diverges logarithmically,
may favor a Pomeranchuck instability (PI), rather than a gapped state.
In that case, the redistribution of the electronic density generates
a deformation of the Fermi surface, which lowers the lattice $C_{3v}$
point group, instead of breaking the $\mathbb{Z}_{2}$ sublattice
symmetry. In the extended Hubbard model at high doping, the
PI is favored by the repulsion between nearest neighbor sites, which
renormalizes the kinetic energy at the mean field level, and competes
with the on-site repulsion, which favors a ferromagnetic state when
the Stoner criterion is satisfied \cite{Val08}.

When coated with metallic atoms that have a strong tendency
to hybridize with the carbon $p_{z}$ orbitals, graphene can induce strong itinerant 
ferromagnetism in the metallic bands \cite{Uch08a}.

\subsection{Local magnetic moments}

For massless Dirac particles, the formation of localized states is 
usually harder than in usual Fermi systems due to the Klein paradox,
in which the fermions can easily tunnel through a barrier regardless of 
its height. Defects such as vacancies, where a carbon atom is knocked
out from the plane, have been shown to generate localized states in
graphene \cite{Per06, VLSG05}, 
and were  recently observed in STM experiments \cite{UBGG10}. 
Vacancies have also been found to host local magnetic states
 \cite{CCWF10, Yaz07}. 

Short range interacting impurities can  generate local resonances,
which are quasi-localized states. At half-filling, the energy of the
resonance, $\varepsilon_{0}$, is given by \cite{Skr06,Weh07} \begin{equation}
U_{0}=\frac{D^{2}}{\varepsilon_{0}\ln\left|\varepsilon_{0}^{2}/(D^{2}-\varepsilon_{0}^{2})\right|}\,,\label{eq:Uo}\end{equation}
where $U_{0}$ is the scattering potential of the impurity
and $D$ is the bandwidth.
 The resonance induces accumulation of LDOS at the Fermi level
around the impurity, $\rho(r,\omega)$, which decays as $1/r$ \cite{Ben05}, 
whereas the Friedel oscillations decay as $1/r^{2}$
for intracone scattering and as $1/r$ for intercone scattering \cite{Ben08}.

\begin{figure}[b]
\centering
\includegraphics[width=0.43\textwidth]{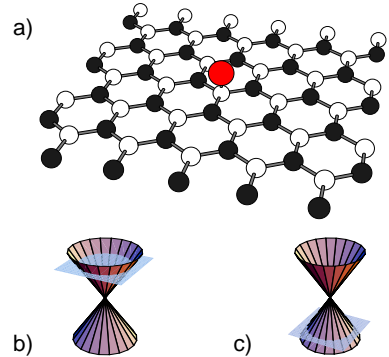}
\caption{{\small  (color on line) (a) Honeycomb lattice with an impurity
atom. Black: sublattice A; White: sublattice B. Intersection of the
Dirac cone spectrum, $E(\mathbf{k})=\pm v|\mathbf{k}|$, with the localized
level $E_{f}=\varepsilon_{0}$: (b) $\varepsilon_{0}>0$, 
(c) $\varepsilon_{0}<0$. }}
\label{bruno1}
\end{figure}

Besides defects, zigzag edges also lead to local magnetism in the
presence of interactions (for a more detailed discussion, see Sec.
\ref{Paco6}). In bulk graphene, a simple way to generate localized magnetic
states is provided by the adsorption of adatoms with inner shell electrons.
On the lattice, the adatoms can stay in different locations relative
to the two sublattices in graphene. Transition metals are usually
more stable sitting in the hollow site, at the center of the honeycomb
hexagon \cite{Cha08}, whereas simple molecules and atoms
such as hydrogen (H) tend to hybridize more strongly with the carbons,
sitting on top of them and generating a large local moment \cite{Yaz07}. 
In particular, H adsorption creates a midgap state \cite{Bou08,WehlingPRL} and distorts locally the
$sp^{2}$ carbon bonds, which acquire $sp^{3}$ character \cite{Eli09}. 
This distortion can induce a strong local enhancement
of the spin-orbit coupling up to $\approx7$ meV, as in diamond, 
and generate a strong local magnetic anisotropy \cite{Cas09}. 
Adatoms can also form local moments from substitutional defects 
on single and double vacancies in graphene \cite{Kra09, Ven09}.

The heuristic criterion that describes the formation of a local magnetic
moment is addressed at the mean field level by the Anderson impurity
model \cite{And61}. In the top carbon case, assuming that the
adatom sits on a carbon (see Fig.~\ref{bruno1}), say on sublattice $B$, the hybridization
Hamiltonian is $H_{V}=V\sum_{\sigma}[f_{\sigma}^{\dagger}b_{\sigma}(0)+h.c.],$
where $f_{\sigma}$ ($f_{\sigma}^{\dagger}$) annihilates (creates)
an electron with spin $\sigma=\uparrow,\downarrow$ at the impurity.
 In momentum space, this translates into:
\begin{equation}
\mathcal{H}_{V}=V\sum_{\mathbf{p},\sigma}(f_{\sigma}^{\dagger}b_{\mathbf{p},\sigma}+b_{\mathbf{p},\sigma}^{\dagger}f_{\sigma})\, .\label{eq:Hv}\end{equation}
If $n_{\sigma}=\langle f_{\sigma}^{\dagger}f_{\sigma}\rangle$ is
the occupation of the localized level for a given spin, the effective
Hamiltonian of the level is \begin{equation}
\mathcal{H}_{f}=\sum_{\sigma}\varepsilon_{\sigma}f_{\sigma}^{\dagger}f_{\sigma}\,,\label{eq:H_f}\end{equation}
with $\varepsilon_{\sigma}=\varepsilon_{0}+Un_{-\sigma}$, after a proper
mean field decomposition of the Hubbard term $H_{U}=Uf_{\uparrow}^{\dagger}f_{\uparrow}f_{\downarrow}^{\dagger}f_{\downarrow}$,
which accounts for the charging energy  $U$ to doubly occupy the level.
The hybridized level becomes magnetic when  $n_{\uparrow}\neq n_{\downarrow}$.
The occupation is derived self-consistently by integrating the $f$-electron 
DOS from the bottom of the graphene band up to the Fermi level $\mu$,
\begin{equation}
n_{\sigma}=-\frac{1}{\pi}\mbox{Im}\int_{-\infty}^{\mu}\mbox{d}\omega\,\frac{1}{\omega-\varepsilon_{\sigma}-\Sigma_{ff}(\omega)}\,,\label{MFh2}\end{equation}
where $\Sigma_{ff}(\omega)$ is the self-energy of the localized electrons.
In the cone approximation of the spectrum in graphene, for the top
carbon case, \begin{equation}
\Sigma_{ff}(\omega)=\omega\left[1-Z^{-1}(\omega)\right]-i\Delta|\omega|\theta(D-\vert\omega\vert)\,,\label{G_bar_R}\end{equation}
where $\Delta=\pi V^{2}/D^{2}$ is the dimensionless hybridization,
$D$ is the effective band width, and \begin{equation}
Z^{-1}(\omega)=1+\frac{V^{2}}{D^{2}}\ln\!\left|1-\frac{D^{2}}{\omega^{2}}\right|\label{eq:Z}\end{equation}
gives the quasiparticle residue, $Z(\omega)$, which vanishes logarithmically
at the Dirac points ($\omega\to0$). 

\begin{figure}[tb]
\centering
\includegraphics[width=0.47\textwidth]{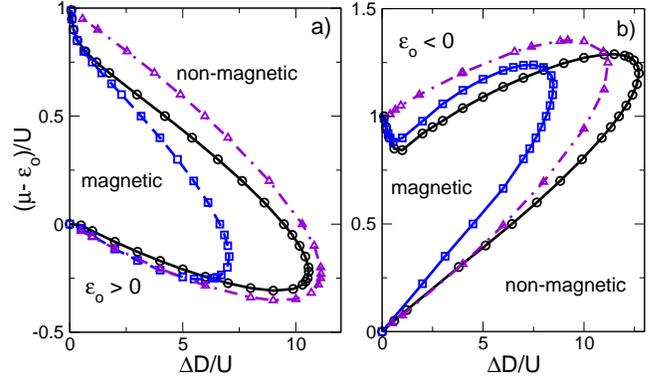}
\vspace{0.7cm}
\caption{Boundary between magnetic and non-magnetic impurity states in the
scaling variables $x=\Delta D/U$ and $y=(\mu-\varepsilon_{0})/U$ for
$\varepsilon_{0}>0$ (a) and $\varepsilon_{0}<0$ (b). $|\varepsilon_{0}|/D=0.029,\,0.043,\,0.029$
and $V/D=0.14,\,0.14,\,0.04$ for circles, squares and triangles,
respectively. The upturn close to $y=1$ and $x\to0$ on panel b)
signals a crossover to the Fermi liquid regime $\mu,U\gg|\varepsilon_{0}|>0$,
where the Dirac points are physically irrelevant. This feature is
not visible in this scale when $V$ is very small (triangles) \cite{Uch08b}.}
\label{fig:Boundary-between-magnetic}
\end{figure}

Because of the vanishing DOS, the level broadening {[}given by $\mbox{Im}\Sigma_{ff}^{R}(\omega)$]
scales linearly with the energy around the Dirac points \cite{Gon98, Skr06, Uch08b, Zha01}. The DOS induced around the bare
level, $\varepsilon_{\sigma}$, does not decay like a Lorentzian as in
usual metals, but shows a long tail proportional to $1/\omega$. This
tail induces several peculiar features in the magnetic states. For
instance, a local moment is allowed to exist when the bare level is
empty ($\varepsilon_{0}<\mu$) or doubly occupied ($\varepsilon_{0}+U>\mu$)
(see Fig.~\ref{fig:Boundary-between-magnetic}). The presence of the
Dirac point also breaks the symmetry around the line $\mu-\varepsilon_{0}=U/2$,
and makes the scaling of the curves shown in Fig.~\ref{fig:Boundary-between-magnetic}
non-universal. Furthermore, there is  a physical asymmetry between
the cases where the level is above ($\varepsilon_{0}>0$) or below ($\varepsilon_{0}<0$)
the Dirac point. When $\varepsilon_{0}=0$, as in the case of a vacancy,
the level decouples from the bath and becomes magnetic for any $\mu>0$,
regardless  of the value of $U$ \cite{Per06, Uch08b}. 

Since the chemical potential in graphene can be tuned, the formation
of local magnetic states can be controlled by the application of a
gate voltage \cite{Uch08b}. The low density of states
around the localized level also makes the formation of local moments
in graphene much easier than in usual metallic hosts. As a result 
the adatoms can achieve high magnetic moments at relatively small
$U$ \cite{Cor09, Uch08b}. 

\begin{figure}[b]
\centering
\includegraphics[width=0.5\textwidth]{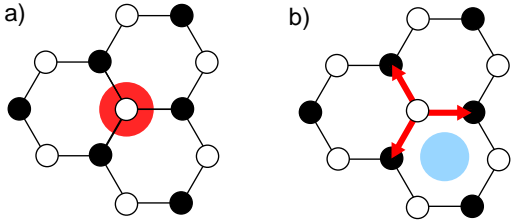}
\caption{{\small Two adatom configurations in graphene: a) the adatom (red
circle) sits on top of a carbon atom, and b) the adatom (blue
circle) sits at the center of the honeycomb hexagon, hybridizing equally
with the two sublattices. Red arrows: nearest neighbor vectors.}}
\label{bruno2}
\end{figure}

The formation of local moments is also affected by the specific location
of the adatom in the lattice (Fig.~\ref{bruno2}). For instance, when the adatom sits in
the center of the honeycomb hexagon ($H$-site), the tight-binding hybridization
Hamiltonian is \cite{Uch09b} \begin{equation}
H_{V}=\sum_{\sigma,i}\left[V_{a,i}a_{\sigma}^{\dagger}(\mathbf{a}_{i})+V_{b,i}b_{\sigma}^{\dagger}(-\mathbf{a}_{i})\right]f_{\sigma}(0)
 + {\mbox{h.c.}}
\,,\label{eq:HV00}\end{equation}
where $\mathbf{a}_{i}$ $(i=1,2,3$) are the three nearest neighbor
vectors of the honeycomb lattice, and $V_{x,i}$ ($x=a,b$) is the
hybridization strength of the adatom with each of the nearest surrounding
carbon atoms. In momentum representation, \begin{equation}
H_{V}=\sum_{\mathbf{p}\sigma}\left(V_{a,\mathbf{p}}^{*}a_{\mathbf{p}\sigma}^{\dagger}+V_{b,\mathbf{p}}b_{\mathbf{p}\sigma}^{\dagger}\right)f_{\sigma}+
{\mbox{h.c.}}\,,\label{Hv2}\end{equation}
where \begin{equation}
V_{x,\mathbf{p}}=\sum_{i=1}^{3}V_{x,i}\,\mbox{e}^{i\mathbf{p}\cdot\mathbf{a}_{i}}\,.\label{eq:Vp}\end{equation}
The top carbon case is recovered
 by setting $V_{a,\mathbf{p}}\equiv V$ and $V_{b,\mathbf{p}}= 0$ or
 vice-versa.
For $s$-wave orbitals, $V_{x,i}\equiv V$, whereas for in-plane $f$-wave orbitals the hybridization is anti-symmetric in the two sublattices, $V_{a,i}=-V_{b,i}\equiv V$. In the case of substitutional impurities ($S$-sites), either $V_{a,i}=0$ or $V_{b,i}=0$. The quantum interference between the different hybridization paths of the electrons can modify
the energy scaling of the level broadening in Eq. (\ref{G_bar_R})
\cite{Uch09b}, and can also
change the shape of the Fano resonances in scanning tunneling spectroscopy
(STS) measurements, allowing a clear identification of the adatom
position with an STS tip \cite{Sah10, Uch09b, Weh10a}. 

\subsection{Kondo effect}

The formation of a Kondo screening cloud around a magnetic moment
is described by the Anderson Hamiltonian (\ref{eq:Hv}) in the strong
coupling limit, $U\to\infty$, where the valence fluctuations are
suppressed and the local moment becomes a good quantum number. In
the standard mean field approach, the spin $1/2$ fermionic fields
are replaced by fermionic fields with larger degeneracy, $N>m$, which
corresponds to an SU($N$) extension of the problem, with a corresponding
Kondo Hamiltonian \cite{Coc69} \begin{equation}
\mathcal{H}_{K}=J_{K}\sum_{m m^{\prime}}
\sum_{\mathbf{k}\mathbf{k}^{\prime}}\psi_{\mathbf{k},m}^{\dagger}f_{m^{\prime}}^{\dagger}f_{m}\psi_{\mathbf{k}^{\prime},m^{\prime}}\,,\label{eq:H_K}\end{equation}
where $J_{K}\sim V^{2}/|\varepsilon_{0}-\mu|$ is the Kondo coupling,
$\psi_{m}$ ($\psi_{m}^{\dagger})$ are annihilation (creation)
operators of the itinerant electrons, and the local $f$ fields are
constrained to a fixed occupancy. At the mean field level, which is
asymptotically exact at large $N$, the Kondo order parameter can
be extracted either from the standard slave boson approach to the
Anderson model   \cite{Col83, New87}, or else
by an equivalent path integral approach starting from the Kondo Hamiltonian
(\ref{eq:H_K}) \cite{Rea83}. 

The application of these methods to semi-metals with a vanishing DOS,
$\rho(\omega)=\rho_{0}|\omega|^{r}$, with $r>0$, resulted in the
prediction of a Kondo quantum critical point (QCP)
at half-filling ($\mu=0$). In that case, a Kondo cloud
is expected for $J_{K}>J_{K}^{c}=r/(\rho_{0}D^{r})$, below the Kondo
temperature \cite{Wit90} \begin{equation}
T_{K}\approx|J_{K}-J_{K}^{c}|^{\nu}\,,\label{eq:Tk}\end{equation}
where $\nu=1/r$, and $D$ is the ultraviolet cut-off.
Since the scaling dimension of the hybridization $V$ in the
    Anderson model is dim$[V]=(1-r)/2$, the case $r=1$ acts as an upper
    critical scaling dimension in the problem, where the scaling is
    marginal \cite{Voj04}. In the marginal case, the Kondo temperature
    may have an additional logarithmic scaling with the coupling, upon implementation 
    of an ultraviolet cut-off smoothly connected to the metallic case ($r=0$)  \cite{Cas96}.
Away from half-filling, there is a crossover to the usual Fermi liquid
case in the weak coupling regime, $J_{K}<J_{K}^{c}$, where \cite{Wit90} 
\begin{equation}
T_{K}\approx \mu\,\mbox{exp}[r^{-1}(D/\mu)^{r}(1-J_{K}^{c}/J_{K})-1/r]\,.\label{eq:Tk2}\end{equation}

Further studies based on NRG techniques
\cite{Fri04, Gon98, Voj01a} predicted a variety of fixed points. At half-filling, in the
particle-hole symmetric case, $\varepsilon_{0}=-U/2$, the Kondo problem
has a metallic Kondo screened fixed point at $r=0$, which evolves
into a strong coupling fixed point for $0<r\leq1/2$. In this case,
the strong ($J_{K}>J_{K}^{c}$) and weak coupling ($J_{K}<J_{K}^{c})$
regimes  are separated by a symmetric quantum critical point
(SCP), whereas for $r>1/2$ the local moment remains unscreened for
all initial values of the Kondo coupling \cite{Che95}.
In the particle-hole asymmetric case ($\mu=0$, $U\neq-2\varepsilon_{0}$),
for $r>r^{*}\approx0.375$, the weak and strong coupling regimes are
separated by an asymmetric critical point (ACP). For $r<r^{*}$, the
particle-hole symmetry is dynamically restored \cite{Gon98, Fri04}. 

\begin{figure}
\centering
\includegraphics[width=0.4\textwidth]{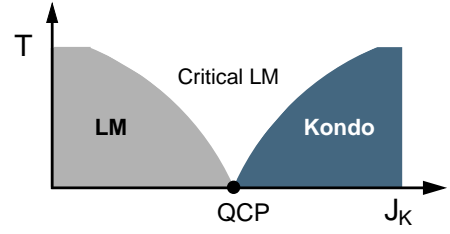}
\caption{{\small Schematic phase diagram around the Kondo QCP at half filling:
temperature }\emph{\small vs}{\small{} Kondo coupling. LM: local moment
phase, where the Kondo cloud is suppressed.
In the critical LM phase, quantum critical fluctuations dominate \cite{Ing02}.
\label{fig:Schematic-phase-diagram} }}
\end{figure}

The phase diagram around the QCP is schematically shown in 
Fig.~\ref{fig:Schematic-phase-diagram}.
The critical local moment fluctuations were studied by \onlinecite{Ing02},
 who found  linear $\omega/T$ scaling of the dynamical
spin susceptibility at the critical point for $0<r<1$. In the marginal
case, $r=1$, there are logarithmic corrections to scaling \cite{Cas97}. 
The Kondo problem for gapless excitations was
also extensively studied in the context of magnetic impurities
in $d$-wave superconductors \cite{Bor92, Cas96, Cas97, Pol01, Pol02, Voj01b, 
Zha01, Zhu00}. For a review, see \onlinecite{Bal06}.

In the graphene case, where $r=1$, the Dirac fermions in the
    bath have an additional pseudospin structure, which motivated 
    several proposals for multichannel Kondo physics \cite{Cas96, Del10,
    Sen08, Zhe10}. The Kondo resonance in
    graphene has been calculated with NRG by \onlinecite{Cor09}. 
 At half filling, the local DOS around the impurity can be spontaneously 
enhanced by the formation of midgap states due to the scattering potential of the
impurity \cite{HG07}, frustrating the Kondo QCP. 

At finite doping, the Kondo temperature has an exponential
    dependence with the DOS at weak coupling, allowing the Kondo cloud
    to be tuned by gating  \cite{Sen08}.  In the crossover regime, at
    $J=J_c$, the scaling of the Kondo temperature with doping becomes
    power law,  $T_{K}\propto|\mu|^{x}$. Recent NRG calculations in
    graphene have found a particle-hole asymmetric scaling of the Kondo
    temperature with  doping, $T_{K}\propto|\mu|^{x}$, where  $x=1$ for
    $\mu>0$ and $x=2.6$ for $\mu<0$ \cite{Voj10}, in contradiction with
    the mean field and poor man scaling analysis for the marginal case
    \cite{Voj10}.  In the
    presence of Landau level quantization, the
    Kondo temperature has reentrant behavior as a function of the
    chemical potential \cite{Dor07}.

Looking at the problem on the lattice, ab initio
    calculations on Cobalt have found that the interplay of spin and
    orbital degrees of freedom can give rise to an SU$(4)$ Kondo effect
    in graphene when the spin orbit coupling is strong enough
    \cite{Weh10b}. Another  ab initio calculation accounting for dynamic
    correlations, also on Co, has identified the possibility of a spin
    $3/2$ Kondo effect, involving multiple orbitals
    \cite{Jac10}. From a tight-binding perspective, for a spin $1/2$
    impurity, the hybridization Hamiltonian (\ref{Hv2}) can be written
    in the diagonal basis

\begin{equation}
\mathcal{H}_{V}=V\sum_{\alpha=\pm}\sum_{\mathbf{p},\sigma}\left[\Theta_{\alpha,\mathbf{p}}c_{\alpha,\mathbf{p}\sigma}^{\dagger}f_{\sigma}+
{\mbox{h.c.}}\right],\label{eq:HV01}\end{equation}
where $
c_{\pm,\mathbf{k}\sigma}=(1/\sqrt{2})[b_{\mathbf{k}\sigma} \pm
 (\phi_{\mathbf{k}}^{*}/|\phi_{\mathbf{k}}|)a_{\mathbf{k}\sigma}]$
are the fermionic operators that diagonalize the graphene Hamiltonian
(\ref{Ho}), $\phi_{\mathbf{k}}=\sum_{i=1}^{3}\mbox{e}^{i\mathbf{a}_{i}\cdot\mathbf{k}}$
is the tight-binding hopping matrix element defined by Eq. (\ref{phi}),
and $\alpha=\pm$ labels the conduction and valence bands. $\Theta$
is a phase factor, which accounts for the symmetry and position of
the localized orbital with respect to the sublattices
 \cite{Uch09b}, \begin{equation}
\Theta_{\alpha,\mathbf{p}}=\frac{1}{\sqrt{2}V}\left(V_{b,\mathbf{p}}+\alpha V_{a,\mathbf{p}}^{*}\frac{\phi_{\mathbf{p}}^{*}}{|\phi_{\mathbf{p}}|}\right)\, , \label{eq:Theta}\end{equation}
where $V_{x,\mathbf{p}}$ is the hybridization as defined in Eq. (\ref{eq:Vp}).

\begin{figure}[t]
\centering
\includegraphics[width=0.35\textwidth]{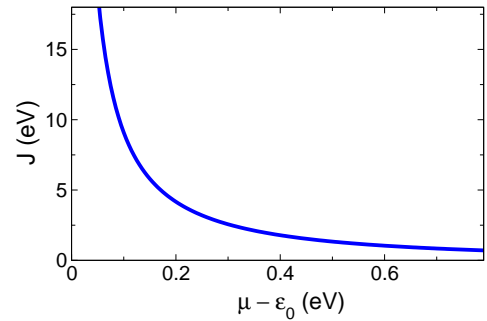}
\caption{{\small  Kondo coupling vs. chemical potential in graphene for  $U=1$ eV and $V=1$ eV. The Kondo coupling can be controlled by gate voltage  across the weak ($J\ll J_{c}$) and strong coupling ($J\gtrsim J_{c}$) Kondo regimes, where $J_{c}$ is the critical coupling at half-filling.
\label{fig:J-mu} }}
\end{figure}

As in metals, the Anderson Hamiltonian in graphene can be mapped into the spin exchange Hamiltonian by a canonical transformation \cite{Sch66}. In the large $U$ limit, the spin exchange Hamiltonian between the
magnetic adatom and the graphene electrons  is \cite{Uch10} \begin{equation}
\mathcal{H}_{e}=-J\sum_{\mathbf{k}\mathbf{k}^{\prime}}\sum_{\alpha\alpha^{\prime}}\Theta_{\alpha,\mathbf{k}}^{*}\Theta_{\alpha^{\prime},\mathbf{k}^{\prime}}\,\mathbf{S}\cdot c_{\alpha^{\prime},\sigma^{\prime},\mathbf{k}^{\prime}}^{\dagger}{\bm \sigma}c_{\alpha,\sigma,\mathbf{k}}\,,\label{eq:He}\end{equation}
where ${\bm \sigma}=(\sigma_{1},\sigma_{2},\sigma_{3})$ are the Pauli
matrices, $\mathbf{S}=\frac{1}{2}f_{\sigma}^{\dagger}{\bm \sigma}f_{\sigma^{\prime}}$
is the localized spin, and \begin{equation}
J(\mu)\approx\frac{V^{2}U}{(\varepsilon_{0}-\mu)(\varepsilon_{0}+U-\mu)}<0\,,\label{eq:J}\end{equation}
is the exchange coupling defined at the Fermi level, $\mu$. 
Within the tight-binding description, we
    realize that the determinant of the exchange coupling matrix in Eq.
(\ref{eq:He}) is identically zero, $\mbox{det}[\hat{J}_{\alpha\alpha^{\prime}}]\equiv0$, and
hence the exchange Hamiltonian (\ref{eq:He}) can be rotated into a new
basis where one of the hybridization channels is decoupled from the
bath \cite{Pul01}.
The eigenvalues in the new diagonal basis are
$J_{u,\mathbf{k},\mathbf{k}^{\prime}}=J\sum_{\alpha}\Theta_{\alpha,\mathbf{k}}^{*}
\Theta_{\alpha,\mathbf{k}^{\prime}}$
and $J_{v}=0$, implying that the one-level exchange Hamiltonian (\ref{eq:He})
maps into the problem of a \emph{single} channel Kondo Hamitonian,
$
\mathcal{H}_{e}=-2\sum_{\mathbf{k}}J_{u,\mathbf{k}\mathbf{k}^{\prime}}\mathbf{S}\cdot\mathbf{s}_{\mathbf{k},\mathbf{k}^{\prime}}$, 
where $\mathbf{s}$ is the itinerant spin, in spite of the implicit
valley degeneracy. A multi-channel description of the one-level problem 
is nevertheless possible for example in graphene 
quantum dots, in the continuum limit, where valley and angular momentum channels become good quantum numbers. 

Unlike the situation in 
metals, the exchange coupling in graphene can be controlled by gating \cite{Jac10, Uch10},  as shown in Fig.~\ref{fig:J-mu}, in particular when the chemical potential is brought to
    the proximity of the localized level, where the Kondo coupling
    becomes resonant. 
 This effect opens the possibility of tuning $J$ to the vicinity of the critical coupling that sets the crossover between the weak and strong coupling regimes.  In this region, at finite doping, quantum criticality is reminiscent of the frustrated QCP at $\mu=0$.  Since the width of the Kondo peak in the spectral function is set by the Kondo temperature only, the gating effect permits measuring the quantum critical scaling of the Kondo temperature with  doping \cite{Uch10,Voj10} directly with STM 
 probes \cite{Sah10, Uch09b, Weh10a, Zhu09}.

\subsection{RKKY interaction}

The Ruderman-Kittel-Kasuya-Yosida (RKKY) interaction between two local spins is obtained by integrating out the itinerant fermions in Eq. (\ref{eq:He}), which gives $
\mathcal{H}_{RKKY}=-J^2\chi_{ij}\,\mathbf{S}_{i}\cdot\mathbf{S}_{j}$,
where $\chi_{ij}$ is a two point correlation function, with  $i,j$
indexing the positions of the local spins. In momentum space \cite{Bre07, Uch10, Sar07},
 \begin{equation}
\chi^{xy}(\mathbf{q})=\sum_{\mathbf{k}\alpha\alpha^{\prime}}\mathcal{M}_{\alpha\alpha^{\prime},\mathbf{k},\mathbf{q}}^{xy}\frac{f[E_{\alpha^{\prime}}(\mathbf{k}+\mathbf{q})]-f[E_{\alpha}(\mathbf{k})]}{E_{\alpha}(\mathbf{k})-E_{\alpha^{\prime}}(\mathbf{k}+\mathbf{q})}\,,\label{eq:chiSpin}\end{equation}
where (omitting the $\alpha\alpha^{\prime}$ labels for simplicity)
\begin{equation}
\mathcal{M}_{\mathbf{k},\mathbf{q}}^{xy}=\Theta_{\alpha,\mathbf{k}}^{
*x}\Theta_{\alpha,\mathbf{k}}^{y}\Theta_{\alpha^{\prime},\mathbf{k}
+\mathbf{q}}^{x}\Theta_{\alpha^{\prime},\mathbf{k}+\mathbf{q}}^{*y}\,,
\label{eq:M1}
\end{equation}
with $x,y=A,\, B,\, H,\,S$ etc, indexing the position of the spins on
the lattice, $E_{\alpha}(\mathbf{k})=\alpha|\phi_{\mathbf{k}}|-\mu$,
and $f$ is the Fermi distribution. $\mathcal{M}_{\mathbf{k},\mathbf{q}}^{AA}=\mathcal{M}_{\mathbf{k},\mathbf{q}}^{BB}=1/4$
for spins on the same sublattice whereas 
\begin{equation}
\mathcal{M}_{\mathbf{k},\mathbf{q}}^{AB}=\frac{1}{4}\alpha\alpha^{
\prime}\frac{\phi_{\mathbf{k}}\phi_{\mathbf{k}+\mathbf{q}}^{*}}{|\phi_
{\mathbf{k}}||\phi_{\mathbf{k}+\mathbf{q}}|}\label{eq:M2}
\end{equation}
for spins on opposite sublattices. In the continuum limit, where the
spectrum is linearized around the Dirac points, $\mathcal{M}_{\mathbf{k},\mathbf{q}}^{AB}=\frac{1}{4}\alpha\alpha^{\prime}\mbox{e}^{i\theta_{\mathbf{k},\mathbf{k}+\mathbf{q}}}$,
where $\theta$ is the angle between $\mathbf{k}$ and $\mathbf{k}+\mathbf{q}$
\cite{Bre07}. 

At half-filling, $k_{F}=0$, the Fermi surface collapses into points
and the RKKY interaction is mediated by interband transitions, which
polarize the vacuum as in QED. In this case, the Friedel oscillations
disappear and the sign of the interaction is ferromagnetic for spins
on the same sublattice and anti-ferromagnetic for spins in opposite sublattices
\cite{Bre07, Sar07}. In the overdoped regime, at $\mu=t$, the nesting among
the Van Hove singularities in graphene reverses the sign of the RKKY
interaction compared to the $\mu=0$ case \cite{Uch10}.

At long distances, the spatial
decay of the RKKY is $r^{-3}$ when $\mu$ is at the neutrality point \cite{Bre07,
Che06, Sar07, VLSG05, Wun07}. Away from half filling, the Friedel
oscillations are restored by the intraband transitions and the RKKY
interaction decays at $r\gg1/k_{F}$ as $1/r^{2}$, similarly to the 2DEG
case \cite{Bre07, Wun07}. For $H$
or $S$ site spins formed in $C_{3v}$ symmetric orbitals, the RKKY
interaction decays with a fast power law $1/r^{7}$ at half filling
\cite{Uch10}. In carbon nanotubes, the RKKY interaction decays as $1/r$ for top carbon spins and
as $1/r^{5}$ for $H$ site spins in isotropic orbitals \cite{Kir08}. 

When distributed regularly on top of graphene, magnetic adatoms such
as  hydrogen (H) can form macroscopic magnetic
states at room temperature \cite{Zho09}. In the disordered case, H atoms in particular can cluster
on top of graphene due to rippling. On top of a ripple, the $sp^{2}$
carbon (C) bonds are spontaneously stretched by the curvature and
acquire $sp^{3}$ character. Contrary to the perfectly flat case,
the adsorption of H atoms on top of the hills helps to stabilize the
ripples \cite{Bou09}. The interplay between the
correlations due to the ripples and the RKKY interaction among the
H spins can generate magnetoresistance hysteresis loops and  a
variety of magnetic spin textures \cite{Rap09}.

\subsection{Superconductivity}
The observation of proximity induced superconductivity in
    graphene junctions has stirred a lot excitement in the field of
    mesoscopics \cite{Hee07}. The Dirac nature of the quasiparticles
    gives rise to  ballistic  transport on a micron scale and
    allows graphene to sustain supercurrents in long junctions, the
    size of the coherence length in the superconducting metallic leads 
    \cite{Du08, Mia07, Hee07, Oje09}.
The experimental realization of the proximity effect motivated
theoretical
studies of the differential conductance (DC) in normal-superconductor
(NS) interfaces in graphene \cite{Bee06, Bur08}, graphene nanoribbons \cite{Rai09}, 
and in graphene
normal-insulator-superconductor (NIS) junctions \cite{Bha06}. 
 Due to the Dirac nature of the spectrum, at half-filling,
    the Andreev conversion of an electron into a hole at the interface
    between a normal and a superconducting region involves specular
    reflection rather than retro reflection \cite{Bee06}.
The specular Andreev reflection leads to the presence of Andreev
modes in SNS junctions that propagate along the graphene edges at
the interface with the superconductor \cite{Tit07}.
The Josephson current in graphene SNS junctions was studied by \onlinecite{Tit06}, 
followed by \onlinecite{Ber09,Mog06, Mai07},  and \onlinecite{Bla08}. 
Possible applications involving the  proximity effect in graphene
include proposals for valley sensors \cite{Akh07},
current switches \cite{Lin08, Lut08}, and a spin current filter \cite{Gre07}. 
A review on Andreev and Klein tunneling processes in graphene can be
found in \onlinecite{Bee08}.

These experimental developments in transport motivated  a surge of 
    interest in the possibility of making graphene  an intrinsic
    superconductor. 
Graphene parent compounds, such as the graphite intercalated materials
CaC$_{6}$ and KC$_{8}$, are low temperature superconductors, although
neither graphite nor alkaline metals alone superconduct \cite{Csa05, Han65, Wel05}. 
Even though 
intrinsic superconductivity has not been observed in the single layer
so far, a few different superconducting mechanisms have been proposed.
One possibility is a plasmon mediated mechanism in graphene coated
with metallic adatoms, in which the plasmons of the metallic band
mediate the attraction between the graphene electrons \cite{Uch07}. 
When isolated islands of metallic atoms are adsorbed
on top of graphene, superconductivity can also be induced by proximity
effect \cite{Fei08}. Another possibility is the
Kohn-Luttinger mechanism, which explores the proximity of the Fermi
surface to the Van-Hove singularities in the high doping regime \cite{Koh65}. 
In this scenario, the superconductivity
can be mediated by a purely electronic mechanism, when the interactions
become attractive along a specific direction of the BZ near the Van-Hove singularity 
\cite{Gon08}. The superconductivity  can also be mediated  by in plane or out of plane 
 flexural phonons \cite{Loz10}.
 In graphene, strong doping regimes can be currently achieved
by  chemical adsorption of alkaline metals, such as potassium
\cite{Gru09b, McC07, Uch08a}, or with metal contacts \cite{Gio08}. 

Alternative proposals include edge state superconductivity, induced
by the large DOS at the edges \cite{Sas07}, or strong
correlations, which so far  have not  been observed in graphene. As
in the cuprates, the antiferromagnetic attraction between spin singlets
on nearest neighbor sites has been proposed as a possible pairing
channel in graphene, provided the on site Hubbard repulsion is strong
enough to suppress the local fluctuations \cite{Pat10}.
\onlinecite{GGV01} considered the possible competition
between ferromagnetic and superconducting states in graphene sheets
through a renormalization group analysis accounting for Coulomb interactions.
A recent functional renormalization group calculation has proposed
the possibility of a strongly correlated SDW state that gives way
to a singlet superconducting instability in the $d$-wave channel,
or else a CDW solution that allows a triplet pairing instability in
the $f$-wave channel \cite{Hon08}. 
In two-layer graphene, the possibility of excitonic pairing 
of electrons in one layer with  holes in the
other one has been considered \cite{Kha08, Min08}. 

\begin{figure}
\centering
\includegraphics[width=0.5\textwidth]{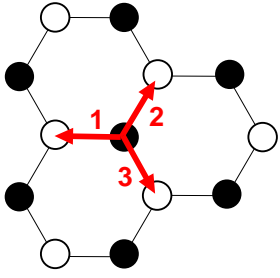}
\caption{{\small Superconducting order parameter $\Delta_{1,j}=\Delta_{1}\mbox{e}^{i\theta_{j}}$
($j=1,\,2,\,3)$, with phases along the three different bond directions in the
lattice. }}
\label{bruno3}
\end{figure}

Regardless of the microscopic origin, the superconducting state in graphene
can be analyzed based on the symmetries of the order parameter in the honeycomb lattice. 
On the lattice, the electrons in graphene carry spin, angular momentum
and sublattice quantum numbers. There are four possible pairing channels:
singlet/triplet spin channels, and same/opposite
sublattices. In the singlet case, if we restrict the analysis to nearest
neighbor site interactions only, two competing order
parameters can be identified: \begin{equation}
\Delta_{0}=g_{0}\langle a_{i\uparrow}a_{j\downarrow}\rangle=g_{0}\langle b_{i\uparrow}b_{j\downarrow}\rangle,\label{eq:Delta0}\end{equation}
which corresponds to an $s$-wave state, and $\Delta_{1}$, defined
as \begin{equation}
\Delta_{1,ij}=g_{1}\langle a_{i\uparrow}b_{j\downarrow}-a_{i\downarrow}b_{j\uparrow}\rangle\label{eq:Delta10}\end{equation}
 for nearest neighbors and zero otherwise, where $g_{0}$ and $g_{1}$
are the coupling strengths. In momentum space, the latter state is
described by \begin{equation}
\Delta_{1,\mathbf{k}}=\sum_{i=1}^{3}\Delta_{1,i}\,\mbox{e}^{i\mathbf{a}_{i}\cdot\mathbf{k}}\,,\label{eq:delta1}\end{equation}
where $\Delta_{1,i}\equiv\Delta_{1}(\mathbf{a}_{i})$  are the
real space pairing amplitudes along the three different bond directions
in the honeycomb lattice (see Fig.~\ref{bruno3}). In the simplest case the pairing amplitudes
are the same,  $\Delta_{1,i}\equiv\Delta_{1}$, and $\Delta_{1}$ is real,
giving  \begin{equation}
\Delta_{1,\mathbf{k}}=\Delta_{1}\phi_{\mathbf{k}}\,,\label{eq:Delta1Iso}\end{equation}
where $\phi_{\mathbf{k}}=\sum_{i=1}^{3}\mbox{e}^{i\mathbf{k}\cdot\mathbf{a}_{i}}$
gives the hopping matrix element in the single particle tight-binding
spectrum \cite{Uch07}. This order parameter represents
the pairing between electronic states in opposite sides of the BZ,
and preserves all the physical symmetries of the honeycomb lattice,
including point group and time-reversal symmetry, $\Delta_{1,\mathbf{k}}=\Delta_{1,-\mathbf{k}}^{*}$,
where the momentum \textbf{$\mathbf{k}$ }is measured with respect
to the center of the BZ, at the $\Gamma$ point. In real space, this
order parameter (OP) has  \emph{extended} $s$-wave symmetry. If expanded around
the Fermi surface centered at the Dirac point $\mathbf{K}$, from
the perspective of the quasiparticle excitations near the Fermi level,
\begin{equation}
\Delta_{1,\mathbf{K}+\mathbf{p}}=\Delta_{1}\mbox{e}^{i\theta}(p_{x}+ip_{y})\label{eq:Delta1k}\end{equation}
describes a $p+ip$ state in one valley and $p-ip$ in the opposite
one \cite{Uch07}. This state is therefore a $p+ip$
state with additional valley degeneracy. Unlike the case of conventional $p+ip$ superconductivity, 
the time reversal
operation involves an additional exchange of valleys, preserving  the TRS
of this state, and we shall refer to it as $p+ip$.

\begin{figure}
\centering
\includegraphics[width=0.4\textwidth]{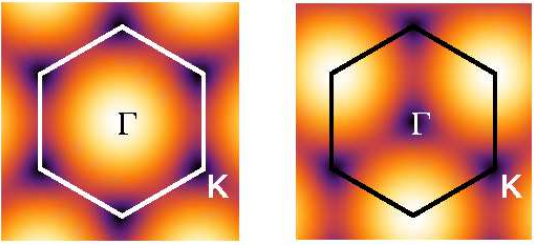}
\caption{{\small (color online) Order parameter (OP) amplitude, $|\Delta_{1,\mathbf{k}}|$,
in the BZ: (left panel) $\Delta_{1,j}=\Delta_{1}$ with $j=1,2,3$
indexing the three different bond directions of the crystal {[}see
Eq. (\ref{eq:delta1})] and (right panel) $\Delta_{1,j}=\mbox{e}^{i2\pi j/3}$,
which describes a flux phase. Light colors represent higher
amplitude. Dirac points are located at the $K$ points, at the edges
of the BZ. In all dark spots, the OP has $p+ip$ symmetry around the
respective high symmetry points. In the three light spots on the right
panel, the OP has $s$-wave symmetry around the $K^{\prime}$ points.
\label{fig:Order parameter}}}
\end{figure}

Another possible paring symmetry is the state \cite{Bla07, Jia08}
 \begin{equation}
\Delta_{1,j}=\Delta_{1}\mbox{e}^{i(2\pi/3)j}\,,\label{eq:delta12}\end{equation}
$j=1,2,3$, which describes on the lattice a real space pairing wavefunction
with $d_{x^{2}-y^{2}}+id_{xy}$-wave symmetry, breaking TRS. This
broken symmetry is caused by the circulation of plaquette current
loops, which amounts to  global circulation of current along the edges.
The low energy description of this state around the Dirac points is
a combination of $s$-wave in one valley and  $p+ip$ state in the
opposite valley \cite{Jia08}, as shown in Fig.~\ref{fig:Order parameter}.
At the mean field level, this state was shown to have lower energy
than the pure $p+ip$ state \cite{Bla07}. Due
to the broken TRS, disorder and quantum fluctuations,
which are paramount in a 2D system, may strongly inhibit the coherence
of the $d+id$ state. Other alternatives are the degenerate states
with $d_{x^{2}-y^{2}}$ and $d_{xy}$-wave symmetries, represented
by the $\Delta_{1,i}$ pairing amplitudes $(2,-1,-1)$ and $(0,1,-1)$,
respectively \cite{Bla07}. These states conserve TRS
but lower the crystal point group symmetry. 

In the spin triplet channel, the OP is a superposition of $S_{z}=-1,0,+1$
states. Since on-site pairing is forbidden by the Pauli principle,
for nearest neighbors interaction  the triplet superconducting states
are $\Delta_{ij,\sigma\sigma}^{t}=\langle a_{i\sigma}b_{j\sigma}\rangle$,
with $\sigma=\uparrow,\downarrow$ for $S_{z}=\pm1$, and $\Delta_{0,\uparrow\downarrow}^{t}=\langle a_{i\uparrow}b_{j\downarrow}+a_{i\downarrow}b_{j\uparrow}\rangle$,
in the $S_{z}=0$ channel. The OP in this case is commonly defined
as a 2$\times$2 tensor, \begin{equation}
\Delta_{ij}=i\sigma_{2}{\bm \sigma}\cdot\mathbf{d}_{ij}\,,\label{eq:triplet}\end{equation}
where the Pauli matrices act  in spin space, and $\mathbf{d}_{ij}=-\mathbf{d}_{ji}$
is an anti-symmetric tensor, violating parity. The case where the
OP $\mathbf{d}$ has a single vector component describes the spinless
fermionic case, discussed by \onlinecite{Ber09}. The possibility
of spin triplet states beyond nearest neighbors in the $S_{z}=0$
channel was recently examined in a variational cluster approximation
calculation \cite{Sah09}. Another possibility is
a  Kekule superconducting  state in the triplet channel, which
breaks the translational symmetry of the lattice and allows the presence
of topological excitations \cite{Roy10}.

At the level of nearest neighbor sites, the electron-electron interaction
can be decomposed into an effective local Hubbard term, 
\begin{equation}
\mathcal{H}_{I}^{0} = 
 \frac{g_{0}}{2}\sum_{i\sigma}\left(a_{i\sigma}^{\dagger}a_{i\sigma}a_{i-\sigma}^{\dagger}a_{i-\sigma} 
+b_{i\sigma}^{\dagger}b_{i\sigma}b_{i-\sigma}^{\dagger}b_{i-\sigma}\right),\label{HI0}
\end{equation}
and a non-local part,\begin{eqnarray}
\mathcal{H}_{I}^{1} & = & g_{1}\sum_{\langle ij\rangle}\sum_{\sigma\sigma^{\prime}}a_{i\sigma}^{\dagger}a_{i\sigma}b_{j\sigma^{\prime}}^{\dagger}b_{j\sigma^{\prime}}.\label{HI1}\end{eqnarray}

In the singlet pairing channel, the non-local term can be decomposed
into $\mathcal{H}_{I}^{1}=g_{1}\sum_{\langle ij\rangle}\left(-\mathcal{B}_{ij}^{\dagger}\mathcal{B}_{ij}+\mathcal{D}_{ij}^{\dagger}\mathcal{D}_{ij}\right)$,
plus one body terms that can be absorbed into the chemical potential
$\mu$. $\mathcal{D}_{ij}=a_{i\uparrow}b_{j\downarrow}-a_{i\downarrow}b_{j\uparrow}$
is a standard singlet pair operator 
 and $\mathcal{B}_{ij}=\sum_{\sigma}a_{i\sigma}^{\dagger}b_{j\sigma}$
is a bond operator. Decomposition of the interaction at the mean field
level with $\langle\mathcal{B}_{ij}\rangle=0$ results in the graphene
tight-binding Hamiltonian for the superconducting phase, $\mathcal{H}^{s}=\sum_{\mathbf{k}}\Psi_{\mathbf{k}}^{\dagger}\hat{\mathcal{H}}^{s}\Psi_{\mathbf{k}}+E_{0}$,
where \begin{equation}
E_{0}=-|\Delta_{0}|^{2}/g_{0}-3\Delta_{1}^{2}/g_{1}\,,\end{equation}
and 
 \begin{equation}
\hat{\mathcal{H}}_{\mathbf{k}}^{S}=\left(\begin{array}{cccc}
-\mu & -t\phi_{\mathbf{k}} & \Delta_{0} & \Delta_{1,\mathbf{k}}\\
-t\phi_{\mathbf{k}}^{*} & -\mu & \Delta_{1,-\mathbf{k}} & \Delta_{0}\\
\Delta_{0}^{*} & \Delta_{1,-\mathbf{k}}^{*} & \mu & t\phi_{\mathbf{k}}\\
\Delta_{1,\mathbf{k}}^{*} & \Delta_{0}^{*} & t\phi_{\mathbf{k}}^{*} & \mu\end{array}\right)\label{eq:H_s}\end{equation}
is the  Bogoliubov-de Gennes matrix written in the sublattice and Nambu basis         $\Psi_{\mathbf{k}}=(a_{\mathbf{k}\uparrow},b_{\mathbf{k}\uparrow},a_{-\mathbf{k}\downarrow}^{\dagger},b_{-\mathbf{k}\downarrow}^{\dagger})$.

\begin{figure}[t]
\centering
\includegraphics[width=0.45\textwidth]{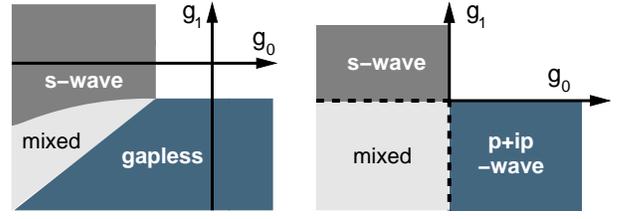}
\caption{{\small Phase diagram between the $s$-wave and effective $p+ip$
phases in the spin singlet channel. On
the left: $\mu=0$ case, which is quantum critical. Right: $\mu\neq0$
case. Continuous  lines represent  second order transitions, and  dashed
lines represent   first order transitions \cite{Uch07}.}}
\label{fig:Phase-diagram}
\end{figure}

The Hamiltonian (\ref{eq:H_s}) can be diagonalized in a basis of 
Bogoliubov quasiparticles: $H^{s}=\sum_{\mathbf{k}\alpha s}E_{\mathbf{k},\alpha,s}\hat{n}_{\mathbf{k},\alpha,s}^{B}+E_{0}$,
where $\hat{n}^{B}$ is the  quasiparticle number operator and $s,\alpha=\pm1$. 
In the isotropic case, $\Delta_{1,\mathbf{k}}=\Delta_{1}\phi_{\mathbf{k}}$,
the spectrum is $E_{\mathbf{k},\alpha,s}=\alpha E_{\mathbf{k},s}$,
with \cite{Uch07} \begin{equation}
E_{\mathbf{k},s}=\sqrt{\left(t|\phi_{\mathbf{k}}|+s\mu\right)^{2}+\left(|\Delta_{0}|+s\Delta_{1}|\phi_{\mathbf{k}}|\right)^{2}}\,,\label{eq:E_k}\end{equation}
where the phase of the OP $\Delta_{0}$ is locked in with $\Delta_{1}$,
 and $\Delta_{1}$ is real. The electronic gap described by the
spectrum (\ref{eq:E_k}) is \begin{equation}
E_{g}=2|t\Delta_{0}-\mu\Delta_{1}|/\sqrt{t^{2}+\Delta_{1}^{2}}\,.\end{equation}
In the $p+ip$ state ($\Delta_{0}=0,\,\Delta_{1}\neq0$), $E_{g}$
is proportional to the deviation of the chemical potential away from
half-filling, and at $\mu=0$ this state becomes quantum critical
and gapless. The instability in this case translates into the renormalization
of the Fermi velocity, where $\bar{t}=t\sqrt{1+\Delta_{1}^{2}}$ is
the renormalized hopping amplitude, instead of the opening of a gap
\cite{Uch07}. Minimization of the free energy \begin{equation}
F=-T\sum_{\mathbf{k},s}\ln\left[2+2\mbox{cosh}\!\left(E_{\mathbf{k,}s}/T\right)\right]+E_{0}\,,\label{eq:(F)}\end{equation}
 with respect to $\Delta_{0}$ and $\Delta_{1}$ 
gives a set of two coupled BCS-like equations, and leads to the
phase diagram shown in Fig.~\ref{fig:Phase-diagram}. At half-filling,
$\mu=0$, the emergence of superconductivity is controlled by quantum
critical lines in the parameters $g_{0}$ and $g_{1}$, with critical
values $g_{0}^{c}=-\pi v^{2}/D$ and $g_{1}^{c}=-4\pi v^{4}/D^{3}$,
in the linear cone approximation, where $D$ is an ultraviolet cut-off
and $v$ is the Fermi velocity near the Dirac point \cite{Cas01, Uch07, Mar06, Zha06}. 
For finite $\mu$, there is a crossover to the standard Fermi liquid case at weak coupling, 
as shown in Fig.~\ref{fig:Phase-diagram}. 

When $\Delta_{1,j}=\Delta_{1}\mbox{e}^{i(2\pi/3)j}$ {[}see Eq. (\ref{eq:delta1})],
the electronic wavefunctions collect different phases along the different
bond links, which gives rise to a current flow, and the $d+id$ state 
cannot coexist with an isotropic TRS $s$-wave state. 
The gap properties of the $d+id$ state and the differential
conductance in SN junctions were derived by \onlinecite{Jia08}.
The Josephson current for this state in SNS junctions was calculated
by \onlinecite{Lin09}.

\begin{figure}
\centering
\includegraphics[width=0.35\textwidth]{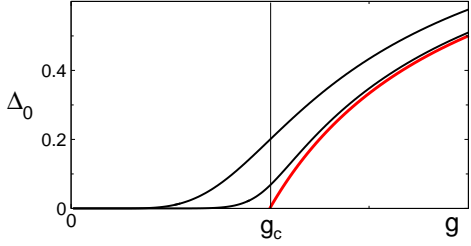}
\caption{{\small Dependence of the gap, normalized by the band cut-off $D$,
on $g$ in the weak $(g<g_{c})$ and strong coupling $(g>g_{c}$)
sectors for $\mu=0$ (red line), $\mu/D=0.1,$ and $0.3$. The model has 
 a QCP at half filling \cite{Uch05}. }}
\label{fig:Superconductor-Gap}
\end{figure}

In the $s$-wave state (we assume $\Delta_{0}$ to be real),
the gap variation with the coupling at half filling, near the quantum
critical point $g_{0}^{c}=-\pi v^{2}/D$, is  \cite{Cas01} \begin{equation}
\Delta_{0}=D(1-g_{0}^{c}/g_{0})\,.\label{eq:Delta}\end{equation}
Away from half-filling, the gap  crosses
over to \cite{Uch05} \begin{equation}
\Delta_{0}=2|\mu|\mbox{exp}\left[D(1-g_{0}^{c}/g_{0})/|\mu|-1\right]\label{eq:delta2}\end{equation}
for $|\mu|\gg\Delta_{0}$, which corresponds to the weak coupling
BCS limit, where $g\ll g_{c}$, as shown in Fig. \ref{fig:Superconductor-Gap}.
The $|\mu|/\Delta_{0}\ll1$ limit corresponds to the strong coupling
regime ($g>g_{c}$), and the intermediate coupling region near $g\sim g_{c}$
sets the crossover scale between the two regimes at finite $\mu$.
Non-equilibrium effects in the presence of a dissipative
    environment may also lead to a dissipation driven quantum phase
    transition away from half filling \cite{Tak08}.

At mean field level, the critical temperature at $\mu=0$ is $T_{c}=\Delta_{0}/2\ln4$,
whereas in the opposite limit, $|\mu|\gg\Delta_{0}$, $T_{c}=\gamma\Delta_{0}/\pi$,
as in the BCS case, where $\ln\gamma\approx0.577$ is the Euler constant
\cite{Uch05}. 
Of course, in two dimensions there is no true long range
    order. The superconducting transition is of Kosterlitz-Thouless (KT)
    type and  coherence is actually lost at much lower temperatures
    due to the role of thermal fluctuations, which unbind vortex and
    anti-vortex pairs above the KT transition temperature, at
    $T_{KT}<T_c$. The mean field result  indicates  the onset of
    critical fluctuations where the amplitude of the Cooper pairs is
    completely destroyed, although the phase coherence is suppressed
    much earlier, at $T_{KT}$. The KT fluctuations of the SC order parameter
    have been considered by  \onlinecite{Lok09}, without accounting, nevertheless,
    for the chiral nature of the quasiparticles in graphene.  

 Zero field thermodynamic
properties, such as the specific heat at fixed volume, $
C_{V}=-T(\partial^{2}F/\partial T^{2})_{V}$, 
can be extracted from the free energy (\ref{eq:(F)}). For an isotropic
condensate of Dirac fermions, the jump of the specific heat at the
phase transition, normalized by the specific heat on the normal side,
is \cite{Uch05} \begin{equation}
\delta C_{V}=2(\ln4)^{2}/[9\zeta(3)] \approx 0.35,\label{eq:Cv}\end{equation}
at half-filling. In the $|\mu|/\Delta_{0}\gg1$ limit, the jump grows
to the standard BCS value $
 \delta C_{V}=12/[7\zeta(3)] \approx 1.43$.

 The Meissner effect in graphene, which describes the
    expulsion of an external magnetic field by the circulation of
    diamagnetic supercurrents,  has been recently examined by
    \onlinecite{Kop08} and \onlinecite{Uch09a}. 
In the presence of vortices, the Bogoliubov de Gennes equations for
    Dirac fermions in 2+1 dimensions allow the presence of zero energy 
    modes \cite{Jac81} which are bound to 
    the vortex cores. For a vortex with vorticity $n$ (the winding 
    number of the OP), $\Delta_{0}=|\Delta_{n}(r)|\mbox{e}^{in\phi}$, 
    with $(r,\phi)$ as cylindrical coordinates. The physical solutions
    allowed by the boundary 
    conditions at the center of the vortex and at infinity result in
    $n$ zero modes at 
    half filling \cite{Gha07}. The subgap spectrum and the 
    wavefunctions in the vortex core have been derived by
    \onlinecite{Ber09,Ser08b}.
Away from half filling, for \emph{odd} vorticity $n$, there is only
one energy branch that crosses zero energy for zero angular momentum.
For $n$ \emph{even}, no subgap branch intersects  zero energy,
and no exact zero modes exist \cite{Ber09, Kha09}. Because of the fermionic degeneracy in the valleys,
the topological zero modes do not lead to fractional statistics
under vortex exchange, as in conventional $p+ip$ superconductors, unless
additional interactions that  lift the fermionic degeneracy are included \cite{Her10}. 
Vortex zero modes for excitonic condensates in bilayers have been discussed
by \onlinecite{Ser08c}.

\section{Interactions at boundaries and lattice defects}
\label{Paco6}
\subsection{Surface states}
The vanishing density of states of graphene at the neutrality point
implies that localized states can exist at the Dirac energy, much in
the same way as localized states appear inside a forbidden energy
gap in semiconductors and insulators. In order for these states to
be normalizable, special boundary conditions are required. These
conditions imply the breaking of the translational symmetry of the
lattice, so that they can only exist near edges or defects.

\begin{figure}
\includegraphics[width=0.9\columnwidth]{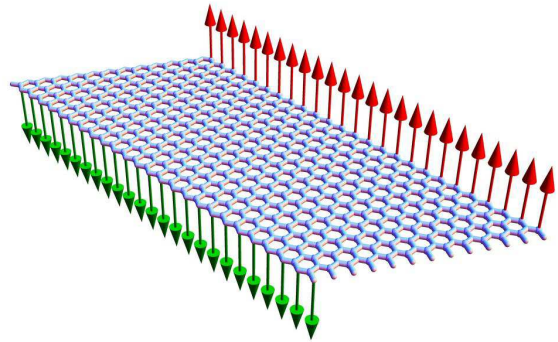}
\caption[fig]{(Color online). Sketch of the magnetization at the zigzag edges of a graphene ribbon.}
 \label{fig_ribbon}
\end{figure}

The most extensively studied examples are the surface states which
exist at graphene zigzag edges, where the lattice is abruptly
terminated  \cite{FWNK96,Netal96}. Such edges have been observed in
graphene flakes \cite{Getal09d,Jetal09}, and also in
graphite \cite{Netal05b}. As the localized states form an energy band
of zero width, the local density of states at the Dirac energy near
a zigzag edge changes from zero to infinity, and the electron
compressibility becomes divergent. Interactions of arbitrarily small
strength lead to instabilities when the Fermi energy lies at the
Dirac point. A mean field analysis showed that a short range Hubbard interaction 
can lead to a ferromagnetic ground state \cite{H01,HE02}. In zigzag
ribbons with two edges, the spins at the two edges are aligned
antiferromagnetically, see Fig.~\ref{fig_ribbon}. These early theoretical results, based on the tight
binding approximation, were later confirmed by calculations based on
the Local Density Approximation \cite{SCL06,PCMH07}. The
ferromagnetic order remained when the dangling bonds at the zigzag
edges where saturated by hydrogen, which probably is closer to the
actual experimental situation. The optimization of the atomic
positions at the edges leads to reconstructed phases with gaps, 
where the spin up and spin down bands do not overlap near the gap,
suggesting a half metallic phase \cite{SCL06}. Other phases, with
ferroelectric properties \cite{F08} or canted moments have been
studied \cite{JM09}. A sketch of the magnetization induced near a
zigzag edge of a graphene ribbon is shown in Fig.~\ref{fig_ribbon}.
Recent experiments \cite{ET09,Jetal10} confirm the existence of
magnetic moments at graphene edges.

The effects of the electron-electron interaction on the midgap
states has also been studied beyond the mean field approximation.
The calculations show that the ferromagnetic phase is stable when
the  band of localized states is half filled. Both a local onsite
interaction or the long range exchange effect lead to this phase. At
very low fillings, electrons tend to form a charge density wave
state, similar to a Wigner crystal \cite{WSG08,WSSG08}. More complex
correlated states are possible at other fillings. The fact that the
midgap states at a zigzag edge resemble the wavefunctions of Landau
levels, in that the momentum parallel to the edge and the spatial
extension are coupled, leads to the intriguing possibility of states
similar to the Laughlin wavefunctions which describe the Fractional
Quantum Hall Effect \cite{WSG08}.

At long distances, straight graphene edges of arbitrary orientation
other than armchair can support midgap states, as zigzag
edges \cite{AB08}. Hence, local magnetic moments can be a generic
property of abrupt graphene edges. Zigzag edges and vacancies in
bilayer (Bernal) graphene also give rise to midgap states, at least
when only the direct nearest neighbor interlayer hopping is
included \cite{CPLNG08}, and magnetic moments can be formed at the
edges of bilayer graphene \cite{SMMB08}. Models which include other
interlayer hoppings lead to sharp resonances near edges and
vacancies. These results suggest that moderate interactions can produce 
 local moments in graphene bilayers or in three dimensional
graphite. The combination of the Zeeman field associated with
 magnetic ordering, and the spin orbit coupling can lead to phases 
characterized by quantized spin currents at the edges \cite{SF10}.
\subsection{States at vacancies and cracks}
\begin{figure}
\includegraphics[width=\columnwidth]{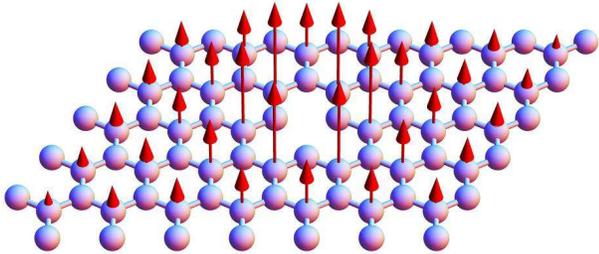}
\caption[fig]{(Color online). Sketch of the magnetization induced near a vacancy.}
 \label{fig_vacancy}
\end{figure}
Midgap states can occur in other situations where the translational
symmetry of the lattice is broken. Similarly to the case of surface
states at a zigzag edge, interactions will lead to the spin
polarization of these states. The simplest situation where the
existence of a partially localized midgap state can be demonstrated
is a lattice vacancy \cite{Per06,Pereira:2008-1}. This analysis can be
extended to
multilayer samples \cite{CLV10}.

The existence of these states has been confirmed by STM spectroscopy
on vacancies in irradiated graphite \cite{UBGG10}. It can be expected
that interactions lead to the
formation of a magnetic moment around the vacancy. The formation of
local moments near vacancies is consistent with the observation of
ferromagnetism in irradiated graphite
\cite{Eetal03,Betal07,Oetal07b,Retal10,CCWF10}.  Absorption of
hydrogen leads to similar effects to those of a vacancy, including
the formation of magnetic moments \cite{Y08}. Other dopants, like
carbon atoms and NO$_2$, also lead to the formation of
spins \cite{Letal03,Wetal08b}.

A sketch of the magnetization induced near a graphene vacancy is shown
in Fig.~\ref{fig_vacancy}. The moment associated with the localized
level around the vacancy is coupled to the extended states, leading to
the possibility of the Kondo effect. Some differences between usual
magnetic impurities and the situations described here can be expected:
i) The vacancy or adatom modifies significantly the electronic density
of states, rendering invalid perturbative treatments which relate the
magnitude of the exchange coupling to the unperturbed electronic
structure. The phase shift induced in the conduction band remains
significant, even near the Dirac energy \cite{HG07}. ii) The localized
state is orthogonal to the extended states. Hence, the coupling
between the local moment and the conduction band does not take place
via virtual hops between the two types of states. Instead, it can be
expected that the electron-electron interaction favors a ferromagnetic
alignment of the local moment and the spins of the conduction
electrons.

Spins at different vacancies interact ferro- or
antiferromagnetically \cite{Bre07,PFB08}, depending on whether the
vacancies occupy the same or different sublattices. At half filling,
the RKKY interaction mediated by the $\pi$ band decays as $1/ |
{\bf r} - {\bf r'} |^3$, and it goes to the $1/ | {\bf r}
- {\bf r'} |^2$ dependence typical of a two dimensional electron
gas at finite carrier concentrations \cite{Che06}. Voids or cracks
can be considered an intermediate case between vacancies and
edges \cite{VLSG05}. They also support localized spins at the
boundaries.

\subsection{Midgap States  and Random Gauge Fields}
\label{sec:PacoDefects}
Midgap states in bulk graphene can also be induced by magnetic
fields (see below), or by strains which mimic the effect of a
magnetic field \cite{GKV08}. These states have been analyzed using
the tight binding approximation \cite{GKV08}, or by means of the
Local Density Functional method \cite{Wetal08}. Corrugations and
wrinkles also induce midgap states in  graphene
\cite{KP08,Pereira:2010}. The presence of these states enhances the
effects of the interactions. Mean field
calculations suggest the formation of magnetic moments, which will
order ferro- or antiferromagnetically \cite{GKV08,GHL08}.

A random strain distribution leads to a random gauge field acting on
the electrons. The changes in the electronic density of states
induced by a random gauge field have been studied by RG
 techniques \cite{LFSG94,HL02}. Related problems arise at the
transition between plateaus in the Quantum Hall Effect, and in d-wave
superconductors.
It can be shown that, above a
certain disorder strength, a random gauge field leads to a divergent
density of states at the Dirac energy \cite{RH01,HL02}. This
divergence leads to a vanishing electron compressibility, and 
 enhances the effects of interactions in the same way as
the midgap states considered earlier.  A random gauge field, ${\bf A} ({\bf r})$, can be characterized by a dimensionless number, $\Delta$,
\begin{equation}
\langle A_{\mu}({\bf r}) A_{\nu}({\bf r'})\rangle = \Delta \delta_{\mu \nu} \delta^{(2)}({\bf r}-{\bf r'}).
\label{gauge}
\end{equation}
If the gauge potential is assumed to arise from random corrugations of
average height $h$ and length $\ell$, then $\Delta \sim h^4 / ( a^2
\ell^2 )$, where $a$ is the lattice constant \cite{GKV08,GHL08}. A
similar parameter can be defined if the gauge potential is due to
topological defects, such as dislocations \cite{GGV01}. The regime
$\Delta \sim 1$ corresponds to ripples large enough to accommodate
midgap states, leading to a divergence in the density of states.
The changes in the density of states induced by a gauge field can be written as a logarithmic renormalization of the Fermi velocity
\begin{equation}
v \rightarrow v \left[ 1 - c \Delta \log \left( \frac{\Lambda}{|{\bf k}|} \right) \right],
\label{ren_gauge}
\end{equation}
where $c$ is a numerical constant, and $\Lambda$ is a high momentum cutoff of the order of the inverse of the lattice constant.

The scaling towards lower Fermi velocities in eq.~\ref{ren_gauge} can be combined with the RG
 analysis of the long range
Coulomb interaction \cite{Stauber05,FL06,FL06b,FA08}. Disorder tends
to increase the density of states near the Dirac energy, while
interactions lead to the opposite effect. To lowest order,
this analysis leads to a line of fixed points characterized by a
finite disorder and finite interactions,
 as discussed in Sec.~\ref{subsubsec:Weak-coupling}, see Fig.~\ref{figdisorder}. The temperature and
frequency dependence of properties such as the conductivity or the
specific heat acquire anomalous exponents \cite{HJV08}. For high
disorder, $\Delta \gtrsim 1$, it can be shown that a gapped state is
more stable than the gapless density of states expected in the absence
of interaction effects \cite{GHL08}.

Certain strain configurations lead to effects similar to those induced
by a constant magnetic field \cite{GKG10}. 
The possible ways in which the degeneracies of
these states are lifted by the interactions have been
studied \cite{H08}, and new phases, with properties similar to those
of topological insulators may exist. It is worth noting that STM
experiments suggest the existence of very large effective fields due
to strains, $B_{eff} \sim 300$T, in small graphene bubbles under high
strains \cite{Letal10b}. The effects of  electron-electron
interactions in this regime remain unexplored.
\begin{figure}
\includegraphics[width=0.8\columnwidth]{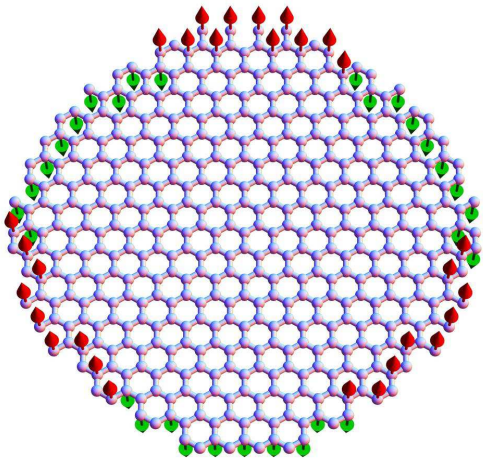}
\caption[fig]{(Color online). Sketch of the magnetization induced at the edges of a quantum dot.}
 \label{fig_dot}
\end{figure}

\section{Interaction effects in mesoscopic systems}
\subsection{Magnetism in quantum dots}
Mesoscopic samples have a large ratio between the perimeter and the
area. Midgap states localized at the edges can have a significant
weight in the total density of states, and interaction effects are
enhanced. Early calculations for planar carbon
molecules  \cite{SB87,TMDM98} showed gaps associated with the
electron-electron interaction, and magnetic moments at the edges. A
large magnetic moment can be found in triangular graphene
flakes  \cite{FP07}, where the three boundaries have the zigzag
orientation, and the carbon atoms at the edges belong to the same
sublattice.

\begin{figure}
\includegraphics[width=1.5\columnwidth,angle=-90]{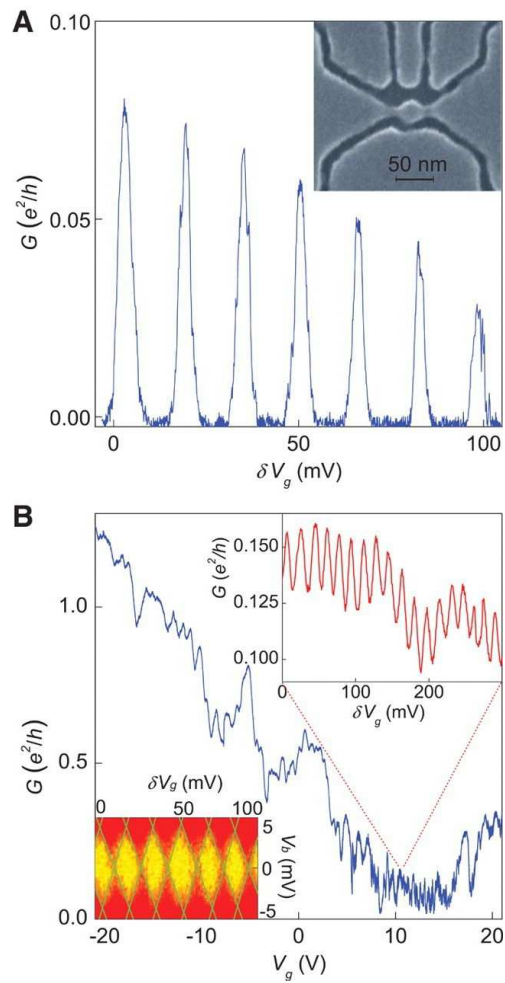}
\caption[fig]{(Color online). Single energy peaks and Coulomb diamonds
in a graphene quantum dot, see \cite{Petal08}.}
 \label{fig_dot_exp}
\end{figure}
\begin{figure}
\includegraphics[width=0.7\columnwidth]{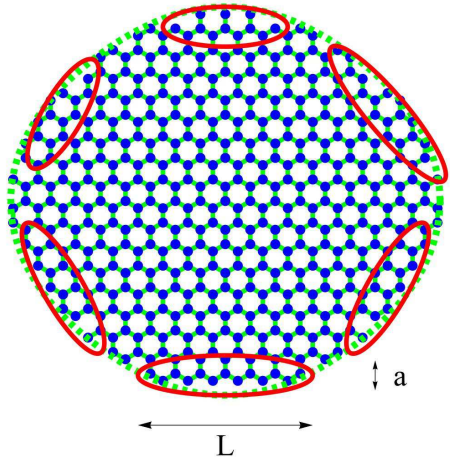}
\caption[fig]{(Color online). Sketch of the extension of edge states in a graphene quantum dot.}
 \label{fig_dot_blockade}
\end{figure}

As mentioned previously, edges of arbitrary orientations, except the
armchair direction, support midgap states  \cite{AB08}. Hence, local
moments and magnetism can be expected in graphene quantum dots of
any shape, provided that the termination at the edges is abrupt.
Model results suggest that this is the case, and the orientation of
the moments at the edges depends on the type of sublattice at the
edge \cite{FP07}, as sketched in Fig.~\ref{fig_dot}. Away from half
filling, correlated states with
unsaturated magnetization, and charge density wave states are also
possible \cite{WSG08,RYL09}. The charging of a quantum dot leads to a
substantial rearrangement of the electronic levels, in a similar way
to the well studied orthogonality catastrophe in
metals \cite{A67,WSG08}. The conductance can acquire a non trivial
voltage or temperature dependence, as in a Luttinger
liquid \cite{KF92}.

 A simple estimate of the number of magnetic moments in a quantum dot
can be obtained by assuming that the average density of edge states
is of order $\rho_{edge} \approx c \times[  R / ( a W )]$,
where $c \sim 1$ is a numerical constant, $R$ is the radius of the
dot, $a$ is the lattice spacing, and $W$ is the bandwidth of the band
of edge states \cite{WAG10}. The Coulomb interaction within each
state, which leads to the formation of local moments is $E_c \approx
e^2 / R \times \log ( R / a )$, see below. Naturally, one has to replace $e^2 \rightarrow e^2/\epsilon_0$
 in all formulas, but we do not write the dielectric constant explicitly in this section. 
The states which are spin
polarized are those whose distance from the Fermi energy is less than
$E_c$. This condition, combined with the estimate for
 $\rho_{edge}$, gives a maximum number of magnetic moments
within the dot, $N \approx E_c \rho_{edge} \approx c \times [ e^2 / ( a W
) ] \times \log ( R / a )$. This number is not too large. For $W \sim
0.3 - 0.5$eV and $R \sim 100$nm we obtain $N \sim 10 - 20$. The total
magnetic moment of the dot depends on the sign of the couplings
between the edge spins, see Fig.~\ref{fig_dot}.

Experimentally, there is evidence which suggests the formation of local
moments in small graphene flakes, of dimensions $10-50$nm
\cite{Setal10}.

\begin{figure}
\includegraphics[width=\columnwidth]{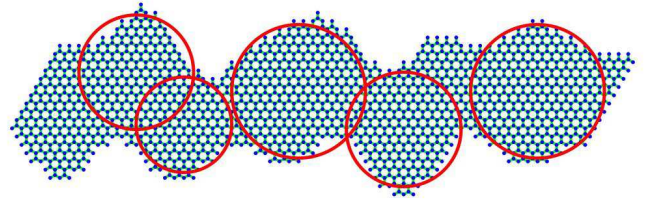}
\caption[fig]{(Color online). Sketch of a graphene ribbon with disordered edges as a series of quantum dots.}
 \label{fig_blockade}
\end{figure}
\begin{figure}
\includegraphics[width=0.9\columnwidth]{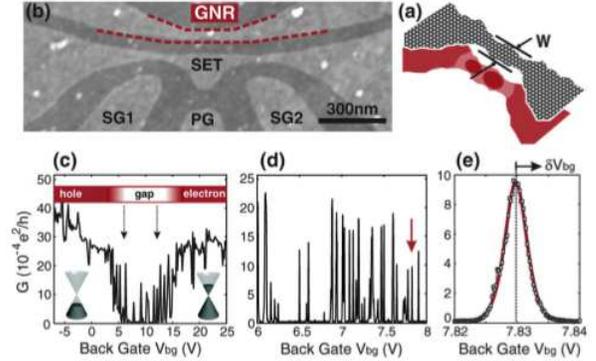}
\caption[fig]{(Color online). Graphene point contact coupled to a
quantum dot, see \cite{Setal09}.}
 \label{fig_constriction}
\end{figure}
\subsection{Charging effects. Coulomb blockade}
Graphene quantum dots of many shapes and dimensions are being
extensively
studied 
\cite{Betal05,Oetal07,HSSTYG07,HOZK07,WDM07,ACP07,Petal08,Setal08,
Setal08b,MB09,Getal09,Metal09b}.
Single electron effects have been observed in many of them. Experiments show clear evidence of charging effects in graphene
quantum dots, as evidenced in the diamond patterns formed by the
resonances in the conductance through the dot as a function of gate
and bias
voltages 
\cite{Petal08,Setal08,Setal08b,Setal09b,Metal09,MB09,Getal09,Metal09b,
RL09,Metal09c}, see Fig.~\ref{fig_dot_exp}.

The electrostatic interaction between electrons leads to Coulomb
blockade, which modulates the energy difference between levels, and
induces non Ohmic features in the conductance through the dot. In a
graphene quantum dot of dimension $R$, the electrostatic energy
required to add a unit of charge scales as $e^2 / R$. The mean level
spacing between extended states in a ballistic dot scales as
$v/R$. As the dimensionless parameter $\alpha = e^2/(\epsilon_0 v)$ in graphene is of
order unity, the energy scales associated with charging and
confinement effects are comparable. The edge states discussed earlier
can lead to charging energies larger than those for extended states.
Assuming that these states are delocalized along the perimeter of the
ribbon, over a scale $L \sim R$, see Fig.~\ref{fig_dot_blockade}, and
 width $a$ comparable to the lattice spacing, see
Fig.~\ref{fig_dot_blockade}, the charging energy becomes $(e^2/R)
\times \log ( R/a )$ \cite{WAG10}.

Charging effects can also modify the transport properties of narrow
graphene ribbons. Irregularities in the edges may induce the
formation of constrictions and quantum dots,  as sketched in Fig.~\ref{fig_blockade}, where charging effects
will lead to a transport gap. In a nanoribbon of width $W$, the
typical size of these dots will  also be  $W$, and the transport gap
will be of order $e^2 / W$. In the absence of charging effects, a
ribbon will have confined subbands, separated by gaps of order
$v/W$. Hence, the  similarity between the energy scales arising
from quantum confinement and charging effects,  which exists in a quantum dot,
also exists in a graphene ribbon. An experimental realization of an all
graphene circuit with a point contact coupled to a quantum dot
\cite{Setal09} is shown in Fig.~\ref{fig_constriction}. This setup can
be used to count the passage of charges through the quantum dot.

Experiments in graphene nanoribbons are compatible with the
relevance of charging effects \cite{HOZK07,TCGA09,HKB09}. Some
observations can be explained by a model of dots formed in the
ribbon connected through many channels with the rest of the
structure. Such a strongly coupled dot always shows  Coulomb blockade
effects, unless there is a perfect transmission through one or more
of the channels. The effective charging energy, however, is strongly
renormalized by the coupling between the dot and the rest of the
system \cite{SGN07}, $E_c \approx e^2 / W e^{-g}$, where $g$ is the
conductance, in dimensionless units, of the junction between the dot
and the electrodes. In general, $g \sim \langle | T |^2 \rangle \times
k_F W$, where $T$ is the transmission amplitude of a given channel.

The electron-electron interactions can be studied in mesoscopic
samples through their effect on the magnetoconductance at low magnetic
fields. These experiments probe the phase coherence of electrons at
low temperatures. This quantum effect is suppressed due to the
dephasing induced by the interactions. Electronic quantum coherence
also gives rise to the universal conductance fluctuations observed in
disordered metals, which are also reduced by the dephasing due to
interactions. The dephasing length shows a temperature dependence
consistent with the expected behavior in a dirty metal, $\ell_\phi
\sim  ( g \hbar v ) / [ T \log ( g ) ]$, where $g$ is the
conductivity in dimensionless units \cite{TKSG10} (see
also \cite{Cetal10}). This dependence is replaced by a $\ell_\phi
\propto T^{-2}$ in high mobility samples \cite{TKSG10}, as expected in
a clean Fermi liquid. Experiments that tune the ratio between the
dephasing length and the mean free path \cite{Metal10} show a variety
of regimes, interpolating between weak and strong localization.

\begin{figure}
\includegraphics[width=0.6\columnwidth]{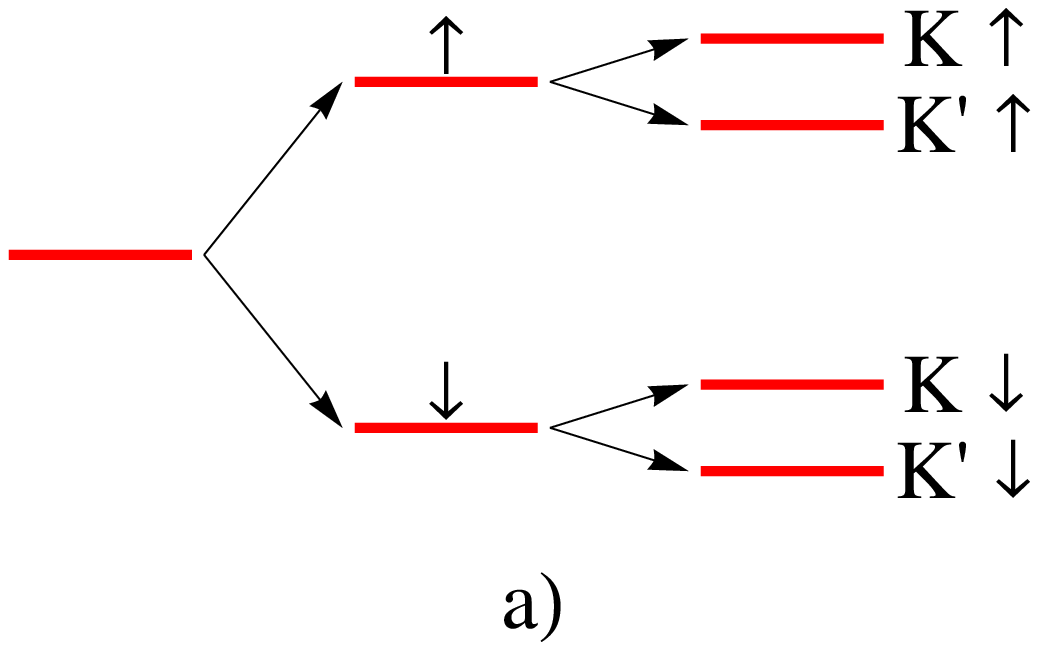}
\includegraphics[width=0.6\columnwidth]{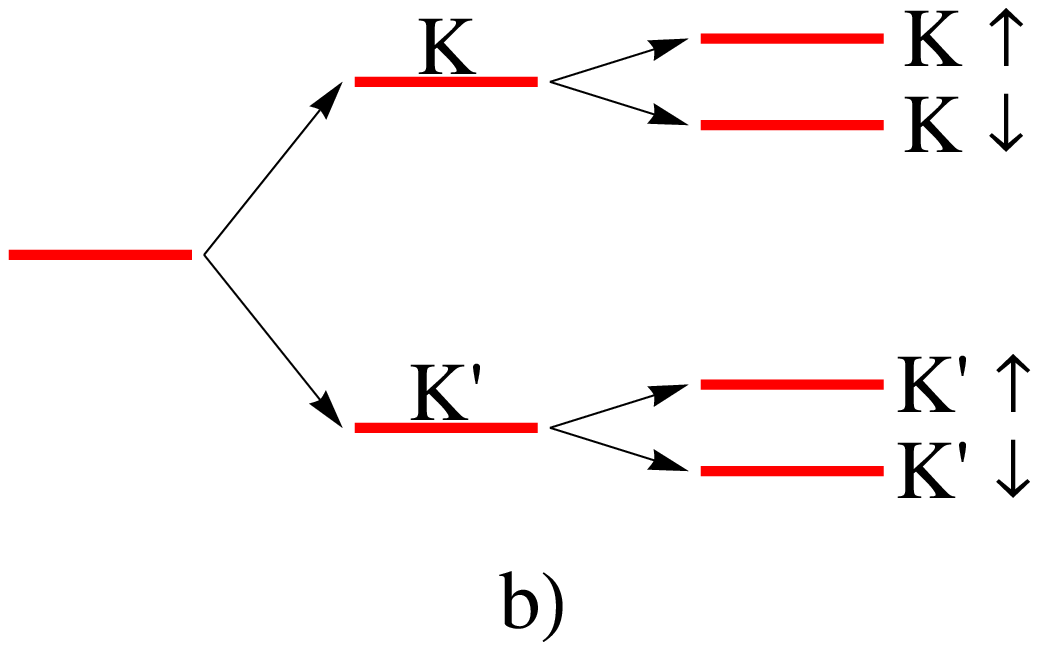}
\caption[fig]{(Color online). Sketch of the successive splittings of the Landau levels as the magnetic field is increased. a) Spin states are split first, and then the valley degeneracy is broken. b) Valley degeneracy is lifted first, followed by the breaking of spin degeneracy.}
 \label{fig_LL}
\end{figure}

\begin{figure}
\includegraphics[width=0.9\columnwidth]{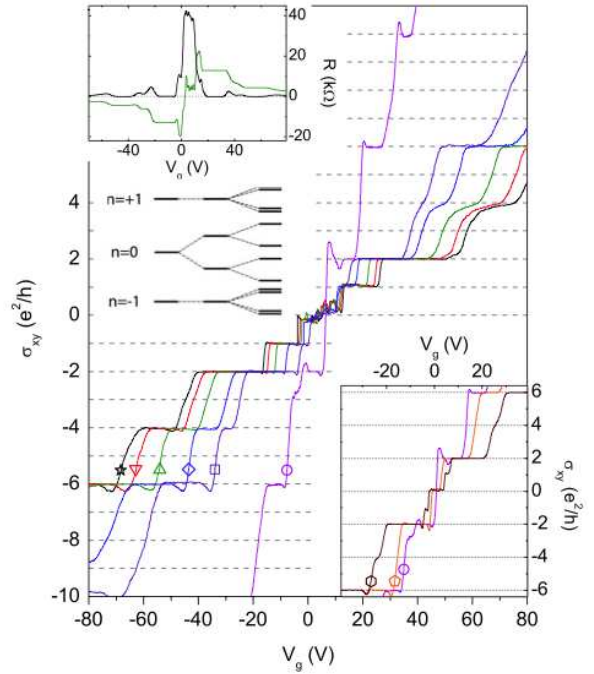}
\caption[fig]{(Color online). Splittings of the Landau levels in
graphene as function of magnetic field, see \cite{Zetal06}.}
 \label{fig_LL_exp}
\end{figure}

\section{Interactions in strong magnetic fields}
A comprehensive review of graphene in magnetic field has  recently appeared \cite{G10},
 and here we only mention some of the main effects. 
The electronic energy bands of graphene in a strong magnetic field
collapse into Landau levels. In the absence of disorder, the
electronic compressibility diverges when the chemical potential
coincides with the energy of a Landau level, and the effects of the
interactions are enhanced, as in other two dimensional metallic
systems. The typical scale of the electronic wavefunctions is the
magnetic length, $\ell_B = \sqrt{\hbar/(e B)} = \sqrt{\Phi_0 / ( 2 \pi B )}$, where $B$ is
the applied field and $\Phi_0$ is the quantum unit of flux. The
separation between levels is of order $v/ \ell_B$, while the
relevant scale for interaction effects is $e^2 / \ell_B$.

\begin{figure}
\includegraphics[width=0.8\columnwidth,angle=-90]{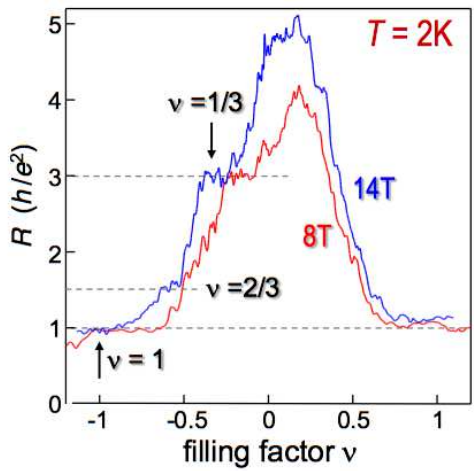}
\caption[fig]{(Color online). Resistance of a suspended graphene sample as a function of carrier density for two different magnetic fields. R. V. Gorbachev, D. C. Elias, A. S. Mayorov, A. A. Zhukov, K.  S. Novoselov, A. K. Geim (unpublished).}
 \label{fig_FQHE}
\end{figure}
There are two sets of Landau levels in graphene, one for each
valley. In addition, graphene has the $n=0$ level, which combines
electron and hole features. Hence, interactions can break either the
valley degeneracy or the spin degeneracy. The long range part of the Coulomb interaction is independent of the valley index. The $n=0$ Landau level is localized in a given sublattice, and its degeneracy can be lifted by interactions which break the symmetry
between sublattices, like the coupling to out of plane optical phonons \cite{FL07}. Hence, the removal of the spin and valley degeneracies of the Landau levels due to interactions depends on other energy scales
 \cite{G10}, such as the Zeeman splitting, or the nearest neighbor repulsion, for the case $n=0$. A sketch of the possible
symmetry breaking patterns as a function of magnetic field is shown in
Fig.~\ref{fig_LL}. Early observations of splittings between Landau
levels are shown in Fig.~\ref{fig_LL_exp}, see \cite{Zetal06}.

It is usually assumed
that the Zeeman splitting is much smaller than the other energy
scales. Calculations suggest that the spin degeneracy is lifted
first, leading to excitations with combined spin and valley
indices \cite{NM06,AF06,YSM06,GMC06,Aetal07,SN08,WIFB08,Getal09c}. The fourfold spin and valley degeneracy when the Zeeman coupling is neglected gives a new $SU(4)$ symmetry, which may lead to new features, not observable in other two dimensional electron gases \cite{GR07,HCJ07}.
 The formation of Landau levels favors the
excitonic transition which can also exist in the absence of a
magnetic field  \cite{GMSS06}. The spin split $n=0$ level leads to
spin polarized edge states \cite{FB06,ALL06,Aetal07,SFP09} where the orientation of the spin depends on the sign of the current, as in topological insulators \cite{QZ08,HK10}.

A magnetic field oriented parallel to the plane does not give rise to
Landau levels. In neutral graphene, it leads to  metallic states
with electrons and holes polarized in opposite directions,  providing
 another route towards an excitonic transition 
\cite{Kharzeev07}.

Experiments show that, indeed, the spin and valley degeneracies of
Landau levels in graphene are
lifted \cite{Zetal06,Getal07,JZSK07,Getal09b}. The opening of a gap in
the $n=0$ level in graphene has been extensively studied, and a
metal insulator transition with critical features consistent with a
Berezinskii-Kosterlitz-Thouless transition has been
reported \cite{CLO08,CLO09,Aetal09}.

The most striking manifestation of the interactions in the presence
of a strong magnetic field is the Fractional Quantum Hall Effect.
Early theoretical calculations showed that the FQHE could be stable
in graphene \cite{CGP06,AC06,TLCJ06}. The conditions for the FQHE are
the existence of sharp Landau levels and sufficiently strong electron
electron interactions. The analysis of FQHE states in graphene can be
done in a similar way to that of  a two dimensional electron gas. The
main difference is a change in the pseudopotentials which describe the
interactions between electrons in a given Landau level, because the
wavefunctions in graphene and in a two dimensional electron gas
differ.

This Fractional Quantum Hall Effect was extensively, but
unsuccessfully, sought in samples deposited on SiO$_2$. Suspended
samples, which showed a much higher electron mobility, did not
exhibit the FQHE, using the standard experimental four terminal
setup. The observation of the IQHE in suspended bilayer graphene
using a two terminal setup \cite{FMY09} led quickly to the discovery
of the FQHE in single layer graphene \cite{DSDLA09,BGSSK09}, using
the same technique. More recently, four terminal measurements in high mobility suspended samples
  \cite{GZCBK10}, and also samples deposited on a new substrate, boron nitride \cite{Detal10,Detal10b}, also show the FQHE. In two terminal measurements, the existence of the FQHE is inferred from plateaus of the longitudinal resistance at carrier densities which correspond to fractional fillings of Landau levels, see Fig.~\ref{fig_FQHE}. The $\nu = 1/3$ state turns out to be more
robust than in other materials, like GaAs, which exhibit the FQHE, and it can be observed at temperatures greater than 10K. Fractional plateaus at
$\nu = 2/3$ and $\nu = 1/2$ have also been reported.
 Theoretical calculations suggest that the so called Moore-Read ground state at fillings with even denominators, which leads to the existence of non Abelian anyonic quasiparticles, is not favored in
graphene \cite{WMC10}.

\section{Interactions in bilayers}

\begin{figure}
  \centering
  \includegraphics[scale=0.32]{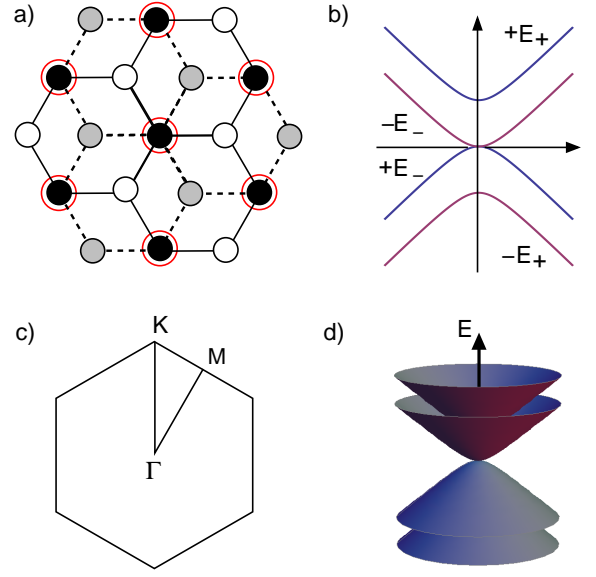}
  \caption{
  a) Top view of a graphene bilayer; white and black  circles:
  top layer carbon atoms; gray and red: bottom layer. 
  b) four-band spectrum of the bilayer, $\pm E_{\gamma}(\mathbf{p})$,
  with $\gamma=\pm$ as shown in Eq. \eqref{eq:spectrumBi}, near the
  corner of the Brillouin zone. 
  c) Brillouin zone with high symmetry points. 
  d) Illustration of the four band spectrum around the $K$
  point.
  }
  \label{bilayer}
\end{figure}

Bilayers are the building blocks for 3D stacks
of graphene, such as graphite. In a bilayer one
has two parallel graphene sheets, separated by an equilibrium
distance similar to the interlayer distance of graphite
(3.35\AA) \cite{Dresselhaus81}. 
The relative position of the two graphene layers is not unique, and
this leads to different stacking arrangements of the bilayer, and
even more possibilities for multilayers, or graphite. 
The most stable configuration seems to be the so-called 
Bernal $AB$ stacking, in which the two layers are rotated by
60$^\text{o}$. As a consequence, one of the
the sublattices in the lower layer (say, sublattice $A$) is vertically
aligned with one of the sublattices of the upper layer (say,
sublattice $B$) [see Fig.~\ref{bilayer}(a)]. Notice that this
particular rotation leads to a breaking of sublattice symmetry between
layers. 
As a first approximation, the electronic coupling between the layers
can be described in terms of the hopping of electrons between the
nearest neighbor atoms in different layers with an energy $t_\perp$
(also known as $\gamma_1 \approx 0.39$ eV in the  graphite literature
\cite{CastroNeto:09}. Another possible
arrangement between the layers is the fully aligned configuration,
also called $AA$ stacking. In both $AB$ and $AA$ stacking, the unit
cell is comprised of 4 atoms, and has the same 2D extension as the
unit cell of a single layer; this implies that the  Brillouin zone
is precisely the same as in monolayer graphene. 

Notice, however, that these configurations are just a few of an infinite
series of commensurate structures between two layers, the so-called
{\it twisted bilayer graphene} \cite{San07}. The problem of
commensurate and incommensurate structures always appears when two
crystalline materials are superimposed, as in the case of bilayers.
For commensurate structures, the angle between the layers is not
arbitrary but follows a well defined sequence \cite{San07}. Obviously,
different angles lead to different broken symmetries and hence to
different electronic states. When the angle of rotation is 60 degrees,
as in the case of the Bernal structure, the sublattices are
nonequivalent, which leads to a broken sublattice symmetry and hence
to a putative gap opening. For other angles, there is no broken
sublattice symmetry but the unit cell is enlarged as the rotation
angle becomes smaller. In this case the massless Dirac dispersion has
to be preserved for symmetry reasons \cite{San07,Mele10,Li10}. From
this perspective, the Bernal configuration is an exception. The
twisted bilayer graphene presents a very rich physics of its own that
we will not cover in this review. Instead, we will focus on the Bernal
configuration which is the most studied case. 

We will start from the minimal tight-binding model for Bernal
bilayers, which includes a basis with two additional layer
flavors (denoted by an overbar),
\begin{equation}
\Psi_{\mathbf{k},\sigma}=(a_{\mathbf{k},\sigma},b_{\mathbf{k},\sigma},
\bar{b}_{\mathbf{k},\sigma},\bar{a}_{\mathbf{k},\sigma})\,,\label{
eq:PsiBi}
\end{equation}
with $\sigma=\uparrow,\downarrow$ representing the spin.
The resulting Bloch Hamiltonian is then a $4\times4$
matrix with two sublattice, and two layer degrees of freedom,
\begin{equation}
\mathcal{H}_{B}=\sum_{\mathbf{k}\sigma}\Psi_{\mathbf{k},\sigma}^{\dagger}\left(\begin{array}{cccc}
0 & -t\phi_{\mathbf{k}} & -t_{\perp} & 0\\
-t\phi_{\mathbf{k}}^{*} & 0 & 0 & 0\\
-t_{\perp} & 0 & 0 & -t\phi_{\mathbf{k}}^{*}\\
0 & 0 & -t\phi_{\mathbf{k}} &
0\end{array}\right)\Psi_{\mathbf{k},\sigma},
\label{eq:H_bilayer}
\end{equation}
where $t_{\perp}\approx0.39$eV is the interlayer hopping,
and $t\approx2.8$eV is the in-plane, nearest neighbor, hopping
amplitude. The momentum dependence is contained in $\phi_{\bk}$, which
is the same as for a monolayer \eqref{phi}.
The band structure associated with eq.~\eqref{eq:H_bilayer}
consists of four non-degenerate bands given by
\begin{equation}
  E(\bk) = \pm \frac{1}{2}
  \Bigl(
    t_\perp \pm\, \sqrt{t^2_\perp+4t^2\vert\phi_{\bk}\vert^2}
  \Bigr)
  \label{eq:Ek-bilayer-full}
  .
\end{equation}
An expansion $\bk=\bK+\bp$ around the $\bK$ points of the BZ  when 
$v|\mathbf{p}|\ll t$  shows that the four-band tight-binding spectrum
\eqref{eq:Ek-bilayer-full} resolves into four hyperbolic bands
\cite{Nil06}, as shown in Fig.~\ref{bilayer}(b), and whose form
reads:
\begin{equation}
\pm
E_{\gamma}(\mathbf{p})=\pm\frac{t_{\perp}}{2}\left[1+\gamma\sqrt{
1+4(v|\mathbf{p}|/t_{\perp})^{2}}\right],
\label{eq:spectrumBi}
\end{equation}
with $v\approx 6$ eV$\mbox{\AA}$ being the Fermi velocity (the same
Fermi velocity of a monolayer), and
$\gamma=\pm1$. The Bernal stacking explicitly breaks the sublattice
symmetry in each layer, causing an energy split of $t_{\perp}$
between the two $\gamma=\pm1$ branches, $E_{+}$ and $E_{-}$, at $p=0$
(see Fig.~\ref{bilayer}). Due to a degeneracy at the $K$
points, the two symmetric branches $+E_{-}$ and $-E_{+}$ touch there, 
resulting in a gapless spectrum. Just as in a monolayer, the Fermi
surface of an undoped bilayer reduces to only two points, at $K$ and
$K'$; but now the valence and conduction bands have a finite
curvature and, hence, notwithstanding the absence of a gap, the
effective electronic degrees of freedom are massive, but still chiral.
The degeneracy at $K$ is protected by the $Z_{2}$
symmetry between the two layers only \cite{McC06b}, and can be lifted
with arbitrarily small perturbations, such as the ones induced by a
bias voltage, by polarizing the two sheets \cite{Zha09}, or else by
independently changing the carrier concentration in each layer
\cite{Oht06}.
This property opens the exciting prospect of using graphene bilayers
as materials with a gate-tunable band gap \cite{Cas07, Min07,
Nil07}. 

We stress that the low energy effective theory of bilayers remains
Lorentz invariant, in the following sense. The rotation of $\pi/3$
between layers breaks the sublattice symmetry leading to $2$ pairs of
massive Dirac particles at the $K\,(K')$ point.
Nevertheless, the system remains metallic because two bands, belonging
to different pairs, touch in a point. More explicitly, the
non-interacting bands \eqref{eq:spectrumBi} have the form: 
\begin{subequations}
\begin{align}
  E_1(\bk) &= - E_-(\bk) = -m v^2 + E(\bk), \\
  E_2(\bk) &= + E_-(\bk) = m v^2 - E(\bk), \\
  E_3(\bk) &= + E_+(\bk) = m v^2 + E(\bk),\\
  E_4(\bk) &=- E_+(\bk) = -m v^2 - E(\bk)
  ,
\end{align}
\end{subequations}
where  $E(\bk)=\sqrt{(m v^2)^2 + (v k)^2}$, and $m=t_{\perp}/(2v^2)$. 
Hence, $E_1(\bk)$ and $E_4(\bk)$ [or $E_2(\bk)$ and $E_3(\bk)$]
describe a massive relativistic dispersion with rest energy given
by $m v^2$. Again, the gapless nature of the full spectrum of
this problem is due to an accidental degeneracy of the simplest tight
binding parametrization. Additional hopping terms \cite{CastroNeto:09}
in the Hamiltonian or many body interactions can easily lift this
degeneracy. This implies that the Bernal bilayer problem is 
unstable from the electronic point of view. In contrast, the twisted
bilayer \cite{San07} is stable because it does not rely on this
particular accidental degeneracy.
Just like in the case of monolayer graphene, the
introduction of the instantaneous Coulomb interaction does not
preserve this Lorentz invariance. 
 
At very low energy, below $\Delta_w\approx 1.5\,$meV, additional
trigonal warping effects take place due to the influence of
next-nearest neighbor hopping matrix elements [which we are
neglecting in \eqref{eq:H_bilayer}]. Trigonal warping introduces 
an asymmetry in the conductivity under electron or hole doping
\cite{Li09}, and leads to a remarkable Lifshitz transition at low
densities, whereby the lowest energy bands split into 4 Dirac cones 
\cite{McC06a,Cserti07}. These effects, however, happen at 
very low densities (around 1 electron per flake for typical
1$\mu$m$^2$ samples), and hence are experimentally very challenging.
A detailed description of the spectral properties of
graphene bilayers can be found in \onlinecite{CastroNeto:09}, and
\onlinecite{Nil08}. 

When $\Delta_{w}<v|\mathbf{p}|\ll t_{\perp}$, we recover the so-called
{\it classical limit} of the ``relativistic'' problem. This means
that the presence of the uppermost band is not too relevant, and the
energy disperses quadratically with momentum (the opposite
limit of $v|\mathbf{p}|\gg t_{\perp}$ corresponds to the
``ultra-relativistic'' regime, where the bandstructure is essentially
linear in momentum, like in the monolayer).
In this case the Hamiltonian \eqref{eq:H_bilayer} near the $K$ points
can be projected onto an effective two-band model, written in terms of
the two valleys and a mixed sublattice-layer basis \cite{McC06a}:
\begin{equation}
\tilde{\Psi}_{\mathbf{p},\sigma}=(a_{\mathbf{K}+\mathbf{p},\sigma},\bar{b}_{
\mathbf{K}+\mathbf{p},\sigma},\bar{b}_{-\mathbf{K}+\mathbf{p},\sigma},
a_{-\mathbf{K}+\mathbf{p},\sigma})\,.\label{eq:Psitilde}
\end{equation}
In such a basis, the effective kinetic Hamiltonian is 
\begin{equation}
\mathcal{H}_{B}=\sum_{\mathbf{p}\sigma}\sum_{\alpha=\pm}\tilde{\Psi}_{
\mathbf{p},\sigma}^{\dagger}\frac{p_{\alpha}^{2}}{2m}\left[\tau_{0}
\otimes\sigma_{\alpha}\right]\tilde{\Psi}_{\mathbf{p},\sigma}\,,\label
{eq:Hb}
\end{equation}
where $p_{\pm}=p_{x}\pm ip_{y}$, $\sigma_{\pm}=(\sigma_{1}\pm
i\sigma_{2})/2$ operating in the sublattice basis, 
and $\tau$ operates in the valley space. The resulting energy spectrum
is parabolic,
\begin{equation}
  E(\bp)=\pm\frac{p^2}{2m}
  ,
  \label{eq:Ep2}
\end{equation}
with $m=t_{\perp}/(2v^2)\approx0.054\, m_e$ as the effective
mass of the electron. From now on we will omit the valley indexes
and assume the two component basis $\tilde{\Psi}_{\mathbf{p},\sigma}\to(a_{\mathbf{p},\sigma},\bar{b}_{\mathbf{p},\sigma})$
with a total degeneracy $N=4$ in valley and spin. 

The electronic Green's function in this two band model,
$\hat{G}^{(0)}(\mathbf{k},\tau)=-\langle
T[\tilde{\Psi}_{\mathbf{k}}(\tau)\tilde{\Psi}_{\mathbf{k}}^{\dagger}
(0)]\rangle$,
is given by
$\hat{G}^{(0)}(\mathbf{k},i\omega)=(i\omega-\hat{\mathcal{H}}_{B})^{-1}$
or, equivalently, by %
\begin{equation}
\hat{G}^{(0)}(\mathbf{k},i\omega)=\frac{1}{2}\sum_{s=\pm}\frac{1+s\hat{
\sigma}_{\mathbf{k}}}{i\omega-s|E(\mathbf{k})|}\label{eq:GBi}
\end{equation}
in the chiral representation, where \begin{equation}
\hat{\sigma}_{\mathbf{k}}=\sum_{\alpha=\pm}\frac{k_{\alpha}^{2}}{|\mathbf{k}|^{2}}\sigma_{\alpha}\,.\label{eq:sigma_p2}\end{equation}
Although the fermions are chiral, in bilayers the wavefunctions of
the quasiparticles acquire a $2\pi$ phase when winding around the
$K$ points, rather than a $\pi$-phase, as for Dirac fermions. This
property is an admixture of the behavior of Dirac particles, which
are chiral, with conventional electrons, which disperse
quadratically.
The combination of chirality and a trivial Berry phase has a clear
experimental signature in the suppression of the zero-level plateau
in the quantum Hall effect of the bilayer, whose plateaus are
quantized by integer numbers \cite{McC06a, Nov06}.

\begin{figure}[tb]
  \centering
  \includegraphics[width=\columnwidth]{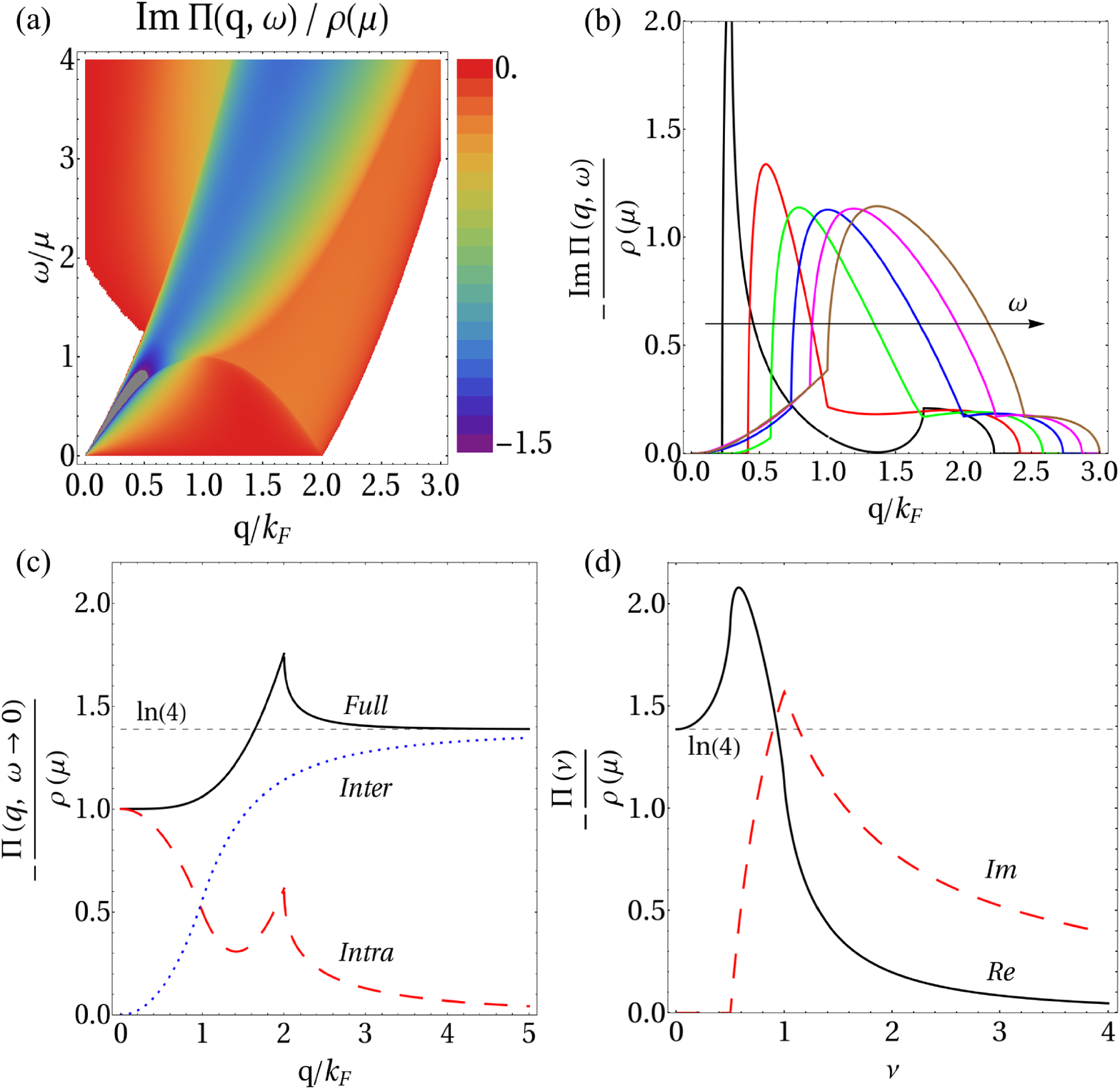}
  \caption{(Color online)
    The polarization $\Pi^{(1)}(q,\omega)$ 
    of bilayer graphene, obtained within the two-band approximation,
    for finite chemical potential, and zero temperature.
    All panels are normalized to the DOS at the Fermi energy, $\mu$.
    Panel (a) shows a density plot of the imaginary part and, in 
    (b), we have cuts of the same at constant frequency, for 
    $\omega/\mu = 0.5,\,1.0,\,1.5,\,2.0,\,2.5,\,3.0$.
    Panel (c) corresponds to the static limit $\Pi^{(1)}(q,0)$
    in eq.~\eqref{eq:Pi2L-Stat},
    and includes the intra-band contribution (dashed), the 
    inter-band contribution (dotted), and the full polarization
    (solid).
    In (d) we represent the real and imaginary parts of the 
    polarization in the undoped case \eqref{eq:piBi} as a function
    of $\nu=2m\omega/q^2$.
  }
  \label{fig:Pol-Bilayer}
\end{figure}

\subsection{Charge polarization}

Within the two band model, the one loop polarization function has the
generic form given in Eq.~(\ref{eq:Pol_E}) for the single layer. The
adaptations for the present case consist in considering the bilayer
spectrum, and a new overlap factor, which, for the bilayer, reads
\begin{equation}
\mathcal{F}_{s,s^{\prime},\mathbf{p},\mathbf{q}}=\frac{1}{2}[1+ss^{
\prime}\cos(2\theta_{\mathbf{p},\mathbf{p}+\mathbf{q}})]
\label{eq:F2}
.
\end{equation}
In this expression $\theta_{\mathbf{p},\mathbf{p}+\mathbf{q}}$ is,
again, the angle between the vectors $\mathbf{p}$ and
$\mathbf{p}+\mathbf{q}$.
Below we shall focus our discussion in terms of the effective
two-band Hamiltonian \eqref{eq:Hb}, and dispersion \eqref{eq:Ep2}.

The polarization function $\Pi^{(1)}(q,\omega)$ at finite density was
obtained by \onlinecite{Hwa08b} in the $T=0$ static limit. The full
dynamical case was calculated by \onlinecite{Sensarma10} at $T=0$, and by
\onlinecite{Lv10} at finite temperature. The finite density result
can be obtained in closed analytical form for $T=0$; but, in
order to avoid reproducing here those lengthy expressions, we simply
present $\Pi^{(1)}(q,\omega)$ graphically in
Figs.~\ref{fig:Pol-Bilayer}(a,b).
The explicit form of the static limit reads 
\cite{Hwa08b,Lv10}
\begin{equation}
  - \frac{\Pi^{(1)}(q,0)}{\rho(\mu)} = 
    g\bigl(\tfrac{q}{k_F}\bigr) 
    - f\bigl(\tfrac{q}{k_F}\bigr) \,\theta(q-2k_F)
  \label{eq:Pi2L-Stat}
\end{equation}
at zero temperature, with
\begin{subequations}
\begin{align}
  f(x) &= \frac{2+x^2}{2x}\sqrt{x^2-4}
    + \ln\left(\frac{x-\sqrt{x^2-4}}{x+\sqrt{x^2-4}}\right)\\
  g(x) &=  \frac{1}{2}\sqrt{4+x^4} 
    - \ln\left(\frac{2+\sqrt{4+x^4}}{4}\right) 
  .
\end{align}
\end{subequations}
The DOS at the Fermi energy, $\rho(\mu) = N m /(2\pi)$,  
is constant  and density independent, by virtue of the parabolic
nature of the low energy approximations \eqref{eq:Hb} and
\eqref{eq:Ep2} [note, however, that the consideration of the
full 4-band spectrum leads to a DOS which is linear in energy; in
this sense, the correction to the DOS that arises from considering
the 4 versus the 2 band model is not negligible \cite{Ando:2007}]. In
this sense the bilayer is similar to the conventional 2DEG. However,
just as in the monolayer, the existence of two symmetric bands
adds an inter-band channel, leading to a rather different
quasiparticle spectrum, in comparison with the 2DEG. This can be seen
by directly comparing Figs.~\ref{fig:Pi-Function-2DEG}(b) and
\ref{fig:Pol-Bilayer}(a).
The behavior of $\Pi^{(1)}(q,0)$ is shown in
Fig.~\ref{fig:Pol-Bilayer}(c), together with its decomposition into
intra- and inter-band contributions, which are respectively associated
with the choice $ss'=1$, or $ss'=-1$ in eq.~\eqref{eq:F2}. As
intuitively expected, the inter-band contribution dominates at large
momenta/small densities, whereas the intra-band transitions dominate
the low momenta/large density regime. Unlike the monolayer, or the
2DEG, the polarization is constant for both $q\ll k_F$ and 
$q\gg k_F$. The former limit makes the bilayer similar  to the
conventional 2DEG and monolayer graphene, while the latter is neither
akin to the 2DEG (for which the polarization decreases rapidly with
$q/k_F$ [Fig.~\ref{fig:Pi-Function-2DEG}(e)]), nor to the monolayer
(for which it increases linearly [Fig.~\ref{fig:Pi-Function}(e)]).
Moreover, at precisely $q=2k_F$, $\Pi^{(1)}(q,0)$ is sharply cusped,
which contrasts with the behavior of a monolayer, whose derivative is
continuous. According to the standard theories of linear response,
this feature at $2k_F$ has important implications for the behavior of
the induced charge, the associated decay of the Friedel oscillations
around charged impurities, the effective RKKY interaction among
magnetic impurities, Kohn's anomaly in the phonon dispersion, etc. 
For example, one  expects qualitative differences between the
resistivity arising from Coulomb scattering in mono- and bilayer
graphene: it should be stronger in the bilayer, and have a more
pronounced temperature dependence \cite{Hwa08b,Lv10}.

At long wavelengths, the RPA screened potential reads
$V^{RPA}(q)=V(q)/[1-V(q)\Pi^{(1)}(q)] \approx 
2\pi e^2 / [\epsilon_0 (q + q_{TF})]$, with a Thomas-Fermi momentum 
$q_{TF} = Nme^2/\epsilon_0$. Notice that $q_{TF}$ is the
same for the bilayer as in the 2DEG, i.e. it is constant (no density
dependence), and also temperature independent \cite{Lv10}. The
temperature independence of $q_{TF}$ at long wavelengths is another
trait that distinguishes this system from both the monolayer and the
2DEG. In real space the statically screened potential decays
asymptotically as $V(r)\propto 1/r^3$  \cite{Hwa08b}.

At half-filling (undoped situation) and zero temperature, the form of
the polarization bubble simplifies further, and can be cast
as
\begin{multline}
\Pi^{(1)}(\nu) = 
-\frac{Nm}{2\pi}\left[\frac{1}{\nu}\ln\!\left(\frac{1+\nu}{1-\nu}
\right)-\frac{1}{2\nu}\ln\!\left(\frac{1+2\nu}{1-2\nu}
\right)\right.
\\
\left.+\ln\!\left(\frac{1-\nu^{2}}{\frac{1}{4}-\nu^{2}}
\right)\right]
\label{eq:piBi}
\end{multline}
\cite{Bar09,Nan10a, Nil06}, where $\nu=2m\omega/q^{2}$ is the only
scaling parameter. This function is plotted in
Fig.~\ref{fig:Pol-Bilayer}(d). It follows at once that the static
limit ($\omega\to0$) is simply
\begin{equation}
\Pi^{(1)}(q,0)=-\frac{N\ln4}{2\pi}m
\label{eq:Pibi3}
,
\end{equation}
consistent with the above discussion when $k_F=0$. Despite the absence
of a Fermi surface at half-filling, the Coulomb interaction among the
quasiparticles is screened due to the finite density of states at
the $K$ points. However, an
important difference here is that $\Pi^{(1)}(q,0)$ is constant for
\emph{all} momenta, unlike traditional 2D systems, and stems from the
presence of the inter-band channel. Hence, the Thomas-Fermi wavevector
is exactly $q_{TF} = Nm\ln(4)e^2/\epsilon_0$ for \emph{all}
wavelengths, and Friedel oscillations are suppressed at half-filling
\cite{Hwa08b}. The additional numerical factor $\ln(4)$ means a
slight increase in the screening strength of undoped bilayer, with
respect to the doped situation. One way to interpret this 
$\ln(4)$ enhancement is the following: the
factor $Nme^2/\epsilon_0$, being exactly the same as in a simple 2DEG,
is attributable to the finite DOS, while the extra $\ln(4)$ 
arises from the virtual inter-band transitions.
In real space, the
statically screened potential of undoped bilayer will decay as
$1/r^3$, which contrasts with the corresponding
behavior in the monolayer, where the decay is $1/r$ (as we saw
before this is due to the fact that, in the RPA, the effect of
interactions in the monolayer is to simply renormalize the background
dielectric constant, keeping the Coulomb form of the potential). 
Inspection of Fig.~\ref{fig:Pol-Bilayer}(d)
reveals that the real part of the RPA dielectric function
$\epsilon_{RPA}(\mathbf{q},\omega)=\epsilon_{0}[1-V(\mathbf{q})\Pi^{(1)}
(\mathbf{q},\omega)]$ will be always nonzero. This means that,
although the lack of a Fermi surface does not prevent
screening in bilayers ($q_{TF}\ne 0$), the formation of zero
temperature infrared plasmons is suppressed at half-filling. 

The screened Coulomb interaction between the layers is $V(q)=2\pi
e^{2}\mbox{e}^{-qd}/[\epsilon_{0}(q+q_{TF}\mbox{e}^{-qd})]$,
where $d=3.35\mbox{ \AA}$ is the interlayer distance. At long wavelengths,
$q\ll t_{\perp}/v<1/d\approx0.3\mbox{\,\AA}^{-1}$, $d$ can be
effectively replaced by zero in first approximation, and the screened
interaction among electrons belonging to the same or different
planes can be treated on the same footing. 

At this point we should pause to point out that the behaviors
discussed so far at large $q$ have to be interpreted within the
restrictions regarding the validity of the two-band approximation.
For example, the fact that in Fig.~\ref{fig:Pol-Bilayer}(c) we see the
polarization becoming constant at $q\gg k_F$
is an artifact of the two-band approximation. In reality, we should
bear in mind that the full dispersion is hyperbolic, and hence
becomes linear at high densities. 
We then expect to recover the linear-in-$q$ dependence of
$\Pi^{(1)}(q,0)$ seen in Fig.~\ref{fig:Pi-Function}(e) for the
monolayer.

For this reason, proper caution is needed when considering the
extrapolation of these results to highly doped bilayers, where
the consideration of the four-band hyperbolic dispersion 
\eqref{eq:spectrumBi} is inevitably required. In terms of electronic
densities, this corresponds to values above $\sim 10^{12}\,$cm$^{-2}$,
for which the two-band model is no longer warranted.
The full dynamical response using the spectrum in
eq.~\eqref{eq:spectrumBi} has been recently derived in closed
analytical form by \onlinecite{Bor09b}. Notwithstanding the lengthy and
cumbersome nature of these analytical results, they afford a more
accurate perspective on the screening response of doped bilayer
graphene, its collective modes, and the crossover between the regimes
of a massive-chiral system at low densities, to a system of weekly
coupled monolayers at higher densities. \onlinecite{Bor09b}'s approach
is ultimately limited by systems of such high densities that $\mu
\approx t$, in which case the full tight-binding dispersion
\eqref{eq:Ek-bilayer-full} is needed, but is beyond closed analytical
approaches.

\subsection{Quasiparticles}

In the two band model, the structure of perturbation theory
for Coulomb interactions is set only by self-energy renormalizations
in the effective mass of the electrons, $m$, and in the quasiparticle
residue, $Z$. 

From the Hamiltonian (\ref{eq:Hb}), the renormalized Green's function
is 
\begin{equation}
\hat{G}(\mathbf{k},\omega)=\frac{1}{\omega-\sum_{\alpha=\pm}k_{\alpha}^{
2}/(2m)\sigma_{\alpha}-\hat{\Sigma}(\mathbf{k},\omega)}
.
\label{eq:Gbi}
\end{equation}
$\hat{\Sigma}(\mathbf{k},\omega)$ is the quasiparticle self-energy
correction, which is described in the
$(a_{\mathbf{k},\sigma},\bar{b}_{\mathbf{k},\sigma})$
basis by a matrix in the form
\begin{equation}
\hat{\Sigma}=\left(\begin{array}{cc}
\Sigma_{0} & \Sigma_{+}\\
\Sigma_{-} & \Sigma_{0}\end{array}\right),\label{eq:Sigma}
\end{equation}
or, equivalently,
$\hat{\Sigma}=\Sigma_{0}\sigma_{0}+\Sigma_{+}\sigma_{+}+\Sigma_{-}
\sigma_{-}$, where $\sigma_{\pm}=(\sigma_x \pm i \sigma_y)/2$. 
By symmetry, $\Sigma_{+}=\Sigma_{-}^{*}$. In a more
conventional form, 
\begin{equation}
\hat{G}(\mathbf{k},\omega)=\frac{Z}{\omega-Z\sum_{\alpha=\pm}[k_{\alpha}^{
2}/(2m)+\Sigma_{\alpha}]\sigma_{\alpha}}
,
\label{eq:Gbi2}
\end{equation}
where $Z^{-1}=1-\partial\Sigma_{0}/\partial\omega$ corresponds to the
quasiparticle residue, and 
\begin{equation}
\frac{m^{*}}{m}=\frac{1-\partial\Sigma_{0}/\partial\omega}{
1+2m \partial\Sigma_{+}/\partial
k_{+}^{2}}
\label{eq:mBi}
\end{equation}
is the mass renormalization. 

\begin{figure}
  \centering
  \includegraphics[width=0.9\columnwidth]{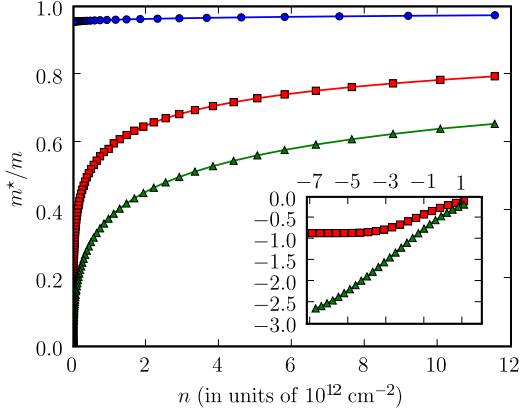}
  \caption{(Color online)
  Mass renormalization for $\alpha=0.5$ in the
  bilayer, calculated with a \emph{static} Thomas-Fermi screened Coulomb
  interaction  $V(q)= e^2/[\epsilon_0(q+\lambda q_{TF})]$, as a
  function  of the  electronic density. Blue
  circles:  $\lambda=1$;  red squares: $\lambda=0.01$; green
  triangles:  $\lambda=10^{-4}$.
  (From \onlinecite{Bor09}). The inset shows $\log_{10}(m^{*}/m)$
  as a function of  $\log_{10}(n)$ for two of the $\lambda$
  values; the mass saturates at a finite value for $n \rightarrow 0$.
  }
  \label{massRenormalization}
\end{figure}

We saw in the previous section that, unlike the monolayer, Coulomb
interactions in the bilayer are screened. The self-energy is given in terms of the bare Green's
function and  the RPA effective interaction by
\begin{equation}
\hat{\Sigma}^{(1)}(\mathbf{q},\omega)=i \int \frac{d^{2}k d \varepsilon }{(2\pi)^{3}}
V^{RPA}(\mathbf{k},\varepsilon)\hat{G}^{(0)}(\mathbf{k}+\mathbf{q},
\varepsilon+\omega),
\label{eq:Sigma1Bi}
\end{equation}
where $V^{RPA}(\mathbf{q},\omega)=V(\mathbf{q})/\epsilon_{RPA}(\mathbf{q},\omega)$
is dressed by the RPA dielectric function. Even if the ratio between
the Coulomb and kinetic energies diverges in the low density limit
(as in a 2DEG) the validity of RPA can be, in principle, justified in
the large $N$ limit.  If only static screening is taken into account (Hartree-Fock-Thomas-Fermi
theory), the
self-energy is frequency independent and, to leading order, the
quasiparticle residue $Z$ does not renormalize. 
Calculations based on the static screening picture for the two-band
model \cite{Bor09}, and also for the four-band model \cite{Kus09}, have
found mass renormalization in the bilayer. The mass 
 decreases ($m^*/m <1$), and the renormalization grows stronger as  the screening is 
suppressed. In Fig.~\ref{massRenormalization} we show this
 renormalization within the two-band model, where the parameter  $\lambda$
 interpolates between the Thomas-Fermi screened potential  ($\lambda=1$)
and the unscreened  Coulomb potential ($\lambda \approx 0$).
As a consequence of the reduced mass, the  charge compressibility
 is also expected to decrease  \cite{Kus08,Borghi10}. 

More recent calculations that account for the full dynamical screening
have found quite different results. When the dynamical RPA polarization
bubble, Eq~\eqref{eq:piBi}, is taken into account,  the self-energy exhibits a strong $\ln^{2}$ leading divergence,
$\mbox{Re}\Sigma_{+}^{(1)}(k.\omega)=
2k_{+}^{2}/(Nm\pi^{2})\ln^{2}(\Lambda/k)$,
 and
$\mbox{Re}\Sigma_{0}^{(1)}(k,\omega)=-4\omega/(N\pi^{2})\ln^{2}[\Lambda/(\sqrt{
m\omega})]$,
 at small energies and momenta \cite{Bar09}.
The ultraviolet momentum scale $\Lambda \sim q_{TF}$ is related to the effective ``Bohr radius", 
$a_0=\epsilon_0/(me^2)$, and we set $\Lambda = 1/a_0$.
 At leading ($\ln^{2}$) order, the
 two terms in the self-energy compensate each other exactly in
 Eq.~\eqref{eq:mBi} and the mass does not renormalize, $m^{*}/m\to1$ at $k\to0$, while 
 the  quasiparticle spectral weight vanishes as $Z \sim \ln^{-2}(\Lambda/k)$.
The RG analysis of the dynamically screened interaction at large $N$ was carried out by
 \onlinecite{Nan10c}, where subleading (single log) contributions were collected.
These were found to cause a (weak) increase of the effective mass
 $m^*/m \approx 1 + [0.56/(N2\pi \ln{4})]\ln{\Lambda}$, and
 consequently an increase of the compressibility.

Once again, the validity of a two band model rests on the assumption
that all relevant energy scales are small compared to
$t_{\perp}\approx0.4$eV. 
However the Coulomb energy $\Lambda_E$ on the scale of $a_0=\epsilon_0/(me^2)$ is substantial for not too
 strong dielectric screening,  
 $\Lambda_E = e^2/(\epsilon_{0}a_0) \approx 1.47/\epsilon_0^2$ eV  \cite{Nan10a,Nan10c}.
Hence, Coulomb interactions can promote electronic transitions
among the four bands, while the two-band model is only justified in the 
limit $\Lambda_E < t_{\perp}$. To what extent the two band model
provides a valid description of the quasiparticles in the presence of
Coulomb interactions is a matter of ongoing discussion.

\subsection{Many-body instabilities}

The finite DOS in the bilayer enhances the possibility of many body
instabilities in comparison with the single layer case. For instance,
the spin polarization tensor in the bilayer is defined in leading
order by Eq. (\ref{eq:Pi_spin}), with the matrix element
 $\hat{\mathcal{A}}^{s}(\mathbf{k})=1+s\sum_{\alpha=\pm}(k_{\alpha}^{2}/\mathbf{k}^2)
\sigma_{\alpha}$.
In matrix form, 
\begin{equation}
\hat{\Pi}^{(1)+-}=\left(\begin{array}{cc}
\Pi_{aa}^{(1)+-} & \Pi_{ab}^{(1)+-}\\
\Pi_{ba}^{(1)+-} &
\Pi_{bb}^{(1)+-}\end{array}\right),\label{eq:PiBiSpin2}
\end{equation}
which leads to one ferromagnetic, and one antiferromagnetic 
eigenstate, $\Pi_{F/AF}=\Pi_{aa}\pm|\Pi_{ab}|$, by symmetry under
exchange of the $a$ and $b$ labels. In bilayers the AF state has a
leading logarithmic divergence with the cut-off, $\Lambda$, at zero
frequency and magnetic field, \cite{Nil06} 
\begin{equation}
\Pi_{AF}^{(1)}(\mathbf{q},0)=\frac{m}{\pi}\ln\!\left(\frac{2\Lambda}{
|\mathbf{q}|}\right)\,,\label{eq:PiBiSpin}
\end{equation}
suggesting (within RPA) a tendency towards an AF instability for any value of the
 Hubbard interaction $U$. 
In addition,  at finite $U$, a first order ferromagnetic
transition can be driven by the Stoner criterion, leading to a
ferromagneto-electric state where the layers have different
magnetization and polarized charge \cite{Cas08}. 

Other possibilities include the emergence of CDW instabilities induced
by the short range part of the Coulomb interaction, \cite{Dah10} 
or else, an excitonic instability at strong local electronic
repulsion \cite{Dil08}. With long range Coulomb interactions,
the inverse electronic compressibility $\kappa^{-1}$ becomes negative at small densities \cite{Kus08}, 
indicating a tendency to Wigner crystallization
\cite{Dahal06}, which is compensated by the positive
compressibility of the lattice. 

Bilayers share  similar features with  one-dimensional (1D) electron systems, such as the
point like Fermi surfaces and the parabolic spectrum.
 In particular,
in biased bilayers, the 1D interface between biased regions confines
chiral modes that propagate as in a strongly interacting Luttinger
liquid  \cite{Kil10}. This
affords the possibility of studying such interacting models
experimentally in appropriately prepared samples of bilayer graphene.

For short-ranged interactions in 2D, the structure of the diagrams in
bilayers and in 1D electron liquids is  quite similar, although the diagrams
compensate each other in a rather different way. The dimensionless
coupling which determines the strength of the interactions is
$Ua^{2}m$, where $U$ is the strength of the local interactions, and
$a$ is the lattice constant. Perturbative renormalization group
calculations in the bilayer have identified distinct leading
instabilities of the electron gas. For different choices of possible
interactions, two different low-temperature broken symmetry phases
have been found: in one case, a ferroelectric gapped phase \cite{Zha10}
induced by the coupling between the different layers; in the other, a
nematic phase \cite{Vaf10, Vafek10}, where each Fermi point splits into two
Dirac points.

The possibility of an excitonic instability has been also
predicted by \onlinecite{Nan10a}, who found that the dynamically
screened Coulomb interaction gives rise to a ferroelectric state that
polarizes the two layers. In the ferroelectric state, the kinetic
energy inflicts an energy cost 
$\delta E_{\mathrm{Kinetic}}\propto\Delta^{2}\ln(\Lambda_E/\Delta)$,
where $\Delta$ is the energy gap.  Finite separation between the
layers generates an additional electrostatic energy cost to polarize
the charge between the layers, which dominates the kinetic energy at
the Hartree level, 
$\delta E_\text{Hartree}\propto\Delta^2\ln^2(\Lambda_E/\Delta)$
\cite{McCa07}.
The excitonic instability is induced by the exchange term,
which is parametrically larger than the Hartree
  term by the factor $(a_0/d)$, where $d$ is the interlayer distance \cite{Nan10a}.
 The existence of a ferroelectric
state has  nevertheless  been disputed by independent RG calculations,
that also accounted for the dynamically screened Coulomb interactions,
and infrared trigonal warping effects \cite{Lem10}. The spontaneous
symmetry breaking found in this work leads to a Lifshitz transition
consistent with the  nematic state found by \onlinecite{Vaf10}, rather
than the opening of a gap.
 
In the Quantum Hall (QH) state, two terminal measurements of the
conductivity
in clean suspended samples have found an insulating state at the
$\nu=0$
filling factor \cite{FMY09}, rather than the metallic QH state 
previously found in supported samples \cite{Nov06}. Further theoretical works
predicted the possibility of a zero field excitonic QH state, which
spontaneously breaks time reversal symmetry, and can evolve into a
ferromagnetic QH state at finite magnetic field \cite{Nan10b}. 
In biased bilayers, a chiral anomaly has been predicted in
the QHE, splitting the degeneracy of valley quantum numbers \cite{Nak09}. 
Another predicted effect resulting from interactions
in the QH state is the formation of charge 2$e$ skyrmions at even
filling factors \cite{Aba09}.

\section{Conclusions}

As we have seen, the understanding of the many-body problem in graphene has evolved quite
 rapidly in only a few years. The case of monolayer graphene in the weak coupling regime (which means, graphene embedded in an environment with large dielectric constant) is quite clear, namely, although Lorentz invariance is explicitly broken because of the Coulomb interactions, the effective low energy theory is still Lorentz invariant with well defined quasiparticles. Nevertheless, these quasiparticles have a renormalized {\it speed of light} that grows logarithmically in the infrared, while their spectral weight 
 decreases slowly in the same limit. This situation can be contrasted with the conventional Fermi liquid picture where all the physical constants (the so-called Landau parameters) and spectral weight are finite in the infrared (that is, at the Fermi surface). Hence, these logarithmic renormalizations are weak enough, even in the presence of strong Coulomb interactions, and a {\it Dirac liquid} picture is preserved. 

In the strong coupling regime (that is, graphene in vacuum), many-body instabilities are possible albeit depending on a delicate balance of energy scales. This occurs because the renormalizations of quasiparticle properties also depend on details of the cut-off procedure in the ultraviolet (as it is shown by the f-sum rule). While mean-field theories have predicted instabilities towards phases with broken chiral symmetry and superconducting quasi-long range order (because of the 2D nature of the material), and earlier Monte Carlo studies on a hyper-cubic lattice suggest the presence of instabilities \cite{Drut09a,Drut09b,Drut09c}, simulations  of interacting electrons
on the honeycomb lattice have still to be performed in order to address these issues, since the strong coupling regime cannot be reached by perturbative methods. This remains, at this point in time, as an important open problem in many-body graphene physics. 

The Coulomb impurity problem in graphene shares many of the issues of the many-body problem but can be studied in much more detail because the 2D hydrogen problem in graphene was solved exactly. In the weak coupling regime (the so-called under-critical regime), the Coulomb interaction between a localized charge and the electrons leads to only mild changes in the physical properties due to the explicitly  broken particle-hole symmetry. In the strong coupling (or super-critical) regime, the situation is rather different because of the phenomenon of {\it fall to the center}, that is, the electron states become unstable, with the generation of resonances near the Dirac point. Just like the many-body problem, the critical local charge depends on the dielectric environment, and in vacuum this amazing effect should be observed by local probes even for a single proton sitting on the graphene surface. So far, there is no experimental evidence of such effect, given that it is difficult to study adatoms in suspended samples with local probes, such as scanning tunneling microscopes. In supported samples, because of dielectric screening that brings the system to weak coupling, and of the disorder in the substrate, the study of this problem can be much more elusive.

In analogy to the 2DEG problem, the effect of disorder is rather strong in graphene which again is the effect of dimensionality. The low dimensionality implies strong quantum fluctuations that can easily couple to spatial variations of random scalar (chemical potential) and gauge (hopping) fields. Strong localization is the ultimate fate of any disordered two dimensional system but because the localization length grows very slowly in the infrared limit, the finite size of the samples, or the finite temperature of the system, ends up cutting off the tendency towards Anderson localization and, in practical terms, graphene behaves in a metallic way. 

The problem of magnetism of adatoms in graphene is rather different from the one found in metallic hosts. Due to the strong energy dependence of the density of states (that vanishes at the Dirac point), the  Anderson impurity  problem has features that are unique. Firstly, in analogy with the  strong coupling regime in the many-body and Coulomb impurity problems, the results are sensitive to the ultraviolet regularization. In fact, this is a generic feature of the Dirac spectrum, namely, strong coupling leads to spectral weight transfer from high energies to low energies, that is, to the Dirac point (as discussed in the context of the f-sum rule). Moreover, the damping by Dirac electrons leads to an  anomalously large (and strongly energy dependent) broadening of the adatom energy level. This leads to an unusual situation as compared to the Anderson impurity problem in a metal, namely, that even when the chemical potential is above (below) the energy of the doubly (singly) occupied state, a magnetic moment can emerge. Hence, adatoms that may not be magnetic in a metal (hydrogen or fluorine, for instance), might become magnetic in graphene. 

On the other hand, the Kondo effect that usually suppresses the appearance of magnetic moments in metals because of magnetic ``screening'' (the ultimate consequence of the so-called ``Kondo cloud''), is strongly suppressed in graphene. This suppression has its roots in the low density of states and the sublattice structure. In fact, there is a strong dependence of the hybridization with the position on the lattice (whether it breaks or not the sublattice symmetry). Furthermore, the Kondo effect is very dependent on the chemical potential (that can be easily tuned in graphene by gating). This state of 
affairs reinforces the conclusion that magnetic states of adatoms could be more the norm than the exception in graphene, in a big contrast with the situation  in ordinary metals. Experimentally, there are very few studies of the magnetism of adatoms in graphene. The main problem here is that most of the experiments done so far are in electronic transport. Just like the Kondo problem in metals and semiconductors, the observation of magnetic effects in transport is rather subtle, and requires careful analysis. At this point in time, this is a rather open field in graphene physics. 

A superconducting state in graphene would have dramatic consequences given its low dimensionality and unusual electronic spectrum. While true long range order would not be possible because of its 2D nature, quasi-long range order would have unusual consequences. For one, because of the sublattice structure, there is room for exotic pairing states with even more exotic vortex excitations. The phase space for pairing is rather large due to the spin, sublattice, and valley degeneracies. However, the low density of states  plays a deleterious role here. One way out of this conundrum would be the enhancement of the density of states by either gating or doping with adatoms. These two techniques have their own limitations. Gating is limited by the distance from the gate to the graphene sample,
and by the dielectric breakdown of the spacer that separates the two. Doping inevitably introduces disorder, or can modify the electronic structure of the $\pi$ band too much leading to extrinsic effects. There are, however, serious hopes that come from the fact that intercalated graphite can be
 made to superconduct. An obvious idea would be intercalation of Ca or Yb in the graphene bilayer. So far, intercalation experiments in bilayers have not been performed, and very little is known about how to intercalate atoms or molecules in such systems. Again, this is very much an open field of research.

In addition to the dielectric environment, which  has a strong influence on many-body effects in bulk graphene, finite size effects are also of great importance. It has been understood very early on that zig-zag edges are strongly interacting because of the high density of states they create at the Dirac point. Systems with high density of states are prone to many-body states due to Stoner-like instabilities. However, the many-body physics of finite 2D systems is even more sensitive to disorder (either in the bulk or in the edge) because of the strong boundary condition dependence. In graphene this problem is magnified because the electronic wavefunctions associated with impurity states do not decay exponentially, as they would in a semiconductor with a finite gap, or would be extended, as in a normal metal, but they are quasi-localized (that is, decay like power law). This implies that evanescent waves play an important role in determining the physical properties. Experiments in mesoscopic graphene samples show very clearly these effects through strong oscillations of the electronic conductance and the presence of Coulomb blockade peaks. From theoretical perspective, such  problems are probably the hardest to solve because they involve the direct interplay between Anderson localization and interactions. 
Thus deeper understanding of mesoscopic graphene systems is still necessary, and 
this topic would merit a review of its own.

Magnetic fields also lead to spatial localization due to the presence of Landau levels with a length scale given by the cyclotron length. Hence, this problem shares many of the difficulties of the previous problems with the added complication that the 2D nature of graphene brings a huge degeneracy into play. Once again, the detailed balance between kinetic and Coulomb energies,  and the details in the ultraviolet, determine the fate of the many-body ground state. The fractional quantum Hall effect was only observed recently in suspended two-probe experiments \cite{DSDLA09}, and very little is known about the sequence of FQHE fractions and their nature. It is believed that magnetic fields can generate a plethora of new many-body states, with symmetries that are rather different from the ones found in the 2DEG. But, compared to the 2DEG problem, this field is still in its infancy.

While we have demonstrated the complexity of the many-body problem in monolayer graphene, we have 
not even touched beyond the surface of the many-body problem in bilayer graphene. There is no doubt, at least from the theoretical perspective, that the many-body problem in the bilayer is much richer than in the monolayer. For one, the bilayer  has a finite density of states at neutrality, making it similar to the 2DEG problem. However, unlike the 2DEG, the graphene bilayer is a Lorentz invariant system with a finite ``rest mass'' (that is, it has a hyperbolic dispersion relation) albeit with an accidental degeneracy that makes it a semi-metal (two of the four bands touch at the Dirac point). This accidental degeneracy can be lifted very easily by hopping or interactions, leading to a huge number of possible many-body states with different quantum numbers. Given this richness one can venture saying that bilayer graphene is the ultimate target of many-body theorists in this field. However, it is technically a major challenge given the high dimensionality of the problem,
with its $2^4$-dimensional spinorial structure (spin, valley, sublattice, and plane). 
Moreover, from the experimental perspective
 many details and conditions are still quite uncontrolled, which has led to a few contradictory results,
 and has so far yielded more questions than answers. 
 In fact, both theoretically and experimentally, the graphene bilayer remains very much an open problem. If we now extrapolate from the monolayer to the bilayer, we see that there are problems that have not even been addressed theoretically and experimentally, like the Anderson impurity problem, or the Kondo effect in bilayers, the problem of magnetism, and superconductivity, just to mention some. These are topics for the future, for future generations of physicists to address and marvel.

\section*{Acknowledgments}
We are indebted to our collaborators, friends and colleagues for
their many invaluable contributions, discussions, comments, and
suggestions. In particular we want to explicitly thank
E. Andrei, Y. Barlas, S. Das Sarma, V. Fal'ko, M. M. Fogler, E. Fradkin, A. Geim, M. Goerbig, J. Gonzalez, I. Herbut, 
M. I. Katsnelson, P. Kim,  A. Lanzara,  J. Lopes dos Santos, A. MacDonald, E. Mucciolo,  
J. Nilsson,  K. Novoselov, N. Peres, S. Sachdev,    O. Sushkov, 
  O. Vafek, S. Viola,   M. A. H. Vozmediano, and A. Yacoby. 

A.H.C.N. acknowledges DOE grant DE-FG02-08ER46512 and ONR grant MURI N00014-09-1-1063.
B.U. acknowledges partial support from DOE grant DE-FG02-91ER45439 at
   the University of Illinois. 
F.G. acknowledges financial support by MICINN (Spain) through grants 
FIS2008-00124 and CONSOLIDER CSD2007-00010, and 
by the Comunidad de Madrid, through NANOBIOMAG.
V.N.K. acknowledges  the financial support of the University of Vermont.

\bibliographystyle{apsrmp}
\bibliography{References}

\begin{thebibliography}{471}
\expandafter\ifx\csname natexlab\endcsname\relax\def\natexlab#1{#1}\fi
\expandafter\ifx\csname bibnamefont\endcsname\relax
  \def\bibnamefont#1{#1}\fi
\expandafter\ifx\csname bibfnamefont\endcsname\relax
  \def\bibfnamefont#1{#1}\fi
\expandafter\ifx\csname citenamefont\endcsname\relax
  \def\citenamefont#1{#1}\fi
\expandafter\ifx\csname url\endcsname\relax
  \def\url#1{\texttt{#1}}\fi
\expandafter\ifx\csname urlprefix\endcsname\relax\def\urlprefix{URL }\fi
\providecommand{\bibinfo}[2]{#2}
\providecommand{\eprint}[2][]{\url{#2}}

\bibitem[{\citenamefont{Abanin} \emph{et~al.}(2006)\citenamefont{Abanin, Lee,
  and Levitov}}]{ALL06}
\bibinfo{author}{\bibnamefont{Abanin}, \bibfnamefont{D.~A.}},
  \bibinfo{author}{\bibfnamefont{P.~A.} \bibnamefont{Lee}}, and
  \bibinfo{author}{\bibfnamefont{L.~S.} \bibnamefont{Levitov}},
  \bibinfo{year}{2006}, \bibinfo{journal}{Phys. Rev. Lett.}
  \textbf{\bibinfo{volume}{96}}, \bibinfo{pages}{176803}.

\bibitem[{\citenamefont{Abanin} \emph{et~al.}(2007)\citenamefont{Abanin,
  Novoselov, Zeitler, Lee, Geim, and Levitov}}]{Aetal07}
\bibinfo{author}{\bibnamefont{Abanin}, \bibfnamefont{D.~A.}},
  \bibinfo{author}{\bibfnamefont{K.~S.} \bibnamefont{Novoselov}},
  \bibinfo{author}{\bibfnamefont{U.}~\bibnamefont{Zeitler}},
  \bibinfo{author}{\bibfnamefont{P.~A.} \bibnamefont{Lee}},
  \bibinfo{author}{\bibfnamefont{A.~K.} \bibnamefont{Geim}}, and
  \bibinfo{author}{\bibfnamefont{L.~S.} \bibnamefont{Levitov}},
  \bibinfo{year}{2007}, \bibinfo{journal}{Phys. Rev. Lett.}
  \textbf{\bibinfo{volume}{98}}, \bibinfo{pages}{196806}.

\bibitem[{\citenamefont{Abanin} \emph{et~al.}(2009)\citenamefont{Abanin,
  Parameswaran, and Sondhi}}]{Aba09}
\bibinfo{author}{\bibnamefont{Abanin}, \bibfnamefont{D.~A.}},
  \bibinfo{author}{\bibfnamefont{S.~A.} \bibnamefont{Parameswaran}}, and
  \bibinfo{author}{\bibfnamefont{S.~L.} \bibnamefont{Sondhi}},
  \bibinfo{year}{2009}, \bibinfo{journal}{Phys. Rev. Lett.}
  \textbf{\bibinfo{volume}{103}}, \bibinfo{pages}{076802}.

\bibitem[{\citenamefont{Abergel} \emph{et~al.}(2009)\citenamefont{Abergel,
  Pietil{\"a}inen, and Chakraborty}}]{Abergel09}
\bibinfo{author}{\bibnamefont{Abergel}, \bibfnamefont{D.~S.~L.}},
  \bibinfo{author}{\bibfnamefont{P.}~\bibnamefont{Pietil{\"a}inen}}, and
  \bibinfo{author}{\bibfnamefont{T.}~\bibnamefont{Chakraborty}},
  \bibinfo{year}{2009}, \bibinfo{journal}{Phys. Rev. B}
  \textbf{\bibinfo{volume}{80}}, \bibinfo{pages}{081408}.

\bibitem[{\citenamefont{Abramowitz and Stegun}(1964)}]{Abramowitz:1964}
\bibinfo{author}{\bibnamefont{Abramowitz}, \bibfnamefont{M.}}, and
  \bibinfo{author}{\bibfnamefont{I.~A.} \bibnamefont{Stegun}},
  \bibinfo{year}{1964}, \emph{\bibinfo{title}{Handbook of Mathematical
  Functions}} (\bibinfo{publisher}{Dover}, \bibinfo{address}{New York}).

\bibitem[{\citenamefont{Abrikosov and Beneslavskii}(1971)}]{Abr71}
\bibinfo{author}{\bibnamefont{Abrikosov}, \bibfnamefont{A.~A.}}, and
  \bibinfo{author}{\bibfnamefont{S.~D.} \bibnamefont{Beneslavskii}},
  \bibinfo{year}{1971}, \bibinfo{journal}{Sov. Phys. JETP}
  \textbf{\bibinfo{volume}{32}}, \bibinfo{pages}{699}.

\bibitem[{\citenamefont{Adam} \emph{et~al.}(2007)\citenamefont{Adam, Hwang,
  Galitski, and {Das Sarma}}}]{Adam:2007}
\bibinfo{author}{\bibnamefont{Adam}, \bibfnamefont{S.}},
  \bibinfo{author}{\bibfnamefont{E.~H.} \bibnamefont{Hwang}},
  \bibinfo{author}{\bibfnamefont{V.~M.} \bibnamefont{Galitski}}, and
  \bibinfo{author}{\bibfnamefont{S.}~\bibnamefont{{Das Sarma}}},
  \bibinfo{year}{2007}, \bibinfo{journal}{Proc. Nat. Acad. Sci.}
  \textbf{\bibinfo{volume}{104}}, \bibinfo{pages}{18392}.

\bibitem[{\citenamefont{Akhmerov and Beenakker}(2007)}]{Akh07}
\bibinfo{author}{\bibnamefont{Akhmerov}, \bibfnamefont{A.~R.}}, and
  \bibinfo{author}{\bibfnamefont{C.~W.~J.} \bibnamefont{Beenakker}},
  \bibinfo{year}{2007}, \bibinfo{journal}{Phys. Rev. Lett.}
  \textbf{\bibinfo{volume}{98}}, \bibinfo{pages}{157003}.

\bibitem[{\citenamefont{Akhmerov and Beenakker}(2008)}]{AB08}
\bibinfo{author}{\bibnamefont{Akhmerov}, \bibfnamefont{A.~R.}}, and
  \bibinfo{author}{\bibfnamefont{C.~W.~J.} \bibnamefont{Beenakker}},
  \bibinfo{year}{2008}, \bibinfo{journal}{Phys. Rev. B}
  \textbf{\bibinfo{volume}{77}}, \bibinfo{pages}{085423}.

\bibitem[{\citenamefont{Aleiner and Efetov}(2006)}]{Aleiner06}
\bibinfo{author}{\bibnamefont{Aleiner}, \bibfnamefont{I.~L.}}, and
  \bibinfo{author}{\bibfnamefont{K.~B.} \bibnamefont{Efetov}},
  \bibinfo{year}{2006}, \bibinfo{journal}{Phys. Rev. Lett.}
  \textbf{\bibinfo{volume}{97}}, \bibinfo{pages}{236801}.

\bibitem[{\citenamefont{Aleiner} \emph{et~al.}(2007)\citenamefont{Aleiner,
  Kharzeev, and Tsvelik}}]{Kharzeev07}
\bibinfo{author}{\bibnamefont{Aleiner}, \bibfnamefont{I.~L.}},
  \bibinfo{author}{\bibfnamefont{D.~E.} \bibnamefont{Kharzeev}}, and
  \bibinfo{author}{\bibfnamefont{A.~M.} \bibnamefont{Tsvelik}},
  \bibinfo{year}{2007}, \bibinfo{journal}{Phys. Rev. B}
  \textbf{\bibinfo{volume}{76}}, \bibinfo{pages}{195415}.

\bibitem[{\citenamefont{Alicea and Fisher}(2006)}]{AF06}
\bibinfo{author}{\bibnamefont{Alicea}, \bibfnamefont{J.}}, and
  \bibinfo{author}{\bibfnamefont{M.~P.} \bibnamefont{Fisher}},
  \bibinfo{year}{2006}, \bibinfo{journal}{Phys. Rev. B}
  \textbf{\bibinfo{volume}{74}}, \bibinfo{pages}{075422}.

\bibitem[{\citenamefont{Amado} \emph{et~al.}(2009)\citenamefont{Amado, Diez,
  {L{\'o}pez-Romero}, Rossella, Caridad, Bellani, and Maude}}]{Aetal09}
\bibinfo{author}{\bibnamefont{Amado}, \bibfnamefont{M.}},
  \bibinfo{author}{\bibfnamefont{E.}~\bibnamefont{Diez}},
  \bibinfo{author}{\bibfnamefont{D.}~\bibnamefont{{L{\'o}pez-Romero}}},
  \bibinfo{author}{\bibfnamefont{F.}~\bibnamefont{Rossella}},
  \bibinfo{author}{\bibfnamefont{J.~M.} \bibnamefont{Caridad}},
  \bibinfo{author}{\bibfnamefont{V.}~\bibnamefont{Bellani}}, and
  \bibinfo{author}{\bibfnamefont{D.~K.} \bibnamefont{Maude}},
  \bibinfo{year}{2009}, \bibinfo{journal}{New. J. Phys.}
  \textbf{\bibinfo{volume}{12}}, \bibinfo{pages}{053004}.

\bibitem[{\citenamefont{Anderson}(1961)}]{And61}
\bibinfo{author}{\bibnamefont{Anderson}, \bibfnamefont{P.~W.}},
  \bibinfo{year}{1961}, \bibinfo{journal}{Phys. Rev.}
  \textbf{\bibinfo{volume}{124}}, \bibinfo{pages}{41}.

\bibitem[{\citenamefont{Anderson}(1967)}]{A67}
\bibinfo{author}{\bibnamefont{Anderson}, \bibfnamefont{P.~W.}},
  \bibinfo{year}{1967}, \bibinfo{journal}{Phys. Rev. Lett.}
  \textbf{\bibinfo{volume}{18}}, \bibinfo{pages}{1049}.

\bibitem[{\citenamefont{Ando}(2006)}]{Ando:2006}
\bibinfo{author}{\bibnamefont{Ando}, \bibfnamefont{T.}}, \bibinfo{year}{2006},
  \bibinfo{journal}{J. Phys. Soc. Jpn.} \textbf{\bibinfo{volume}{75}},
  \bibinfo{pages}{074716}.

\bibitem[{\citenamefont{Ando}(2007)}]{Ando:2007}
\bibinfo{author}{\bibnamefont{Ando}, \bibfnamefont{T.}}, \bibinfo{year}{2007},
  \bibinfo{journal}{J. Phys. Soc. Jpn.} \textbf{\bibinfo{volume}{76}},
  \bibinfo{pages}{104711}.

\bibitem[{\citenamefont{Ando} \emph{et~al.}(1982)\citenamefont{Ando, Fowler,
  and Stern}}]{Ando:1982}
\bibinfo{author}{\bibnamefont{Ando}, \bibfnamefont{T.}},
  \bibinfo{author}{\bibfnamefont{A.~B.} \bibnamefont{Fowler}}, and
  \bibinfo{author}{\bibfnamefont{F.}~\bibnamefont{Stern}},
  \bibinfo{year}{1982}, \bibinfo{journal}{Rev. Mod. Phys.}
  \textbf{\bibinfo{volume}{54}}, \bibinfo{pages}{437}.

\bibitem[{\citenamefont{Apalkov and Chakraborty}(2006)}]{AC06}
\bibinfo{author}{\bibnamefont{Apalkov}, \bibfnamefont{V.~M.}}, and
  \bibinfo{author}{\bibfnamefont{T.}~\bibnamefont{Chakraborty}},
  \bibinfo{year}{2006}, \bibinfo{journal}{Phys. Rev. Lett.}
  \textbf{\bibinfo{volume}{97}}, \bibinfo{pages}{126801}.

\bibitem[{\citenamefont{Appelquist}
  \emph{et~al.}(1988)\citenamefont{Appelquist, Bostwick, Karabali, T., and
  Wijewardhana}}]{App88}
\bibinfo{author}{\bibnamefont{Appelquist}, \bibfnamefont{T.}},
  \bibinfo{author}{\bibfnamefont{M.}~\bibnamefont{Bostwick}},
  \bibinfo{author}{\bibfnamefont{D.}~\bibnamefont{Karabali}},
  \bibinfo{author}{\bibnamefont{T.}}, and \bibinfo{author}{\bibfnamefont{L.~R.}
  \bibnamefont{Wijewardhana}}, \bibinfo{year}{1988}, \bibinfo{journal}{Phys.
  Rev. D} \textbf{\bibinfo{volume}{60}}, \bibinfo{pages}{2575}.

\bibitem[{\citenamefont{Appelquist}
  \emph{et~al.}(1986)\citenamefont{Appelquist, Bostwick, Karabali, and
  Wijewardhana}}]{Appel86}
\bibinfo{author}{\bibnamefont{Appelquist}, \bibfnamefont{T.~W.}},
  \bibinfo{author}{\bibfnamefont{M.}~\bibnamefont{Bostwick}},
  \bibinfo{author}{\bibfnamefont{D.}~\bibnamefont{Karabali}}, and
  \bibinfo{author}{\bibfnamefont{L.~C.~R.} \bibnamefont{Wijewardhana}},
  \bibinfo{year}{1986}, \bibinfo{journal}{Phys. Rev. D}
  \textbf{\bibinfo{volume}{33}}, \bibinfo{pages}{3704}.

\bibitem[{\citenamefont{Avouris} \emph{et~al.}(2007)\citenamefont{Avouris,
  Chen, and Perebeinos}}]{ACP07}
\bibinfo{author}{\bibnamefont{Avouris}, \bibfnamefont{P.}},
  \bibinfo{author}{\bibfnamefont{Z.}~\bibnamefont{Chen}}, and
  \bibinfo{author}{\bibfnamefont{V.}~\bibnamefont{Perebeinos}},
  \bibinfo{year}{2007}, \bibinfo{journal}{Nature Nanotechnology}
  \textbf{\bibinfo{volume}{2}}, \bibinfo{pages}{605}.

\bibitem[{\citenamefont{Balatsky} \emph{et~al.}(2006)\citenamefont{Balatsky,
  Vekhter, and Zhu}}]{Bal06}
\bibinfo{author}{\bibnamefont{Balatsky}, \bibfnamefont{A.~V.}},
  \bibinfo{author}{\bibfnamefont{I.}~\bibnamefont{Vekhter}}, and
  \bibinfo{author}{\bibfnamefont{J.~X.} \bibnamefont{Zhu}},
  \bibinfo{year}{2006}, \bibinfo{journal}{Rev. Mod. Phys.}
  \textbf{\bibinfo{volume}{78}}, \bibinfo{pages}{373}.

\bibitem[{\citenamefont{Barlas} \emph{et~al.}(2007)\citenamefont{Barlas,
  Pereg-Barnea, Polini, Asgari, and MacDonald}}]{Bar07}
\bibinfo{author}{\bibnamefont{Barlas}, \bibfnamefont{Y.}},
  \bibinfo{author}{\bibfnamefont{T.}~\bibnamefont{Pereg-Barnea}},
  \bibinfo{author}{\bibfnamefont{M.}~\bibnamefont{Polini}},
  \bibinfo{author}{\bibfnamefont{R.}~\bibnamefont{Asgari}}, and
  \bibinfo{author}{\bibfnamefont{A.~H.} \bibnamefont{MacDonald}},
  \bibinfo{year}{2007}, \bibinfo{journal}{\prl} \textbf{\bibinfo{volume}{98}},
  \bibinfo{pages}{236601}.

\bibitem[{\citenamefont{Barlas and Yang}(2009)}]{Bar09}
\bibinfo{author}{\bibnamefont{Barlas}, \bibfnamefont{Y.}}, and
  \bibinfo{author}{\bibfnamefont{K.}~\bibnamefont{Yang}}, \bibinfo{year}{2009},
  \bibinfo{journal}{Phys. Rev. B} \textbf{\bibinfo{volume}{80}},
  \bibinfo{pages}{161408}.

\bibitem[{\citenamefont{Barraza-Lopez}
  \emph{et~al.}(2010)\citenamefont{Barraza-Lopez, {Vanevi\ifmmode
  \acute{c}\else {\'c}\fi{}}, Kindermann, and Chou}}]{BarrazaLopez:2010}
\bibinfo{author}{\bibnamefont{Barraza-Lopez}, \bibfnamefont{S.}},
  \bibinfo{author}{\bibfnamefont{M.}~\bibnamefont{{Vanevi\ifmmode
  \acute{c}\else {\'c}\fi{}}}},
  \bibinfo{author}{\bibfnamefont{M.}~\bibnamefont{Kindermann}}, and
  \bibinfo{author}{\bibfnamefont{M.~Y.} \bibnamefont{Chou}},
  \bibinfo{year}{2010}, \bibinfo{journal}{Phys. Rev. Lett.}
  \textbf{\bibinfo{volume}{104}}, \bibinfo{pages}{076807}.

\bibitem[{\citenamefont{Barzola-Quiquia}
  \emph{et~al.}(2007)\citenamefont{Barzola-Quiquia, Esquinazi, Rothermel,
  Spemann, Butz, and Garc{\'i}a}}]{Betal07}
\bibinfo{author}{\bibnamefont{Barzola-Quiquia}, \bibfnamefont{J.}},
  \bibinfo{author}{\bibfnamefont{P.}~\bibnamefont{Esquinazi}},
  \bibinfo{author}{\bibfnamefont{M.}~\bibnamefont{Rothermel}},
  \bibinfo{author}{\bibfnamefont{D.}~\bibnamefont{Spemann}},
  \bibinfo{author}{\bibfnamefont{T.}~\bibnamefont{Butz}}, and
  \bibinfo{author}{\bibfnamefont{N.}~\bibnamefont{Garc{\'i}a}},
  \bibinfo{year}{2007}, \bibinfo{journal}{Phys. Rev. B}
  \textbf{\bibinfo{volume}{76}}, \bibinfo{pages}{161403}.

\bibitem[{\citenamefont{Baym}(1969)}]{Baym:69}
\bibinfo{author}{\bibnamefont{Baym}, \bibfnamefont{G.}}, \bibinfo{year}{1969},
  \emph{\bibinfo{title}{Lectures on Quantum Mechanics}}
  (\bibinfo{publisher}{Addison-Wesley}, \bibinfo{address}{Reading, MA}).

\bibitem[{\citenamefont{Baym and Pethick}(1991)}]{Baym:1991}
\bibinfo{author}{\bibnamefont{Baym}, \bibfnamefont{G.}}, and
  \bibinfo{author}{\bibfnamefont{C.}~\bibnamefont{Pethick}},
  \bibinfo{year}{1991}, \emph{\bibinfo{title}{Landau Fermi-Liquid Theory}}
  (\bibinfo{publisher}{John Wiley}, \bibinfo{address}{New York}).

\bibitem[{\citenamefont{Beenakker}(2006)}]{Bee06}
\bibinfo{author}{\bibnamefont{Beenakker}, \bibfnamefont{C.~W.~J.}},
  \bibinfo{year}{2006}, \bibinfo{journal}{Phys. Rev. Lett.}
  \textbf{\bibinfo{volume}{97}}, \bibinfo{pages}{067007}.

\bibitem[{\citenamefont{Beenakker}(2008)}]{Bee08}
\bibinfo{author}{\bibnamefont{Beenakker}, \bibfnamefont{C.~W.~J.}},
  \bibinfo{year}{2008}, \bibinfo{journal}{Rev. Mod. Phys.}
  \textbf{\bibinfo{volume}{80}}, \bibinfo{pages}{1337}.

\bibitem[{\citenamefont{Bena}(2008)}]{Ben08}
\bibinfo{author}{\bibnamefont{Bena}, \bibfnamefont{C.}}, \bibinfo{year}{2008},
  \bibinfo{journal}{Phys. Rev. Lett.} \textbf{\bibinfo{volume}{100}},
  \bibinfo{pages}{076601}.

\bibitem[{\citenamefont{Bena and Kivelson}(2006)}]{Ben05}
\bibinfo{author}{\bibnamefont{Bena}, \bibfnamefont{C.}}, and
  \bibinfo{author}{\bibfnamefont{S.}~\bibnamefont{Kivelson}},
  \bibinfo{year}{2006}, \bibinfo{journal}{Phys. Rev. B}
  \textbf{\bibinfo{volume}{72}}, \bibinfo{pages}{125432}.

\bibitem[{\citenamefont{Bercx} \emph{et~al.}(2009)\citenamefont{Bercx, Lang,
  and Assaad}}]{Berc09}
\bibinfo{author}{\bibnamefont{Bercx}, \bibfnamefont{M.}},
  \bibinfo{author}{\bibfnamefont{T.~C.} \bibnamefont{Lang}}, and
  \bibinfo{author}{\bibfnamefont{F.~F.} \bibnamefont{Assaad}},
  \bibinfo{year}{2009}, \bibinfo{journal}{Phys. Rev. B}
  \textbf{\bibinfo{volume}{80}}, \bibinfo{pages}{045412}.

\bibitem[{\citenamefont{Bergman and Hur}(2009)}]{Ber09}
\bibinfo{author}{\bibnamefont{Bergman}, \bibfnamefont{D.}}, and
  \bibinfo{author}{\bibfnamefont{K.~L.} \bibnamefont{Hur}},
  \bibinfo{year}{2009}, \bibinfo{journal}{Phys. Rev. B}
  \textbf{\bibinfo{volume}{79}}, \bibinfo{pages}{185420}.

\bibitem[{\citenamefont{Bhattacharjee and Sengupta}(2006)}]{Bha06}
\bibinfo{author}{\bibnamefont{Bhattacharjee}, \bibfnamefont{S.}}, and
  \bibinfo{author}{\bibfnamefont{K.}~\bibnamefont{Sengupta}},
  \bibinfo{year}{2006}, \bibinfo{journal}{Phys. Rev. Lett.}
  \textbf{\bibinfo{volume}{97}}, \bibinfo{pages}{217001}.

\bibitem[{\citenamefont{Biswas} \emph{et~al.}(2007)\citenamefont{Biswas,
  Sachdev, and Son}}]{Biswas:2007}
\bibinfo{author}{\bibnamefont{Biswas}, \bibfnamefont{R.~R.}},
  \bibinfo{author}{\bibfnamefont{S.}~\bibnamefont{Sachdev}}, and
  \bibinfo{author}{\bibfnamefont{D.~T.} \bibnamefont{Son}},
  \bibinfo{year}{2007}, \bibinfo{journal}{Phys. Rev. B}
  \textbf{\bibinfo{volume}{76}}, \bibinfo{pages}{205122}.

\bibitem[{\citenamefont{Black-Schaffer and Doniach}(2007)}]{Bla07}
\bibinfo{author}{\bibnamefont{Black-Schaffer}, \bibfnamefont{A.~M.}}, and
  \bibinfo{author}{\bibfnamefont{S.}~\bibnamefont{Doniach}},
  \bibinfo{year}{2007}, \bibinfo{journal}{Phys. Rev. B}
  \textbf{\bibinfo{volume}{75}}, \bibinfo{pages}{134512}.

\bibitem[{\citenamefont{Black-Schaffer and Doniach}(2008)}]{Bla08}
\bibinfo{author}{\bibnamefont{Black-Schaffer}, \bibfnamefont{A.~M.}}, and
  \bibinfo{author}{\bibfnamefont{S.}~\bibnamefont{Doniach}},
  \bibinfo{year}{2008}, \bibinfo{journal}{Phys. Rev. B}
  \textbf{\bibinfo{volume}{78}}, \bibinfo{pages}{024504}.

\bibitem[{\citenamefont{Bloch}(1928)}]{Bloch:28}
\bibinfo{author}{\bibnamefont{Bloch}, \bibfnamefont{F.}}, \bibinfo{year}{1928},
  \bibinfo{journal}{Z. Physik} \textbf{\bibinfo{volume}{52}},
  \bibinfo{pages}{555}.

\bibitem[{\citenamefont{Bolotin} \emph{et~al.}(2009)\citenamefont{Bolotin,
  Ghahari, Shulman, Stormer, and Kim}}]{BGSSK09}
\bibinfo{author}{\bibnamefont{Bolotin}, \bibfnamefont{K.~I.}},
  \bibinfo{author}{\bibfnamefont{F.}~\bibnamefont{Ghahari}},
  \bibinfo{author}{\bibfnamefont{M.~D.} \bibnamefont{Shulman}},
  \bibinfo{author}{\bibfnamefont{H.~L.} \bibnamefont{Stormer}}, and
  \bibinfo{author}{\bibfnamefont{P.}~\bibnamefont{Kim}}, \bibinfo{year}{2009},
  \bibinfo{journal}{Nature} \textbf{\bibinfo{volume}{462}},
  \bibinfo{pages}{196}.

\bibitem[{\citenamefont{Borghi} \emph{et~al.}(2010)\citenamefont{Borghi,
  Polini, Asgari, and MacDonald}}]{Borghi10}
\bibinfo{author}{\bibnamefont{Borghi}, \bibfnamefont{G.}},
  \bibinfo{author}{\bibfnamefont{M.}~\bibnamefont{Polini}},
  \bibinfo{author}{\bibfnamefont{R.}~\bibnamefont{Asgari}}, and
  \bibinfo{author}{\bibfnamefont{A.}~\bibnamefont{MacDonald}},
  \bibinfo{year}{2010}, \bibinfo{journal}{Phys. Rev. B}
  \textbf{\bibinfo{volume}{82}}, \bibinfo{pages}{155403}.

\bibitem[{\citenamefont{Borghi}
  \emph{et~al.}(2009{\natexlab{a}})\citenamefont{Borghi, Polini, Asgari, and
  MacDonald}}]{Bor09}
\bibinfo{author}{\bibnamefont{Borghi}, \bibfnamefont{G.}},
  \bibinfo{author}{\bibfnamefont{M.}~\bibnamefont{Polini}},
  \bibinfo{author}{\bibfnamefont{R.}~\bibnamefont{Asgari}}, and
  \bibinfo{author}{\bibfnamefont{A.~H.} \bibnamefont{MacDonald}},
  \bibinfo{year}{2009}{\natexlab{a}}, \bibinfo{journal}{Solid State Commun.}
  \textbf{\bibinfo{volume}{149}}, \bibinfo{pages}{1117}.

\bibitem[{\citenamefont{Borghi}
  \emph{et~al.}(2009{\natexlab{b}})\citenamefont{Borghi, Polini, Asgari, and
  MacDonald}}]{Bor09b}
\bibinfo{author}{\bibnamefont{Borghi}, \bibfnamefont{G.}},
  \bibinfo{author}{\bibfnamefont{M.}~\bibnamefont{Polini}},
  \bibinfo{author}{\bibfnamefont{R.}~\bibnamefont{Asgari}}, and
  \bibinfo{author}{\bibfnamefont{A.~H.} \bibnamefont{MacDonald}},
  \bibinfo{year}{2009}{\natexlab{b}}, \bibinfo{journal}{Phys. Rev. B}
  \textbf{\bibinfo{volume}{80}}, \bibinfo{pages}{241402}.

\bibitem[{\citenamefont{Borkowski and Hirschfeld}(1992)}]{Bor92}
\bibinfo{author}{\bibnamefont{Borkowski}, \bibfnamefont{L.~S.}}, and
  \bibinfo{author}{\bibfnamefont{P.~J.} \bibnamefont{Hirschfeld}},
  \bibinfo{year}{1992}, \bibinfo{journal}{Phys. Rev. B}
  \textbf{\bibinfo{volume}{46}}, \bibinfo{pages}{9274}.

\bibitem[{\citenamefont{Bostwick} \emph{et~al.}(2007)\citenamefont{Bostwick,
  Ohta, Seyller, Horn, and Rotenberg}}]{Bos07}
\bibinfo{author}{\bibnamefont{Bostwick}, \bibfnamefont{A.}},
  \bibinfo{author}{\bibfnamefont{T.}~\bibnamefont{Ohta}},
  \bibinfo{author}{\bibfnamefont{T.}~\bibnamefont{Seyller}},
  \bibinfo{author}{\bibfnamefont{K.}~\bibnamefont{Horn}}, and
  \bibinfo{author}{\bibfnamefont{E.}~\bibnamefont{Rotenberg}},
  \bibinfo{year}{2007}, \bibinfo{journal}{Nature Physics}
  \textbf{\bibinfo{volume}{3}}, \bibinfo{pages}{36}.

\bibitem[{\citenamefont{Bostwick} \emph{et~al.}(2010)\citenamefont{Bostwick,
  Speck, Seyller, Horn, Polini, Asgari, MacDonald, and Rotenberg}}]{Bos10}
\bibinfo{author}{\bibnamefont{Bostwick}, \bibfnamefont{A.}},
  \bibinfo{author}{\bibfnamefont{F.}~\bibnamefont{Speck}},
  \bibinfo{author}{\bibfnamefont{T.}~\bibnamefont{Seyller}},
  \bibinfo{author}{\bibfnamefont{K.}~\bibnamefont{Horn}},
  \bibinfo{author}{\bibfnamefont{M.}~\bibnamefont{Polini}},
  \bibinfo{author}{\bibfnamefont{R.}~\bibnamefont{Asgari}},
  \bibinfo{author}{\bibfnamefont{A.~H.} \bibnamefont{MacDonald}}, and
  \bibinfo{author}{\bibfnamefont{E.}~\bibnamefont{Rotenberg}},
  \bibinfo{year}{2010}, \bibinfo{journal}{Science}
  \textbf{\bibinfo{volume}{328}}, \bibinfo{pages}{999}.

\bibitem[{\citenamefont{Boukhvalov and Katsnelson}(2009)}]{Bou09}
\bibinfo{author}{\bibnamefont{Boukhvalov}, \bibfnamefont{D.~W.}}, and
  \bibinfo{author}{\bibfnamefont{M.~I.} \bibnamefont{Katsnelson}},
  \bibinfo{year}{2009}, \bibinfo{journal}{J. Phys. Chem. C}
  \textbf{\bibinfo{volume}{113}}, \bibinfo{pages}{14176}.

\bibitem[{\citenamefont{Boukhvalov}
  \emph{et~al.}(2008)\citenamefont{Boukhvalov, Katsnelson, and
  Lichtenstein}}]{Bou08}
\bibinfo{author}{\bibnamefont{Boukhvalov}, \bibfnamefont{D.~W.}},
  \bibinfo{author}{\bibfnamefont{M.~I.} \bibnamefont{Katsnelson}}, and
  \bibinfo{author}{\bibfnamefont{A.~I.} \bibnamefont{Lichtenstein}},
  \bibinfo{year}{2008}, \bibinfo{journal}{Phys. Rev. B}
  \textbf{\bibinfo{volume}{77}}, \bibinfo{pages}{035427}.

\bibitem[{\citenamefont{Brar} \emph{et~al.}(2010)\citenamefont{Brar,
  Wickenburg, Panlasigui, Park, Wehling, Zhang, Decker, Girit, Balatsky, Louie,
  Zettl, and Crommie}}]{Brar10}
\bibinfo{author}{\bibnamefont{Brar}, \bibfnamefont{V.~W.}},
  \bibinfo{author}{\bibfnamefont{S.}~\bibnamefont{Wickenburg}},
  \bibinfo{author}{\bibfnamefont{M.}~\bibnamefont{Panlasigui}},
  \bibinfo{author}{\bibfnamefont{C.-H.} \bibnamefont{Park}},
  \bibinfo{author}{\bibfnamefont{T.~O.} \bibnamefont{Wehling}},
  \bibinfo{author}{\bibfnamefont{Y.}~\bibnamefont{Zhang}},
  \bibinfo{author}{\bibfnamefont{R.}~\bibnamefont{Decker}},
  \bibinfo{author}{\bibfnamefont{C.}~\bibnamefont{Girit}},
  \bibinfo{author}{\bibfnamefont{A.~V.} \bibnamefont{Balatsky}},
  \bibinfo{author}{\bibfnamefont{S.~G.} \bibnamefont{Louie}},
  \bibinfo{author}{\bibfnamefont{A.}~\bibnamefont{Zettl}}, and
  \bibinfo{author}{\bibfnamefont{M.~F.} \bibnamefont{Crommie}},
  \bibinfo{year}{2010}, \bibinfo{journal}{Phys. Rev. Lett.}
  \textbf{\bibinfo{volume}{104}}, \bibinfo{pages}{036805}.

\bibitem[{\citenamefont{Brey} \emph{et~al.}(2007)\citenamefont{Brey, Fertig,
  and {Das Sarma}}}]{Bre07}
\bibinfo{author}{\bibnamefont{Brey}, \bibfnamefont{L.}},
  \bibinfo{author}{\bibfnamefont{H.~A.} \bibnamefont{Fertig}}, and
  \bibinfo{author}{\bibfnamefont{S.}~\bibnamefont{{Das Sarma}}},
  \bibinfo{year}{2007}, \bibinfo{journal}{Phys. Rev. Lett.}
  \textbf{\bibinfo{volume}{99}}, \bibinfo{pages}{116802}.

\bibitem[{\citenamefont{Brown} \emph{et~al.}(1975)\citenamefont{Brown, Cahn,
  and McLerran}}]{Brown:1975}
\bibinfo{author}{\bibnamefont{Brown}, \bibfnamefont{L.~S.}},
  \bibinfo{author}{\bibfnamefont{R.~N.} \bibnamefont{Cahn}}, and
  \bibinfo{author}{\bibfnamefont{L.~D.} \bibnamefont{McLerran}},
  \bibinfo{year}{1975}, \bibinfo{journal}{Phys. Rev. D}
  \textbf{\bibinfo{volume}{12}}, \bibinfo{pages}{581}.

\bibitem[{\citenamefont{Bunch} \emph{et~al.}(2005)\citenamefont{Bunch, Yaish,
  Brink, Bolotin, and McEuen}}]{Betal05}
\bibinfo{author}{\bibnamefont{Bunch}, \bibfnamefont{J.~S.}},
  \bibinfo{author}{\bibfnamefont{Y.}~\bibnamefont{Yaish}},
  \bibinfo{author}{\bibfnamefont{M.}~\bibnamefont{Brink}},
  \bibinfo{author}{\bibfnamefont{K.}~\bibnamefont{Bolotin}}, and
  \bibinfo{author}{\bibfnamefont{P.~L.} \bibnamefont{McEuen}},
  \bibinfo{year}{2005}, \bibinfo{journal}{Nano Lett.}
  \textbf{\bibinfo{volume}{5}}, \bibinfo{pages}{287}.

\bibitem[{\citenamefont{Burset} \emph{et~al.}(2008)\citenamefont{Burset,
  Yeyati, and Mart{\'i}n-Rodero}}]{Bur08}
\bibinfo{author}{\bibnamefont{Burset}, \bibfnamefont{P.}},
  \bibinfo{author}{\bibfnamefont{A.~L.} \bibnamefont{Yeyati}}, and
  \bibinfo{author}{\bibfnamefont{A.}~\bibnamefont{Mart{\'i}n-Rodero}},
  \bibinfo{year}{2008}, \bibinfo{journal}{Phys. Rev. B}
  \textbf{\bibinfo{volume}{77}}, \bibinfo{pages}{205425}.

\bibitem[{\citenamefont{Calandra and Mauri}(2007)}]{Cal07}
\bibinfo{author}{\bibnamefont{Calandra}, \bibfnamefont{M.}}, and
  \bibinfo{author}{\bibfnamefont{F.}~\bibnamefont{Mauri}},
  \bibinfo{year}{2007}, \bibinfo{journal}{Phys. Rev. B}
  \textbf{\bibinfo{volume}{76}}, \bibinfo{pages}{205111}.

\bibitem[{\citenamefont{Case}(1960)}]{Case:1960}
\bibinfo{author}{\bibnamefont{Case}, \bibfnamefont{K.~M.}},
  \bibinfo{year}{1960}, \bibinfo{journal}{Phys. Rev.}
  \textbf{\bibinfo{volume}{80}}, \bibinfo{pages}{797}.

\bibitem[{\citenamefont{Cassanello and Fradkin}(1996)}]{Cas96}
\bibinfo{author}{\bibnamefont{Cassanello}, \bibfnamefont{C.~R.}}, and
  \bibinfo{author}{\bibfnamefont{E.}~\bibnamefont{Fradkin}},
  \bibinfo{year}{1996}, \bibinfo{journal}{Phys. Rev. B}
  \textbf{\bibinfo{volume}{53}}, \bibinfo{pages}{15079}.

\bibitem[{\citenamefont{Cassanello and Fradkin}(1997)}]{Cas97}
\bibinfo{author}{\bibnamefont{Cassanello}, \bibfnamefont{C.~R.}}, and
  \bibinfo{author}{\bibfnamefont{E.}~\bibnamefont{Fradkin}},
  \bibinfo{year}{1997}, \bibinfo{journal}{Phys. Rev. B}
  \textbf{\bibinfo{volume}{56}}, \bibinfo{pages}{11246}.

\bibitem[{\citenamefont{Castro} \emph{et~al.}(2010)\citenamefont{Castro,
  L{\'o}pez-Sancho, and Vozmediano}}]{CLV10}
\bibinfo{author}{\bibnamefont{Castro}, \bibfnamefont{E.~V.}},
  \bibinfo{author}{\bibfnamefont{M.~P.} \bibnamefont{L{\'o}pez-Sancho}}, and
  \bibinfo{author}{\bibfnamefont{M.~A.~H.} \bibnamefont{Vozmediano}},
  \bibinfo{year}{2010}, \bibinfo{journal}{Phys. Rev. Lett.}
  \textbf{\bibinfo{volume}{104}}, \bibinfo{pages}{036802}.

\bibitem[{\citenamefont{Castro} \emph{et~al.}(2007)\citenamefont{Castro,
  Novoselov, Morozov, Peres, {Lopes dos Santos}, Nilsson, Guinea, Geim, and
  {Castro Neto}}}]{Cas07}
\bibinfo{author}{\bibnamefont{Castro}, \bibfnamefont{E.~V.}},
  \bibinfo{author}{\bibfnamefont{K.~S.} \bibnamefont{Novoselov}},
  \bibinfo{author}{\bibfnamefont{S.~V.} \bibnamefont{Morozov}},
  \bibinfo{author}{\bibfnamefont{N.~R.} \bibnamefont{Peres}},
  \bibinfo{author}{\bibfnamefont{J.~M.~B.} \bibnamefont{{Lopes dos Santos}}},
  \bibinfo{author}{\bibfnamefont{J.}~\bibnamefont{Nilsson}},
  \bibinfo{author}{\bibfnamefont{F.}~\bibnamefont{Guinea}},
  \bibinfo{author}{\bibfnamefont{A.~K.} \bibnamefont{Geim}}, and
  \bibinfo{author}{\bibfnamefont{A.~H.} \bibnamefont{{Castro Neto}}},
  \bibinfo{year}{2007}, \bibinfo{journal}{Phys. Rev. Lett.}
  \textbf{\bibinfo{volume}{99}}, \bibinfo{pages}{216802}.

\bibitem[{\citenamefont{Castro}
  \emph{et~al.}(2008{\natexlab{a}})\citenamefont{Castro, Peres, {Lopes dos
  Santos}, {Castro Neto}, and Guinea}}]{CPLNG08}
\bibinfo{author}{\bibnamefont{Castro}, \bibfnamefont{E.~V.}},
  \bibinfo{author}{\bibfnamefont{N.~M.~R.} \bibnamefont{Peres}},
  \bibinfo{author}{\bibfnamefont{J.~M.~B.} \bibnamefont{{Lopes dos Santos}}},
  \bibinfo{author}{\bibfnamefont{A.~H.} \bibnamefont{{Castro Neto}}}, and
  \bibinfo{author}{\bibfnamefont{F.}~\bibnamefont{Guinea}},
  \bibinfo{year}{2008}{\natexlab{a}}, \bibinfo{journal}{Phys. Rev. Lett.}
  \textbf{\bibinfo{volume}{100}}, \bibinfo{pages}{026802}.

\bibitem[{\citenamefont{Castro}
  \emph{et~al.}(2008{\natexlab{b}})\citenamefont{Castro, Peres, Stauber, and
  Silva}}]{Cas08}
\bibinfo{author}{\bibnamefont{Castro}, \bibfnamefont{E.~V.}},
  \bibinfo{author}{\bibfnamefont{N.~M.~R.} \bibnamefont{Peres}},
  \bibinfo{author}{\bibfnamefont{T.}~\bibnamefont{Stauber}}, and
  \bibinfo{author}{\bibfnamefont{N.~P.} \bibnamefont{Silva}},
  \bibinfo{year}{2008}{\natexlab{b}}, \bibinfo{journal}{Phys. Rev. Lett.}
  \textbf{\bibinfo{volume}{100}}, \bibinfo{pages}{186803}.

\bibitem[{\citenamefont{{Castro Neto}}(2001)}]{Cas01}
\bibinfo{author}{\bibnamefont{{Castro Neto}}, \bibfnamefont{A.~H.}},
  \bibinfo{year}{2001}, \bibinfo{journal}{Phys. Rev. Lett.}
  \textbf{\bibinfo{volume}{86}}, \bibinfo{pages}{4382}.

\bibitem[{\citenamefont{{Castro Neto} and Guinea}(2009)}]{Cas09}
\bibinfo{author}{\bibnamefont{{Castro Neto}}, \bibfnamefont{A.~H.}}, and
  \bibinfo{author}{\bibfnamefont{F.}~\bibnamefont{Guinea}},
  \bibinfo{year}{2009}, \bibinfo{journal}{Phys. Rev. Lett.}
  \textbf{\bibinfo{volume}{103}}, \bibinfo{pages}{026804}.

\bibitem[{\citenamefont{{Castro Neto}} \emph{et~al.}(2006)\citenamefont{{Castro
  Neto}, Guinea, and Peres}}]{CGP06}
\bibinfo{author}{\bibnamefont{{Castro Neto}}, \bibfnamefont{A.~H.}},
  \bibinfo{author}{\bibfnamefont{F.}~\bibnamefont{Guinea}}, and
  \bibinfo{author}{\bibfnamefont{N.~M.} \bibnamefont{Peres}},
  \bibinfo{year}{2006}, \bibinfo{journal}{Phys. Rev. B}
  \textbf{\bibinfo{volume}{73}}, \bibinfo{pages}{205408}.

\bibitem[{\citenamefont{{Castro Neto}}
  \emph{et~al.}(2009{\natexlab{a}})\citenamefont{{Castro Neto}, Guinea, Peres,
  Novoselov, and Geim}}]{CastroNeto:09}
\bibinfo{author}{\bibnamefont{{Castro Neto}}, \bibfnamefont{A.~H.}},
  \bibinfo{author}{\bibfnamefont{F.}~\bibnamefont{Guinea}},
  \bibinfo{author}{\bibfnamefont{N.~M.~R.} \bibnamefont{Peres}},
  \bibinfo{author}{\bibfnamefont{K.~S.} \bibnamefont{Novoselov}}, and
  \bibinfo{author}{\bibfnamefont{A.~K.} \bibnamefont{Geim}},
  \bibinfo{year}{2009}{\natexlab{a}}, \bibinfo{journal}{Rev. Mod. Phys.}
  \textbf{\bibinfo{volume}{81}}, \bibinfo{pages}{109}.

\bibitem[{\citenamefont{{Castro Neto}}
  \emph{et~al.}(2009{\natexlab{b}})\citenamefont{{Castro Neto}, Kotov, Nilsson,
  Pereira, Peres, and Uchoa}}]{CastroNeto:2009}
\bibinfo{author}{\bibnamefont{{Castro Neto}}, \bibfnamefont{A.~H.}},
  \bibinfo{author}{\bibfnamefont{V.~N.} \bibnamefont{Kotov}},
  \bibinfo{author}{\bibfnamefont{J.}~\bibnamefont{Nilsson}},
  \bibinfo{author}{\bibfnamefont{V.~M.} \bibnamefont{Pereira}},
  \bibinfo{author}{\bibfnamefont{N.~M.~R.} \bibnamefont{Peres}}, and
  \bibinfo{author}{\bibfnamefont{B.}~\bibnamefont{Uchoa}},
  \bibinfo{year}{2009}{\natexlab{b}}, \bibinfo{journal}{Solid State Commun.}
  \textbf{\bibinfo{volume}{149}}, \bibinfo{pages}{1094}.

\bibitem[{\citenamefont{{Castro Neto}} \emph{et~al.}(2007)\citenamefont{{Castro
  Neto}, Nilsson, Guinea, and Peres}}]{Nil07}
\bibinfo{author}{\bibnamefont{{Castro Neto}}, \bibfnamefont{A.~H.}},
  \bibinfo{author}{\bibfnamefont{J.}~\bibnamefont{Nilsson}},
  \bibinfo{author}{\bibfnamefont{F.}~\bibnamefont{Guinea}}, and
  \bibinfo{author}{\bibfnamefont{N.~M.~R.} \bibnamefont{Peres}},
  \bibinfo{year}{2007}, \bibinfo{journal}{Phys. Rev. B}
  \textbf{\bibinfo{volume}{76}}, \bibinfo{pages}{165416}.

\bibitem[{\citenamefont{Ceni}(2001)}]{Cen01}
\bibinfo{author}{\bibnamefont{Ceni}, \bibfnamefont{R.}}, \bibinfo{year}{2001},
  \bibinfo{journal}{Nucl. Phys. A} \textbf{\bibinfo{volume}{696}},
  \bibinfo{pages}{605}.

\bibitem[{\citenamefont{Ceperley}(1978)}]{ceperley:78}
\bibinfo{author}{\bibnamefont{Ceperley}, \bibfnamefont{D.}},
  \bibinfo{year}{1978}, \bibinfo{journal}{Phys. Rev. B}
  \textbf{\bibinfo{volume}{18}}, \bibinfo{pages}{3126}.

\bibitem[{\citenamefont{Chamon}
  \emph{et~al.}(2008{\natexlab{a}})\citenamefont{Chamon, Hou, Jackiw, Mudry,
  Pi, and Semenoff}}]{Cha08b}
\bibinfo{author}{\bibnamefont{Chamon}, \bibfnamefont{C.}},
  \bibinfo{author}{\bibfnamefont{C.-Y.} \bibnamefont{Hou}},
  \bibinfo{author}{\bibfnamefont{R.}~\bibnamefont{Jackiw}},
  \bibinfo{author}{\bibfnamefont{C.}~\bibnamefont{Mudry}},
  \bibinfo{author}{\bibfnamefont{S.-Y.} \bibnamefont{Pi}}, and
  \bibinfo{author}{\bibfnamefont{G.}~\bibnamefont{Semenoff}},
  \bibinfo{year}{2008}{\natexlab{a}}, \bibinfo{journal}{Phys. Rev. B}
  \textbf{\bibinfo{volume}{77}}, \bibinfo{pages}{235431}.

\bibitem[{\citenamefont{Chamon}
  \emph{et~al.}(2008{\natexlab{b}})\citenamefont{Chamon, Hou, Jackiw, Mudry,
  S.-Y.Pi, and Schnyder}}]{Cha08a}
\bibinfo{author}{\bibnamefont{Chamon}, \bibfnamefont{C.}},
  \bibinfo{author}{\bibfnamefont{C.-Y.} \bibnamefont{Hou}},
  \bibinfo{author}{\bibfnamefont{R.}~\bibnamefont{Jackiw}},
  \bibinfo{author}{\bibfnamefont{C.}~\bibnamefont{Mudry}},
  \bibinfo{author}{\bibnamefont{S.-Y.Pi}}, and
  \bibinfo{author}{\bibfnamefont{A.~P.} \bibnamefont{Schnyder}},
  \bibinfo{year}{2008}{\natexlab{b}}, \bibinfo{journal}{Phys. Rev. Lett.}
  \textbf{\bibinfo{volume}{100}}, \bibinfo{pages}{110405}.

\bibitem[{\citenamefont{Chan} \emph{et~al.}(2008)\citenamefont{Chan, Neaton,
  and Cohen}}]{Cha08}
\bibinfo{author}{\bibnamefont{Chan}, \bibfnamefont{K.~T.}},
  \bibinfo{author}{\bibfnamefont{J.~B.} \bibnamefont{Neaton}}, and
  \bibinfo{author}{\bibfnamefont{M.~L.} \bibnamefont{Cohen}},
  \bibinfo{year}{2008}, \bibinfo{journal}{Phys. Rev. B}
  \textbf{\bibinfo{volume}{77}}, \bibinfo{pages}{235430}.

\bibitem[{\citenamefont{Checkelsky}
  \emph{et~al.}(2008)\citenamefont{Checkelsky, Li, and Ong}}]{CLO08}
\bibinfo{author}{\bibnamefont{Checkelsky}, \bibfnamefont{J.~G.}},
  \bibinfo{author}{\bibfnamefont{L.}~\bibnamefont{Li}}, and
  \bibinfo{author}{\bibfnamefont{N.~P.} \bibnamefont{Ong}},
  \bibinfo{year}{2008}, \bibinfo{journal}{Phys. Rev. Lett.}
  \textbf{\bibinfo{volume}{100}}, \bibinfo{pages}{206801}.

\bibitem[{\citenamefont{Checkelsky}
  \emph{et~al.}(2009)\citenamefont{Checkelsky, Li, and Ong}}]{CLO09}
\bibinfo{author}{\bibnamefont{Checkelsky}, \bibfnamefont{J.~G.}},
  \bibinfo{author}{\bibfnamefont{L.}~\bibnamefont{Li}}, and
  \bibinfo{author}{\bibfnamefont{N.~P.} \bibnamefont{Ong}},
  \bibinfo{year}{2009}, \bibinfo{journal}{Phys. Rev. B}
  \textbf{\bibinfo{volume}{79}}, \bibinfo{pages}{115434}.

\bibitem[{\citenamefont{Cheianov and Fal'ko}(2006)}]{Che06}
\bibinfo{author}{\bibnamefont{Cheianov}, \bibfnamefont{V.~V.}}, and
  \bibinfo{author}{\bibfnamefont{V.~I.} \bibnamefont{Fal'ko}},
  \bibinfo{year}{2006}, \bibinfo{journal}{Phys. Rev. Lett.}
  \textbf{\bibinfo{volume}{97}}, \bibinfo{pages}{226801}.

\bibitem[{\citenamefont{Chen}
  \emph{et~al.}(2009{\natexlab{a}})\citenamefont{Chen, Xia, Ferry, and
  Tao}}]{Chen:2009-1}
\bibinfo{author}{\bibnamefont{Chen}, \bibfnamefont{F.}},
  \bibinfo{author}{\bibfnamefont{J.}~\bibnamefont{Xia}},
  \bibinfo{author}{\bibfnamefont{D.~K.} \bibnamefont{Ferry}}, and
  \bibinfo{author}{\bibfnamefont{N.}~\bibnamefont{Tao}},
  \bibinfo{year}{2009}{\natexlab{a}}, \bibinfo{journal}{Nano Lett.}
  \textbf{\bibinfo{volume}{9}}, \bibinfo{pages}{2571}.

\bibitem[{\citenamefont{Chen}
  \emph{et~al.}(2009{\natexlab{b}})\citenamefont{Chen, Xia, and
  Tao}}]{Chen:2009-2}
\bibinfo{author}{\bibnamefont{Chen}, \bibfnamefont{F.}},
  \bibinfo{author}{\bibfnamefont{J.}~\bibnamefont{Xia}}, and
  \bibinfo{author}{\bibfnamefont{N.}~\bibnamefont{Tao}},
  \bibinfo{year}{2009}{\natexlab{b}}, \bibinfo{journal}{Nano Lett.}
  \textbf{\bibinfo{volume}{9}}, \bibinfo{pages}{1621}.

\bibitem[{\citenamefont{Chen} \emph{et~al.}(2011)\citenamefont{Chen, Cullen,
  Williams, and Fuhrer}}]{CCWF10}
\bibinfo{author}{\bibnamefont{Chen}, \bibfnamefont{J.-H.}},
  \bibinfo{author}{\bibfnamefont{W.~G.} \bibnamefont{Cullen}},
  \bibinfo{author}{\bibfnamefont{E.~D.} \bibnamefont{Williams}}, and
  \bibinfo{author}{\bibfnamefont{M.~S.} \bibnamefont{Fuhrer}},
  \bibinfo{year}{2011}, \bibinfo{journal}{Nature Physics}
  \textbf{\bibinfo{volume}{7}}, \bibinfo{pages}{535}.

\bibitem[{\citenamefont{Chen} \emph{et~al.}(2008)\citenamefont{Chen, Jang,
  Adam, Fuhrer, Williams, and Ishigami}}]{Chen:2008}
\bibinfo{author}{\bibnamefont{Chen}, \bibfnamefont{J.~H.}},
  \bibinfo{author}{\bibfnamefont{C.}~\bibnamefont{Jang}},
  \bibinfo{author}{\bibfnamefont{S.}~\bibnamefont{Adam}},
  \bibinfo{author}{\bibfnamefont{M.~S.} \bibnamefont{Fuhrer}},
  \bibinfo{author}{\bibfnamefont{E.~D.} \bibnamefont{Williams}}, and
  \bibinfo{author}{\bibfnamefont{M.}~\bibnamefont{Ishigami}},
  \bibinfo{year}{2008}, \bibinfo{journal}{Nature Physics}
  \textbf{\bibinfo{volume}{4}}, \bibinfo{pages}{377}.

\bibitem[{\citenamefont{Chen and Jayaprakash}(1995)}]{Che95}
\bibinfo{author}{\bibnamefont{Chen}, \bibfnamefont{K.}}, and
  \bibinfo{author}{\bibfnamefont{C.}~\bibnamefont{Jayaprakash}},
  \bibinfo{year}{1995}, \bibinfo{journal}{J. Phys.: Condens. Mattter}
  \textbf{\bibinfo{volume}{7}}, \bibinfo{pages}{L491}.

\bibitem[{\citenamefont{Chen} \emph{et~al.}(2010)\citenamefont{Chen, Bae,
  Chialvo, Dirks, Bezryadin, and Mason}}]{Cetal10}
\bibinfo{author}{\bibnamefont{Chen}, \bibfnamefont{Y.-F.}},
  \bibinfo{author}{\bibfnamefont{M.-H.} \bibnamefont{Bae}},
  \bibinfo{author}{\bibfnamefont{C.}~\bibnamefont{Chialvo}},
  \bibinfo{author}{\bibfnamefont{T.}~\bibnamefont{Dirks}},
  \bibinfo{author}{\bibfnamefont{A.}~\bibnamefont{Bezryadin}}, and
  \bibinfo{author}{\bibfnamefont{N.}~\bibnamefont{Mason}},
  \bibinfo{year}{2010}, \bibinfo{journal}{J. Phys.: Condens. Matter}
  \textbf{\bibinfo{volume}{22}}, \bibinfo{pages}{205301}.

\bibitem[{\citenamefont{Coleman}(1983)}]{Col83}
\bibinfo{author}{\bibnamefont{Coleman}, \bibfnamefont{P.}},
  \bibinfo{year}{1983}, \bibinfo{journal}{Phys. Rev. B}
  \textbf{\bibinfo{volume}{28}}, \bibinfo{pages}{5255}.

\bibitem[{\citenamefont{Cooper}(1956)}]{Cooper:1956}
\bibinfo{author}{\bibnamefont{Cooper}, \bibfnamefont{L.~N.}},
  \bibinfo{year}{1956}, \bibinfo{journal}{Phys. Rev.}
  \textbf{\bibinfo{volume}{104}}, \bibinfo{pages}{1189}.

\bibitem[{\citenamefont{Coqblin and Schrieffer}(1969)}]{Coc69}
\bibinfo{author}{\bibnamefont{Coqblin}, \bibfnamefont{B.}}, and
  \bibinfo{author}{\bibfnamefont{J.~R.} \bibnamefont{Schrieffer}},
  \bibinfo{year}{1969}, \bibinfo{journal}{Phys. Rev.}
  \textbf{\bibinfo{volume}{185}}, \bibinfo{pages}{847}.

\bibitem[{\citenamefont{Cornaglia} \emph{et~al.}(2009)\citenamefont{Cornaglia,
  Usaj, and Balseiro}}]{Cor09}
\bibinfo{author}{\bibnamefont{Cornaglia}, \bibfnamefont{P.~S.}},
  \bibinfo{author}{\bibfnamefont{G.}~\bibnamefont{Usaj}}, and
  \bibinfo{author}{\bibfnamefont{C.~A.} \bibnamefont{Balseiro}},
  \bibinfo{year}{2009}, \bibinfo{journal}{Phys. Rev. Lett.}
  \textbf{\bibinfo{volume}{102}}, \bibinfo{pages}{046801}.

\bibitem[{\citenamefont{Csanyi} \emph{et~al.}(2005)\citenamefont{Csanyi,
  Littlewood, Nevidomskyy, and Simons}}]{Csa05}
\bibinfo{author}{\bibnamefont{Csanyi}, \bibfnamefont{G.}},
  \bibinfo{author}{\bibfnamefont{P.~B.} \bibnamefont{Littlewood}},
  \bibinfo{author}{\bibfnamefont{A.~H.} \bibnamefont{Nevidomskyy}}, and
  \bibinfo{author}{\bibfnamefont{C.~P. B.~D.} \bibnamefont{Simons}},
  \bibinfo{year}{2005}, \bibinfo{journal}{Nature Physics}
  \textbf{\bibinfo{volume}{1}}, \bibinfo{pages}{42}.

\bibitem[{\citenamefont{Cserti} \emph{et~al.}(2007)\citenamefont{Cserti,
  Csord\'as, and D\'avid}}]{Cserti07}
\bibinfo{author}{\bibnamefont{Cserti}, \bibfnamefont{J.}},
  \bibinfo{author}{\bibfnamefont{A.}~\bibnamefont{Csord\'as}}, and
  \bibinfo{author}{\bibfnamefont{G.}~\bibnamefont{D\'avid}},
  \bibinfo{year}{2007}, \bibinfo{journal}{Phys. Rev. Lett.}
  \textbf{\bibinfo{volume}{99}}, \bibinfo{pages}{066802}.

\bibitem[{\citenamefont{Dahal} \emph{et~al.}(2006)\citenamefont{Dahal,
  Joglekar, Bedell, and Balatsky}}]{Dahal06}
\bibinfo{author}{\bibnamefont{Dahal}, \bibfnamefont{H.~P.}},
  \bibinfo{author}{\bibfnamefont{Y.~N.} \bibnamefont{Joglekar}},
  \bibinfo{author}{\bibfnamefont{K.~S.} \bibnamefont{Bedell}}, and
  \bibinfo{author}{\bibfnamefont{A.~V.} \bibnamefont{Balatsky}},
  \bibinfo{year}{2006}, \bibinfo{journal}{Phys. Rev. B}
  \textbf{\bibinfo{volume}{74}}, \bibinfo{pages}{233405}.

\bibitem[{\citenamefont{Dahal} \emph{et~al.}(2010)\citenamefont{Dahal, Wehling,
  Bedell, Zhu, and Balatsky}}]{Dah10}
\bibinfo{author}{\bibnamefont{Dahal}, \bibfnamefont{H.~P.}},
  \bibinfo{author}{\bibfnamefont{T.~O.} \bibnamefont{Wehling}},
  \bibinfo{author}{\bibfnamefont{K.~S.} \bibnamefont{Bedell}},
  \bibinfo{author}{\bibfnamefont{J.-X.} \bibnamefont{Zhu}}, and
  \bibinfo{author}{\bibfnamefont{A.}~\bibnamefont{Balatsky}},
  \bibinfo{year}{2010}, \bibinfo{journal}{Physica B}
  \textbf{\bibinfo{volume}{405}}, \bibinfo{pages}{2241}.

\bibitem[{\citenamefont{{Das Sarma}} \emph{et~al.}(2011)\citenamefont{{Das
  Sarma}, Adam, Hwang, and Rossi}}]{DasSarmaRMP}
\bibinfo{author}{\bibnamefont{{Das Sarma}}, \bibfnamefont{S.}},
  \bibinfo{author}{\bibfnamefont{S.}~\bibnamefont{Adam}},
  \bibinfo{author}{\bibfnamefont{E.~H.} \bibnamefont{Hwang}}, and
  \bibinfo{author}{\bibfnamefont{E.}~\bibnamefont{Rossi}},
  \bibinfo{year}{2011}, \bibinfo{journal}{Rev. Mod. Phys.}
  \textbf{\bibinfo{volume}{83}}, \bibinfo{pages}{407}.

\bibitem[{\citenamefont{{Das Sarma}} \emph{et~al.}(2007)\citenamefont{{Das
  Sarma}, Hwang, and Tse}}]{DSar07}
\bibinfo{author}{\bibnamefont{{Das Sarma}}, \bibfnamefont{S.}},
  \bibinfo{author}{\bibfnamefont{E.~H.} \bibnamefont{Hwang}}, and
  \bibinfo{author}{\bibfnamefont{W.-K.} \bibnamefont{Tse}},
  \bibinfo{year}{2007}, \bibinfo{journal}{Phys. Rev. B}
  \textbf{\bibinfo{volume}{75}}, \bibinfo{pages}{121406(R)}.

\bibitem[{\citenamefont{{de Juan}} \emph{et~al.}(2010)\citenamefont{{de Juan},
  Grushin, and Vozmediano}}]{Vozm10}
\bibinfo{author}{\bibnamefont{{de Juan}}, \bibfnamefont{F.}},
  \bibinfo{author}{\bibfnamefont{A.~G.} \bibnamefont{Grushin}}, and
  \bibinfo{author}{\bibfnamefont{M.~A.~H.} \bibnamefont{Vozmediano}},
  \bibinfo{year}{2010}, \bibinfo{journal}{Phys. Rev. B}
  \textbf{\bibinfo{volume}{82}}, \bibinfo{pages}{125409}.

\bibitem[{\citenamefont{Dean} \emph{et~al.}(2011)\citenamefont{Dean, Young,
  Cadden-Zimansky, Wang, Ren, Watanabe, Taniguchi, Kim, Hone, and
  Shepard}}]{Detal10b}
\bibinfo{author}{\bibnamefont{Dean}, \bibfnamefont{C.~R.}},
  \bibinfo{author}{\bibfnamefont{A.~F.} \bibnamefont{Young}},
  \bibinfo{author}{\bibfnamefont{P.}~\bibnamefont{Cadden-Zimansky}},
  \bibinfo{author}{\bibfnamefont{L.}~\bibnamefont{Wang}},
  \bibinfo{author}{\bibfnamefont{H.}~\bibnamefont{Ren}},
  \bibinfo{author}{\bibfnamefont{K.}~\bibnamefont{Watanabe}},
  \bibinfo{author}{\bibfnamefont{T.}~\bibnamefont{Taniguchi}},
  \bibinfo{author}{\bibfnamefont{P.}~\bibnamefont{Kim}},
  \bibinfo{author}{\bibfnamefont{J.}~\bibnamefont{Hone}}, and
  \bibinfo{author}{\bibfnamefont{K.~L.} \bibnamefont{Shepard}},
  \bibinfo{year}{2011}, \bibinfo{journal}{Nature Physics}
  \textbf{\bibinfo{volume}{7}}, \bibinfo{pages}{693}.

\bibitem[{\citenamefont{Dean} \emph{et~al.}(2010)\citenamefont{Dean, Young,
  Meric, C.~Lee, Sorgenfrei, Watanabe, Taniguchi, Kim, Shepard, and
  Hone}}]{Detal10}
\bibinfo{author}{\bibnamefont{Dean}, \bibfnamefont{C.~R.}},
  \bibinfo{author}{\bibfnamefont{A.~F.} \bibnamefont{Young}},
  \bibinfo{author}{\bibfnamefont{I.}~\bibnamefont{Meric}},
  \bibinfo{author}{\bibfnamefont{L.~W.} \bibnamefont{C.~Lee}},
  \bibinfo{author}{\bibfnamefont{S.}~\bibnamefont{Sorgenfrei}},
  \bibinfo{author}{\bibfnamefont{K.}~\bibnamefont{Watanabe}},
  \bibinfo{author}{\bibfnamefont{T.}~\bibnamefont{Taniguchi}},
  \bibinfo{author}{\bibfnamefont{P.}~\bibnamefont{Kim}},
  \bibinfo{author}{\bibfnamefont{K.~L.} \bibnamefont{Shepard}}, and
  \bibinfo{author}{\bibfnamefont{J.}~\bibnamefont{Hone}}, \bibinfo{year}{2010},
  \bibinfo{journal}{Nature Nanotechnology} \textbf{\bibinfo{volume}{5}},
  \bibinfo{pages}{722}.

\bibitem[{\citenamefont{Dell'Anna}(2010)}]{Del10}
\bibinfo{author}{\bibnamefont{Dell'Anna}, \bibfnamefont{L.}},
  \bibinfo{year}{2010}, \bibinfo{journal}{J. Stat. Mech.} ,
  \bibinfo{pages}{P01007}.

\bibitem[{\citenamefont{Dillenschneider and Han}(2008)}]{Dil08}
\bibinfo{author}{\bibnamefont{Dillenschneider}, \bibfnamefont{R.}}, and
  \bibinfo{author}{\bibfnamefont{J.~H.} \bibnamefont{Han}},
  \bibinfo{year}{2008}, \bibinfo{journal}{Phys. Rev. B}
  \textbf{\bibinfo{volume}{78}}, \bibinfo{pages}{045401}.

\bibitem[{\citenamefont{DiVincenzo and Mele}(1984)}]{DiVincenzo:1984}
\bibinfo{author}{\bibnamefont{DiVincenzo}, \bibfnamefont{D.}}, and
  \bibinfo{author}{\bibfnamefont{E.}~\bibnamefont{Mele}}, \bibinfo{year}{1984},
  \bibinfo{journal}{Phys. Rev. B} \textbf{\bibinfo{volume}{29}},
  \bibinfo{pages}{1685}.

\bibitem[{\citenamefont{Dora and Thalmeier}(2007)}]{Dor07}
\bibinfo{author}{\bibnamefont{Dora}, \bibfnamefont{B.}}, and
  \bibinfo{author}{\bibfnamefont{P.}~\bibnamefont{Thalmeier}},
  \bibinfo{year}{2007}, \bibinfo{journal}{Phys. Rev. B}
  \textbf{\bibinfo{volume}{76}}, \bibinfo{pages}{115407}.

\bibitem[{\citenamefont{Dresselhaus and Dresselhaus}(1981)}]{Dresselhaus81}
\bibinfo{author}{\bibnamefont{Dresselhaus}, \bibfnamefont{M.~S.}}, and
  \bibinfo{author}{\bibfnamefont{G.}~\bibnamefont{Dresselhaus}},
  \bibinfo{year}{1981}, \bibinfo{journal}{Adv. Phys.}
  \textbf{\bibinfo{volume}{30}}, \bibinfo{pages}{139}.

\bibitem[{\citenamefont{Drude}(1900{\natexlab{a}})}]{Drude1:1900}
\bibinfo{author}{\bibnamefont{Drude}, \bibfnamefont{P.}},
  \bibinfo{year}{1900}{\natexlab{a}}, \bibinfo{journal}{Annalen der Physik}
  \textbf{\bibinfo{volume}{1}}, \bibinfo{pages}{566}.

\bibitem[{\citenamefont{Drude}(1900{\natexlab{b}})}]{Drude2:1900}
\bibinfo{author}{\bibnamefont{Drude}, \bibfnamefont{P.}},
  \bibinfo{year}{1900}{\natexlab{b}}, \bibinfo{journal}{Annalen der Physik}
  \textbf{\bibinfo{volume}{3}}, \bibinfo{pages}{369}.

\bibitem[{\citenamefont{Drut and L{\"a}hde}(2009{\natexlab{a}})}]{Drut09c}
\bibinfo{author}{\bibnamefont{Drut}, \bibfnamefont{J.~E.}}, and
  \bibinfo{author}{\bibfnamefont{T.~A.} \bibnamefont{L{\"a}hde}},
  \bibinfo{year}{2009}{\natexlab{a}}, \bibinfo{journal}{Phys. Rev. B}
  \textbf{\bibinfo{volume}{79}}, \bibinfo{pages}{241405(R)}.

\bibitem[{\citenamefont{Drut and L{\"a}hde}(2009{\natexlab{b}})}]{Drut09a}
\bibinfo{author}{\bibnamefont{Drut}, \bibfnamefont{J.~E.}}, and
  \bibinfo{author}{\bibfnamefont{T.~A.} \bibnamefont{L{\"a}hde}},
  \bibinfo{year}{2009}{\natexlab{b}}, \bibinfo{journal}{Phys. Rev. Lett.}
  \textbf{\bibinfo{volume}{102}}, \bibinfo{pages}{026802}.

\bibitem[{\citenamefont{Drut and L{\"a}hde}(2009{\natexlab{c}})}]{Drut09b}
\bibinfo{author}{\bibnamefont{Drut}, \bibfnamefont{J.~E.}}, and
  \bibinfo{author}{\bibfnamefont{T.~A.} \bibnamefont{L{\"a}hde}},
  \bibinfo{year}{2009}{\natexlab{c}}, \bibinfo{journal}{Phys. Rev. B}
  \textbf{\bibinfo{volume}{79}}, \bibinfo{pages}{165425}.

\bibitem[{\citenamefont{Du} \emph{et~al.}(2008)\citenamefont{Du, Skachko, and
  Andrei}}]{Du08}
\bibinfo{author}{\bibnamefont{Du}, \bibfnamefont{X.}},
  \bibinfo{author}{\bibfnamefont{I.}~\bibnamefont{Skachko}}, and
  \bibinfo{author}{\bibfnamefont{E.~Y.} \bibnamefont{Andrei}},
  \bibinfo{year}{2008}, \bibinfo{journal}{Phys. Rev. B}
  \textbf{\bibinfo{volume}{77}}, \bibinfo{pages}{184507}.

\bibitem[{\citenamefont{Du} \emph{et~al.}(2009)\citenamefont{Du, Skachko,
  Duerr, Luican, and Andrei}}]{DSDLA09}
\bibinfo{author}{\bibnamefont{Du}, \bibfnamefont{X.}},
  \bibinfo{author}{\bibfnamefont{I.}~\bibnamefont{Skachko}},
  \bibinfo{author}{\bibfnamefont{F.}~\bibnamefont{Duerr}},
  \bibinfo{author}{\bibfnamefont{A.}~\bibnamefont{Luican}}, and
  \bibinfo{author}{\bibfnamefont{E.~Y.} \bibnamefont{Andrei}},
  \bibinfo{year}{2009}, \bibinfo{journal}{Nature}
  \textbf{\bibinfo{volume}{462}}, \bibinfo{pages}{192}.

\bibitem[{\citenamefont{Eberlein} \emph{et~al.}(2008)\citenamefont{Eberlein,
  Bangert, Nair, Jones, M.Grass, Bleloch, Novoselov, Geim, and
  Briddon}}]{Ebe08}
\bibinfo{author}{\bibnamefont{Eberlein}, \bibfnamefont{T.}},
  \bibinfo{author}{\bibfnamefont{U.}~\bibnamefont{Bangert}},
  \bibinfo{author}{\bibfnamefont{R.~R.} \bibnamefont{Nair}},
  \bibinfo{author}{\bibfnamefont{R.}~\bibnamefont{Jones}},
  \bibinfo{author}{\bibnamefont{M.Grass}},
  \bibinfo{author}{\bibfnamefont{A.~L.} \bibnamefont{Bleloch}},
  \bibinfo{author}{\bibfnamefont{K.~S.} \bibnamefont{Novoselov}},
  \bibinfo{author}{\bibfnamefont{A.}~\bibnamefont{Geim}}, and
  \bibinfo{author}{\bibfnamefont{P.~R.} \bibnamefont{Briddon}},
  \bibinfo{year}{2008}, \bibinfo{journal}{Phys. Rev. B}
  \textbf{\bibinfo{volume}{77}}, \bibinfo{pages}{233406}.

\bibitem[{\citenamefont{Elias} \emph{et~al.}(2011)\citenamefont{Elias,
  Gorbachev, Mayorov, Morozov, Zhukov, Blake, Ponomarenko, Grigorieva,
  Novoselov, Guinea, and Geim}}]{Elias11}
\bibinfo{author}{\bibnamefont{Elias}, \bibfnamefont{D.~C.}},
  \bibinfo{author}{\bibfnamefont{R.~V.} \bibnamefont{Gorbachev}},
  \bibinfo{author}{\bibfnamefont{A.~S.} \bibnamefont{Mayorov}},
  \bibinfo{author}{\bibfnamefont{S.~V.} \bibnamefont{Morozov}},
  \bibinfo{author}{\bibfnamefont{A.~A.} \bibnamefont{Zhukov}},
  \bibinfo{author}{\bibfnamefont{P.}~\bibnamefont{Blake}},
  \bibinfo{author}{\bibfnamefont{L.~A.} \bibnamefont{Ponomarenko}},
  \bibinfo{author}{\bibfnamefont{I.~V.} \bibnamefont{Grigorieva}},
  \bibinfo{author}{\bibfnamefont{K.~S.} \bibnamefont{Novoselov}},
  \bibinfo{author}{\bibfnamefont{F.}~\bibnamefont{Guinea}}, and
  \bibinfo{author}{\bibfnamefont{A.~K.} \bibnamefont{Geim}},
  \bibinfo{year}{2011}, \bibinfo{journal}{Nature Physics}
  \textbf{\bibinfo{volume}{7}}, \bibinfo{pages}{701}.

\bibitem[{\citenamefont{Elias} \emph{et~al.}(2009)\citenamefont{Elias, Nair,
  Mohiuddin, Morozov, Blake, Halsall, Ferrari, Boukhvalov, Katsnelson, Geim,
  and Novoselov}}]{Eli09}
\bibinfo{author}{\bibnamefont{Elias}, \bibfnamefont{D.~C.}},
  \bibinfo{author}{\bibfnamefont{R.~R.} \bibnamefont{Nair}},
  \bibinfo{author}{\bibfnamefont{T.~M.~G.} \bibnamefont{Mohiuddin}},
  \bibinfo{author}{\bibfnamefont{S.}~\bibnamefont{Morozov}},
  \bibinfo{author}{\bibfnamefont{P.}~\bibnamefont{Blake}},
  \bibinfo{author}{\bibfnamefont{M.~P.} \bibnamefont{Halsall}},
  \bibinfo{author}{\bibfnamefont{A.~C.} \bibnamefont{Ferrari}},
  \bibinfo{author}{\bibfnamefont{D.~W.} \bibnamefont{Boukhvalov}},
  \bibinfo{author}{\bibfnamefont{M.~I.} \bibnamefont{Katsnelson}},
  \bibinfo{author}{\bibfnamefont{A.~K.} \bibnamefont{Geim}}, and
  \bibinfo{author}{\bibfnamefont{K.~S.} \bibnamefont{Novoselov}},
  \bibinfo{year}{2009}, \bibinfo{journal}{Science}
  \textbf{\bibinfo{volume}{323}}, \bibinfo{pages}{610}.

\bibitem[{\citenamefont{Enoki and Takai}(2009)}]{ET09}
\bibinfo{author}{\bibnamefont{Enoki}, \bibfnamefont{T.}}, and
  \bibinfo{author}{\bibfnamefont{K.}~\bibnamefont{Takai}},
  \bibinfo{year}{2009}, \bibinfo{journal}{Solid State Commun.}
  \textbf{\bibinfo{volume}{149}}, \bibinfo{pages}{1144}.

\bibitem[{\citenamefont{Esquinazi} \emph{et~al.}(2003)\citenamefont{Esquinazi,
  Spemann, H{\"o}hne, Setzer, Han, and Butz}}]{Eetal03}
\bibinfo{author}{\bibnamefont{Esquinazi}, \bibfnamefont{P.}},
  \bibinfo{author}{\bibfnamefont{D.}~\bibnamefont{Spemann}},
  \bibinfo{author}{\bibfnamefont{R.}~\bibnamefont{H{\"o}hne}},
  \bibinfo{author}{\bibfnamefont{A.}~\bibnamefont{Setzer}},
  \bibinfo{author}{\bibfnamefont{K.-H.} \bibnamefont{Han}}, and
  \bibinfo{author}{\bibfnamefont{T.}~\bibnamefont{Butz}}, \bibinfo{year}{2003},
  \bibinfo{journal}{Phys. Rev. Lett.} \textbf{\bibinfo{volume}{91}},
  \bibinfo{pages}{227201}.

\bibitem[{\citenamefont{Feigel'man}
  \emph{et~al.}(2008)\citenamefont{Feigel'man, Skvortov, and Tikhonov}}]{Fei08}
\bibinfo{author}{\bibnamefont{Feigel'man}, \bibfnamefont{M.~V.}},
  \bibinfo{author}{\bibfnamefont{M.~A.} \bibnamefont{Skvortov}}, and
  \bibinfo{author}{\bibfnamefont{K.~S.} \bibnamefont{Tikhonov}},
  \bibinfo{year}{2008}, \bibinfo{journal}{Pris'ma v ZhETP}
  \textbf{\bibinfo{volume}{88}}, \bibinfo{pages}{780}.

\bibitem[{\citenamefont{Feldman} \emph{et~al.}(2009)\citenamefont{Feldman,
  Martin, and Yacoby}}]{FMY09}
\bibinfo{author}{\bibnamefont{Feldman}, \bibfnamefont{B.~E.}},
  \bibinfo{author}{\bibfnamefont{J.}~\bibnamefont{Martin}}, and
  \bibinfo{author}{\bibfnamefont{A.}~\bibnamefont{Yacoby}},
  \bibinfo{year}{2009}, \bibinfo{journal}{Nature Physics}
  \textbf{\bibinfo{volume}{5}}, \bibinfo{pages}{890}.

\bibitem[{\citenamefont{Fermi}(1927)}]{Fermi:1927}
\bibinfo{author}{\bibnamefont{Fermi}, \bibfnamefont{E.}}, \bibinfo{year}{1927},
  \bibinfo{journal}{Rend. Accad. Naz. Lincei} \textbf{\bibinfo{volume}{6}},
  \bibinfo{pages}{602}.

\bibitem[{\citenamefont{{Fern{\'a}ndez-Rossier}}(2008)}]{F08}
\bibinfo{author}{\bibnamefont{{Fern{\'a}ndez-Rossier}}, \bibfnamefont{J.}},
  \bibinfo{year}{2008}, \bibinfo{journal}{Phys. Rev. B}
  \textbf{\bibinfo{volume}{77}}, \bibinfo{pages}{075430}.

\bibitem[{\citenamefont{Fern{\'a}ndez-Rossier and Palacios}(2007)}]{FP07}
\bibinfo{author}{\bibnamefont{Fern{\'a}ndez-Rossier}, \bibfnamefont{J.}}, and
  \bibinfo{author}{\bibfnamefont{J.~J.} \bibnamefont{Palacios}},
  \bibinfo{year}{2007}, \bibinfo{journal}{Phys. Rev. Lett.}
  \textbf{\bibinfo{volume}{99}}, \bibinfo{pages}{177204}.

\bibitem[{\citenamefont{Fertig and Brey}(2006)}]{FB06}
\bibinfo{author}{\bibnamefont{Fertig}, \bibfnamefont{H.~A.}}, and
  \bibinfo{author}{\bibfnamefont{L.}~\bibnamefont{Brey}}, \bibinfo{year}{2006},
  \bibinfo{journal}{Phys. Rev. Lett.} \textbf{\bibinfo{volume}{97}},
  \bibinfo{pages}{116805}.

\bibitem[{\citenamefont{Fetter}(1974)}]{Fet74}
\bibinfo{author}{\bibnamefont{Fetter}, \bibfnamefont{A.}},
  \bibinfo{year}{1974}, \bibinfo{journal}{Ann. Phys.}
  \textbf{\bibinfo{volume}{88}}, \bibinfo{pages}{1}.

\bibitem[{\citenamefont{Fetter and Walecka}(1971)}]{Fetter:1971}
\bibinfo{author}{\bibnamefont{Fetter}, \bibfnamefont{A.~L.}}, and
  \bibinfo{author}{\bibfnamefont{J.~D.} \bibnamefont{Walecka}},
  \bibinfo{year}{1971}, \emph{\bibinfo{title}{Quantum Theory of Many-Particle
  systems}}, International Series in Pure and Applied Physics
  (\bibinfo{publisher}{McGraw-Hill Book Company}).

\bibitem[{\citenamefont{Fogler} \emph{et~al.}(2007)\citenamefont{Fogler,
  Novikov, and Shklovskii}}]{Fogler:2007}
\bibinfo{author}{\bibnamefont{Fogler}, \bibfnamefont{M.}},
  \bibinfo{author}{\bibfnamefont{D.}~\bibnamefont{Novikov}}, and
  \bibinfo{author}{\bibfnamefont{B.}~\bibnamefont{Shklovskii}},
  \bibinfo{year}{2007}, \bibinfo{journal}{Phys. Rev. B}
  \textbf{\bibinfo{volume}{76}}, \bibinfo{pages}{233402}.

\bibitem[{\citenamefont{Foster and Aleiner}(2008)}]{FA08}
\bibinfo{author}{\bibnamefont{Foster}, \bibfnamefont{M.~S.}}, and
  \bibinfo{author}{\bibfnamefont{I.~L.} \bibnamefont{Aleiner}},
  \bibinfo{year}{2008}, \bibinfo{journal}{Phys. Rev. B}
  \textbf{\bibinfo{volume}{77}}, \bibinfo{pages}{195413}.

\bibitem[{\citenamefont{Foster and Ludwig}(2006{\natexlab{a}})}]{FL06}
\bibinfo{author}{\bibnamefont{Foster}, \bibfnamefont{M.~S.}}, and
  \bibinfo{author}{\bibfnamefont{A.~W.~W.} \bibnamefont{Ludwig}},
  \bibinfo{year}{2006}{\natexlab{a}}, \bibinfo{journal}{Phys. Rev. B}
  \textbf{\bibinfo{volume}{73}}, \bibinfo{pages}{155104}.

\bibitem[{\citenamefont{Foster and Ludwig}(2006{\natexlab{b}})}]{FL06b}
\bibinfo{author}{\bibnamefont{Foster}, \bibfnamefont{M.~S.}}, and
  \bibinfo{author}{\bibfnamefont{A.~W.~W.} \bibnamefont{Ludwig}},
  \bibinfo{year}{2006}{\natexlab{b}}, \bibinfo{journal}{Phys. Rev. B}
  \textbf{\bibinfo{volume}{74}}, \bibinfo{pages}{241102(R)}.

\bibitem[{\citenamefont{Fradkin}(1986)}]{Fradkin86}
\bibinfo{author}{\bibnamefont{Fradkin}, \bibfnamefont{E.}},
  \bibinfo{year}{1986}, \bibinfo{journal}{\prb} \textbf{\bibinfo{volume}{33}},
  \bibinfo{pages}{3263}.

\bibitem[{\citenamefont{Fritz} \emph{et~al.}(2008)\citenamefont{Fritz,
  J.Schmalian, M{\"u}ller, and Sachdev}}]{Fri08}
\bibinfo{author}{\bibnamefont{Fritz}, \bibfnamefont{L.}},
  \bibinfo{author}{\bibnamefont{J.Schmalian}},
  \bibinfo{author}{\bibfnamefont{M.}~\bibnamefont{M{\"u}ller}}, and
  \bibinfo{author}{\bibfnamefont{S.}~\bibnamefont{Sachdev}},
  \bibinfo{year}{2008}, \bibinfo{journal}{\prb} \textbf{\bibinfo{volume}{78}},
  \bibinfo{pages}{085416}.

\bibitem[{\citenamefont{Fritz and Vojta}(2004)}]{Fri04}
\bibinfo{author}{\bibnamefont{Fritz}, \bibfnamefont{L.}}, and
  \bibinfo{author}{\bibfnamefont{M.}~\bibnamefont{Vojta}},
  \bibinfo{year}{2004}, \bibinfo{journal}{Phys. Rev. B}
  \textbf{\bibinfo{volume}{70}}, \bibinfo{pages}{214427}.

\bibitem[{\citenamefont{Fuchs and Lederer}(2007)}]{FL07}
\bibinfo{author}{\bibnamefont{Fuchs}, \bibfnamefont{J.}}, and
  \bibinfo{author}{\bibfnamefont{P.}~\bibnamefont{Lederer}},
  \bibinfo{year}{2007}, \bibinfo{journal}{Phys. Rev. Lett.}
  \textbf{\bibinfo{volume}{98}}, \bibinfo{pages}{016803}.

\bibitem[{\citenamefont{Fujita} \emph{et~al.}(1996)\citenamefont{Fujita,
  Wakabayashi, Nakada, and Kusakabe}}]{FWNK96}
\bibinfo{author}{\bibnamefont{Fujita}, \bibfnamefont{M.}},
  \bibinfo{author}{\bibfnamefont{K.}~\bibnamefont{Wakabayashi}},
  \bibinfo{author}{\bibfnamefont{K.}~\bibnamefont{Nakada}}, and
  \bibinfo{author}{\bibfnamefont{K.}~\bibnamefont{Kusakabe}},
  \bibinfo{year}{1996}, \bibinfo{journal}{J. Phys. Soc. Jpn.}
  \textbf{\bibinfo{volume}{65}}, \bibinfo{pages}{1920}.

\bibitem[{\citenamefont{Furukawa}(2001)}]{Fur01}
\bibinfo{author}{\bibnamefont{Furukawa}, \bibfnamefont{N.}},
  \bibinfo{year}{2001}, \bibinfo{journal}{J. Phys. Soc. Jpn.}
  \textbf{\bibinfo{volume}{70}}, \bibinfo{pages}{1483}.

\bibitem[{\citenamefont{Gamayun} \emph{et~al.}(2009)\citenamefont{Gamayun,
  Gorbar, and Gusynin}}]{Gamayun:2009}
\bibinfo{author}{\bibnamefont{Gamayun}, \bibfnamefont{O.~V.}},
  \bibinfo{author}{\bibfnamefont{E.~V.} \bibnamefont{Gorbar}}, and
  \bibinfo{author}{\bibfnamefont{V.~P.} \bibnamefont{Gusynin}},
  \bibinfo{year}{2009}, \bibinfo{journal}{Phys. Rev. B}
  \textbf{\bibinfo{volume}{80}}, \bibinfo{pages}{165429}.

\bibitem[{\citenamefont{Gamayun} \emph{et~al.}(2010)\citenamefont{Gamayun,
  Gorbar, and Gusynin}}]{Gorbar10}
\bibinfo{author}{\bibnamefont{Gamayun}, \bibfnamefont{O.~V.}},
  \bibinfo{author}{\bibfnamefont{E.~V.} \bibnamefont{Gorbar}}, and
  \bibinfo{author}{\bibfnamefont{V.~P.} \bibnamefont{Gusynin}},
  \bibinfo{year}{2010}, \bibinfo{journal}{Phys. Rev. B}
  \textbf{\bibinfo{volume}{81}}, \bibinfo{pages}{075429}.

\bibitem[{\citenamefont{Gangadharaiah}
  \emph{et~al.}(2008)\citenamefont{Gangadharaiah, Farid, and
  Mishchenko}}]{Gan08}
\bibinfo{author}{\bibnamefont{Gangadharaiah}, \bibfnamefont{S.}},
  \bibinfo{author}{\bibfnamefont{A.~M.} \bibnamefont{Farid}}, and
  \bibinfo{author}{\bibfnamefont{E.~G.} \bibnamefont{Mishchenko}},
  \bibinfo{year}{2008}, \bibinfo{journal}{Phys. Rev. Lett.}
  \textbf{\bibinfo{volume}{100}}, \bibinfo{pages}{166802}.

\bibitem[{\citenamefont{Geim and Novoselov}(2007)}]{Geim07}
\bibinfo{author}{\bibnamefont{Geim}, \bibfnamefont{A.~K.}}, and
  \bibinfo{author}{\bibfnamefont{K.~S.} \bibnamefont{Novoselov}},
  \bibinfo{year}{2007}, \bibinfo{journal}{Nature Materials}
  \textbf{\bibinfo{volume}{6}}, \bibinfo{pages}{183}.

\bibitem[{\citenamefont{Ghaemi and Wilczek}(2012)}]{Gha07}
\bibinfo{author}{\bibnamefont{Ghaemi}, \bibfnamefont{P.}}, and
  \bibinfo{author}{\bibfnamefont{F.}~\bibnamefont{Wilczek}},
  \bibinfo{year}{2012}, \bibinfo{journal}{Physica Scripta}
  \textbf{\bibinfo{volume}{T146}}, \bibinfo{pages}{014019}.

\bibitem[{\citenamefont{Ghahari} \emph{et~al.}(2011)\citenamefont{Ghahari,
  Zhao, Cadden-Zimansky, Bolotin, and Kim}}]{GZCBK10}
\bibinfo{author}{\bibnamefont{Ghahari}, \bibfnamefont{F.}},
  \bibinfo{author}{\bibfnamefont{Y.}~\bibnamefont{Zhao}},
  \bibinfo{author}{\bibfnamefont{P.}~\bibnamefont{Cadden-Zimansky}},
  \bibinfo{author}{\bibfnamefont{K.}~\bibnamefont{Bolotin}}, and
  \bibinfo{author}{\bibfnamefont{P.}~\bibnamefont{Kim}}, \bibinfo{year}{2011},
  \bibinfo{journal}{Phys. Rev. Lett.} \textbf{\bibinfo{volume}{106}},
  \bibinfo{pages}{046801}.

\bibitem[{\citenamefont{Ghosal} \emph{et~al.}(2007)\citenamefont{Ghosal,
  Goswami, and Chakravarty}}]{Sudip07}
\bibinfo{author}{\bibnamefont{Ghosal}, \bibfnamefont{A.}},
  \bibinfo{author}{\bibfnamefont{P.}~\bibnamefont{Goswami}}, and
  \bibinfo{author}{\bibfnamefont{S.}~\bibnamefont{Chakravarty}},
  \bibinfo{year}{2007}, \bibinfo{journal}{Phys. Rev. B}
  \textbf{\bibinfo{volume}{75}}, \bibinfo{pages}{115123}.

\bibitem[{\citenamefont{Giesbers} \emph{et~al.}(2007)\citenamefont{Giesbers,
  Zeitler, Katsnelson, Ponomarenko, Mohiuddin, and Maan}}]{Getal07}
\bibinfo{author}{\bibnamefont{Giesbers}, \bibfnamefont{A.~J.}},
  \bibinfo{author}{\bibfnamefont{U.}~\bibnamefont{Zeitler}},
  \bibinfo{author}{\bibfnamefont{M.~I.} \bibnamefont{Katsnelson}},
  \bibinfo{author}{\bibfnamefont{L.~A.} \bibnamefont{Ponomarenko}},
  \bibinfo{author}{\bibfnamefont{T.~M.} \bibnamefont{Mohiuddin}}, and
  \bibinfo{author}{\bibfnamefont{J.~C.} \bibnamefont{Maan}},
  \bibinfo{year}{2007}, \bibinfo{journal}{Phys. Rev. Lett.}
  \textbf{\bibinfo{volume}{99}}, \bibinfo{pages}{206803}.

\bibitem[{\citenamefont{Giesbers} \emph{et~al.}(2009)\citenamefont{Giesbers,
  Zeitler, Ponomarenko, Yang, Novoselov, Geim, and Maan}}]{Getal09b}
\bibinfo{author}{\bibnamefont{Giesbers}, \bibfnamefont{A.~J.~M.}},
  \bibinfo{author}{\bibfnamefont{U.}~\bibnamefont{Zeitler}},
  \bibinfo{author}{\bibfnamefont{L.~A.} \bibnamefont{Ponomarenko}},
  \bibinfo{author}{\bibfnamefont{R.}~\bibnamefont{Yang}},
  \bibinfo{author}{\bibfnamefont{K.~S.} \bibnamefont{Novoselov}},
  \bibinfo{author}{\bibfnamefont{A.~K.} \bibnamefont{Geim}}, and
  \bibinfo{author}{\bibfnamefont{J.~C.} \bibnamefont{Maan}},
  \bibinfo{year}{2009}, \bibinfo{journal}{Phys. Rev. B}
  \textbf{\bibinfo{volume}{80}}, \bibinfo{pages}{241411}.

\bibitem[{\citenamefont{Giovannetti}
  \emph{et~al.}(2008)\citenamefont{Giovannetti, Khomyakov, Brocks, M.Karpan,
  {van den Brink}, and Kelly}}]{Gio08}
\bibinfo{author}{\bibnamefont{Giovannetti}, \bibfnamefont{G.}},
  \bibinfo{author}{\bibfnamefont{P.~A.} \bibnamefont{Khomyakov}},
  \bibinfo{author}{\bibfnamefont{G.}~\bibnamefont{Brocks}},
  \bibinfo{author}{\bibfnamefont{V.}~\bibnamefont{M.Karpan}},
  \bibinfo{author}{\bibfnamefont{J.}~\bibnamefont{{van den Brink}}}, and
  \bibinfo{author}{\bibfnamefont{P.~J.} \bibnamefont{Kelly}},
  \bibinfo{year}{2008}, \bibinfo{journal}{Phys. Rev. Lett.}
  \textbf{\bibinfo{volume}{101}}, \bibinfo{pages}{026803}.

\bibitem[{\citenamefont{Girit} \emph{et~al.}(2009)\citenamefont{Girit, Meyer,
  Erni, Rossell, Kisielowski, Yang, Park, Crommie, Cohen, Louie, and
  Zettl}}]{Getal09d}
\bibinfo{author}{\bibnamefont{Girit}, \bibfnamefont{C.~{\"O}.}},
  \bibinfo{author}{\bibfnamefont{J.~C.} \bibnamefont{Meyer}},
  \bibinfo{author}{\bibfnamefont{R.}~\bibnamefont{Erni}},
  \bibinfo{author}{\bibfnamefont{M.~D.} \bibnamefont{Rossell}},
  \bibinfo{author}{\bibfnamefont{C.}~\bibnamefont{Kisielowski}},
  \bibinfo{author}{\bibfnamefont{L.}~\bibnamefont{Yang}},
  \bibinfo{author}{\bibfnamefont{C.-H. C.-H.} \bibnamefont{Park}},
  \bibinfo{author}{\bibfnamefont{M.~F.} \bibnamefont{Crommie}},
  \bibinfo{author}{\bibfnamefont{M.~L.} \bibnamefont{Cohen}},
  \bibinfo{author}{\bibfnamefont{S.~G.} \bibnamefont{Louie}}, and
  \bibinfo{author}{\bibfnamefont{A.}~\bibnamefont{Zettl}},
  \bibinfo{year}{2009}, \bibinfo{journal}{Science}
  \textbf{\bibinfo{volume}{323}}, \bibinfo{pages}{1705}.

\bibitem[{\citenamefont{Goerbig}(2011)}]{G10}
\bibinfo{author}{\bibnamefont{Goerbig}, \bibfnamefont{M.~O.}},
  \bibinfo{year}{2011}, \bibinfo{journal}{Rev. Mod. Phys.}
  \textbf{\bibinfo{volume}{83}}, \bibinfo{pages}{1193}.

\bibitem[{\citenamefont{Goerbig} \emph{et~al.}(2006)\citenamefont{Goerbig,
  Moessner, and Doucot}}]{GMC06}
\bibinfo{author}{\bibnamefont{Goerbig}, \bibfnamefont{M.~O.}},
  \bibinfo{author}{\bibfnamefont{R.}~\bibnamefont{Moessner}}, and
  \bibinfo{author}{\bibfnamefont{B.}~\bibnamefont{Doucot}},
  \bibinfo{year}{2006}, \bibinfo{journal}{Phys. Rev. B}
  \textbf{\bibinfo{volume}{74}}, \bibinfo{pages}{161407}.

\bibitem[{\citenamefont{Goerbig and Regnault}(2007)}]{GR07}
\bibinfo{author}{\bibnamefont{Goerbig}, \bibfnamefont{M.~O.}}, and
  \bibinfo{author}{\bibfnamefont{N.}~\bibnamefont{Regnault}},
  \bibinfo{year}{2007}, \bibinfo{journal}{Phys. Rev. B}
  \textbf{\bibinfo{volume}{75}}, \bibinfo{pages}{241405}.

\bibitem[{\citenamefont{Goldman and Drake}(1982)}]{Gol82}
\bibinfo{author}{\bibnamefont{Goldman}, \bibfnamefont{S.}}, and
  \bibinfo{author}{\bibfnamefont{G.}~\bibnamefont{Drake}},
  \bibinfo{year}{1982}, \bibinfo{journal}{Phys. Rev. A}
  \textbf{\bibinfo{volume}{25}}, \bibinfo{pages}{2877}.

\bibitem[{\citenamefont{Gonz{\'a}lez}
  \emph{et~al.}(1994)\citenamefont{Gonz{\'a}lez, J., Guinea, and
  Vozmediano}}]{Gon94}
\bibinfo{author}{\bibnamefont{Gonz{\'a}lez}},
  \bibinfo{author}{\bibnamefont{J.}},
  \bibinfo{author}{\bibfnamefont{F.}~\bibnamefont{Guinea}}, and
  \bibinfo{author}{\bibfnamefont{M.~A.~H.} \bibnamefont{Vozmediano}},
  \bibinfo{year}{1994}, \bibinfo{journal}{Nucl. Phys. B}
  \textbf{\bibinfo{volume}{424}}, \bibinfo{pages}{595}.

\bibitem[{\citenamefont{Gonz{\'a}lez}
  \emph{et~al.}(1996)\citenamefont{Gonz{\'a}lez, J., Guinea, and
  Vozmediano}}]{Gon96}
\bibinfo{author}{\bibnamefont{Gonz{\'a}lez}},
  \bibinfo{author}{\bibnamefont{J.}},
  \bibinfo{author}{\bibfnamefont{F.}~\bibnamefont{Guinea}}, and
  \bibinfo{author}{\bibfnamefont{M.~A.~H.} \bibnamefont{Vozmediano}},
  \bibinfo{year}{1996}, \bibinfo{journal}{\prl} \textbf{\bibinfo{volume}{77}},
  \bibinfo{pages}{3589}.

\bibitem[{\citenamefont{Gonz{\'a}lez}
  \emph{et~al.}(1999)\citenamefont{Gonz{\'a}lez, J., Guinea, and
  Vozmediano}}]{Gon99}
\bibinfo{author}{\bibnamefont{Gonz{\'a}lez}},
  \bibinfo{author}{\bibnamefont{J.}},
  \bibinfo{author}{\bibfnamefont{F.}~\bibnamefont{Guinea}}, and
  \bibinfo{author}{\bibfnamefont{M.~A.~H.} \bibnamefont{Vozmediano}},
  \bibinfo{year}{1999}, \bibinfo{journal}{\prb} \textbf{\bibinfo{volume}{59}},
  \bibinfo{pages}{R2474}.

\bibitem[{\citenamefont{Gonz{\'a}lez}(2008)}]{Gon08}
\bibinfo{author}{\bibnamefont{Gonz{\'a}lez}, \bibfnamefont{J.}},
  \bibinfo{year}{2008}, \bibinfo{journal}{Phys. Rev. B}
  \textbf{\bibinfo{volume}{78}}, \bibinfo{pages}{205431}.

\bibitem[{\citenamefont{Gonz{\'a}lez}(2010)}]{Gonzalez10}
\bibinfo{author}{\bibnamefont{Gonz{\'a}lez}, \bibfnamefont{J.}},
  \bibinfo{year}{2010}, \bibinfo{journal}{Phys. Rev. B}
  \textbf{\bibinfo{volume}{82}}, \bibinfo{pages}{155404}.

\bibitem[{\citenamefont{Gonz{\'a}lez}
  \emph{et~al.}(2001)\citenamefont{Gonz{\'a}lez, Guinea, and
  Vozmediano}}]{GGV01}
\bibinfo{author}{\bibnamefont{Gonz{\'a}lez}, \bibfnamefont{J.}},
  \bibinfo{author}{\bibfnamefont{F.}~\bibnamefont{Guinea}}, and
  \bibinfo{author}{\bibfnamefont{M.~A.~H.} \bibnamefont{Vozmediano}},
  \bibinfo{year}{2001}, \bibinfo{journal}{Phys. Rev. B}
  \textbf{\bibinfo{volume}{63}}, \bibinfo{pages}{134421}.

\bibitem[{\citenamefont{Gonz{\'a}lez-Buxton and Ingersent}(1998)}]{Gon98}
\bibinfo{author}{\bibnamefont{Gonz{\'a}lez-Buxton}, \bibfnamefont{C.}}, and
  \bibinfo{author}{\bibfnamefont{K.}~\bibnamefont{Ingersent}},
  \bibinfo{year}{1998}, \bibinfo{journal}{Phys. Rev. B}
  \textbf{\bibinfo{volume}{57}}, \bibinfo{pages}{14254}.

\bibitem[{\citenamefont{Gorbar} \emph{et~al.}(2002)\citenamefont{Gorbar,
  Gusynin, Miransky, and Shovkovy}}]{Gorbar02}
\bibinfo{author}{\bibnamefont{Gorbar}, \bibfnamefont{E.~V.}},
  \bibinfo{author}{\bibfnamefont{V.~P.} \bibnamefont{Gusynin}},
  \bibinfo{author}{\bibfnamefont{V.~A.} \bibnamefont{Miransky}}, and
  \bibinfo{author}{\bibfnamefont{I.~A.} \bibnamefont{Shovkovy}},
  \bibinfo{year}{2002}, \bibinfo{journal}{Phys. Rev. B}
  \textbf{\bibinfo{volume}{66}}, \bibinfo{pages}{045108}.

\bibitem[{\citenamefont{Greenbaum} \emph{et~al.}(2007)\citenamefont{Greenbaum,
  Das, Schwiete, and Silvestrov}}]{Gre07}
\bibinfo{author}{\bibnamefont{Greenbaum}, \bibfnamefont{D.}},
  \bibinfo{author}{\bibfnamefont{S.}~\bibnamefont{Das}},
  \bibinfo{author}{\bibfnamefont{G.}~\bibnamefont{Schwiete}}, and
  \bibinfo{author}{\bibfnamefont{P.~G.} \bibnamefont{Silvestrov}},
  \bibinfo{year}{2007}, \bibinfo{journal}{Phys. Rev. B}
  \textbf{\bibinfo{volume}{75}}, \bibinfo{pages}{195437}.

\bibitem[{\citenamefont{Greiner} \emph{et~al.}(1985)\citenamefont{Greiner,
  M{\"u}ller, Rafelski, Greiner, Muller, and Rafelski}}]{Greiner:1985}
\bibinfo{author}{\bibnamefont{Greiner}, \bibfnamefont{W.}},
  \bibinfo{author}{\bibfnamefont{B.}~\bibnamefont{M{\"u}ller}},
  \bibinfo{author}{\bibfnamefont{J.}~\bibnamefont{Rafelski}},
  \bibinfo{author}{\bibfnamefont{W.}~\bibnamefont{Greiner}},
  \bibinfo{author}{\bibfnamefont{B.}~\bibnamefont{Muller}}, and
  \bibinfo{author}{\bibfnamefont{J.}~\bibnamefont{Rafelski}},
  \bibinfo{year}{1985}, \emph{\bibinfo{title}{Quantum Electrodynamics of Strong
  Fields}} (\bibinfo{publisher}{Springer}).

\bibitem[{\citenamefont{Gross and Neveu}(1974)}]{Gro74}
\bibinfo{author}{\bibnamefont{Gross}, \bibfnamefont{D.~J.}}, and
  \bibinfo{author}{\bibfnamefont{A.}~\bibnamefont{Neveu}},
  \bibinfo{year}{1974}, \bibinfo{journal}{Phys. Rev. D}
  \textbf{\bibinfo{volume}{10}}, \bibinfo{pages}{3235}.

\bibitem[{\citenamefont{Gr{\"u}neis}
  \emph{et~al.}(2009)\citenamefont{Gr{\"u}neis, Attaccalite, Rubio, Vyalikh,
  Molodtsov, Fink, Follath, Eberhardt, B{\"u}chner, and Pichler}}]{Gru09b}
\bibinfo{author}{\bibnamefont{Gr{\"u}neis}, \bibfnamefont{A.}},
  \bibinfo{author}{\bibfnamefont{C.}~\bibnamefont{Attaccalite}},
  \bibinfo{author}{\bibfnamefont{A.}~\bibnamefont{Rubio}},
  \bibinfo{author}{\bibfnamefont{D.~V.} \bibnamefont{Vyalikh}},
  \bibinfo{author}{\bibfnamefont{S.~L.} \bibnamefont{Molodtsov}},
  \bibinfo{author}{\bibfnamefont{J.}~\bibnamefont{Fink}},
  \bibinfo{author}{\bibfnamefont{R.}~\bibnamefont{Follath}},
  \bibinfo{author}{\bibfnamefont{W.}~\bibnamefont{Eberhardt}},
  \bibinfo{author}{\bibfnamefont{B.}~\bibnamefont{B{\"u}chner}}, and
  \bibinfo{author}{\bibfnamefont{T.}~\bibnamefont{Pichler}},
  \bibinfo{year}{2009}, \bibinfo{journal}{Phys. Rev. B}
  \textbf{\bibinfo{volume}{80}}, \bibinfo{pages}{075431}.

\bibitem[{\citenamefont{Gr{\"u}neis and Vyalikh}(2008)}]{Grneis:2008}
\bibinfo{author}{\bibnamefont{Gr{\"u}neis}, \bibfnamefont{A.}}, and
  \bibinfo{author}{\bibfnamefont{D.~V.} \bibnamefont{Vyalikh}},
  \bibinfo{year}{2008}, \bibinfo{journal}{Physical Review B}
  \textbf{\bibinfo{volume}{77}}, \bibinfo{pages}{193401}.

\bibitem[{\citenamefont{Gruner}(1994)}]{Gruner:94}
\bibinfo{author}{\bibnamefont{Gruner}, \bibfnamefont{G.}},
  \bibinfo{year}{1994}, \emph{\bibinfo{title}{Density waves in solids}}
  (\bibinfo{publisher}{Perseus}, \bibinfo{address}{Cambridge, MA}).

\bibitem[{\citenamefont{Grushin} \emph{et~al.}(2009)\citenamefont{Grushin,
  Valenzuela, and Vozmediano}}]{Grushin09}
\bibinfo{author}{\bibnamefont{Grushin}, \bibfnamefont{A.~G.}},
  \bibinfo{author}{\bibfnamefont{B.}~\bibnamefont{Valenzuela}}, and
  \bibinfo{author}{\bibfnamefont{M.~A.~H.} \bibnamefont{Vozmediano}},
  \bibinfo{year}{2009}, \bibinfo{journal}{Phys. Rev. B}
  \textbf{\bibinfo{volume}{80}}, \bibinfo{pages}{155417}.

\bibitem[{\citenamefont{Guettinger}
  \emph{et~al.}(2008)\citenamefont{Guettinger, Stampfer, Hellmueller, Molitor,
  Ihn, and Ensslin}}]{Setal08}
\bibinfo{author}{\bibnamefont{Guettinger}, \bibfnamefont{J.}},
  \bibinfo{author}{\bibfnamefont{C.}~\bibnamefont{Stampfer}},
  \bibinfo{author}{\bibfnamefont{S.}~\bibnamefont{Hellmueller}},
  \bibinfo{author}{\bibfnamefont{F.}~\bibnamefont{Molitor}},
  \bibinfo{author}{\bibfnamefont{T.}~\bibnamefont{Ihn}}, and
  \bibinfo{author}{\bibfnamefont{K.}~\bibnamefont{Ensslin}},
  \bibinfo{year}{2008}, \bibinfo{journal}{Appl. Phys. Lett.}
  \textbf{\bibinfo{volume}{93}}, \bibinfo{pages}{212102}.

\bibitem[{\citenamefont{Guinea}(2007)}]{Gui07}
\bibinfo{author}{\bibnamefont{Guinea}, \bibfnamefont{F.}},
  \bibinfo{year}{2007}, \bibinfo{journal}{Phys. Rev. B}
  \textbf{\bibinfo{volume}{75}}, \bibinfo{pages}{235433}.

\bibitem[{\citenamefont{Guinea}
  \emph{et~al.}(2008{\natexlab{a}})\citenamefont{Guinea, Horowitz, and
  Doussal}}]{GHL08}
\bibinfo{author}{\bibnamefont{Guinea}, \bibfnamefont{F.}},
  \bibinfo{author}{\bibfnamefont{B.}~\bibnamefont{Horowitz}}, and
  \bibinfo{author}{\bibfnamefont{P.~L.} \bibnamefont{Doussal}},
  \bibinfo{year}{2008}{\natexlab{a}}, \bibinfo{journal}{Phys. Rev. B}
  \textbf{\bibinfo{volume}{77}}, \bibinfo{pages}{205421}.

\bibitem[{\citenamefont{Guinea} \emph{et~al.}(2010)\citenamefont{Guinea,
  Katsnelson, and Geim}}]{GKG10}
\bibinfo{author}{\bibnamefont{Guinea}, \bibfnamefont{F.}},
  \bibinfo{author}{\bibfnamefont{M.~I.} \bibnamefont{Katsnelson}}, and
  \bibinfo{author}{\bibfnamefont{A.~K.} \bibnamefont{Geim}},
  \bibinfo{year}{2010}, \bibinfo{journal}{Nature Physics}
  \textbf{\bibinfo{volume}{6}}, \bibinfo{pages}{30}.

\bibitem[{\citenamefont{Guinea}
  \emph{et~al.}(2008{\natexlab{b}})\citenamefont{Guinea, Katsnelson, and
  Vozmediano}}]{GKV08}
\bibinfo{author}{\bibnamefont{Guinea}, \bibfnamefont{F.}},
  \bibinfo{author}{\bibfnamefont{M.~I.} \bibnamefont{Katsnelson}}, and
  \bibinfo{author}{\bibfnamefont{M.~A.~H.} \bibnamefont{Vozmediano}},
  \bibinfo{year}{2008}{\natexlab{b}}, \bibinfo{journal}{Phys. Rev. B}
  \textbf{\bibinfo{volume}{77}}, \bibinfo{pages}{075422}.

\bibitem[{\citenamefont{Gupta and Sen}(2008)}]{Gupta:2008}
\bibinfo{author}{\bibnamefont{Gupta}, \bibfnamefont{K.}}, and
  \bibinfo{author}{\bibfnamefont{S.}~\bibnamefont{Sen}}, \bibinfo{year}{2008},
  \bibinfo{journal}{Phys. Rev. B} \textbf{\bibinfo{volume}{78}},
  \bibinfo{pages}{205429}.

\bibitem[{\citenamefont{Gupta} \emph{et~al.}(2010)\citenamefont{Gupta,
  Samsarov, and Sen}}]{Gupta:2010}
\bibinfo{author}{\bibnamefont{Gupta}, \bibfnamefont{K.~S.}},
  \bibinfo{author}{\bibfnamefont{A.}~\bibnamefont{Samsarov}}, and
  \bibinfo{author}{\bibfnamefont{S.}~\bibnamefont{Sen}}, \bibinfo{year}{2010},
  \bibinfo{journal}{Eur. Phys. J. B} \textbf{\bibinfo{volume}{73}},
  \bibinfo{pages}{389}.

\bibitem[{\citenamefont{Gupta and Sen}(2009)}]{Gupta:2009}
\bibinfo{author}{\bibnamefont{Gupta}, \bibfnamefont{K.~S.}}, and
  \bibinfo{author}{\bibfnamefont{S.}~\bibnamefont{Sen}}, \bibinfo{year}{2009},
  \bibinfo{journal}{Mod. Phys. Lett. A} \textbf{\bibinfo{volume}{24}},
  \bibinfo{pages}{99}.

\bibitem[{\citenamefont{Gusynin} \emph{et~al.}(2006)\citenamefont{Gusynin,
  Miransky, Sharapov, and Shovkovy}}]{GMSS06}
\bibinfo{author}{\bibnamefont{Gusynin}, \bibfnamefont{V.~P.}},
  \bibinfo{author}{\bibfnamefont{V.~A.} \bibnamefont{Miransky}},
  \bibinfo{author}{\bibfnamefont{S.~G.} \bibnamefont{Sharapov}}, and
  \bibinfo{author}{\bibfnamefont{I.~A.} \bibnamefont{Shovkovy}},
  \bibinfo{year}{2006}, \bibinfo{journal}{Phys. Rev. B}
  \textbf{\bibinfo{volume}{74}}, \bibinfo{pages}{195429}.

\bibitem[{\citenamefont{Gusynin} \emph{et~al.}(2009)\citenamefont{Gusynin,
  Miransky, Sharapov, Shovkovy, and Wyenberg}}]{Getal09c}
\bibinfo{author}{\bibnamefont{Gusynin}, \bibfnamefont{V.~P.}},
  \bibinfo{author}{\bibfnamefont{V.~A.} \bibnamefont{Miransky}},
  \bibinfo{author}{\bibfnamefont{S.~G.} \bibnamefont{Sharapov}},
  \bibinfo{author}{\bibfnamefont{I.~A.} \bibnamefont{Shovkovy}}, and
  \bibinfo{author}{\bibfnamefont{C.~M.} \bibnamefont{Wyenberg}},
  \bibinfo{year}{2009}, \bibinfo{journal}{Phys. Rev. B}
  \textbf{\bibinfo{volume}{79}}, \bibinfo{pages}{115431}.

\bibitem[{\citenamefont{Gusynin} \emph{et~al.}(2007)\citenamefont{Gusynin,
  Sharapov, and Carbotte}}]{Gus07}
\bibinfo{author}{\bibnamefont{Gusynin}, \bibfnamefont{V.~P.}},
  \bibinfo{author}{\bibfnamefont{S.~G.} \bibnamefont{Sharapov}}, and
  \bibinfo{author}{\bibfnamefont{J.~P.} \bibnamefont{Carbotte}},
  \bibinfo{year}{2007}, \bibinfo{journal}{Int. J. Mod. Phys. B}
  \textbf{\bibinfo{volume}{21}}, \bibinfo{pages}{4611}.

\bibitem[{\citenamefont{{G{\"u}ttinger}}
  \emph{et~al.}(2009)\citenamefont{{G{\"u}ttinger}, Stampfer, Libisch, Frey,
  Burgdorfer, Ihn, and Ensslin}}]{Getal09}
\bibinfo{author}{\bibnamefont{{G{\"u}ttinger}}, \bibfnamefont{J.}},
  \bibinfo{author}{\bibfnamefont{C.}~\bibnamefont{Stampfer}},
  \bibinfo{author}{\bibfnamefont{F.}~\bibnamefont{Libisch}},
  \bibinfo{author}{\bibfnamefont{T.}~\bibnamefont{Frey}},
  \bibinfo{author}{\bibfnamefont{J.}~\bibnamefont{Burgdorfer}},
  \bibinfo{author}{\bibfnamefont{T.}~\bibnamefont{Ihn}}, and
  \bibinfo{author}{\bibfnamefont{K.}~\bibnamefont{Ensslin}},
  \bibinfo{year}{2009}, \bibinfo{journal}{Phys. Rev. Lett.}
  \textbf{\bibinfo{volume}{103}}, \bibinfo{pages}{046810}.

\bibitem[{\citenamefont{Haldane}(1988)}]{Hal88}
\bibinfo{author}{\bibnamefont{Haldane}, \bibfnamefont{F.~D.~M.}},
  \bibinfo{year}{1988}, \bibinfo{journal}{Phys. Rev. Lett.}
  \textbf{\bibinfo{volume}{61}}, \bibinfo{pages}{2015}.

\bibitem[{\citenamefont{Han} \emph{et~al.}(2010)\citenamefont{Han, Brant, and
  Kim}}]{HKB09}
\bibinfo{author}{\bibnamefont{Han}, \bibfnamefont{M.~Y.}},
  \bibinfo{author}{\bibfnamefont{J.~C.} \bibnamefont{Brant}}, and
  \bibinfo{author}{\bibfnamefont{P.}~\bibnamefont{Kim}}, \bibinfo{year}{2010},
  \bibinfo{journal}{Phys. Rev. Lett.} \textbf{\bibinfo{volume}{104}},
  \bibinfo{pages}{056801}.

\bibitem[{\citenamefont{Han} \emph{et~al.}(2007)\citenamefont{Han,
  {\"O}zyilmaz, Zhang, and Kim}}]{HOZK07}
\bibinfo{author}{\bibnamefont{Han}, \bibfnamefont{M.~Y.}},
  \bibinfo{author}{\bibfnamefont{B.}~\bibnamefont{{\"O}zyilmaz}},
  \bibinfo{author}{\bibfnamefont{Y.}~\bibnamefont{Zhang}}, and
  \bibinfo{author}{\bibfnamefont{P.}~\bibnamefont{Kim}}, \bibinfo{year}{2007},
  \bibinfo{journal}{Phys. Rev. Lett.} \textbf{\bibinfo{volume}{98}},
  \bibinfo{pages}{206805}.

\bibitem[{\citenamefont{Hands and Strouthos}(2008)}]{Hands08}
\bibinfo{author}{\bibnamefont{Hands}, \bibfnamefont{S.}}, and
  \bibinfo{author}{\bibfnamefont{C.}~\bibnamefont{Strouthos}},
  \bibinfo{year}{2008}, \bibinfo{journal}{Phys. Rev. B}
  \textbf{\bibinfo{volume}{78}}, \bibinfo{pages}{165423}.

\bibitem[{\citenamefont{Hanish} \emph{et~al.}(1995)\citenamefont{Hanish,
  Kleine, Ritzl, and M{\"u}ller-Hartmann}}]{Han95}
\bibinfo{author}{\bibnamefont{Hanish}, \bibfnamefont{T.}},
  \bibinfo{author}{\bibfnamefont{B.}~\bibnamefont{Kleine}},
  \bibinfo{author}{\bibfnamefont{A.}~\bibnamefont{Ritzl}}, and
  \bibinfo{author}{\bibfnamefont{E.}~\bibnamefont{M{\"u}ller-Hartmann}},
  \bibinfo{year}{1995}, \bibinfo{journal}{Ann. Phys.}
  \textbf{\bibinfo{volume}{4}}, \bibinfo{pages}{303}.

\bibitem[{\citenamefont{Hannay} \emph{et~al.}(1965)\citenamefont{Hannay,
  Gaballe, Matthias, Andres, Schimidt, and MacNair}}]{Han65}
\bibinfo{author}{\bibnamefont{Hannay}, \bibfnamefont{N.~B.}},
  \bibinfo{author}{\bibfnamefont{T.~H.} \bibnamefont{Gaballe}},
  \bibinfo{author}{\bibfnamefont{B.~T.} \bibnamefont{Matthias}},
  \bibinfo{author}{\bibfnamefont{K.}~\bibnamefont{Andres}},
  \bibinfo{author}{\bibfnamefont{P.}~\bibnamefont{Schimidt}}, and
  \bibinfo{author}{\bibfnamefont{D.}~\bibnamefont{MacNair}},
  \bibinfo{year}{1965}, \bibinfo{journal}{Phys. Rev. Lett.}
  \textbf{\bibinfo{volume}{14}}, \bibinfo{pages}{225}.

\bibitem[{\citenamefont{Harigaya}(2001)}]{H01}
\bibinfo{author}{\bibnamefont{Harigaya}, \bibfnamefont{K.}},
  \bibinfo{year}{2001}, \bibinfo{journal}{J. Phys. C.: Condens. Matt.}
  \textbf{\bibinfo{volume}{13}}, \bibinfo{pages}{1295}.

\bibitem[{\citenamefont{Harigaya and Enoki}(2002)}]{HE02}
\bibinfo{author}{\bibnamefont{Harigaya}, \bibfnamefont{K.}}, and
  \bibinfo{author}{\bibfnamefont{T.}~\bibnamefont{Enoki}},
  \bibinfo{year}{2002}, \bibinfo{journal}{Chem. Phys. Lett.}
  \textbf{\bibinfo{volume}{351}}, \bibinfo{pages}{128}.

\bibitem[{\citenamefont{Hasan and Kane}(2010)}]{HK10}
\bibinfo{author}{\bibnamefont{Hasan}, \bibfnamefont{M.~Z.}}, and
  \bibinfo{author}{\bibfnamefont{C.~L.} \bibnamefont{Kane}},
  \bibinfo{year}{2010}, \bibinfo{journal}{Rev. Mod. Phys.}
  \textbf{\bibinfo{volume}{82}}, \bibinfo{pages}{3045}.

\bibitem[{\citenamefont{Hawrylak}(1987)}]{Haw87}
\bibinfo{author}{\bibnamefont{Hawrylak}, \bibfnamefont{P.}},
  \bibinfo{year}{1987}, \bibinfo{journal}{Phys. Rev. Lett.}
  \textbf{\bibinfo{volume}{59}}, \bibinfo{pages}{485}.

\bibitem[{\citenamefont{Heersche} \emph{et~al.}(2007)\citenamefont{Heersche,
  Herrero, Oostinga, Versypen, and Morpurgo}}]{Hee07}
\bibinfo{author}{\bibnamefont{Heersche}, \bibfnamefont{H.~B.}},
  \bibinfo{author}{\bibfnamefont{P.~J.} \bibnamefont{Herrero}},
  \bibinfo{author}{\bibfnamefont{J.~B.} \bibnamefont{Oostinga}},
  \bibinfo{author}{\bibfnamefont{L.~K.} \bibnamefont{Versypen}}, and
  \bibinfo{author}{\bibfnamefont{A.}~\bibnamefont{Morpurgo}},
  \bibinfo{year}{2007}, \bibinfo{journal}{Nature}
  \textbf{\bibinfo{volume}{446}}, \bibinfo{pages}{56}.

\bibitem[{\citenamefont{Hentschel and Guinea}(2007)}]{HG07}
\bibinfo{author}{\bibnamefont{Hentschel}, \bibfnamefont{M.}}, and
  \bibinfo{author}{\bibfnamefont{F.}~\bibnamefont{Guinea}},
  \bibinfo{year}{2007}, \bibinfo{journal}{Phys. Rev. B}
  \textbf{\bibinfo{volume}{76}}, \bibinfo{pages}{115407}.

\bibitem[{\citenamefont{Herbut}(2006)}]{Her06}
\bibinfo{author}{\bibnamefont{Herbut}, \bibfnamefont{I.~F.}},
  \bibinfo{year}{2006}, \bibinfo{journal}{Phys. Rev. Lett.}
  \textbf{\bibinfo{volume}{97}}, \bibinfo{pages}{146401}.

\bibitem[{\citenamefont{Herbut}(2008)}]{H08}
\bibinfo{author}{\bibnamefont{Herbut}, \bibfnamefont{I.~F.}},
  \bibinfo{year}{2008}, \bibinfo{journal}{Phys. Rev. B}
  \textbf{\bibinfo{volume}{78}}, \bibinfo{pages}{205433}.

\bibitem[{\citenamefont{Herbut}(2010)}]{Her10}
\bibinfo{author}{\bibnamefont{Herbut}, \bibfnamefont{I.~F.}},
  \bibinfo{year}{2010}, \bibinfo{journal}{Phys. Rev. Lett.}
  \textbf{\bibinfo{volume}{104}}, \bibinfo{pages}{066404}.

\bibitem[{\citenamefont{Herbut} \emph{et~al.}(2009)\citenamefont{Herbut,
  Juricic, and Roy}}]{Her09}
\bibinfo{author}{\bibnamefont{Herbut}, \bibfnamefont{I.~F.}},
  \bibinfo{author}{\bibfnamefont{V.}~\bibnamefont{Juricic}}, and
  \bibinfo{author}{\bibfnamefont{B.}~\bibnamefont{Roy}}, \bibinfo{year}{2009},
  \bibinfo{journal}{Phys. Rev. B} \textbf{\bibinfo{volume}{79}},
  \bibinfo{pages}{085116}.

\bibitem[{\citenamefont{Herbut} \emph{et~al.}(2008)\citenamefont{Herbut,
  Juricic, and Vafek}}]{HJV08}
\bibinfo{author}{\bibnamefont{Herbut}, \bibfnamefont{I.~F.}},
  \bibinfo{author}{\bibfnamefont{V.}~\bibnamefont{Juricic}}, and
  \bibinfo{author}{\bibfnamefont{O.}~\bibnamefont{Vafek}},
  \bibinfo{year}{2008}, \bibinfo{journal}{Phys. Rev. Lett.}
  \textbf{\bibinfo{volume}{100}}, \bibinfo{pages}{046403}.

\bibitem[{\citenamefont{Hoddeson} \emph{et~al.}(1987)\citenamefont{Hoddeson,
  Baym, and Eckert}}]{Hoddeson:87}
\bibinfo{author}{\bibnamefont{Hoddeson}, \bibfnamefont{L.}},
  \bibinfo{author}{\bibfnamefont{G.}~\bibnamefont{Baym}}, and
  \bibinfo{author}{\bibfnamefont{M.}~\bibnamefont{Eckert}},
  \bibinfo{year}{1987}, \bibinfo{journal}{Rev. Mod. Phys.}
  \textbf{\bibinfo{volume}{59}}, \bibinfo{pages}{287}.

\bibitem[{\citenamefont{Honerkamp}(2008)}]{Hon08}
\bibinfo{author}{\bibnamefont{Honerkamp}, \bibfnamefont{C.}},
  \bibinfo{year}{2008}, \bibinfo{journal}{Phys. Rev. Lett.}
  \textbf{\bibinfo{volume}{100}}, \bibinfo{pages}{146404}.

\bibitem[{\citenamefont{Horowitz and Doussal}(2002)}]{HL02}
\bibinfo{author}{\bibnamefont{Horowitz}, \bibfnamefont{B.}}, and
  \bibinfo{author}{\bibfnamefont{P.~L.} \bibnamefont{Doussal}},
  \bibinfo{year}{2002}, \bibinfo{journal}{Phys. Rev. B}
  \textbf{\bibinfo{volume}{65}}, \bibinfo{pages}{125323}.

\bibitem[{\citenamefont{Hou} \emph{et~al.}(2007)\citenamefont{Hou, Chamon, and
  Mudry}}]{Hou07}
\bibinfo{author}{\bibnamefont{Hou}, \bibfnamefont{C.-Y.}},
  \bibinfo{author}{\bibfnamefont{C.}~\bibnamefont{Chamon}}, and
  \bibinfo{author}{\bibfnamefont{C.}~\bibnamefont{Mudry}},
  \bibinfo{year}{2007}, \bibinfo{journal}{Phys. Rev. Lett.}
  \textbf{\bibinfo{volume}{98}}, \bibinfo{pages}{186809}.

\bibitem[{\citenamefont{Huard} \emph{et~al.}(2008)\citenamefont{Huard, Stander,
  Sulpizio, and Goldhaber-Gordon}}]{Huard:2008}
\bibinfo{author}{\bibnamefont{Huard}, \bibfnamefont{B.}},
  \bibinfo{author}{\bibfnamefont{N.}~\bibnamefont{Stander}},
  \bibinfo{author}{\bibfnamefont{J.~A.} \bibnamefont{Sulpizio}}, and
  \bibinfo{author}{\bibfnamefont{D.}~\bibnamefont{Goldhaber-Gordon}},
  \bibinfo{year}{2008}, \bibinfo{journal}{Phys. Rev. B}
  \textbf{\bibinfo{volume}{78}}, \bibinfo{pages}{121402}.

\bibitem[{\citenamefont{Huard} \emph{et~al.}(2007)\citenamefont{Huard,
  Sulpizio, Stander, Todd, Yang, and Goldhaber-Gordon}}]{HSSTYG07}
\bibinfo{author}{\bibnamefont{Huard}, \bibfnamefont{B.}},
  \bibinfo{author}{\bibfnamefont{J.~A.} \bibnamefont{Sulpizio}},
  \bibinfo{author}{\bibfnamefont{N.}~\bibnamefont{Stander}},
  \bibinfo{author}{\bibfnamefont{K.}~\bibnamefont{Todd}},
  \bibinfo{author}{\bibfnamefont{B.}~\bibnamefont{Yang}}, and
  \bibinfo{author}{\bibfnamefont{D.}~\bibnamefont{Goldhaber-Gordon}},
  \bibinfo{year}{2007}, \bibinfo{journal}{Phys. Rev. Lett.}
  \textbf{\bibinfo{volume}{98}}, \bibinfo{pages}{236803}.

\bibitem[{\citenamefont{Huertas-Herno}
  \emph{et~al.}(2006)\citenamefont{Huertas-Herno, Guinea, and Brataas}}]{Hue06}
\bibinfo{author}{\bibnamefont{Huertas-Herno}, \bibfnamefont{D.}},
  \bibinfo{author}{\bibfnamefont{F.}~\bibnamefont{Guinea}}, and
  \bibinfo{author}{\bibfnamefont{A.}~\bibnamefont{Brataas}},
  \bibinfo{year}{2006}, \bibinfo{journal}{Phys. Rev. B}
  \textbf{\bibinfo{volume}{74}}, \bibinfo{pages}{155426}.

\bibitem[{\citenamefont{Hwang and {Das Sarma}}(2007)}]{Hwa07}
\bibinfo{author}{\bibnamefont{Hwang}, \bibfnamefont{E.~H.}}, and
  \bibinfo{author}{\bibfnamefont{S.}~\bibnamefont{{Das Sarma}}},
  \bibinfo{year}{2007}, \bibinfo{journal}{Phys. Rev. B}
  \textbf{\bibinfo{volume}{75}}, \bibinfo{pages}{205418}.

\bibitem[{\citenamefont{Hwang and {Das Sarma}}(2008{\natexlab{a}})}]{Hwa08b}
\bibinfo{author}{\bibnamefont{Hwang}, \bibfnamefont{E.~H.}}, and
  \bibinfo{author}{\bibfnamefont{S.}~\bibnamefont{{Das Sarma}}},
  \bibinfo{year}{2008}{\natexlab{a}}, \bibinfo{journal}{Phys. Rev. Lett.}
  \textbf{\bibinfo{volume}{101}}, \bibinfo{pages}{156802}.

\bibitem[{\citenamefont{Hwang and {Das Sarma}}(2008{\natexlab{b}})}]{Hwa08}
\bibinfo{author}{\bibnamefont{Hwang}, \bibfnamefont{E.~H.}}, and
  \bibinfo{author}{\bibfnamefont{S.}~\bibnamefont{{Das Sarma}}},
  \bibinfo{year}{2008}{\natexlab{b}}, \bibinfo{journal}{Phys. Rev. B}
  \textbf{\bibinfo{volume}{77}}, \bibinfo{pages}{081412}.

\bibitem[{\citenamefont{Hwang} \emph{et~al.}(2007)\citenamefont{Hwang, Hu, and
  {Das Sarma}}}]{Hwang07}
\bibinfo{author}{\bibnamefont{Hwang}, \bibfnamefont{E.~H.}},
  \bibinfo{author}{\bibfnamefont{B.~Y.-K.} \bibnamefont{Hu}}, and
  \bibinfo{author}{\bibfnamefont{S.}~\bibnamefont{{Das Sarma}}},
  \bibinfo{year}{2007}, \bibinfo{journal}{Phys. Rev. Lett.}
  \textbf{\bibinfo{volume}{99}}, \bibinfo{pages}{226801}.

\bibitem[{\citenamefont{Ingersent and Si}(2002)}]{Ing02}
\bibinfo{author}{\bibnamefont{Ingersent}, \bibfnamefont{K.}}, and
  \bibinfo{author}{\bibfnamefont{Q.}~\bibnamefont{Si}}, \bibinfo{year}{2002},
  \bibinfo{journal}{Phys. Rev. Lett.} \textbf{\bibinfo{volume}{89}},
  \bibinfo{pages}{076403}.

\bibitem[{\citenamefont{Jackiw}(1984)}]{Jac84}
\bibinfo{author}{\bibnamefont{Jackiw}, \bibfnamefont{R.}},
  \bibinfo{year}{1984}, \bibinfo{journal}{Phys. Rev. D}
  \textbf{\bibinfo{volume}{27}}, \bibinfo{pages}{2375}.

\bibitem[{\citenamefont{Jackiw and Rossi}(1981)}]{Jac81}
\bibinfo{author}{\bibnamefont{Jackiw}, \bibfnamefont{R.}}, and
  \bibinfo{author}{\bibfnamefont{P.}~\bibnamefont{Rossi}},
  \bibinfo{year}{1981}, \bibinfo{journal}{Nucl. Phys. B}
  \textbf{\bibinfo{volume}{190}}, \bibinfo{pages}{681}.

\bibitem[{\citenamefont{Jacob and Kotliar}(2010)}]{Jac10}
\bibinfo{author}{\bibnamefont{Jacob}, \bibfnamefont{D.}}, and
  \bibinfo{author}{\bibfnamefont{G.}~\bibnamefont{Kotliar}},
  \bibinfo{year}{2010}, \bibinfo{journal}{Phys. Rev. B}
  \textbf{\bibinfo{volume}{82}}, \bibinfo{pages}{085423}.

\bibitem[{\citenamefont{Jang} \emph{et~al.}(2008)\citenamefont{Jang, Adam,
  Chen, Williams, {Das Sarma}, and Fuhrer}}]{Jang08}
\bibinfo{author}{\bibnamefont{Jang}, \bibfnamefont{C.}},
  \bibinfo{author}{\bibfnamefont{S.}~\bibnamefont{Adam}},
  \bibinfo{author}{\bibfnamefont{J.-H.} \bibnamefont{Chen}},
  \bibinfo{author}{\bibfnamefont{E.~D.} \bibnamefont{Williams}},
  \bibinfo{author}{\bibfnamefont{S.}~\bibnamefont{{Das Sarma}}}, and
  \bibinfo{author}{\bibfnamefont{M.~S.} \bibnamefont{Fuhrer}},
  \bibinfo{year}{2008}, \bibinfo{journal}{Phys. Rev. Lett.}
  \textbf{\bibinfo{volume}{101}}, \bibinfo{pages}{146805}.

\bibitem[{\citenamefont{Jia} \emph{et~al.}(2009)\citenamefont{Jia, Hofmann,
  Meunier, Sumpter, Campos-Delgado, Romo-Herrera, Son, Hsieh, Reina, Kong,
  Terrones, and Dresselhaus}}]{Jetal09}
\bibinfo{author}{\bibnamefont{Jia}, \bibfnamefont{X.}},
  \bibinfo{author}{\bibfnamefont{M.}~\bibnamefont{Hofmann}},
  \bibinfo{author}{\bibfnamefont{V.}~\bibnamefont{Meunier}},
  \bibinfo{author}{\bibfnamefont{B.~G.} \bibnamefont{Sumpter}},
  \bibinfo{author}{\bibfnamefont{J.}~\bibnamefont{Campos-Delgado}},
  \bibinfo{author}{\bibfnamefont{J.-M.} \bibnamefont{Romo-Herrera}},
  \bibinfo{author}{\bibfnamefont{H.}~\bibnamefont{Son}},
  \bibinfo{author}{\bibfnamefont{Y.-P.} \bibnamefont{Hsieh}},
  \bibinfo{author}{\bibfnamefont{A.}~\bibnamefont{Reina}},
  \bibinfo{author}{\bibfnamefont{J.}~\bibnamefont{Kong}},
  \bibinfo{author}{\bibfnamefont{M.}~\bibnamefont{Terrones}}, and
  \bibinfo{author}{\bibfnamefont{M.~S.} \bibnamefont{Dresselhaus}},
  \bibinfo{year}{2009}, \bibinfo{journal}{Science}
  \textbf{\bibinfo{volume}{323}}, \bibinfo{pages}{1701}.

\bibitem[{\citenamefont{Jiang} \emph{et~al.}(2008)\citenamefont{Jiang, Yao,
  Carlson, Chen, and Hu}}]{Jia08}
\bibinfo{author}{\bibnamefont{Jiang}, \bibfnamefont{Y.}},
  \bibinfo{author}{\bibfnamefont{D.-X.} \bibnamefont{Yao}},
  \bibinfo{author}{\bibfnamefont{E.~W.} \bibnamefont{Carlson}},
  \bibinfo{author}{\bibfnamefont{H.-D.} \bibnamefont{Chen}}, and
  \bibinfo{author}{\bibfnamefont{J.-P.} \bibnamefont{Hu}},
  \bibinfo{year}{2008}, \bibinfo{journal}{Phys. Rev. B}
  \textbf{\bibinfo{volume}{77}}, \bibinfo{pages}{235420}.

\bibitem[{\citenamefont{Jiang} \emph{et~al.}(2007)\citenamefont{Jiang, Zhang,
  Stormer, and Kim}}]{JZSK07}
\bibinfo{author}{\bibnamefont{Jiang}, \bibfnamefont{Z.}},
  \bibinfo{author}{\bibfnamefont{Y.}~\bibnamefont{Zhang}},
  \bibinfo{author}{\bibfnamefont{H.~L.} \bibnamefont{Stormer}}, and
  \bibinfo{author}{\bibfnamefont{P.}~\bibnamefont{Kim}}, \bibinfo{year}{2007},
  \bibinfo{journal}{Phys. Rev. Lett.} \textbf{\bibinfo{volume}{99}},
  \bibinfo{pages}{106802}.

\bibitem[{\citenamefont{Joly} \emph{et~al.}(2010)\citenamefont{Joly, Kiguchi,
  Hao, Takai, Enoki, Sumii, Amemiya, Muramatsu, Hayashi, Kim, Endo,
  Campos-Delgado} \emph{et~al.}}]{Jetal10}
\bibinfo{author}{\bibnamefont{Joly}, \bibfnamefont{V.~L.~J.}},
  \bibinfo{author}{\bibfnamefont{M.}~\bibnamefont{Kiguchi}},
  \bibinfo{author}{\bibfnamefont{S.-J.} \bibnamefont{Hao}},
  \bibinfo{author}{\bibfnamefont{K.}~\bibnamefont{Takai}},
  \bibinfo{author}{\bibfnamefont{T.}~\bibnamefont{Enoki}},
  \bibinfo{author}{\bibfnamefont{R.}~\bibnamefont{Sumii}},
  \bibinfo{author}{\bibfnamefont{K.}~\bibnamefont{Amemiya}},
  \bibinfo{author}{\bibfnamefont{H.}~\bibnamefont{Muramatsu}},
  \bibinfo{author}{\bibfnamefont{T.}~\bibnamefont{Hayashi}},
  \bibinfo{author}{\bibfnamefont{Y.~A.} \bibnamefont{Kim}},
  \bibinfo{author}{\bibfnamefont{M.}~\bibnamefont{Endo}},
  \bibinfo{author}{\bibfnamefont{J.}~\bibnamefont{Campos-Delgado}},
  \emph{et~al.}, \bibinfo{year}{2010}, \bibinfo{journal}{Phys. Rev. B}
  \textbf{\bibinfo{volume}{81}}, \bibinfo{pages}{245428}.

\bibitem[{\citenamefont{Jung and MacDonald}(2010)}]{JM09}
\bibinfo{author}{\bibnamefont{Jung}, \bibfnamefont{J.}}, and
  \bibinfo{author}{\bibfnamefont{A.~H.} \bibnamefont{MacDonald}},
  \bibinfo{year}{2010}, \bibinfo{journal}{Phys. Rev. B}
  \textbf{\bibinfo{volume}{81}}, \bibinfo{pages}{195408}.

\bibitem[{\citenamefont{Juricic} \emph{et~al.}(2010)\citenamefont{Juricic,
  Vafek, and Herbut}}]{Juricic11}
\bibinfo{author}{\bibnamefont{Juricic}, \bibfnamefont{V.}},
  \bibinfo{author}{\bibfnamefont{O.}~\bibnamefont{Vafek}}, and
  \bibinfo{author}{\bibfnamefont{I.~F.} \bibnamefont{Herbut}},
  \bibinfo{year}{2010}, \bibinfo{journal}{Phys. Rev. B}
  \textbf{\bibinfo{volume}{82}}, \bibinfo{pages}{235402}.

\bibitem[{\citenamefont{Kane and Mele}(2005)}]{Kan05}
\bibinfo{author}{\bibnamefont{Kane}, \bibfnamefont{C.}}, and
  \bibinfo{author}{\bibfnamefont{E.~J.} \bibnamefont{Mele}},
  \bibinfo{year}{2005}, \bibinfo{journal}{Phys. Rev. Lett.}
  \textbf{\bibinfo{volume}{95}}, \bibinfo{pages}{226801}.

\bibitem[{\citenamefont{Kane and Fisher}(1992)}]{KF92}
\bibinfo{author}{\bibnamefont{Kane}, \bibfnamefont{C.~L.}}, and
  \bibinfo{author}{\bibfnamefont{M.~P.~A.} \bibnamefont{Fisher}},
  \bibinfo{year}{1992}, \bibinfo{journal}{Phys. Rev. Lett.}
  \textbf{\bibinfo{volume}{68}}, \bibinfo{pages}{1220}.

\bibitem[{\citenamefont{Kashuba}(2008)}]{Kashuba08}
\bibinfo{author}{\bibnamefont{Kashuba}, \bibfnamefont{A.~B.}},
  \bibinfo{year}{2008}, \bibinfo{journal}{Phys. Rev. B}
  \textbf{\bibinfo{volume}{78}}, \bibinfo{pages}{085415}.

\bibitem[{\citenamefont{Katsnelson}(2006)}]{Katsnelson:2006}
\bibinfo{author}{\bibnamefont{Katsnelson}, \bibfnamefont{M.}},
  \bibinfo{year}{2006}, \bibinfo{journal}{Phys. Rev. B}
  \textbf{\bibinfo{volume}{74}}, \bibinfo{pages}{201401(R)}.

\bibitem[{\citenamefont{Katsnelson and Prokhorova}(2008)}]{KP08}
\bibinfo{author}{\bibnamefont{Katsnelson}, \bibfnamefont{M.~I.}}, and
  \bibinfo{author}{\bibfnamefont{M.~F.} \bibnamefont{Prokhorova}},
  \bibinfo{year}{2008}, \bibinfo{journal}{Phys. Rev. B}
  \textbf{\bibinfo{volume}{77}}, \bibinfo{pages}{205424}.

\bibitem[{\citenamefont{Khalilov and Ho}(1998)}]{Khalilov:1998}
\bibinfo{author}{\bibnamefont{Khalilov}, \bibfnamefont{V.~R.}}, and
  \bibinfo{author}{\bibfnamefont{C.-L.} \bibnamefont{Ho}},
  \bibinfo{year}{1998}, \bibinfo{journal}{Mod. Phys. Lett. A}
  \textbf{\bibinfo{volume}{13}}, \bibinfo{pages}{615}.

\bibitem[{\citenamefont{Kharitonov and Efetov}(2008)}]{Kha08}
\bibinfo{author}{\bibnamefont{Kharitonov}, \bibfnamefont{M.~Y.}}, and
  \bibinfo{author}{\bibfnamefont{K.~B.} \bibnamefont{Efetov}},
  \bibinfo{year}{2008}, \bibinfo{journal}{Phys. Rev. B}
  \textbf{\bibinfo{volume}{78}}, \bibinfo{pages}{241401(R)}.

\bibitem[{\citenamefont{Khaymovich}
  \emph{et~al.}(2009)\citenamefont{Khaymovich, Kopnin, Melnikov, and
  Shereshevskii}}]{Kha09}
\bibinfo{author}{\bibnamefont{Khaymovich}, \bibfnamefont{M.}},
  \bibinfo{author}{\bibfnamefont{N.~B.} \bibnamefont{Kopnin}},
  \bibinfo{author}{\bibfnamefont{A.~S.} \bibnamefont{Melnikov}}, and
  \bibinfo{author}{\bibfnamefont{I.~A.} \bibnamefont{Shereshevskii}},
  \bibinfo{year}{2009}, \bibinfo{journal}{Phys. Rev. B}
  \textbf{\bibinfo{volume}{79}}, \bibinfo{pages}{224506}.

\bibitem[{\citenamefont{Khveshchenko}(2001{\natexlab{a}})}]{Khv01a}
\bibinfo{author}{\bibnamefont{Khveshchenko}, \bibfnamefont{D.~V.}},
  \bibinfo{year}{2001}{\natexlab{a}}, \bibinfo{journal}{Phys. Rev. Lett.}
  \textbf{\bibinfo{volume}{87}}, \bibinfo{pages}{206401}.

\bibitem[{\citenamefont{Khveshchenko}(2001{\natexlab{b}})}]{Khv01b}
\bibinfo{author}{\bibnamefont{Khveshchenko}, \bibfnamefont{D.~V.}},
  \bibinfo{year}{2001}{\natexlab{b}}, \bibinfo{journal}{Phys. Rev. Lett.}
  \textbf{\bibinfo{volume}{87}}, \bibinfo{pages}{246802}.

\bibitem[{\citenamefont{Khveshchenko}(2006)}]{Khve06b}
\bibinfo{author}{\bibnamefont{Khveshchenko}, \bibfnamefont{D.~V.}},
  \bibinfo{year}{2006}, \bibinfo{journal}{\prb} \textbf{\bibinfo{volume}{74}},
  \bibinfo{pages}{161402(R)}.

\bibitem[{\citenamefont{Khveshchenko}(2009)}]{Khve08}
\bibinfo{author}{\bibnamefont{Khveshchenko}, \bibfnamefont{D.~V.}},
  \bibinfo{year}{2009}, \bibinfo{journal}{J. Phys.: Condens. Matter}
  \textbf{\bibinfo{volume}{21}}, \bibinfo{pages}{075303}.

\bibitem[{\citenamefont{Khveshchenko and Leal}(2004)}]{Khv04}
\bibinfo{author}{\bibnamefont{Khveshchenko}, \bibfnamefont{D.~V.}}, and
  \bibinfo{author}{\bibfnamefont{H.}~\bibnamefont{Leal}}, \bibinfo{year}{2004},
  \bibinfo{journal}{Nucl. Phys. B} \textbf{\bibinfo{volume}{687}},
  \bibinfo{pages}{323}.

\bibitem[{\citenamefont{Killi} \emph{et~al.}(2010)\citenamefont{Killi, Wei,
  Affleck, and Paramekanti}}]{Kil10}
\bibinfo{author}{\bibnamefont{Killi}, \bibfnamefont{M.}},
  \bibinfo{author}{\bibfnamefont{T.}~\bibnamefont{Wei}},
  \bibinfo{author}{\bibfnamefont{I.}~\bibnamefont{Affleck}}, and
  \bibinfo{author}{\bibfnamefont{A.}~\bibnamefont{Paramekanti}},
  \bibinfo{year}{2010}, \bibinfo{journal}{Phys. Rev. Lett.}
  \textbf{\bibinfo{volume}{104}}, \bibinfo{pages}{216406}.

\bibitem[{\citenamefont{Kirwan} \emph{et~al.}(2008)\citenamefont{Kirwan, Rocha,
  Costa, and Ferreira}}]{Kir08}
\bibinfo{author}{\bibnamefont{Kirwan}, \bibfnamefont{D.~F.}},
  \bibinfo{author}{\bibfnamefont{C.~G.} \bibnamefont{Rocha}},
  \bibinfo{author}{\bibfnamefont{A.~T.} \bibnamefont{Costa}}, and
  \bibinfo{author}{\bibfnamefont{M.~S.} \bibnamefont{Ferreira}},
  \bibinfo{year}{2008}, \bibinfo{journal}{Phys. Rev. B}
  \textbf{\bibinfo{volume}{77}}, \bibinfo{pages}{085432}.

\bibitem[{\citenamefont{Kohn and Luttinger}(1965)}]{Koh65}
\bibinfo{author}{\bibnamefont{Kohn}, \bibfnamefont{W.}}, and
  \bibinfo{author}{\bibfnamefont{J.~M.} \bibnamefont{Luttinger}},
  \bibinfo{year}{1965}, \bibinfo{journal}{Phys. Rev. Lett.}
  \textbf{\bibinfo{volume}{15}}, \bibinfo{pages}{524}.

\bibitem[{\citenamefont{Kolezhuk} \emph{et~al.}(2006)\citenamefont{Kolezhuk,
  Sachdev, Biswas, and Chen}}]{Kolezhuk:2006}
\bibinfo{author}{\bibnamefont{Kolezhuk}, \bibfnamefont{A.}},
  \bibinfo{author}{\bibfnamefont{S.}~\bibnamefont{Sachdev}},
  \bibinfo{author}{\bibfnamefont{R.~R.} \bibnamefont{Biswas}}, and
  \bibinfo{author}{\bibfnamefont{P.}~\bibnamefont{Chen}}, \bibinfo{year}{2006},
  \bibinfo{journal}{Phys. Rev. B} \textbf{\bibinfo{volume}{74}},
  \bibinfo{pages}{165114}.

\bibitem[{\citenamefont{Kopnin and Sonin}(2008)}]{Kop08}
\bibinfo{author}{\bibnamefont{Kopnin}, \bibfnamefont{N.~P.}}, and
  \bibinfo{author}{\bibfnamefont{E.~B.} \bibnamefont{Sonin}},
  \bibinfo{year}{2008}, \bibinfo{journal}{Phys. Rev. Lett.}
  \textbf{\bibinfo{volume}{100}}, \bibinfo{pages}{246808}.

\bibitem[{\citenamefont{Kotov}
  \emph{et~al.}(2008{\natexlab{a}})\citenamefont{Kotov, Pereira, and
  Uchoa}}]{Kotov:2008}
\bibinfo{author}{\bibnamefont{Kotov}, \bibfnamefont{V.~N.}},
  \bibinfo{author}{\bibfnamefont{V.~M.} \bibnamefont{Pereira}}, and
  \bibinfo{author}{\bibfnamefont{B.}~\bibnamefont{Uchoa}},
  \bibinfo{year}{2008}{\natexlab{a}}, \bibinfo{journal}{Phys. Rev. B}
  \textbf{\bibinfo{volume}{78}}, \bibinfo{pages}{075433}.

\bibitem[{\citenamefont{Kotov}
  \emph{et~al.}(2008{\natexlab{b}})\citenamefont{Kotov, Uchoa, and {Castro
  Neto}}}]{Kot08a}
\bibinfo{author}{\bibnamefont{Kotov}, \bibfnamefont{V.~N.}},
  \bibinfo{author}{\bibfnamefont{B.}~\bibnamefont{Uchoa}}, and
  \bibinfo{author}{\bibfnamefont{A.~H.} \bibnamefont{{Castro Neto}}},
  \bibinfo{year}{2008}{\natexlab{b}}, \bibinfo{journal}{\prb}
  \textbf{\bibinfo{volume}{78}}, \bibinfo{pages}{035119}.

\bibitem[{\citenamefont{Kotov} \emph{et~al.}(2009)\citenamefont{Kotov, Uchoa,
  and {Castro Neto}}}]{Kot1N}
\bibinfo{author}{\bibnamefont{Kotov}, \bibfnamefont{V.~N.}},
  \bibinfo{author}{\bibfnamefont{B.}~\bibnamefont{Uchoa}}, and
  \bibinfo{author}{\bibfnamefont{A.~H.} \bibnamefont{{Castro Neto}}},
  \bibinfo{year}{2009}, \bibinfo{journal}{Phys. Rev. B}
  \textbf{\bibinfo{volume}{80}}, \bibinfo{pages}{165424}.

\bibitem[{\citenamefont{Kovtun} \emph{et~al.}(2005)\citenamefont{Kovtun, Son,
  and Starinets}}]{Kovtun05}
\bibinfo{author}{\bibnamefont{Kovtun}, \bibfnamefont{P.~K.}},
  \bibinfo{author}{\bibfnamefont{D.~T.} \bibnamefont{Son}}, and
  \bibinfo{author}{\bibfnamefont{A.~O.} \bibnamefont{Starinets}},
  \bibinfo{year}{2005}, \bibinfo{journal}{\prl} \textbf{\bibinfo{volume}{94}},
  \bibinfo{pages}{111601}.

\bibitem[{\citenamefont{Kramberger}
  \emph{et~al.}(2008)\citenamefont{Kramberger, Hambach, Giorgetti, Rummeli,
  Knupfer, Fink, Buchner, Reining, Einarsson, Maruyama, Sottile, Hannewald}
  \emph{et~al.}}]{Kram08}
\bibinfo{author}{\bibnamefont{Kramberger}, \bibfnamefont{C.}},
  \bibinfo{author}{\bibfnamefont{R.}~\bibnamefont{Hambach}},
  \bibinfo{author}{\bibfnamefont{C.}~\bibnamefont{Giorgetti}},
  \bibinfo{author}{\bibfnamefont{M.~H.} \bibnamefont{Rummeli}},
  \bibinfo{author}{\bibfnamefont{M.}~\bibnamefont{Knupfer}},
  \bibinfo{author}{\bibfnamefont{J.}~\bibnamefont{Fink}},
  \bibinfo{author}{\bibfnamefont{B.}~\bibnamefont{Buchner}},
  \bibinfo{author}{\bibfnamefont{L.}~\bibnamefont{Reining}},
  \bibinfo{author}{\bibfnamefont{E.}~\bibnamefont{Einarsson}},
  \bibinfo{author}{\bibfnamefont{S.}~\bibnamefont{Maruyama}},
  \bibinfo{author}{\bibfnamefont{F.}~\bibnamefont{Sottile}},
  \bibinfo{author}{\bibfnamefont{K.}~\bibnamefont{Hannewald}}, \emph{et~al.},
  \bibinfo{year}{2008}, \bibinfo{journal}{Phys. Rev. Lett.}
  \textbf{\bibinfo{volume}{100}}, \bibinfo{pages}{196803}.

\bibitem[{\citenamefont{Krasheninnikov}
  \emph{et~al.}(2009)\citenamefont{Krasheninnikov, Lehtinen, Foster, Pyykko,
  and Nieminen}}]{Kra09}
\bibinfo{author}{\bibnamefont{Krasheninnikov}, \bibfnamefont{A.~V.}},
  \bibinfo{author}{\bibfnamefont{P.~O.} \bibnamefont{Lehtinen}},
  \bibinfo{author}{\bibfnamefont{A.~S.} \bibnamefont{Foster}},
  \bibinfo{author}{\bibfnamefont{P.}~\bibnamefont{Pyykko}}, and
  \bibinfo{author}{\bibfnamefont{R.~M.} \bibnamefont{Nieminen}},
  \bibinfo{year}{2009}, \bibinfo{journal}{Phys. Rev. Lett.}
  \textbf{\bibinfo{volume}{102}}, \bibinfo{pages}{126807}.

\bibitem[{\citenamefont{Kusminskiy}
  \emph{et~al.}(2008)\citenamefont{Kusminskiy, Nilsson, Campbell, and {Castro
  Neto}}}]{Kus08}
\bibinfo{author}{\bibnamefont{Kusminskiy}, \bibfnamefont{S.~V.}},
  \bibinfo{author}{\bibfnamefont{J.}~\bibnamefont{Nilsson}},
  \bibinfo{author}{\bibfnamefont{D.~K.} \bibnamefont{Campbell}}, and
  \bibinfo{author}{\bibfnamefont{A.~H.} \bibnamefont{{Castro Neto}}},
  \bibinfo{year}{2008}, \bibinfo{journal}{Phys. Rev. Lett.}
  \textbf{\bibinfo{volume}{100}}, \bibinfo{pages}{106805}.

\bibitem[{\citenamefont{Kusminskiy}
  \emph{et~al.}(2009)\citenamefont{Kusminskiy, Nilsson, Campbell, and {Castro
  Neto}}}]{Kus09}
\bibinfo{author}{\bibnamefont{Kusminskiy}, \bibfnamefont{S.~V.}},
  \bibinfo{author}{\bibfnamefont{J.}~\bibnamefont{Nilsson}},
  \bibinfo{author}{\bibfnamefont{D.~K.} \bibnamefont{Campbell}}, and
  \bibinfo{author}{\bibfnamefont{A.~H.} \bibnamefont{{Castro Neto}}},
  \bibinfo{year}{2009}, \bibinfo{journal}{Europhys. Lett.}
  \textbf{\bibinfo{volume}{85}}, \bibinfo{pages}{58005}.

\bibitem[{\citenamefont{Laitenberger and Palmer}(1996)}]{Lait96}
\bibinfo{author}{\bibnamefont{Laitenberger}, \bibfnamefont{P.}}, and
  \bibinfo{author}{\bibfnamefont{R.~E.} \bibnamefont{Palmer}},
  \bibinfo{year}{1996}, \bibinfo{journal}{Phys. Rev. Lett.}
  \textbf{\bibinfo{volume}{76}}, \bibinfo{pages}{1952}.

\bibitem[{\citenamefont{Landau and Lifshitz}(1981)}]{Landau-QM:1981}
\bibinfo{author}{\bibnamefont{Landau}, \bibfnamefont{L.~D.}}, and
  \bibinfo{author}{\bibfnamefont{E.~M.} \bibnamefont{Lifshitz}},
  \bibinfo{year}{1981}, \emph{\bibinfo{title}{Quantum Mechanics:
  Non-Relativistic Theory}} (\bibinfo{publisher}{Pergamon Press}).

\bibitem[{\citenamefont{Laughlin}(1983)}]{Laughlin:83}
\bibinfo{author}{\bibnamefont{Laughlin}, \bibfnamefont{R.~B.}},
  \bibinfo{year}{1983}, \bibinfo{journal}{Phys. Rev. Lett.}
  \textbf{\bibinfo{volume}{50}}, \bibinfo{pages}{1395}.

\bibitem[{\citenamefont{Lee}(1993)}]{Lee93}
\bibinfo{author}{\bibnamefont{Lee}, \bibfnamefont{P.~A.}},
  \bibinfo{year}{1993}, \bibinfo{journal}{\prl} \textbf{\bibinfo{volume}{71}},
  \bibinfo{pages}{1887}.

\bibitem[{\citenamefont{Lehtinen} \emph{et~al.}(2003)\citenamefont{Lehtinen,
  Foster, Ayuela, Krasheninnikov, Nordlund, and Nieminen}}]{Letal03}
\bibinfo{author}{\bibnamefont{Lehtinen}, \bibfnamefont{P.~O.}},
  \bibinfo{author}{\bibfnamefont{A.~S.} \bibnamefont{Foster}},
  \bibinfo{author}{\bibfnamefont{A.}~\bibnamefont{Ayuela}},
  \bibinfo{author}{\bibfnamefont{A.}~\bibnamefont{Krasheninnikov}},
  \bibinfo{author}{\bibfnamefont{K.}~\bibnamefont{Nordlund}}, and
  \bibinfo{author}{\bibfnamefont{R.~M.} \bibnamefont{Nieminen}},
  \bibinfo{year}{2003}, \bibinfo{journal}{Phys. Rev. Lett.}
  \textbf{\bibinfo{volume}{91}}, \bibinfo{pages}{017202}.

\bibitem[{\citenamefont{Lemonik} \emph{et~al.}(2010)\citenamefont{Lemonik,
  Aleiner, Toke, and Fal'ko}}]{Lem10}
\bibinfo{author}{\bibnamefont{Lemonik}, \bibfnamefont{Y.}},
  \bibinfo{author}{\bibfnamefont{I.~L.} \bibnamefont{Aleiner}},
  \bibinfo{author}{\bibfnamefont{C.}~\bibnamefont{Toke}}, and
  \bibinfo{author}{\bibfnamefont{V.~I.} \bibnamefont{Fal'ko}},
  \bibinfo{year}{2010}, \bibinfo{journal}{Phys. Rev. B}
  \textbf{\bibinfo{volume}{82}}, \bibinfo{pages}{201408}.

\bibitem[{\citenamefont{Levy} \emph{et~al.}(2010)\citenamefont{Levy, Burke,
  Meaker, Panlasigui, Zettl, Guinea, {Castro Neto}, and Crommie}}]{Letal10b}
\bibinfo{author}{\bibnamefont{Levy}, \bibfnamefont{N.}},
  \bibinfo{author}{\bibfnamefont{S.~A.} \bibnamefont{Burke}},
  \bibinfo{author}{\bibfnamefont{K.~L.} \bibnamefont{Meaker}},
  \bibinfo{author}{\bibfnamefont{M.}~\bibnamefont{Panlasigui}},
  \bibinfo{author}{\bibfnamefont{A.}~\bibnamefont{Zettl}},
  \bibinfo{author}{\bibfnamefont{F.}~\bibnamefont{Guinea}},
  \bibinfo{author}{\bibfnamefont{A.~H.} \bibnamefont{{Castro Neto}}}, and
  \bibinfo{author}{\bibfnamefont{M.~F.} \bibnamefont{Crommie}},
  \bibinfo{year}{2010}, \bibinfo{journal}{Science}
  \textbf{\bibinfo{volume}{329}}, \bibinfo{pages}{544}.

\bibitem[{\citenamefont{Li} \emph{et~al.}(2009{\natexlab{a}})\citenamefont{Li,
  Luican, and Andrei}}]{Li08b}
\bibinfo{author}{\bibnamefont{Li}, \bibfnamefont{G.}},
  \bibinfo{author}{\bibfnamefont{A.}~\bibnamefont{Luican}}, and
  \bibinfo{author}{\bibfnamefont{E.~Y.} \bibnamefont{Andrei}},
  \bibinfo{year}{2009}{\natexlab{a}}, \bibinfo{journal}{Phys. Rev. Lett.}
  \textbf{\bibinfo{volume}{102}}, \bibinfo{pages}{176804}.

\bibitem[{\citenamefont{Li} \emph{et~al.}(2010)\citenamefont{Li, Luican, {Lopes
  dos Santos}, {Castro Neto}, Reina, Kong, and Andrei}}]{Li10}
\bibinfo{author}{\bibnamefont{Li}, \bibfnamefont{G.}},
  \bibinfo{author}{\bibfnamefont{A.}~\bibnamefont{Luican}},
  \bibinfo{author}{\bibfnamefont{J.~M.~B.} \bibnamefont{{Lopes dos Santos}}},
  \bibinfo{author}{\bibfnamefont{A.~H.} \bibnamefont{{Castro Neto}}},
  \bibinfo{author}{\bibfnamefont{A.}~\bibnamefont{Reina}},
  \bibinfo{author}{\bibfnamefont{J.}~\bibnamefont{Kong}}, and
  \bibinfo{author}{\bibfnamefont{E.~Y.} \bibnamefont{Andrei}},
  \bibinfo{year}{2010}, \bibinfo{journal}{Nature Physics}
  \textbf{\bibinfo{volume}{6}}, \bibinfo{pages}{109}.

\bibitem[{\citenamefont{Li} \emph{et~al.}(2008)\citenamefont{Li, Henriksen,
  Jiang, Hao, Martin, Kim, Stormer, and Basov}}]{Li08a}
\bibinfo{author}{\bibnamefont{Li}, \bibfnamefont{Z.~Q.}},
  \bibinfo{author}{\bibfnamefont{E.~A.} \bibnamefont{Henriksen}},
  \bibinfo{author}{\bibfnamefont{Z.}~\bibnamefont{Jiang}},
  \bibinfo{author}{\bibfnamefont{Z.}~\bibnamefont{Hao}},
  \bibinfo{author}{\bibfnamefont{M.~C.} \bibnamefont{Martin}},
  \bibinfo{author}{\bibfnamefont{P.}~\bibnamefont{Kim}},
  \bibinfo{author}{\bibfnamefont{H.~L.} \bibnamefont{Stormer}}, and
  \bibinfo{author}{\bibfnamefont{D.~N.} \bibnamefont{Basov}},
  \bibinfo{year}{2008}, \bibinfo{journal}{Nature Physics}
  \textbf{\bibinfo{volume}{4}}, \bibinfo{pages}{532}.

\bibitem[{\citenamefont{Li} \emph{et~al.}(2009{\natexlab{b}})\citenamefont{Li,
  Henriksen, Jiang, Hao, Martin, Kim, Stormer, and Basov}}]{Li09}
\bibinfo{author}{\bibnamefont{Li}, \bibfnamefont{Z.~Q.}},
  \bibinfo{author}{\bibfnamefont{E.~A.} \bibnamefont{Henriksen}},
  \bibinfo{author}{\bibfnamefont{Z.}~\bibnamefont{Jiang}},
  \bibinfo{author}{\bibfnamefont{Z.}~\bibnamefont{Hao}},
  \bibinfo{author}{\bibfnamefont{M.~C.} \bibnamefont{Martin}},
  \bibinfo{author}{\bibfnamefont{P.}~\bibnamefont{Kim}},
  \bibinfo{author}{\bibfnamefont{H.~L.} \bibnamefont{Stormer}}, and
  \bibinfo{author}{\bibfnamefont{D.~N.} \bibnamefont{Basov}},
  \bibinfo{year}{2009}{\natexlab{b}}, \bibinfo{journal}{Phys. Rev. Lett.}
  \textbf{\bibinfo{volume}{102}}, \bibinfo{pages}{037403}.

\bibitem[{\citenamefont{Lieb}(1981)}]{Lieb:1981}
\bibinfo{author}{\bibnamefont{Lieb}, \bibfnamefont{E.~H.}},
  \bibinfo{year}{1981}, \bibinfo{journal}{Rev. Mod. Phys.}
  \textbf{\bibinfo{volume}{53}}, \bibinfo{pages}{603}.

\bibitem[{\citenamefont{Lin}(2006)}]{Lin:2006}
\bibinfo{author}{\bibnamefont{Lin}, \bibfnamefont{D.-H.}},
  \bibinfo{year}{2006}, \bibinfo{journal}{Phys. Rev. A}
  \textbf{\bibinfo{volume}{73}}, \bibinfo{pages}{044701}.

\bibitem[{\citenamefont{Lin and Shung}(1996)}]{Lin96}
\bibinfo{author}{\bibnamefont{Lin}, \bibfnamefont{M.~F.}}, and
  \bibinfo{author}{\bibfnamefont{K.~W.~K.} \bibnamefont{Shung}},
  \bibinfo{year}{1996}, \bibinfo{journal}{Phys. Rev. B}
  \textbf{\bibinfo{volume}{53}}, \bibinfo{pages}{1109}.

\bibitem[{\citenamefont{Linder} \emph{et~al.}(2009)\citenamefont{Linder,
  Black-Schaffer, Yokoyama, Doniach, and Sudb\o{}}}]{Lin09}
\bibinfo{author}{\bibnamefont{Linder}, \bibfnamefont{J.}},
  \bibinfo{author}{\bibfnamefont{A.~M.} \bibnamefont{Black-Schaffer}},
  \bibinfo{author}{\bibfnamefont{T.}~\bibnamefont{Yokoyama}},
  \bibinfo{author}{\bibfnamefont{S.}~\bibnamefont{Doniach}}, and
  \bibinfo{author}{\bibfnamefont{A.}~\bibnamefont{Sudb\o{}}},
  \bibinfo{year}{2009}, \bibinfo{journal}{Phys. Rev. B}
  \textbf{\bibinfo{volume}{80}}, \bibinfo{pages}{094522}.

\bibitem[{\citenamefont{Linder} \emph{et~al.}(2008)\citenamefont{Linder,
  Yokoyama, Huertas-Herno, and Sudb\o{}}}]{Lin08}
\bibinfo{author}{\bibnamefont{Linder}, \bibfnamefont{J.}},
  \bibinfo{author}{\bibfnamefont{T.}~\bibnamefont{Yokoyama}},
  \bibinfo{author}{\bibfnamefont{D.}~\bibnamefont{Huertas-Herno}}, and
  \bibinfo{author}{\bibfnamefont{A.}~\bibnamefont{Sudb\o{}}},
  \bibinfo{year}{2008}, \bibinfo{journal}{Phys. Rev. Lett.}
  \textbf{\bibinfo{volume}{100}}, \bibinfo{pages}{187004}.

\bibitem[{\citenamefont{Lindhard}(1954)}]{Lindhard:54}
\bibinfo{author}{\bibnamefont{Lindhard}, \bibfnamefont{J.}},
  \bibinfo{year}{1954}, \bibinfo{journal}{Det Kgl Danske Vid. Selskab,
  Matematisk-fysiske Meddelelser} \textbf{\bibinfo{volume}{28}}.

\bibitem[{\citenamefont{Liu} \emph{et~al.}(2009)\citenamefont{Liu, Li, and
  Cheng}}]{Liu09}
\bibinfo{author}{\bibnamefont{Liu}, \bibfnamefont{G.-Z.}},
  \bibinfo{author}{\bibfnamefont{W.}~\bibnamefont{Li}}, and
  \bibinfo{author}{\bibfnamefont{G.}~\bibnamefont{Cheng}},
  \bibinfo{year}{2009}, \bibinfo{journal}{Phys. Rev. B}
  \textbf{\bibinfo{volume}{79}}, \bibinfo{pages}{205429(R)}.

\bibitem[{\citenamefont{Liu} \emph{et~al.}(2008)\citenamefont{Liu, Willis,
  Emtsev, and Seyller}}]{Liu08}
\bibinfo{author}{\bibnamefont{Liu}, \bibfnamefont{Y.}},
  \bibinfo{author}{\bibfnamefont{R.~F.} \bibnamefont{Willis}},
  \bibinfo{author}{\bibfnamefont{K.~V.} \bibnamefont{Emtsev}}, and
  \bibinfo{author}{\bibfnamefont{T.}~\bibnamefont{Seyller}},
  \bibinfo{year}{2008}, \bibinfo{journal}{Phys. Rev. B}
  \textbf{\bibinfo{volume}{78}}, \bibinfo{pages}{201403(R)}.

\bibitem[{\citenamefont{Loktev and Turkowski}(2009)}]{Lok09}
\bibinfo{author}{\bibnamefont{Loktev}, \bibfnamefont{V.~M.}}, and
  \bibinfo{author}{\bibfnamefont{V.}~\bibnamefont{Turkowski}},
  \bibinfo{year}{2009}, \bibinfo{journal}{Phys. Rev. B}
  \textbf{\bibinfo{volume}{79}}, \bibinfo{pages}{233402}.

\bibitem[{\citenamefont{{Lopes dos Santos}}
  \emph{et~al.}(2007)\citenamefont{{Lopes dos Santos}, Peres, and {Castro
  Neto}}}]{San07}
\bibinfo{author}{\bibnamefont{{Lopes dos Santos}}, \bibfnamefont{J.~M.~B.}},
  \bibinfo{author}{\bibfnamefont{N.~M.~R.} \bibnamefont{Peres}}, and
  \bibinfo{author}{\bibfnamefont{A.~H.} \bibnamefont{{Castro Neto}}},
  \bibinfo{year}{2007}, \bibinfo{journal}{Phys. Rev. Lett.}
  \textbf{\bibinfo{volume}{99}}, \bibinfo{pages}{256802}.

\bibitem[{\citenamefont{Lozovik and Sokolik}(2010)}]{Loz10}
\bibinfo{author}{\bibnamefont{Lozovik}, \bibfnamefont{Y.~E.}}, and
  \bibinfo{author}{\bibfnamefont{A.~A.} \bibnamefont{Sokolik}},
  \bibinfo{year}{2010}, \bibinfo{journal}{Phys. Lett. A}
  \textbf{\bibinfo{volume}{374}}, \bibinfo{pages}{2785}.

\bibitem[{\citenamefont{Ludwig} \emph{et~al.}(1994)\citenamefont{Ludwig,
  Fisher, Shankar, and Grinstein}}]{LFSG94}
\bibinfo{author}{\bibnamefont{Ludwig}, \bibfnamefont{A.~W.}},
  \bibinfo{author}{\bibfnamefont{M.~P.~A.} \bibnamefont{Fisher}},
  \bibinfo{author}{\bibfnamefont{R.}~\bibnamefont{Shankar}}, and
  \bibinfo{author}{\bibfnamefont{G.}~\bibnamefont{Grinstein}},
  \bibinfo{year}{1994}, \bibinfo{journal}{Phys. Rev. B}
  \textbf{\bibinfo{volume}{50}}, \bibinfo{pages}{7526}.

\bibitem[{\citenamefont{Lundqvist}(1967)}]{Lundqvist}
\bibinfo{author}{\bibnamefont{Lundqvist}, \bibfnamefont{B.}},
  \bibinfo{year}{1967}, \bibinfo{journal}{Zeitschrift fur Physik B Condensed
  Matter} \textbf{\bibinfo{volume}{6}}, \bibinfo{pages}{193}.

\bibitem[{\citenamefont{Lutchyn} \emph{et~al.}(2008)\citenamefont{Lutchyn,
  Galitski, Refael, and {Das Sarma}}}]{Lut08}
\bibinfo{author}{\bibnamefont{Lutchyn}, \bibfnamefont{R.~M.}},
  \bibinfo{author}{\bibfnamefont{V.}~\bibnamefont{Galitski}},
  \bibinfo{author}{\bibfnamefont{G.}~\bibnamefont{Refael}}, and
  \bibinfo{author}{\bibfnamefont{S.}~\bibnamefont{{Das Sarma}}},
  \bibinfo{year}{2008}, \bibinfo{journal}{Phys. Rev. Lett.}
  \textbf{\bibinfo{volume}{101}}, \bibinfo{pages}{106402}.

\bibitem[{\citenamefont{Lv and Wan}(2010)}]{Lv10}
\bibinfo{author}{\bibnamefont{Lv}, \bibfnamefont{M.}}, and
  \bibinfo{author}{\bibfnamefont{S.}~\bibnamefont{Wan}}, \bibinfo{year}{2010},
  \bibinfo{journal}{Phys. Rev. B} \textbf{\bibinfo{volume}{81}},
  \bibinfo{pages}{195409}.

\bibitem[{\citenamefont{Mahan}(2000)}]{Mahan}
\bibinfo{author}{\bibnamefont{Mahan}, \bibfnamefont{G.~D.}},
  \bibinfo{year}{2000}, \emph{\bibinfo{title}{Many-Particle Physics}}
  (\bibinfo{publisher}{Plenum}, \bibinfo{address}{New York}).

\bibitem[{\citenamefont{Maiti and Sengupta}(2007)}]{Mai07}
\bibinfo{author}{\bibnamefont{Maiti}, \bibfnamefont{M.}}, and
  \bibinfo{author}{\bibfnamefont{K.}~\bibnamefont{Sengupta}},
  \bibinfo{year}{2007}, \bibinfo{journal}{Phys. Rev. B}
  \textbf{\bibinfo{volume}{76}}, \bibinfo{pages}{054513}.

\bibitem[{\citenamefont{Marino and Nunes}(2006)}]{Mar06}
\bibinfo{author}{\bibnamefont{Marino}, \bibfnamefont{E.~C.}}, and
  \bibinfo{author}{\bibfnamefont{L.~H. C.~M.} \bibnamefont{Nunes}},
  \bibinfo{year}{2006}, \bibinfo{journal}{Nucl. Phys. B}
  \textbf{\bibinfo{volume}{741}}, \bibinfo{pages}{404}.

\bibitem[{\citenamefont{Marinopoulus}
  \emph{et~al.}(2004)\citenamefont{Marinopoulus, Reining, Rubio, and
  Olevano}}]{Mar04}
\bibinfo{author}{\bibnamefont{Marinopoulus}, \bibfnamefont{A.~G.}},
  \bibinfo{author}{\bibfnamefont{L.}~\bibnamefont{Reining}},
  \bibinfo{author}{\bibfnamefont{A.}~\bibnamefont{Rubio}}, and
  \bibinfo{author}{\bibfnamefont{V.}~\bibnamefont{Olevano}},
  \bibinfo{year}{2004}, \bibinfo{journal}{Phys. Rev. B}
  \textbf{\bibinfo{volume}{69}}, \bibinfo{pages}{245419}.

\bibitem[{\citenamefont{Martelo} \emph{et~al.}(1997)\citenamefont{Martelo,
  Dzierzawa, Siffert, and Baeriswyl}}]{Mar97}
\bibinfo{author}{\bibnamefont{Martelo}, \bibfnamefont{L.~M.}},
  \bibinfo{author}{\bibfnamefont{M.}~\bibnamefont{Dzierzawa}},
  \bibinfo{author}{\bibfnamefont{L.}~\bibnamefont{Siffert}}, and
  \bibinfo{author}{\bibfnamefont{D.}~\bibnamefont{Baeriswyl}},
  \bibinfo{year}{1997}, \bibinfo{journal}{Z. Phys. B}
  \textbf{\bibinfo{volume}{103}}, \bibinfo{pages}{335}.

\bibitem[{\citenamefont{Martin} \emph{et~al.}(2008)\citenamefont{Martin,
  Akerman, Ulbricht, Lohmann, Smet, {von Klitzing}, and Yacoby}}]{Mar08}
\bibinfo{author}{\bibnamefont{Martin}, \bibfnamefont{J.}},
  \bibinfo{author}{\bibfnamefont{N.}~\bibnamefont{Akerman}},
  \bibinfo{author}{\bibfnamefont{G.}~\bibnamefont{Ulbricht}},
  \bibinfo{author}{\bibfnamefont{T.}~\bibnamefont{Lohmann}},
  \bibinfo{author}{\bibfnamefont{J.~H.} \bibnamefont{Smet}},
  \bibinfo{author}{\bibfnamefont{K.}~\bibnamefont{{von Klitzing}}}, and
  \bibinfo{author}{\bibfnamefont{A.}~\bibnamefont{Yacoby}},
  \bibinfo{year}{2008}, \bibinfo{journal}{Nature Physics}
  \textbf{\bibinfo{volume}{4}}, \bibinfo{pages}{144}.

\bibitem[{\citenamefont{Martinazzo}
  \emph{et~al.}(2010)\citenamefont{Martinazzo, Casolo, and
  Tantardini}}]{Martinazzo:2009}
\bibinfo{author}{\bibnamefont{Martinazzo}, \bibfnamefont{R.}},
  \bibinfo{author}{\bibfnamefont{S.}~\bibnamefont{Casolo}}, and
  \bibinfo{author}{\bibfnamefont{G.~F.} \bibnamefont{Tantardini}},
  \bibinfo{year}{2010}, \bibinfo{journal}{Phys. Rev. B}
  \textbf{\bibinfo{volume}{81}}, \bibinfo{pages}{245420}.

\bibitem[{\citenamefont{McCann}(2006)}]{McC06b}
\bibinfo{author}{\bibnamefont{McCann}, \bibfnamefont{E.}},
  \bibinfo{year}{2006}, \bibinfo{journal}{Phys. Rev. B}
  \textbf{\bibinfo{volume}{74}}, \bibinfo{pages}{161403(R)}.

\bibitem[{\citenamefont{McCann} \emph{et~al.}(2007)\citenamefont{McCann,
  Abergel, and Falko}}]{McCa07}
\bibinfo{author}{\bibnamefont{McCann}, \bibfnamefont{E.}},
  \bibinfo{author}{\bibfnamefont{D.~S.~L.} \bibnamefont{Abergel}}, and
  \bibinfo{author}{\bibfnamefont{V.~I.} \bibnamefont{Falko}},
  \bibinfo{year}{2007}, \bibinfo{journal}{Solid State Commun.}
  \textbf{\bibinfo{volume}{143}}, \bibinfo{pages}{110}.

\bibitem[{\citenamefont{McCann and Fal'ko}(2006)}]{McC06a}
\bibinfo{author}{\bibnamefont{McCann}, \bibfnamefont{E.}}, and
  \bibinfo{author}{\bibfnamefont{V.~I.} \bibnamefont{Fal'ko}},
  \bibinfo{year}{2006}, \bibinfo{journal}{Phys. Rev. Lett.}
  \textbf{\bibinfo{volume}{96}}, \bibinfo{pages}{086805}.

\bibitem[{\citenamefont{McChesney} \emph{et~al.}(2010)\citenamefont{McChesney,
  Bostwick, Ohta, Seyller, Horn, Gonzalez, and Rotenberg}}]{McC07}
\bibinfo{author}{\bibnamefont{McChesney}, \bibfnamefont{J.~L.}},
  \bibinfo{author}{\bibfnamefont{A.}~\bibnamefont{Bostwick}},
  \bibinfo{author}{\bibfnamefont{T.}~\bibnamefont{Ohta}},
  \bibinfo{author}{\bibfnamefont{T.}~\bibnamefont{Seyller}},
  \bibinfo{author}{\bibfnamefont{K.}~\bibnamefont{Horn}},
  \bibinfo{author}{\bibfnamefont{J.}~\bibnamefont{Gonzalez}}, and
  \bibinfo{author}{\bibfnamefont{E.}~\bibnamefont{Rotenberg}},
  \bibinfo{year}{2010}, \bibinfo{journal}{Phys. Rev. Let.}
  \textbf{\bibinfo{volume}{104}}, \bibinfo{pages}{136803}.

\bibitem[{\citenamefont{Mele}(2010)}]{Mele10}
\bibinfo{author}{\bibnamefont{Mele}, \bibfnamefont{E.~J.}},
  \bibinfo{year}{2010}, \bibinfo{journal}{Phys. Rev. B}
  \textbf{\bibinfo{volume}{81}}, \bibinfo{pages}{161405(R)}.

\bibitem[{\citenamefont{Meng} \emph{et~al.}(2010)\citenamefont{Meng, Lang,
  Wessel, Assaad, and Muramatsu}}]{Meng}
\bibinfo{author}{\bibnamefont{Meng}, \bibfnamefont{Z.~Y.}},
  \bibinfo{author}{\bibfnamefont{T.~C.} \bibnamefont{Lang}},
  \bibinfo{author}{\bibfnamefont{S.}~\bibnamefont{Wessel}},
  \bibinfo{author}{\bibfnamefont{F.~F.} \bibnamefont{Assaad}}, and
  \bibinfo{author}{\bibfnamefont{A.}~\bibnamefont{Muramatsu}},
  \bibinfo{year}{2010}, \bibinfo{journal}{Nature}
  \textbf{\bibinfo{volume}{464}}, \bibinfo{pages}{847}.

\bibitem[{\citenamefont{Miao} \emph{et~al.}(2007)\citenamefont{Miao, Wijeratne,
  Zhang, Coskun, W.Bao, and Lau}}]{Mia07}
\bibinfo{author}{\bibnamefont{Miao}, \bibfnamefont{F.}},
  \bibinfo{author}{\bibfnamefont{S.}~\bibnamefont{Wijeratne}},
  \bibinfo{author}{\bibfnamefont{Y.}~\bibnamefont{Zhang}},
  \bibinfo{author}{\bibfnamefont{U.~C.} \bibnamefont{Coskun}},
  \bibinfo{author}{\bibnamefont{W.Bao}}, and
  \bibinfo{author}{\bibfnamefont{C.~N.} \bibnamefont{Lau}},
  \bibinfo{year}{2007}, \bibinfo{journal}{Science}
  \textbf{\bibinfo{volume}{317}}, \bibinfo{pages}{1530}.

\bibitem[{\citenamefont{Mil'shtein and Strakhovenko}(1982)}]{Milshtein:1982}
\bibinfo{author}{\bibnamefont{Mil'shtein}, \bibfnamefont{A.~I.}}, and
  \bibinfo{author}{\bibfnamefont{V.~M.} \bibnamefont{Strakhovenko}},
  \bibinfo{year}{1982}, \bibinfo{journal}{Phys. Lett. A}
  \textbf{\bibinfo{volume}{90}}, \bibinfo{pages}{447}.

\bibitem[{\citenamefont{Min} \emph{et~al.}(2008)\citenamefont{Min, Bistritzer,
  Su, and MacDonald}}]{Min08}
\bibinfo{author}{\bibnamefont{Min}, \bibfnamefont{H.}},
  \bibinfo{author}{\bibfnamefont{R.}~\bibnamefont{Bistritzer}},
  \bibinfo{author}{\bibfnamefont{J.-J.} \bibnamefont{Su}}, and
  \bibinfo{author}{\bibfnamefont{A.~H.} \bibnamefont{MacDonald}},
  \bibinfo{year}{2008}, \bibinfo{journal}{Phys. Rev. B}
  \textbf{\bibinfo{volume}{78}}, \bibinfo{pages}{121401}.

\bibitem[{\citenamefont{Min} \emph{et~al.}(2006)\citenamefont{Min, Hill,
  Sinitsyn, Sahu, Kleinman, and MacDonald}}]{Min06}
\bibinfo{author}{\bibnamefont{Min}, \bibfnamefont{H.}},
  \bibinfo{author}{\bibfnamefont{J.~E.} \bibnamefont{Hill}},
  \bibinfo{author}{\bibfnamefont{N.~A.} \bibnamefont{Sinitsyn}},
  \bibinfo{author}{\bibfnamefont{B.~R.} \bibnamefont{Sahu}},
  \bibinfo{author}{\bibfnamefont{L.}~\bibnamefont{Kleinman}}, and
  \bibinfo{author}{\bibfnamefont{A.~H.} \bibnamefont{MacDonald}},
  \bibinfo{year}{2006}, \bibinfo{journal}{Phys. Rev. B}
  \textbf{\bibinfo{volume}{74}}, \bibinfo{pages}{165310}.

\bibitem[{\citenamefont{Min} \emph{et~al.}(2007)\citenamefont{Min, Sahu,
  Banerjee, and MacDonald}}]{Min07}
\bibinfo{author}{\bibnamefont{Min}, \bibfnamefont{H.}},
  \bibinfo{author}{\bibfnamefont{B.}~\bibnamefont{Sahu}},
  \bibinfo{author}{\bibfnamefont{S.~K.} \bibnamefont{Banerjee}}, and
  \bibinfo{author}{\bibfnamefont{A.~H.} \bibnamefont{MacDonald}},
  \bibinfo{year}{2007}, \bibinfo{journal}{Phys. Rev. B}
  \textbf{\bibinfo{volume}{75}}, \bibinfo{pages}{155115}.

\bibitem[{\citenamefont{Mishchenko}(2007)}]{Mi07}
\bibinfo{author}{\bibnamefont{Mishchenko}, \bibfnamefont{E.~G.}},
  \bibinfo{year}{2007}, \bibinfo{journal}{\prl} \textbf{\bibinfo{volume}{98}},
  \bibinfo{pages}{216801}.

\bibitem[{\citenamefont{Mishchenko}(2008)}]{Mi08}
\bibinfo{author}{\bibnamefont{Mishchenko}, \bibfnamefont{E.~G.}},
  \bibinfo{year}{2008}, \bibinfo{journal}{Europhys. Lett.}
  \textbf{\bibinfo{volume}{83}}, \bibinfo{pages}{17005}.

\bibitem[{\citenamefont{Moghaddam and Zareyan}(2006)}]{Mog06}
\bibinfo{author}{\bibnamefont{Moghaddam}, \bibfnamefont{A.~G.}}, and
  \bibinfo{author}{\bibfnamefont{M.}~\bibnamefont{Zareyan}},
  \bibinfo{year}{2006}, \bibinfo{journal}{Phys. Rev. B}
  \textbf{\bibinfo{volume}{74}}, \bibinfo{pages}{241403(R)}.

\bibitem[{\citenamefont{Molitor}
  \emph{et~al.}(2009{\natexlab{a}})\citenamefont{Molitor, Droscher,
  {G{\"u}ttinger}, Jacobsen, Stampfer, Ihn, and Ensslin}}]{Metal09b}
\bibinfo{author}{\bibnamefont{Molitor}, \bibfnamefont{F.}},
  \bibinfo{author}{\bibfnamefont{S.}~\bibnamefont{Droscher}},
  \bibinfo{author}{\bibfnamefont{J.}~\bibnamefont{{G{\"u}ttinger}}},
  \bibinfo{author}{\bibfnamefont{A.}~\bibnamefont{Jacobsen}},
  \bibinfo{author}{\bibfnamefont{C.}~\bibnamefont{Stampfer}},
  \bibinfo{author}{\bibfnamefont{T.}~\bibnamefont{Ihn}}, and
  \bibinfo{author}{\bibfnamefont{K.}~\bibnamefont{Ensslin}},
  \bibinfo{year}{2009}{\natexlab{a}}, \bibinfo{journal}{Appl. Phys. Lett.}
  \textbf{\bibinfo{volume}{94}}, \bibinfo{pages}{222107}.

\bibitem[{\citenamefont{Molitor}
  \emph{et~al.}(2009{\natexlab{b}})\citenamefont{Molitor, Jacobsen, Stampfer,
  {G{\"u}ttinger}, Ihn, and Ensslin}}]{Metal09}
\bibinfo{author}{\bibnamefont{Molitor}, \bibfnamefont{F.}},
  \bibinfo{author}{\bibfnamefont{A.}~\bibnamefont{Jacobsen}},
  \bibinfo{author}{\bibfnamefont{C.}~\bibnamefont{Stampfer}},
  \bibinfo{author}{\bibfnamefont{J.}~\bibnamefont{{G{\"u}ttinger}}},
  \bibinfo{author}{\bibfnamefont{T.}~\bibnamefont{Ihn}}, and
  \bibinfo{author}{\bibfnamefont{K.}~\bibnamefont{Ensslin}},
  \bibinfo{year}{2009}{\natexlab{b}}, \bibinfo{journal}{Phys. Rev. B}
  \textbf{\bibinfo{volume}{79}}, \bibinfo{pages}{075426}.

\bibitem[{\citenamefont{Monteverde}
  \emph{et~al.}(2010)\citenamefont{Monteverde, Ojeda-Aristizabal, Weil,
  Bennaceur, Ferrier, Gu\'eron, Glattli, Bouchiat, Fuchs, and
  Maslov}}]{Monteverde}
\bibinfo{author}{\bibnamefont{Monteverde}, \bibfnamefont{M.}},
  \bibinfo{author}{\bibfnamefont{C.}~\bibnamefont{Ojeda-Aristizabal}},
  \bibinfo{author}{\bibfnamefont{R.}~\bibnamefont{Weil}},
  \bibinfo{author}{\bibfnamefont{K.}~\bibnamefont{Bennaceur}},
  \bibinfo{author}{\bibfnamefont{M.}~\bibnamefont{Ferrier}},
  \bibinfo{author}{\bibfnamefont{S.}~\bibnamefont{Gu\'eron}},
  \bibinfo{author}{\bibfnamefont{C.}~\bibnamefont{Glattli}},
  \bibinfo{author}{\bibfnamefont{H.}~\bibnamefont{Bouchiat}},
  \bibinfo{author}{\bibfnamefont{J.~N.} \bibnamefont{Fuchs}}, and
  \bibinfo{author}{\bibfnamefont{D.~L.} \bibnamefont{Maslov}},
  \bibinfo{year}{2010}, \bibinfo{journal}{Phys. Rev. Lett.}
  \textbf{\bibinfo{volume}{104}}, \bibinfo{pages}{126801}.

\bibitem[{\citenamefont{Moriyama} \emph{et~al.}(2009)\citenamefont{Moriyama,
  Tsuya, Watanabe, Uji, Shimizu, Mori, Yamaguchi, and Ishibashi}}]{Metal09c}
\bibinfo{author}{\bibnamefont{Moriyama}, \bibfnamefont{S.}},
  \bibinfo{author}{\bibfnamefont{D.}~\bibnamefont{Tsuya}},
  \bibinfo{author}{\bibfnamefont{E.}~\bibnamefont{Watanabe}},
  \bibinfo{author}{\bibfnamefont{S.}~\bibnamefont{Uji}},
  \bibinfo{author}{\bibfnamefont{M.}~\bibnamefont{Shimizu}},
  \bibinfo{author}{\bibfnamefont{T.}~\bibnamefont{Mori}},
  \bibinfo{author}{\bibfnamefont{T.}~\bibnamefont{Yamaguchi}}, and
  \bibinfo{author}{\bibfnamefont{K.}~\bibnamefont{Ishibashi}},
  \bibinfo{year}{2009}, \bibinfo{journal}{Nano Lett.}
  \textbf{\bibinfo{volume}{9}}, \bibinfo{pages}{2891}.

\bibitem[{\citenamefont{Moser and Bachtold}(2009)}]{MB09}
\bibinfo{author}{\bibnamefont{Moser}, \bibfnamefont{J.}}, and
  \bibinfo{author}{\bibfnamefont{A.}~\bibnamefont{Bachtold}},
  \bibinfo{year}{2009}, \bibinfo{journal}{Appl. Phys. Lett.}
  \textbf{\bibinfo{volume}{95}}, \bibinfo{pages}{173506}.

\bibitem[{\citenamefont{Moser} \emph{et~al.}(2010)\citenamefont{Moser, Tao,
  Roche, Alzina, {Sotomayor Torres}, and Bachtold}}]{Metal10}
\bibinfo{author}{\bibnamefont{Moser}, \bibfnamefont{J.}},
  \bibinfo{author}{\bibfnamefont{H.}~\bibnamefont{Tao}},
  \bibinfo{author}{\bibfnamefont{S.}~\bibnamefont{Roche}},
  \bibinfo{author}{\bibfnamefont{F.}~\bibnamefont{Alzina}},
  \bibinfo{author}{\bibfnamefont{C.~M.} \bibnamefont{{Sotomayor Torres}}}, and
  \bibinfo{author}{\bibfnamefont{A.}~\bibnamefont{Bachtold}},
  \bibinfo{year}{2010}, \bibinfo{journal}{Phys. Rev. B}
  \textbf{\bibinfo{volume}{81}}, \bibinfo{pages}{205445}.

\bibitem[{\citenamefont{Mott}(1949)}]{Mott:49}
\bibinfo{author}{\bibnamefont{Mott}, \bibfnamefont{N.~F.}},
  \bibinfo{year}{1949}, \bibinfo{journal}{Proc. Phys. Soc. London, Ser. A}
  \textbf{\bibinfo{volume}{62}}, \bibinfo{pages}{416}.

\bibitem[{\citenamefont{M{\"u}ller and Rafelski}(1975)}]{Muller:1975}
\bibinfo{author}{\bibnamefont{M{\"u}ller}, \bibfnamefont{B.}}, and
  \bibinfo{author}{\bibfnamefont{J.}~\bibnamefont{Rafelski}},
  \bibinfo{year}{1975}, \bibinfo{journal}{Phys. Rev. Lett.}
  \textbf{\bibinfo{volume}{34}}, \bibinfo{pages}{349}.

\bibitem[{\citenamefont{M{\"u}ller}
  \emph{et~al.}(2008)\citenamefont{M{\"u}ller, Fritz, and Sachdev}}]{Mu08}
\bibinfo{author}{\bibnamefont{M{\"u}ller}, \bibfnamefont{M.}},
  \bibinfo{author}{\bibfnamefont{L.}~\bibnamefont{Fritz}}, and
  \bibinfo{author}{\bibfnamefont{S.}~\bibnamefont{Sachdev}},
  \bibinfo{year}{2008}, \bibinfo{journal}{\prb} \textbf{\bibinfo{volume}{78}},
  \bibinfo{pages}{115406}.

\bibitem[{\citenamefont{M\"uller} \emph{et~al.}(2009)\citenamefont{M\"uller,
  Schmalian, and Fritz}}]{Viscosity09}
\bibinfo{author}{\bibnamefont{M\"uller}, \bibfnamefont{M.}},
  \bibinfo{author}{\bibfnamefont{J.}~\bibnamefont{Schmalian}}, and
  \bibinfo{author}{\bibfnamefont{L.}~\bibnamefont{Fritz}},
  \bibinfo{year}{2009}, \bibinfo{journal}{Phys. Rev. Lett.}
  \textbf{\bibinfo{volume}{103}}, \bibinfo{pages}{025301}.

\bibitem[{\citenamefont{Nair} \emph{et~al.}(2008)\citenamefont{Nair, Blake,
  Grigorenko, Novoselov, Booth, Stauber, Peres, and Geim}}]{Nair08}
\bibinfo{author}{\bibnamefont{Nair}, \bibfnamefont{R.~R.}},
  \bibinfo{author}{\bibfnamefont{P.}~\bibnamefont{Blake}},
  \bibinfo{author}{\bibfnamefont{A.~N.} \bibnamefont{Grigorenko}},
  \bibinfo{author}{\bibfnamefont{K.~S.} \bibnamefont{Novoselov}},
  \bibinfo{author}{\bibfnamefont{T.~J.} \bibnamefont{Booth}},
  \bibinfo{author}{\bibfnamefont{T.}~\bibnamefont{Stauber}},
  \bibinfo{author}{\bibfnamefont{N.~M.~R.} \bibnamefont{Peres}}, and
  \bibinfo{author}{\bibfnamefont{A.~K.} \bibnamefont{Geim}},
  \bibinfo{year}{2008}, \bibinfo{journal}{Science}
  \textbf{\bibinfo{volume}{320}}, \bibinfo{pages}{1308}.

\bibitem[{\citenamefont{Nakada} \emph{et~al.}(1996)\citenamefont{Nakada,
  Fujita, Dresselhaus, and Dresselhaus}}]{Netal96}
\bibinfo{author}{\bibnamefont{Nakada}, \bibfnamefont{K.}},
  \bibinfo{author}{\bibfnamefont{M.}~\bibnamefont{Fujita}},
  \bibinfo{author}{\bibfnamefont{G.}~\bibnamefont{Dresselhaus}}, and
  \bibinfo{author}{\bibfnamefont{M.~S.} \bibnamefont{Dresselhaus}},
  \bibinfo{year}{1996}, \bibinfo{journal}{Phys. Rev. B}
  \textbf{\bibinfo{volume}{54}}, \bibinfo{pages}{17954}.

\bibitem[{\citenamefont{Nakamura} \emph{et~al.}(2009)\citenamefont{Nakamura,
  Castro, and Dora}}]{Nak09}
\bibinfo{author}{\bibnamefont{Nakamura}, \bibfnamefont{M.}},
  \bibinfo{author}{\bibfnamefont{E.~V.} \bibnamefont{Castro}}, and
  \bibinfo{author}{\bibfnamefont{B.}~\bibnamefont{Dora}}, \bibinfo{year}{2009},
  \bibinfo{journal}{Phys. Rev. Lett.} \textbf{\bibinfo{volume}{103}},
  \bibinfo{pages}{266804}.

\bibitem[{\citenamefont{Nandkishore and Levitov}(2010a)}]{Nan10a}
\bibinfo{author}{\bibnamefont{Nandkishore}, \bibfnamefont{R.}}, and
  \bibinfo{author}{\bibfnamefont{L.~S.} \bibnamefont{Levitov}},
  \bibinfo{year}{2010a}, \bibinfo{journal}{Phys. Rev. Lett.}
  \textbf{\bibinfo{volume}{104}}, \bibinfo{pages}{156803}.

\bibitem[{\citenamefont{Nandkishore and Levitov}(2010b)}]{Nan10b}
\bibinfo{author}{\bibnamefont{Nandkishore}, \bibfnamefont{R.}}, and
  \bibinfo{author}{\bibfnamefont{L.~S.} \bibnamefont{Levitov}},
  \bibinfo{year}{2010b}, \eprint{arXiv:1002.1966}.

\bibitem[{\citenamefont{Nandkishore and Levitov}(2010c)}]{Nan10c}
\bibinfo{author}{\bibnamefont{Nandkishore}, \bibfnamefont{R.}}, and
  \bibinfo{author}{\bibfnamefont{L.~S.} \bibnamefont{Levitov}},
  \bibinfo{year}{2010c}, \bibinfo{journal}{Phys. Rev. B}
  \textbf{\bibinfo{volume}{82}}, \bibinfo{pages}{115431}.

\bibitem[{\citenamefont{Newns and Read}(1987)}]{New87}
\bibinfo{author}{\bibnamefont{Newns}, \bibfnamefont{D.~M.}}, and
  \bibinfo{author}{\bibfnamefont{N.}~\bibnamefont{Read}}, \bibinfo{year}{1987},
  \bibinfo{journal}{Adv. Phys.} \textbf{\bibinfo{volume}{36}},
  \bibinfo{pages}{799}.

\bibitem[{\citenamefont{Ni} \emph{et~al.}(2010)\citenamefont{Ni, Ponomarenko,
  Nair, Yang, Anissimova, Grigorieva, Schedin, Blake, Shen, Hill, Novoselov,
  and Geim}}]{Ni:2010}
\bibinfo{author}{\bibnamefont{Ni}, \bibfnamefont{Z.~H.}},
  \bibinfo{author}{\bibfnamefont{L.~A.} \bibnamefont{Ponomarenko}},
  \bibinfo{author}{\bibfnamefont{R.~R.} \bibnamefont{Nair}},
  \bibinfo{author}{\bibfnamefont{R.}~\bibnamefont{Yang}},
  \bibinfo{author}{\bibfnamefont{S.}~\bibnamefont{Anissimova}},
  \bibinfo{author}{\bibfnamefont{I.~V.} \bibnamefont{Grigorieva}},
  \bibinfo{author}{\bibfnamefont{F.}~\bibnamefont{Schedin}},
  \bibinfo{author}{\bibfnamefont{P.}~\bibnamefont{Blake}},
  \bibinfo{author}{\bibfnamefont{Z.~X.} \bibnamefont{Shen}},
  \bibinfo{author}{\bibfnamefont{E.~H.} \bibnamefont{Hill}},
  \bibinfo{author}{\bibfnamefont{K.~S.} \bibnamefont{Novoselov}}, and
  \bibinfo{author}{\bibfnamefont{A.~K.} \bibnamefont{Geim}},
  \bibinfo{year}{2010}, \bibinfo{journal}{Nano Letters}
  \textbf{\bibinfo{volume}{10}}, \bibinfo{pages}{3868}.

\bibitem[{\citenamefont{Niimi} \emph{et~al.}(2005)\citenamefont{Niimi, Matsui,
  Kambara, Tagami, Tsukada, and Fukuyama}}]{Netal05b}
\bibinfo{author}{\bibnamefont{Niimi}, \bibfnamefont{Y.}},
  \bibinfo{author}{\bibfnamefont{T.}~\bibnamefont{Matsui}},
  \bibinfo{author}{\bibfnamefont{H.}~\bibnamefont{Kambara}},
  \bibinfo{author}{\bibfnamefont{K.}~\bibnamefont{Tagami}},
  \bibinfo{author}{\bibfnamefont{M.}~\bibnamefont{Tsukada}}, and
  \bibinfo{author}{\bibfnamefont{H.}~\bibnamefont{Fukuyama}},
  \bibinfo{year}{2005}, \bibinfo{journal}{Appl. Phys. Lett.}
  \textbf{\bibinfo{volume}{241}}, \bibinfo{pages}{43}.

\bibitem[{\citenamefont{Nilsson} \emph{et~al.}(2008)\citenamefont{Nilsson,
  {Castro Neto}, Guinea, and Peres}}]{Nil08}
\bibinfo{author}{\bibnamefont{Nilsson}, \bibfnamefont{J.}},
  \bibinfo{author}{\bibfnamefont{A.~H.} \bibnamefont{{Castro Neto}}},
  \bibinfo{author}{\bibfnamefont{F.}~\bibnamefont{Guinea}}, and
  \bibinfo{author}{\bibfnamefont{N.~M.~R.} \bibnamefont{Peres}},
  \bibinfo{year}{2008}, \bibinfo{journal}{Phys. Rev. B}
  \textbf{\bibinfo{volume}{78}}, \bibinfo{pages}{045405}.

\bibitem[{\citenamefont{Nilsson} \emph{et~al.}(2006)\citenamefont{Nilsson,
  {Castro Neto}, Peres, and Guinea}}]{Nil06}
\bibinfo{author}{\bibnamefont{Nilsson}, \bibfnamefont{J.}},
  \bibinfo{author}{\bibfnamefont{A.~H.} \bibnamefont{{Castro Neto}}},
  \bibinfo{author}{\bibfnamefont{N.~M.~R.} \bibnamefont{Peres}}, and
  \bibinfo{author}{\bibfnamefont{F.}~\bibnamefont{Guinea}},
  \bibinfo{year}{2006}, \bibinfo{journal}{Phys. Rev. B}
  \textbf{\bibinfo{volume}{73}}, \bibinfo{pages}{214418}.

\bibitem[{\citenamefont{Nomura and MacDonald}(2007)}]{Nomura:2007}
\bibinfo{author}{\bibnamefont{Nomura}, \bibfnamefont{K.}}, and
  \bibinfo{author}{\bibfnamefont{A.}~\bibnamefont{MacDonald}},
  \bibinfo{year}{2007}, \bibinfo{journal}{Phys. Rev. Lett.}
  \textbf{\bibinfo{volume}{98}}, \bibinfo{pages}{076602}.

\bibitem[{\citenamefont{Nomura and MacDonald}(2006)}]{NM06}
\bibinfo{author}{\bibnamefont{Nomura}, \bibfnamefont{K.}}, and
  \bibinfo{author}{\bibfnamefont{A.~H.} \bibnamefont{MacDonald}},
  \bibinfo{year}{2006}, \bibinfo{journal}{Phys. Rev. Lett.}
  \textbf{\bibinfo{volume}{96}}, \bibinfo{pages}{256602}.

\bibitem[{\citenamefont{Nouchi and Tanigaki}(2010)}]{Nouchi:2010}
\bibinfo{author}{\bibnamefont{Nouchi}, \bibfnamefont{R.}}, and
  \bibinfo{author}{\bibfnamefont{K.}~\bibnamefont{Tanigaki}},
  \bibinfo{year}{2010}, \bibinfo{journal}{Appl. Phys. Lett.}
  \textbf{\bibinfo{volume}{96}}, \bibinfo{pages}{253503}.

\bibitem[{\citenamefont{Novikov}(2007{\natexlab{a}})}]{Novikov:2007}
\bibinfo{author}{\bibnamefont{Novikov}, \bibfnamefont{D.}},
  \bibinfo{year}{2007}{\natexlab{a}}, \bibinfo{journal}{Phys. Rev. B}
  \textbf{\bibinfo{volume}{76}}, \bibinfo{pages}{245435}.

\bibitem[{\citenamefont{Novikov}(2007{\natexlab{b}})}]{Novikov:2007-1}
\bibinfo{author}{\bibnamefont{Novikov}, \bibfnamefont{D.~S.}},
  \bibinfo{year}{2007}{\natexlab{b}}, \bibinfo{journal}{Appl. Phys. Lett.}
  \textbf{\bibinfo{volume}{91}}, \bibinfo{pages}{102102}.

\bibitem[{\citenamefont{Novoselov} \emph{et~al.}(2005)\citenamefont{Novoselov,
  Geim, Morozov, Jiang, Katsnelson, Grigorieva, Dubonos, and
  Firsov}}]{Novoselov:2005}
\bibinfo{author}{\bibnamefont{Novoselov}, \bibfnamefont{K.~S.}},
  \bibinfo{author}{\bibfnamefont{A.~K.} \bibnamefont{Geim}},
  \bibinfo{author}{\bibfnamefont{S.~V.} \bibnamefont{Morozov}},
  \bibinfo{author}{\bibfnamefont{D.}~\bibnamefont{Jiang}},
  \bibinfo{author}{\bibfnamefont{M.~I.} \bibnamefont{Katsnelson}},
  \bibinfo{author}{\bibfnamefont{I.~V.} \bibnamefont{Grigorieva}},
  \bibinfo{author}{\bibfnamefont{S.~V.} \bibnamefont{Dubonos}}, and
  \bibinfo{author}{\bibfnamefont{A.~A.} \bibnamefont{Firsov}},
  \bibinfo{year}{2005}, \bibinfo{journal}{Nature}
  \textbf{\bibinfo{volume}{438}}, \bibinfo{pages}{197}.

\bibitem[{\citenamefont{Novoselov}
  \emph{et~al.}(2004{\natexlab{a}})\citenamefont{Novoselov, Geim, Morozov,
  Jiang, Zhang, Dubonos, Grigorieva, and Firsov}}]{Novoselov:2004}
\bibinfo{author}{\bibnamefont{Novoselov}, \bibfnamefont{K.~S.}},
  \bibinfo{author}{\bibfnamefont{A.~K.} \bibnamefont{Geim}},
  \bibinfo{author}{\bibfnamefont{S.~V.} \bibnamefont{Morozov}},
  \bibinfo{author}{\bibfnamefont{D.}~\bibnamefont{Jiang}},
  \bibinfo{author}{\bibfnamefont{Y.}~\bibnamefont{Zhang}},
  \bibinfo{author}{\bibfnamefont{S.~V.} \bibnamefont{Dubonos}},
  \bibinfo{author}{\bibfnamefont{I.~V.} \bibnamefont{Grigorieva}}, and
  \bibinfo{author}{\bibfnamefont{A.~A.} \bibnamefont{Firsov}},
  \bibinfo{year}{2004}{\natexlab{a}}, \bibinfo{journal}{Science}
  \textbf{\bibinfo{volume}{306}}, \bibinfo{pages}{666}.

\bibitem[{\citenamefont{Novoselov}
  \emph{et~al.}(2004{\natexlab{b}})\citenamefont{Novoselov, Jiang, Schedin,
  Booth, Khotkevich, Morozov, and Geim}}]{Nov04b}
\bibinfo{author}{\bibnamefont{Novoselov}, \bibfnamefont{K.~S.}},
  \bibinfo{author}{\bibfnamefont{D.}~\bibnamefont{Jiang}},
  \bibinfo{author}{\bibfnamefont{F.}~\bibnamefont{Schedin}},
  \bibinfo{author}{\bibfnamefont{T.~J.} \bibnamefont{Booth}},
  \bibinfo{author}{\bibfnamefont{V.~V.} \bibnamefont{Khotkevich}},
  \bibinfo{author}{\bibfnamefont{S.~V.} \bibnamefont{Morozov}}, and
  \bibinfo{author}{\bibfnamefont{A.~K.} \bibnamefont{Geim}},
  \bibinfo{year}{2004}{\natexlab{b}}, \bibinfo{journal}{Proc. Nat. Acad. Sci.}
  \textbf{\bibinfo{volume}{102}}, \bibinfo{pages}{10451}.

\bibitem[{\citenamefont{Novoselov} \emph{et~al.}(2006)\citenamefont{Novoselov,
  McCann, Morozov, Fal'ko, Katsnelson, Zeitler, Jiang, Schedin, and
  Geim}}]{Nov06}
\bibinfo{author}{\bibnamefont{Novoselov}, \bibfnamefont{K.~S.}},
  \bibinfo{author}{\bibfnamefont{E.}~\bibnamefont{McCann}},
  \bibinfo{author}{\bibfnamefont{S.~V.} \bibnamefont{Morozov}},
  \bibinfo{author}{\bibfnamefont{V.~I.} \bibnamefont{Fal'ko}},
  \bibinfo{author}{\bibfnamefont{M.~I.} \bibnamefont{Katsnelson}},
  \bibinfo{author}{\bibfnamefont{U.}~\bibnamefont{Zeitler}},
  \bibinfo{author}{\bibfnamefont{D.}~\bibnamefont{Jiang}},
  \bibinfo{author}{\bibfnamefont{F.}~\bibnamefont{Schedin}}, and
  \bibinfo{author}{\bibfnamefont{A.~K.} \bibnamefont{Geim}},
  \bibinfo{year}{2006}, \bibinfo{journal}{Nature Physics}
  \textbf{\bibinfo{volume}{2}}, \bibinfo{pages}{177}.

\bibitem[{\citenamefont{Nozi{\`e}res}(1964)}]{Noz64}
\bibinfo{author}{\bibnamefont{Nozi{\`e}res}, \bibfnamefont{P.}},
  \bibinfo{year}{1964}, \emph{\bibinfo{title}{The theory of interacting Fermi
  systems}} (\bibinfo{publisher}{Benjamin}, \bibinfo{address}{NewYork}).

\bibitem[{\citenamefont{Ohldag} \emph{et~al.}(2007)\citenamefont{Ohldag,
  Tyliszczak, H{\"o}hne, Spemann, Esquinazi, Ungureanu, and Butz}}]{Oetal07b}
\bibinfo{author}{\bibnamefont{Ohldag}, \bibfnamefont{H.}},
  \bibinfo{author}{\bibfnamefont{T.}~\bibnamefont{Tyliszczak}},
  \bibinfo{author}{\bibfnamefont{R.}~\bibnamefont{H{\"o}hne}},
  \bibinfo{author}{\bibfnamefont{D.}~\bibnamefont{Spemann}},
  \bibinfo{author}{\bibfnamefont{P.}~\bibnamefont{Esquinazi}},
  \bibinfo{author}{\bibfnamefont{M.}~\bibnamefont{Ungureanu}}, and
  \bibinfo{author}{\bibfnamefont{T.}~\bibnamefont{Butz}}, \bibinfo{year}{2007},
  \bibinfo{journal}{Phys. Rev. Lett.} \textbf{\bibinfo{volume}{98}},
  \bibinfo{pages}{187204}.

\bibitem[{\citenamefont{Ohta} \emph{et~al.}(2006)\citenamefont{Ohta, Bostwick,
  Seyller, Horn, and Rotenberg}}]{Oht06}
\bibinfo{author}{\bibnamefont{Ohta}, \bibfnamefont{T.}},
  \bibinfo{author}{\bibfnamefont{A.}~\bibnamefont{Bostwick}},
  \bibinfo{author}{\bibfnamefont{T.}~\bibnamefont{Seyller}},
  \bibinfo{author}{\bibfnamefont{K.}~\bibnamefont{Horn}}, and
  \bibinfo{author}{\bibfnamefont{E.}~\bibnamefont{Rotenberg}},
  \bibinfo{year}{2006}, \bibinfo{journal}{Science}
  \textbf{\bibinfo{volume}{313}}, \bibinfo{pages}{951}.

\bibitem[{\citenamefont{Ojeda-Aristizabal}
  \emph{et~al.}(2009)\citenamefont{Ojeda-Aristizabal, Ferrier, Gu{\'e}ron, and
  Bouchiat}}]{Oje09}
\bibinfo{author}{\bibnamefont{Ojeda-Aristizabal}, \bibfnamefont{C.}},
  \bibinfo{author}{\bibfnamefont{M.}~\bibnamefont{Ferrier}},
  \bibinfo{author}{\bibfnamefont{S.}~\bibnamefont{Gu{\'e}ron}}, and
  \bibinfo{author}{\bibfnamefont{H.}~\bibnamefont{Bouchiat}},
  \bibinfo{year}{2009}, \bibinfo{journal}{Phys. Rev. B}
  \textbf{\bibinfo{volume}{79}}, \bibinfo{pages}{165436}.

\bibitem[{\citenamefont{Ouyang} \emph{et~al.}(2002)\citenamefont{Ouyang, Huang,
  and Lieber}}]{Ouyang:2002}
\bibinfo{author}{\bibnamefont{Ouyang}, \bibfnamefont{M.}},
  \bibinfo{author}{\bibfnamefont{J.-L.} \bibnamefont{Huang}}, and
  \bibinfo{author}{\bibfnamefont{C.~M.} \bibnamefont{Lieber}},
  \bibinfo{year}{2002}, \bibinfo{journal}{Phys. Rev. Lett.}
  \textbf{\bibinfo{volume}{88}}, \bibinfo{pages}{066804}.

\bibitem[{\citenamefont{{\"O}zyilmaz}
  \emph{et~al.}(2007)\citenamefont{{\"O}zyilmaz, Jarillo-Herrero, Efetov,
  Abanin, Levitov, and Kim}}]{Oetal07}
\bibinfo{author}{\bibnamefont{{\"O}zyilmaz}, \bibfnamefont{B.}},
  \bibinfo{author}{\bibfnamefont{P.}~\bibnamefont{Jarillo-Herrero}},
  \bibinfo{author}{\bibfnamefont{D.}~\bibnamefont{Efetov}},
  \bibinfo{author}{\bibfnamefont{D.~A.} \bibnamefont{Abanin}},
  \bibinfo{author}{\bibfnamefont{L.~S.} \bibnamefont{Levitov}}, and
  \bibinfo{author}{\bibfnamefont{P.}~\bibnamefont{Kim}}, \bibinfo{year}{2007},
  \bibinfo{journal}{Phys. Rev. Lett.} \textbf{\bibinfo{volume}{99}},
  \bibinfo{pages}{166804}.

\bibitem[{\citenamefont{Paiva} \emph{et~al.}(2005)\citenamefont{Paiva,
  Scalettar, Zheng, Singh, and Oitmaa}}]{Pai05}
\bibinfo{author}{\bibnamefont{Paiva}, \bibfnamefont{M.}},
  \bibinfo{author}{\bibfnamefont{R.~T.} \bibnamefont{Scalettar}},
  \bibinfo{author}{\bibfnamefont{W.}~\bibnamefont{Zheng}},
  \bibinfo{author}{\bibfnamefont{R.~R.~P.} \bibnamefont{Singh}}, and
  \bibinfo{author}{\bibfnamefont{J.}~\bibnamefont{Oitmaa}},
  \bibinfo{year}{2005}, \bibinfo{journal}{Phys. Rev. B}
  \textbf{\bibinfo{volume}{72}}, \bibinfo{pages}{085123}.

\bibitem[{\citenamefont{Palacios} \emph{et~al.}(2008)\citenamefont{Palacios,
  {Fern{\'a}ndez-Rossier}, and Brey}}]{PFB08}
\bibinfo{author}{\bibnamefont{Palacios}, \bibfnamefont{J.~J.}},
  \bibinfo{author}{\bibfnamefont{J.}~\bibnamefont{{Fern{\'a}ndez-Rossier}}},
  and \bibinfo{author}{\bibfnamefont{L.}~\bibnamefont{Brey}},
  \bibinfo{year}{2008}, \bibinfo{journal}{Phys. Rev. B}
  \textbf{\bibinfo{volume}{77}}, \bibinfo{pages}{195428}.

\bibitem[{\citenamefont{Park} \emph{et~al.}(2009)\citenamefont{Park, Giustino,
  Spataru, Cohen, and Louie}}]{Park09}
\bibinfo{author}{\bibnamefont{Park}, \bibfnamefont{C.-H.}},
  \bibinfo{author}{\bibfnamefont{F.}~\bibnamefont{Giustino}},
  \bibinfo{author}{\bibfnamefont{C.~D.} \bibnamefont{Spataru}},
  \bibinfo{author}{\bibfnamefont{M.~L.} \bibnamefont{Cohen}}, and
  \bibinfo{author}{\bibfnamefont{S.~G.} \bibnamefont{Louie}},
  \bibinfo{year}{2009}, \bibinfo{journal}{Phys. Rev. Lett.}
  \textbf{\bibinfo{volume}{102}}, \bibinfo{pages}{076803}.

\bibitem[{\citenamefont{Pathak} \emph{et~al.}(2010)\citenamefont{Pathak,
  Shenoy, and Baskaran}}]{Pat10}
\bibinfo{author}{\bibnamefont{Pathak}, \bibfnamefont{S.}},
  \bibinfo{author}{\bibfnamefont{V.~B.} \bibnamefont{Shenoy}}, and
  \bibinfo{author}{\bibfnamefont{G.}~\bibnamefont{Baskaran}},
  \bibinfo{year}{2010}, \bibinfo{journal}{Phys. Rev. B}
  \textbf{\bibinfo{volume}{81}}, \bibinfo{pages}{085431}.

\bibitem[{\citenamefont{Pereira} \emph{et~al.}(2010)\citenamefont{Pereira,
  {Castro Neto}, Liang, and Mahadevan}}]{Pereira:2010}
\bibinfo{author}{\bibnamefont{Pereira}, \bibfnamefont{V.~M.}},
  \bibinfo{author}{\bibfnamefont{A.~H.} \bibnamefont{{Castro Neto}}},
  \bibinfo{author}{\bibfnamefont{H.~Y.} \bibnamefont{Liang}}, and
  \bibinfo{author}{\bibfnamefont{L.}~\bibnamefont{Mahadevan}},
  \bibinfo{year}{2010}, \bibinfo{journal}{Phys. Rev. Lett.}
  \textbf{\bibinfo{volume}{105}}, \bibinfo{pages}{156603}.

\bibitem[{\citenamefont{Pereira} \emph{et~al.}(2006)\citenamefont{Pereira,
  Guinea, {Lopes dos Santos}, Peres, and {Castro Neto}}}]{Per06}
\bibinfo{author}{\bibnamefont{Pereira}, \bibfnamefont{V.~M.}},
  \bibinfo{author}{\bibfnamefont{F.}~\bibnamefont{Guinea}},
  \bibinfo{author}{\bibfnamefont{J.~M.~B.} \bibnamefont{{Lopes dos Santos}}},
  \bibinfo{author}{\bibfnamefont{N.~M.~R.} \bibnamefont{Peres}}, and
  \bibinfo{author}{\bibfnamefont{A.~H.} \bibnamefont{{Castro Neto}}},
  \bibinfo{year}{2006}, \bibinfo{journal}{Phys. Rev. Lett.}
  \textbf{\bibinfo{volume}{96}}, \bibinfo{pages}{036801}.

\bibitem[{\citenamefont{Pereira}
  \emph{et~al.}(2008{\natexlab{a}})\citenamefont{Pereira, Kotov, and {Castro
  Neto}}}]{Pereira:2008}
\bibinfo{author}{\bibnamefont{Pereira}, \bibfnamefont{V.~M.}},
  \bibinfo{author}{\bibfnamefont{V.~N.} \bibnamefont{Kotov}}, and
  \bibinfo{author}{\bibfnamefont{A.~H.} \bibnamefont{{Castro Neto}}},
  \bibinfo{year}{2008}{\natexlab{a}}, \bibinfo{journal}{Phys. Rev. B}
  \textbf{\bibinfo{volume}{78}}, \bibinfo{pages}{085101}.

\bibitem[{\citenamefont{Pereira}
  \emph{et~al.}(2008{\natexlab{b}})\citenamefont{Pereira, {Lopes dos Santos},
  and {Castro Neto}}}]{Pereira:2008-1}
\bibinfo{author}{\bibnamefont{Pereira}, \bibfnamefont{V.~M.}},
  \bibinfo{author}{\bibfnamefont{J.~M.~B.} \bibnamefont{{Lopes dos Santos}}},
  and \bibinfo{author}{\bibfnamefont{A.~H.} \bibnamefont{{Castro Neto}}},
  \bibinfo{year}{2008}{\natexlab{b}}, \bibinfo{journal}{Phys. Rev. B}
  \textbf{\bibinfo{volume}{77}}, \bibinfo{pages}{115109}.

\bibitem[{\citenamefont{Pereira} \emph{et~al.}(2007)\citenamefont{Pereira,
  Nilsson, and {Castro Neto}}}]{Pereira:2007}
\bibinfo{author}{\bibnamefont{Pereira}, \bibfnamefont{V.~M.}},
  \bibinfo{author}{\bibfnamefont{J.}~\bibnamefont{Nilsson}}, and
  \bibinfo{author}{\bibfnamefont{A.~H.} \bibnamefont{{Castro Neto}}},
  \bibinfo{year}{2007}, \bibinfo{journal}{Phys. Rev. Lett.}
  \textbf{\bibinfo{volume}{99}}, \bibinfo{pages}{166802}.

\bibitem[{\citenamefont{Perelomov and Popov}(1970)}]{Perelomov:1970}
\bibinfo{author}{\bibnamefont{Perelomov}, \bibfnamefont{A.~M.}}, and
  \bibinfo{author}{\bibfnamefont{V.~S.} \bibnamefont{Popov}},
  \bibinfo{year}{1970}, \bibinfo{journal}{Theor. Mat. Phys.}
  \textbf{\bibinfo{volume}{4}}, \bibinfo{pages}{664}.

\bibitem[{\citenamefont{Peres}(2010)}]{NunoRMP}
\bibinfo{author}{\bibnamefont{Peres}, \bibfnamefont{N.~M.~R.}},
  \bibinfo{year}{2010}, \bibinfo{journal}{Rev. Mod. Phys.}
  \textbf{\bibinfo{volume}{82}}, \bibinfo{pages}{2673}.

\bibitem[{\citenamefont{Peres} \emph{et~al.}(2004)\citenamefont{Peres,
  Ara{\'u}jo, and Bozi}}]{Per04}
\bibinfo{author}{\bibnamefont{Peres}, \bibfnamefont{N.~M.~R.}},
  \bibinfo{author}{\bibfnamefont{M.~A.~N.} \bibnamefont{Ara{\'u}jo}}, and
  \bibinfo{author}{\bibfnamefont{D.}~\bibnamefont{Bozi}}, \bibinfo{year}{2004},
  \bibinfo{journal}{Phys. Rev. B} \textbf{\bibinfo{volume}{70}},
  \bibinfo{pages}{195122}.

\bibitem[{\citenamefont{Peres} \emph{et~al.}(2005)\citenamefont{Peres, Guinea,
  and {Castro Neto}}}]{Peres05}
\bibinfo{author}{\bibnamefont{Peres}, \bibfnamefont{N.~M.~R.}},
  \bibinfo{author}{\bibfnamefont{F.}~\bibnamefont{Guinea}}, and
  \bibinfo{author}{\bibfnamefont{A.~H.} \bibnamefont{{Castro Neto}}},
  \bibinfo{year}{2005}, \bibinfo{journal}{\prb} \textbf{\bibinfo{volume}{72}},
  \bibinfo{pages}{174406}.

\bibitem[{\citenamefont{Pisani} \emph{et~al.}(2007)\citenamefont{Pisani, Chan,
  Montanari, and Harrison}}]{PCMH07}
\bibinfo{author}{\bibnamefont{Pisani}, \bibfnamefont{L.}},
  \bibinfo{author}{\bibfnamefont{J.~A.} \bibnamefont{Chan}},
  \bibinfo{author}{\bibfnamefont{B.}~\bibnamefont{Montanari}}, and
  \bibinfo{author}{\bibfnamefont{N.~M.} \bibnamefont{Harrison}},
  \bibinfo{year}{2007}, \bibinfo{journal}{Phys. Rev. B}
  \textbf{\bibinfo{volume}{75}}, \bibinfo{pages}{064418}.

\bibitem[{\citenamefont{Pisarski}(1984)}]{Pisarski84}
\bibinfo{author}{\bibnamefont{Pisarski}, \bibfnamefont{R.~D.}},
  \bibinfo{year}{1984}, \bibinfo{journal}{Phys. Rev. D}
  \textbf{\bibinfo{volume}{29}}, \bibinfo{pages}{2423}.

\bibitem[{\citenamefont{Polini} \emph{et~al.}(2007)\citenamefont{Polini,
  Asgari, Barlas, Pereg-Barnea, and MacDonald}}]{Pol07}
\bibinfo{author}{\bibnamefont{Polini}, \bibfnamefont{M.}},
  \bibinfo{author}{\bibfnamefont{R.}~\bibnamefont{Asgari}},
  \bibinfo{author}{\bibfnamefont{Y.}~\bibnamefont{Barlas}},
  \bibinfo{author}{\bibfnamefont{T.}~\bibnamefont{Pereg-Barnea}}, and
  \bibinfo{author}{\bibfnamefont{A.~H.} \bibnamefont{MacDonald}},
  \bibinfo{year}{2007}, \bibinfo{journal}{Solid State Commun.}
  \textbf{\bibinfo{volume}{143}}, \bibinfo{pages}{58}.

\bibitem[{\citenamefont{Polini}
  \emph{et~al.}(2008{\natexlab{a}})\citenamefont{Polini, Asgari, Borghi,
  Barlas, Pereg-Barnea, and MacDonald}}]{Pol08a}
\bibinfo{author}{\bibnamefont{Polini}, \bibfnamefont{M.}},
  \bibinfo{author}{\bibfnamefont{R.}~\bibnamefont{Asgari}},
  \bibinfo{author}{\bibfnamefont{G.}~\bibnamefont{Borghi}},
  \bibinfo{author}{\bibfnamefont{Y.}~\bibnamefont{Barlas}},
  \bibinfo{author}{\bibfnamefont{T.}~\bibnamefont{Pereg-Barnea}}, and
  \bibinfo{author}{\bibfnamefont{A.~H.} \bibnamefont{MacDonald}},
  \bibinfo{year}{2008}{\natexlab{a}}, \bibinfo{journal}{\prb}
  \textbf{\bibinfo{volume}{77}}, \bibinfo{pages}{081411(R)}.

\bibitem[{\citenamefont{Polini}
  \emph{et~al.}(2008{\natexlab{b}})\citenamefont{Polini, Tomadin, Asgari, and
  MacDonald}}]{Pol08b}
\bibinfo{author}{\bibnamefont{Polini}, \bibfnamefont{M.}},
  \bibinfo{author}{\bibfnamefont{A.}~\bibnamefont{Tomadin}},
  \bibinfo{author}{\bibfnamefont{R.}~\bibnamefont{Asgari}}, and
  \bibinfo{author}{\bibfnamefont{A.~H.} \bibnamefont{MacDonald}},
  \bibinfo{year}{2008}{\natexlab{b}}, \bibinfo{journal}{Phys. Rev. B}
  \textbf{\bibinfo{volume}{78}}, \bibinfo{pages}{115426}.

\bibitem[{\citenamefont{Polkovnikov}(2002)}]{Pol02}
\bibinfo{author}{\bibnamefont{Polkovnikov}, \bibfnamefont{A.}},
  \bibinfo{year}{2002}, \bibinfo{journal}{Phys. Rev. B}
  \textbf{\bibinfo{volume}{65}}, \bibinfo{pages}{064503}.

\bibitem[{\citenamefont{Polkovnikov}
  \emph{et~al.}(2001)\citenamefont{Polkovnikov, Vojta, and Sachdev}}]{Pol01}
\bibinfo{author}{\bibnamefont{Polkovnikov}, \bibfnamefont{A.}},
  \bibinfo{author}{\bibfnamefont{M.}~\bibnamefont{Vojta}}, and
  \bibinfo{author}{\bibfnamefont{S.}~\bibnamefont{Sachdev}},
  \bibinfo{year}{2001}, \bibinfo{journal}{Phys. Rev. Lett.}
  \textbf{\bibinfo{volume}{86}}, \bibinfo{pages}{296}.

\bibitem[{\citenamefont{Ponomarenko}
  \emph{et~al.}(2008)\citenamefont{Ponomarenko, Schedin, Katsnelson, Yang,
  Hill, Novoselov, and Geim}}]{Petal08}
\bibinfo{author}{\bibnamefont{Ponomarenko}, \bibfnamefont{L.~A.}},
  \bibinfo{author}{\bibfnamefont{F.}~\bibnamefont{Schedin}},
  \bibinfo{author}{\bibfnamefont{M.~I.} \bibnamefont{Katsnelson}},
  \bibinfo{author}{\bibfnamefont{R.}~\bibnamefont{Yang}},
  \bibinfo{author}{\bibfnamefont{E.~W.} \bibnamefont{Hill}},
  \bibinfo{author}{\bibfnamefont{K.~S.} \bibnamefont{Novoselov}}, and
  \bibinfo{author}{\bibfnamefont{A.~K.} \bibnamefont{Geim}},
  \bibinfo{year}{2008}, \bibinfo{journal}{Science}
  \textbf{\bibinfo{volume}{320}}, \bibinfo{pages}{356}.

\bibitem[{\citenamefont{Ponomarenko}
  \emph{et~al.}(2009)\citenamefont{Ponomarenko, Yang, Mohiuddin, Morozov,
  Zhukov, Schedin, Hill, Novoselov, Katsnelson, and Geim}}]{Ponomarenko:2009}
\bibinfo{author}{\bibnamefont{Ponomarenko}, \bibfnamefont{L.~A.}},
  \bibinfo{author}{\bibfnamefont{R.}~\bibnamefont{Yang}},
  \bibinfo{author}{\bibfnamefont{T.~M.} \bibnamefont{Mohiuddin}},
  \bibinfo{author}{\bibfnamefont{S.~M.} \bibnamefont{Morozov}},
  \bibinfo{author}{\bibfnamefont{A.~A.} \bibnamefont{Zhukov}},
  \bibinfo{author}{\bibfnamefont{F.}~\bibnamefont{Schedin}},
  \bibinfo{author}{\bibfnamefont{E.~W.} \bibnamefont{Hill}},
  \bibinfo{author}{\bibfnamefont{K.~S.} \bibnamefont{Novoselov}},
  \bibinfo{author}{\bibfnamefont{M.~I.} \bibnamefont{Katsnelson}}, and
  \bibinfo{author}{\bibfnamefont{A.~K.} \bibnamefont{Geim}},
  \bibinfo{year}{2009}, \bibinfo{journal}{Phys. Rev. Lett.}
  \textbf{\bibinfo{volume}{102}}, \bibinfo{pages}{206603}.

\bibitem[{\citenamefont{Popov}(1971{\natexlab{a}})}]{Popov:1971}
\bibinfo{author}{\bibnamefont{Popov}, \bibfnamefont{V.~S.}},
  \bibinfo{year}{1971}{\natexlab{a}}, \bibinfo{journal}{Sov. J. Nucl. Phys.}
  \textbf{\bibinfo{volume}{12}}, \bibinfo{pages}{235}.

\bibitem[{\citenamefont{Popov}(1971{\natexlab{b}})}]{Popov:1971-1}
\bibinfo{author}{\bibnamefont{Popov}, \bibfnamefont{V.~S.}},
  \bibinfo{year}{1971}{\natexlab{b}}, \bibinfo{journal}{Sov. Phys. JETP}
  \textbf{\bibinfo{volume}{32}}, \bibinfo{pages}{526}.

\bibitem[{\citenamefont{Principi} \emph{et~al.}(2010)\citenamefont{Principi,
  Polini, Vignale, and Katsnelson}}]{Principi10}
\bibinfo{author}{\bibnamefont{Principi}, \bibfnamefont{A.}},
  \bibinfo{author}{\bibfnamefont{M.}~\bibnamefont{Polini}},
  \bibinfo{author}{\bibfnamefont{G.}~\bibnamefont{Vignale}}, and
  \bibinfo{author}{\bibfnamefont{M.~I.} \bibnamefont{Katsnelson}},
  \bibinfo{year}{2010}, \bibinfo{journal}{Phys. Rev. Lett.}
  \textbf{\bibinfo{volume}{104}}, \bibinfo{pages}{225503}.

\bibitem[{\citenamefont{Pustilnik and Glazman}(2001)}]{Pul01}
\bibinfo{author}{\bibnamefont{Pustilnik}, \bibfnamefont{M.}}, and
  \bibinfo{author}{\bibfnamefont{L.~I.} \bibnamefont{Glazman}},
  \bibinfo{year}{2001}, \bibinfo{journal}{Phys. Rev. Lett.}
  \textbf{\bibinfo{volume}{87}}, \bibinfo{pages}{216601}.

\bibitem[{\citenamefont{Qi and Zhang}(2011)}]{QZ08}
\bibinfo{author}{\bibnamefont{Qi}, \bibfnamefont{X.-L.}}, and
  \bibinfo{author}{\bibfnamefont{S.-C.} \bibnamefont{Zhang}},
  \bibinfo{year}{2011}, \bibinfo{journal}{Rev. Mod. Phys.}
  \textbf{\bibinfo{volume}{83}}, \bibinfo{pages}{1057}.

\bibitem[{\citenamefont{Raghu} \emph{et~al.}(2008)\citenamefont{Raghu, Qi,
  Honerkamp, and Zhang}}]{Rag08}
\bibinfo{author}{\bibnamefont{Raghu}, \bibfnamefont{S.}},
  \bibinfo{author}{\bibfnamefont{X.-L.} \bibnamefont{Qi}},
  \bibinfo{author}{\bibfnamefont{C.}~\bibnamefont{Honerkamp}}, and
  \bibinfo{author}{\bibfnamefont{S.-C.} \bibnamefont{Zhang}},
  \bibinfo{year}{2008}, \bibinfo{journal}{Phys. Rev. Lett.}
  \textbf{\bibinfo{volume}{100}}, \bibinfo{pages}{156401}.

\bibitem[{\citenamefont{Rainis} \emph{et~al.}(2009)\citenamefont{Rainis,
  Taddei, Dolcini, Polini, and Fazio}}]{Rai09}
\bibinfo{author}{\bibnamefont{Rainis}, \bibfnamefont{D.}},
  \bibinfo{author}{\bibfnamefont{F.}~\bibnamefont{Taddei}},
  \bibinfo{author}{\bibfnamefont{F.}~\bibnamefont{Dolcini}},
  \bibinfo{author}{\bibfnamefont{M.}~\bibnamefont{Polini}}, and
  \bibinfo{author}{\bibfnamefont{R.}~\bibnamefont{Fazio}},
  \bibinfo{year}{2009}, \bibinfo{journal}{Phys. Rev. B}
  \textbf{\bibinfo{volume}{79}}, \bibinfo{pages}{115131}.

\bibitem[{\citenamefont{Ramos} \emph{et~al.}(2010)\citenamefont{Ramos,
  Barzola-Quiquia, Esquinazi, Mu{\~n}oz-Martin, Climent-Font, and
  Garc{\'i}a-Hern{\'a}ndez}}]{Retal10}
\bibinfo{author}{\bibnamefont{Ramos}, \bibfnamefont{M.~A.}},
  \bibinfo{author}{\bibfnamefont{J.}~\bibnamefont{Barzola-Quiquia}},
  \bibinfo{author}{\bibfnamefont{P.}~\bibnamefont{Esquinazi}},
  \bibinfo{author}{\bibfnamefont{A.}~\bibnamefont{Mu{\~n}oz-Martin}},
  \bibinfo{author}{\bibfnamefont{A.}~\bibnamefont{Climent-Font}}, and
  \bibinfo{author}{\bibfnamefont{M.}~\bibnamefont{Garc{\'i}a-Hern{\'a}ndez}},
  \bibinfo{year}{2010}, \bibinfo{journal}{Phys. Rev. B}
  \textbf{\bibinfo{volume}{81}}, \bibinfo{pages}{214404}.

\bibitem[{\citenamefont{Rappoport} \emph{et~al.}(2009)\citenamefont{Rappoport,
  Uchoa, and {Castro Neto}}}]{Rap09}
\bibinfo{author}{\bibnamefont{Rappoport}, \bibfnamefont{T.~G.}},
  \bibinfo{author}{\bibfnamefont{B.}~\bibnamefont{Uchoa}}, and
  \bibinfo{author}{\bibfnamefont{A.~H.} \bibnamefont{{Castro Neto}}},
  \bibinfo{year}{2009}, \bibinfo{journal}{Phys. Rev. B}
  \textbf{\bibinfo{volume}{80}}, \bibinfo{pages}{245408}.

\bibitem[{\citenamefont{Read and Newns}(1983)}]{Rea83}
\bibinfo{author}{\bibnamefont{Read}, \bibfnamefont{N.}}, and
  \bibinfo{author}{\bibfnamefont{D.~M.} \bibnamefont{Newns}},
  \bibinfo{year}{1983}, \bibinfo{journal}{J. Phys. C}
  \textbf{\bibinfo{volume}{16}}, \bibinfo{pages}{3273}.

\bibitem[{\citenamefont{Reed} \emph{et~al.}(2010)\citenamefont{Reed, Uchoa,
  Joe, Gan, Casa, Fradkin, and Abbamonte}}]{Bruno}
\bibinfo{author}{\bibnamefont{Reed}, \bibfnamefont{J.~P.}},
  \bibinfo{author}{\bibfnamefont{B.}~\bibnamefont{Uchoa}},
  \bibinfo{author}{\bibfnamefont{Y.~I.} \bibnamefont{Joe}},
  \bibinfo{author}{\bibfnamefont{Y.}~\bibnamefont{Gan}},
  \bibinfo{author}{\bibfnamefont{D.}~\bibnamefont{Casa}},
  \bibinfo{author}{\bibfnamefont{E.}~\bibnamefont{Fradkin}}, and
  \bibinfo{author}{\bibfnamefont{P.}~\bibnamefont{Abbamonte}},
  \bibinfo{year}{2010}, \bibinfo{journal}{Science}
  \textbf{\bibinfo{volume}{330}}, \bibinfo{pages}{805}.

\bibitem[{\citenamefont{Ritter and Lyding}(2009)}]{RL09}
\bibinfo{author}{\bibnamefont{Ritter}, \bibfnamefont{K.~A.}}, and
  \bibinfo{author}{\bibfnamefont{J.~W.} \bibnamefont{Lyding}},
  \bibinfo{year}{2009}, \bibinfo{journal}{Nature Materials}
  \textbf{\bibinfo{volume}{8}}, \bibinfo{pages}{235}.

\bibitem[{\citenamefont{Riu and Hatsugai}(2001)}]{RH01}
\bibinfo{author}{\bibnamefont{Riu}, \bibfnamefont{S.}}, and
  \bibinfo{author}{\bibfnamefont{Y.}~\bibnamefont{Hatsugai}},
  \bibinfo{year}{2001}, \bibinfo{journal}{Phys. Rev. B}
  \textbf{\bibinfo{volume}{65}}, \bibinfo{pages}{033301}.

\bibitem[{\citenamefont{Rold{\'a}n}
  \emph{et~al.}(2008)\citenamefont{Rold{\'a}n, L{\'o}pez-Sancho, and
  Guinea}}]{RLG07}
\bibinfo{author}{\bibnamefont{Rold{\'a}n}, \bibfnamefont{R.}},
  \bibinfo{author}{\bibfnamefont{M.~P.} \bibnamefont{L{\'o}pez-Sancho}}, and
  \bibinfo{author}{\bibfnamefont{F.}~\bibnamefont{Guinea}},
  \bibinfo{year}{2008}, \bibinfo{journal}{Phys. Rev. B}
  \textbf{\bibinfo{volume}{77}}, \bibinfo{pages}{115410}.

\bibitem[{\citenamefont{Romanovsky}
  \emph{et~al.}(2009)\citenamefont{Romanovsky, Yannouleas, and
  Landman}}]{RYL09}
\bibinfo{author}{\bibnamefont{Romanovsky}, \bibfnamefont{I.}},
  \bibinfo{author}{\bibfnamefont{C.}~\bibnamefont{Yannouleas}}, and
  \bibinfo{author}{\bibfnamefont{U.}~\bibnamefont{Landman}},
  \bibinfo{year}{2009}, \bibinfo{journal}{Phys. Rev. B}
  \textbf{\bibinfo{volume}{79}}, \bibinfo{pages}{075311}.

\bibitem[{\citenamefont{Roy and Herbut}(2010)}]{Roy10}
\bibinfo{author}{\bibnamefont{Roy}, \bibfnamefont{B.}}, and
  \bibinfo{author}{\bibfnamefont{I.~F.} \bibnamefont{Herbut}},
  \bibinfo{year}{2010}, \bibinfo{journal}{Phys. Rev. B}
  \textbf{\bibinfo{volume}{82}}, \bibinfo{pages}{035429}.

\bibitem[{\citenamefont{Ryu} \emph{et~al.}(2009)\citenamefont{Ryu, Mudry, Hou,
  and Chamon}}]{Ryu09}
\bibinfo{author}{\bibnamefont{Ryu}, \bibfnamefont{S.}},
  \bibinfo{author}{\bibfnamefont{C.}~\bibnamefont{Mudry}},
  \bibinfo{author}{\bibfnamefont{C.-Y.} \bibnamefont{Hou}}, and
  \bibinfo{author}{\bibfnamefont{C.}~\bibnamefont{Chamon}},
  \bibinfo{year}{2009}, \bibinfo{journal}{Phys. Rev. B}
  \textbf{\bibinfo{volume}{80}}, \bibinfo{pages}{205319}.

\bibitem[{\citenamefont{Sabio} \emph{et~al.}(2008)\citenamefont{Sabio, Nilsson,
  and {Castro Neto}}}]{Sab08}
\bibinfo{author}{\bibnamefont{Sabio}, \bibfnamefont{J.}},
  \bibinfo{author}{\bibfnamefont{J.}~\bibnamefont{Nilsson}}, and
  \bibinfo{author}{\bibfnamefont{A.~H.} \bibnamefont{{Castro Neto}}},
  \bibinfo{year}{2008}, \bibinfo{journal}{Phys. Rev. B}
  \textbf{\bibinfo{volume}{78}}, \bibinfo{pages}{075410}.

\bibitem[{\citenamefont{Sabio}
  \emph{et~al.}(2010{\natexlab{a}})\citenamefont{Sabio, Sols, and
  Guinea}}]{Sabio10}
\bibinfo{author}{\bibnamefont{Sabio}, \bibfnamefont{J.}},
  \bibinfo{author}{\bibfnamefont{F.}~\bibnamefont{Sols}}, and
  \bibinfo{author}{\bibfnamefont{F.}~\bibnamefont{Guinea}},
  \bibinfo{year}{2010}{\natexlab{a}}, \bibinfo{journal}{Phys. Rev. B}
  \textbf{\bibinfo{volume}{82}}, \bibinfo{pages}{121413}.

\bibitem[{\citenamefont{Sabio}
  \emph{et~al.}(2010{\natexlab{b}})\citenamefont{Sabio, Sols, and
  Guinea}}]{Sabio:2009}
\bibinfo{author}{\bibnamefont{Sabio}, \bibfnamefont{J.}},
  \bibinfo{author}{\bibfnamefont{F.}~\bibnamefont{Sols}}, and
  \bibinfo{author}{\bibfnamefont{F.}~\bibnamefont{Guinea}},
  \bibinfo{year}{2010}{\natexlab{b}}, \bibinfo{journal}{Phys. Rev. B}
  \textbf{\bibinfo{volume}{81}}, \bibinfo{pages}{045428}.

\bibitem[{\citenamefont{Sachdev}(1999)}]{Sachdev:99}
\bibinfo{author}{\bibnamefont{Sachdev}, \bibfnamefont{S.}},
  \bibinfo{year}{1999}, \emph{\bibinfo{title}{Quantum Phase Transitions}}
  (\bibinfo{publisher}{Cambridge University Press},
  \bibinfo{address}{Cambridge, UK}).

\bibitem[{\citenamefont{Saha} \emph{et~al.}(2010)\citenamefont{Saha, Paul, and
  Sengupta}}]{Sah10}
\bibinfo{author}{\bibnamefont{Saha}, \bibfnamefont{K.}},
  \bibinfo{author}{\bibfnamefont{I.}~\bibnamefont{Paul}}, and
  \bibinfo{author}{\bibfnamefont{K.}~\bibnamefont{Sengupta}},
  \bibinfo{year}{2010}, \bibinfo{journal}{Phys. Rev. B}
  \textbf{\bibinfo{volume}{81}}, \bibinfo{pages}{165446}.

\bibitem[{\citenamefont{Sahebsara and S{\'e}n{\'e}chal}(2009)}]{Sah09}
\bibinfo{author}{\bibnamefont{Sahebsara}, \bibfnamefont{P.}}, and
  \bibinfo{author}{\bibfnamefont{D.}~\bibnamefont{S{\'e}n{\'e}chal}},
  \bibinfo{year}{2009}, \eprint{arXiv:0908.0474}.

\bibitem[{\citenamefont{Sahu} \emph{et~al.}(2008)\citenamefont{Sahu, Min,
  MacDonald, and Banerjee}}]{SMMB08}
\bibinfo{author}{\bibnamefont{Sahu}, \bibfnamefont{B.}},
  \bibinfo{author}{\bibfnamefont{H.}~\bibnamefont{Min}},
  \bibinfo{author}{\bibfnamefont{A.~H.} \bibnamefont{MacDonald}}, and
  \bibinfo{author}{\bibfnamefont{S.~K.} \bibnamefont{Banerjee}},
  \bibinfo{year}{2008}, \bibinfo{journal}{Phys. Rev. B}
  \textbf{\bibinfo{volume}{78}}, \bibinfo{pages}{075404}.

\bibitem[{\citenamefont{Saremi}(2007)}]{Sar07}
\bibinfo{author}{\bibnamefont{Saremi}, \bibfnamefont{S.}},
  \bibinfo{year}{2007}, \bibinfo{journal}{Phys. Rev. B}
  \textbf{\bibinfo{volume}{76}}, \bibinfo{pages}{184430}.

\bibitem[{\citenamefont{Sasakia} \emph{et~al.}(2007)\citenamefont{Sasakia,
  Jiang, Saito, Onari, and Tanaka}}]{Sas07}
\bibinfo{author}{\bibnamefont{Sasakia}, \bibfnamefont{K.}},
  \bibinfo{author}{\bibfnamefont{J.}~\bibnamefont{Jiang}},
  \bibinfo{author}{\bibfnamefont{R.}~\bibnamefont{Saito}},
  \bibinfo{author}{\bibfnamefont{S.}~\bibnamefont{Onari}}, and
  \bibinfo{author}{\bibfnamefont{Y.}~\bibnamefont{Tanaka}},
  \bibinfo{year}{2007}, \bibinfo{journal}{J. Phys. Soc. Jpn.}
  \textbf{\bibinfo{volume}{76}}, \bibinfo{pages}{033702}.

\bibitem[{\citenamefont{Schnez} \emph{et~al.}(2009)\citenamefont{Schnez,
  Molitor, Stampfer, Guettinger, Shorubalko, Ihn, and Ensslin}}]{Setal09b}
\bibinfo{author}{\bibnamefont{Schnez}, \bibfnamefont{S.}},
  \bibinfo{author}{\bibfnamefont{F.}~\bibnamefont{Molitor}},
  \bibinfo{author}{\bibfnamefont{C.}~\bibnamefont{Stampfer}},
  \bibinfo{author}{\bibfnamefont{J.}~\bibnamefont{Guettinger}},
  \bibinfo{author}{\bibfnamefont{I.}~\bibnamefont{Shorubalko}},
  \bibinfo{author}{\bibfnamefont{T.}~\bibnamefont{Ihn}}, and
  \bibinfo{author}{\bibfnamefont{K.}~\bibnamefont{Ensslin}},
  \bibinfo{year}{2009}, \bibinfo{journal}{Appl. Phys. Lett.}
  \textbf{\bibinfo{volume}{94}}, \bibinfo{pages}{012107}.

\bibitem[{\citenamefont{Schrieffer and Wolff}(1966)}]{Sch66}
\bibinfo{author}{\bibnamefont{Schrieffer}, \bibfnamefont{J.~R.}}, and
  \bibinfo{author}{\bibfnamefont{P.~A.} \bibnamefont{Wolff}},
  \bibinfo{year}{1966}, \bibinfo{journal}{Phys. Rev.}
  \textbf{\bibinfo{volume}{149}}, \bibinfo{pages}{491}.

\bibitem[{\citenamefont{Schwinger}(1951)}]{Schwinger:1951}
\bibinfo{author}{\bibnamefont{Schwinger}, \bibfnamefont{J.}},
  \bibinfo{year}{1951}, \bibinfo{journal}{Phys. Rev.}
  \textbf{\bibinfo{volume}{82}}, \bibinfo{pages}{664}.

\bibitem[{\citenamefont{Semenoff}(1984)}]{Sem84}
\bibinfo{author}{\bibnamefont{Semenoff}, \bibfnamefont{G.~W.}},
  \bibinfo{year}{1984}, \bibinfo{journal}{Phys. Rev. Lett.}
  \textbf{\bibinfo{volume}{53}}, \bibinfo{pages}{2449}.

\bibitem[{\citenamefont{Sengupta and Baskaran}(2008)}]{Sen08}
\bibinfo{author}{\bibnamefont{Sengupta}, \bibfnamefont{K.}}, and
  \bibinfo{author}{\bibfnamefont{G.}~\bibnamefont{Baskaran}},
  \bibinfo{year}{2008}, \bibinfo{journal}{Phys. Rev. B}
  \textbf{\bibinfo{volume}{77}}, \bibinfo{pages}{045417}.

\bibitem[{\citenamefont{Sensarma} \emph{et~al.}(2010)\citenamefont{Sensarma,
  Hwang, and Sarma}}]{Sensarma10}
\bibinfo{author}{\bibnamefont{Sensarma}, \bibfnamefont{R.}},
  \bibinfo{author}{\bibfnamefont{E.~H.} \bibnamefont{Hwang}}, and
  \bibinfo{author}{\bibfnamefont{S.~D.} \bibnamefont{Sarma}},
  \bibinfo{year}{2010}, \bibinfo{journal}{Phys. Rev. B}
  \textbf{\bibinfo{volume}{82}}, \bibinfo{pages}{195428}.

\bibitem[{\citenamefont{Sepioni} \emph{et~al.}(2010)\citenamefont{Sepioni,
  Nair, Rablen, Narayanan, Tuna, Winpenny, Geim, and Grigorieva}}]{Setal10}
\bibinfo{author}{\bibnamefont{Sepioni}, \bibfnamefont{M.}},
  \bibinfo{author}{\bibfnamefont{R.~R.} \bibnamefont{Nair}},
  \bibinfo{author}{\bibfnamefont{S.}~\bibnamefont{Rablen}},
  \bibinfo{author}{\bibfnamefont{J.}~\bibnamefont{Narayanan}},
  \bibinfo{author}{\bibfnamefont{F.}~\bibnamefont{Tuna}},
  \bibinfo{author}{\bibfnamefont{R.}~\bibnamefont{Winpenny}},
  \bibinfo{author}{\bibfnamefont{A.~K.} \bibnamefont{Geim}}, and
  \bibinfo{author}{\bibfnamefont{I.~V.} \bibnamefont{Grigorieva}},
  \bibinfo{year}{2010}, \bibinfo{journal}{Phys. Rev. Lett.}
  \textbf{\bibinfo{volume}{105}}, \bibinfo{pages}{207205}.

\bibitem[{\citenamefont{Seradjeh}(2008)}]{Ser08b}
\bibinfo{author}{\bibnamefont{Seradjeh}, \bibfnamefont{B.}},
  \bibinfo{year}{2008}, \bibinfo{journal}{Nucl. Phys. B}
  \textbf{\bibinfo{volume}{805}}, \bibinfo{pages}{182}.

\bibitem[{\citenamefont{Seradjeh and Franz}(2008)}]{Ser08a}
\bibinfo{author}{\bibnamefont{Seradjeh}, \bibfnamefont{B.}}, and
  \bibinfo{author}{\bibfnamefont{M.}~\bibnamefont{Franz}},
  \bibinfo{year}{2008}, \bibinfo{journal}{Phys. Rev. Lett.}
  \textbf{\bibinfo{volume}{101}}, \bibinfo{pages}{146401}.

\bibitem[{\citenamefont{Seradjeh} \emph{et~al.}(2008)\citenamefont{Seradjeh,
  Weber, and Franz}}]{Ser08c}
\bibinfo{author}{\bibnamefont{Seradjeh}, \bibfnamefont{B.}},
  \bibinfo{author}{\bibfnamefont{H.}~\bibnamefont{Weber}}, and
  \bibinfo{author}{\bibfnamefont{M.}~\bibnamefont{Franz}},
  \bibinfo{year}{2008}, \bibinfo{journal}{Phys. Rev. Lett.}
  \textbf{\bibinfo{volume}{101}}, \bibinfo{pages}{246404}.

\bibitem[{\citenamefont{Sheehy and Schmalian}(2007)}]{She07}
\bibinfo{author}{\bibnamefont{Sheehy}, \bibfnamefont{D.~E.}}, and
  \bibinfo{author}{\bibfnamefont{J.}~\bibnamefont{Schmalian}},
  \bibinfo{year}{2007}, \bibinfo{journal}{\prl} \textbf{\bibinfo{volume}{99}},
  \bibinfo{pages}{226803}.

\bibitem[{\citenamefont{Sheehy and Schmalian}(2009)}]{She09}
\bibinfo{author}{\bibnamefont{Sheehy}, \bibfnamefont{D.~E.}}, and
  \bibinfo{author}{\bibfnamefont{J.}~\bibnamefont{Schmalian}},
  \bibinfo{year}{2009}, \bibinfo{journal}{Phys. Rev. B}
  \textbf{\bibinfo{volume}{80}}, \bibinfo{pages}{193411}.

\bibitem[{\citenamefont{Shibata and Nomura}(2008)}]{SN08}
\bibinfo{author}{\bibnamefont{Shibata}, \bibfnamefont{N.}}, and
  \bibinfo{author}{\bibfnamefont{K.}~\bibnamefont{Nomura}},
  \bibinfo{year}{2008}, \bibinfo{journal}{Phys. Rev. B}
  \textbf{\bibinfo{volume}{77}}, \bibinfo{pages}{235426}.

\bibitem[{\citenamefont{Shimshoni} \emph{et~al.}(2009)\citenamefont{Shimshoni,
  Fertig, and Pai}}]{SFP09}
\bibinfo{author}{\bibnamefont{Shimshoni}, \bibfnamefont{E.}},
  \bibinfo{author}{\bibfnamefont{H.~A.} \bibnamefont{Fertig}}, and
  \bibinfo{author}{\bibfnamefont{G.~V.} \bibnamefont{Pai}},
  \bibinfo{year}{2009}, \bibinfo{journal}{Phys. Rev. Lett.}
  \textbf{\bibinfo{volume}{102}}, \bibinfo{pages}{206408}.

\bibitem[{\citenamefont{Shulke} \emph{et~al.}(1988)\citenamefont{Shulke, Bonse,
  Nagasawa, Kaprolat, and Berthold}}]{Shu88}
\bibinfo{author}{\bibnamefont{Shulke}, \bibfnamefont{W.}},
  \bibinfo{author}{\bibfnamefont{U.}~\bibnamefont{Bonse}},
  \bibinfo{author}{\bibfnamefont{H.}~\bibnamefont{Nagasawa}},
  \bibinfo{author}{\bibfnamefont{A.}~\bibnamefont{Kaprolat}}, and
  \bibinfo{author}{\bibfnamefont{A.}~\bibnamefont{Berthold}},
  \bibinfo{year}{1988}, \bibinfo{journal}{Phys. Rev. B}
  \textbf{\bibinfo{volume}{38}}, \bibinfo{pages}{2112}.

\bibitem[{\citenamefont{Shung}(1986{\natexlab{a}})}]{Shu86}
\bibinfo{author}{\bibnamefont{Shung}, \bibfnamefont{K.~W.-K.}},
  \bibinfo{year}{1986}{\natexlab{a}}, \bibinfo{journal}{Phys. Rev. B}
  \textbf{\bibinfo{volume}{34}}, \bibinfo{pages}{979}.

\bibitem[{\citenamefont{Shung}(1986{\natexlab{b}})}]{ShungLi}
\bibinfo{author}{\bibnamefont{Shung}, \bibfnamefont{K.~W.~K.}},
  \bibinfo{year}{1986}{\natexlab{b}}, \bibinfo{journal}{Phys. Rev. B}
  \textbf{\bibinfo{volume}{34}}, \bibinfo{pages}{1264}.

\bibitem[{\citenamefont{Shytov}
  \emph{et~al.}(2007{\natexlab{a}})\citenamefont{Shytov, Katsnelson, and
  Levitov}}]{Shytov:2007}
\bibinfo{author}{\bibnamefont{Shytov}, \bibfnamefont{A.}},
  \bibinfo{author}{\bibfnamefont{M.}~\bibnamefont{Katsnelson}}, and
  \bibinfo{author}{\bibfnamefont{L.}~\bibnamefont{Levitov}},
  \bibinfo{year}{2007}{\natexlab{a}}, \bibinfo{journal}{Phys. Rev. Lett.}
  \textbf{\bibinfo{volume}{99}}, \bibinfo{pages}{246802}.

\bibitem[{\citenamefont{Shytov}
  \emph{et~al.}(2007{\natexlab{b}})\citenamefont{Shytov, Katsnelson, and
  Levitov}}]{Shytov:2007-1}
\bibinfo{author}{\bibnamefont{Shytov}, \bibfnamefont{A.}},
  \bibinfo{author}{\bibfnamefont{M.}~\bibnamefont{Katsnelson}}, and
  \bibinfo{author}{\bibfnamefont{L.}~\bibnamefont{Levitov}},
  \bibinfo{year}{2007}{\natexlab{b}}, \bibinfo{journal}{Phys. Rev. Lett.}
  \textbf{\bibinfo{volume}{99}}, \bibinfo{pages}{236801}.

\bibitem[{\citenamefont{Siegel} \emph{et~al.}(2011)\citenamefont{Siegel, Park,
  Hwang, Deslippe, Fedorov, Louie, and Lanzara}}]{LanzaraMB}
\bibinfo{author}{\bibnamefont{Siegel}, \bibfnamefont{D.~A.}},
  \bibinfo{author}{\bibfnamefont{C.-H.} \bibnamefont{Park}},
  \bibinfo{author}{\bibfnamefont{C.}~\bibnamefont{Hwang}},
  \bibinfo{author}{\bibfnamefont{J.}~\bibnamefont{Deslippe}},
  \bibinfo{author}{\bibfnamefont{A.~V.} \bibnamefont{Fedorov}},
  \bibinfo{author}{\bibfnamefont{S.~G.} \bibnamefont{Louie}}, and
  \bibinfo{author}{\bibfnamefont{A.}~\bibnamefont{Lanzara}},
  \bibinfo{year}{2011}, \bibinfo{journal}{Proceedings of the National Academy
  of Sciences} \textbf{\bibinfo{volume}{108}}, \bibinfo{pages}{11365}.

\bibitem[{\citenamefont{Skrypnyk and Loktev}(2006)}]{Skr06}
\bibinfo{author}{\bibnamefont{Skrypnyk}, \bibfnamefont{Y.~V.}}, and
  \bibinfo{author}{\bibfnamefont{V.~M.} \bibnamefont{Loktev}},
  \bibinfo{year}{2006}, \bibinfo{journal}{Phys. Rev. B}
  \textbf{\bibinfo{volume}{73}}, \bibinfo{pages}{241402(R)}.

\bibitem[{\citenamefont{Sols} \emph{et~al.}(2007)\citenamefont{Sols, Guinea,
  and {Castro Neto}}}]{SGN07}
\bibinfo{author}{\bibnamefont{Sols}, \bibfnamefont{F.}},
  \bibinfo{author}{\bibfnamefont{F.}~\bibnamefont{Guinea}}, and
  \bibinfo{author}{\bibfnamefont{A.~H.} \bibnamefont{{Castro Neto}}},
  \bibinfo{year}{2007}, \bibinfo{journal}{Phys. Rev. Lett.}
  \textbf{\bibinfo{volume}{99}}, \bibinfo{pages}{166803}.

\bibitem[{\citenamefont{Son}(2007)}]{Son07}
\bibinfo{author}{\bibnamefont{Son}, \bibfnamefont{D.~T.}},
  \bibinfo{year}{2007}, \bibinfo{journal}{\prb} \textbf{\bibinfo{volume}{75}},
  \bibinfo{pages}{235423}.

\bibitem[{\citenamefont{Son} \emph{et~al.}(2006)\citenamefont{Son, Cohen, and
  Louie}}]{SCL06}
\bibinfo{author}{\bibnamefont{Son}, \bibfnamefont{Y.-W.}},
  \bibinfo{author}{\bibfnamefont{M.~L.} \bibnamefont{Cohen}}, and
  \bibinfo{author}{\bibfnamefont{S.~G.} \bibnamefont{Louie}},
  \bibinfo{year}{2006}, \bibinfo{journal}{Nature}
  \textbf{\bibinfo{volume}{444}}, \bibinfo{pages}{347}.

\bibitem[{\citenamefont{Sorella and Tosatti}(1992)}]{Sor92}
\bibinfo{author}{\bibnamefont{Sorella}, \bibfnamefont{S.}}, and
  \bibinfo{author}{\bibfnamefont{E.}~\bibnamefont{Tosatti}},
  \bibinfo{year}{1992}, \bibinfo{journal}{Europhys. Lett.}
  \textbf{\bibinfo{volume}{19}}, \bibinfo{pages}{699}.

\bibitem[{\citenamefont{Soriano and {Fern{\'a}ndez-Rossier}}(2010)}]{SF10}
\bibinfo{author}{\bibnamefont{Soriano}, \bibfnamefont{D.}}, and
  \bibinfo{author}{\bibfnamefont{J.}~\bibnamefont{{Fern{\'a}ndez-Rossier}}},
  \bibinfo{year}{2010}, \bibinfo{journal}{Phys. Rev. B}
  \textbf{\bibinfo{volume}{82}}, \bibinfo{pages}{161302}.

\bibitem[{\citenamefont{Sprinkle} \emph{et~al.}(2009)\citenamefont{Sprinkle,
  Siegel, Hu, Hicks, Tejeda, Taleb-Ibrahimi, {Le F{\`e}vre}, Bertran, Vizzini,
  Enriquez, Chiang, Soukiassian} \emph{et~al.}}]{Sprinkle09}
\bibinfo{author}{\bibnamefont{Sprinkle}, \bibfnamefont{M.}},
  \bibinfo{author}{\bibfnamefont{D.}~\bibnamefont{Siegel}},
  \bibinfo{author}{\bibfnamefont{Y.}~\bibnamefont{Hu}},
  \bibinfo{author}{\bibfnamefont{J.}~\bibnamefont{Hicks}},
  \bibinfo{author}{\bibfnamefont{A.}~\bibnamefont{Tejeda}},
  \bibinfo{author}{\bibfnamefont{A.}~\bibnamefont{Taleb-Ibrahimi}},
  \bibinfo{author}{\bibfnamefont{P.}~\bibnamefont{{Le F{\`e}vre}}},
  \bibinfo{author}{\bibfnamefont{F.}~\bibnamefont{Bertran}},
  \bibinfo{author}{\bibfnamefont{S.}~\bibnamefont{Vizzini}},
  \bibinfo{author}{\bibfnamefont{H.}~\bibnamefont{Enriquez}},
  \bibinfo{author}{\bibfnamefont{S.}~\bibnamefont{Chiang}},
  \bibinfo{author}{\bibfnamefont{P.}~\bibnamefont{Soukiassian}}, \emph{et~al.},
  \bibinfo{year}{2009}, \bibinfo{journal}{Phys. Rev. Lett.}
  \textbf{\bibinfo{volume}{103}}, \bibinfo{pages}{226803}.

\bibitem[{\citenamefont{Stampfer} \emph{et~al.}(2009)\citenamefont{Stampfer,
  {G{\"u}ttinger}, Hellmueller, Molitor, Ensslin, and Ihn}}]{Setal09}
\bibinfo{author}{\bibnamefont{Stampfer}, \bibfnamefont{C.}},
  \bibinfo{author}{\bibfnamefont{J.}~\bibnamefont{{G{\"u}ttinger}}},
  \bibinfo{author}{\bibfnamefont{S.}~\bibnamefont{Hellmueller}},
  \bibinfo{author}{\bibfnamefont{F.}~\bibnamefont{Molitor}},
  \bibinfo{author}{\bibfnamefont{K.}~\bibnamefont{Ensslin}}, and
  \bibinfo{author}{\bibfnamefont{T.}~\bibnamefont{Ihn}}, \bibinfo{year}{2009},
  \bibinfo{journal}{Phys. Rev. Lett.} \textbf{\bibinfo{volume}{102}},
  \bibinfo{pages}{056403}.

\bibitem[{\citenamefont{Stampfer} \emph{et~al.}(2008)\citenamefont{Stampfer,
  Schurtenberger, Molitor, {G{\"u}ttinger}, Ihn, and Ensslin}}]{Setal08b}
\bibinfo{author}{\bibnamefont{Stampfer}, \bibfnamefont{C.}},
  \bibinfo{author}{\bibfnamefont{E.}~\bibnamefont{Schurtenberger}},
  \bibinfo{author}{\bibfnamefont{F.}~\bibnamefont{Molitor}},
  \bibinfo{author}{\bibfnamefont{J.}~\bibnamefont{{G{\"u}ttinger}}},
  \bibinfo{author}{\bibfnamefont{T.}~\bibnamefont{Ihn}}, and
  \bibinfo{author}{\bibfnamefont{K.}~\bibnamefont{Ensslin}},
  \bibinfo{year}{2008}, \bibinfo{journal}{Nano Lett.}
  \textbf{\bibinfo{volume}{8}}, \bibinfo{pages}{2378}.

\bibitem[{\citenamefont{Stauber} \emph{et~al.}(2005)\citenamefont{Stauber,
  Guinea, and Vozmediano}}]{Stauber05}
\bibinfo{author}{\bibnamefont{Stauber}, \bibfnamefont{T.}},
  \bibinfo{author}{\bibfnamefont{F.}~\bibnamefont{Guinea}}, and
  \bibinfo{author}{\bibfnamefont{M.~A.~H.} \bibnamefont{Vozmediano}},
  \bibinfo{year}{2005}, \bibinfo{journal}{Phys. Rev. B}
  \textbf{\bibinfo{volume}{71}}, \bibinfo{pages}{041406}.

\bibitem[{\citenamefont{Stein and Brown}(1987)}]{SB87}
\bibinfo{author}{\bibnamefont{Stein}, \bibfnamefont{S.~E.}}, and
  \bibinfo{author}{\bibfnamefont{R.~L.} \bibnamefont{Brown}},
  \bibinfo{year}{1987}, \bibinfo{journal}{J. Am. Chem. Soc.}
  \textbf{\bibinfo{volume}{109}}, \bibinfo{pages}{3721}.

\bibitem[{\citenamefont{Taft and Philipp}(1965)}]{Taf65}
\bibinfo{author}{\bibnamefont{Taft}, \bibfnamefont{E.~A.}}, and
  \bibinfo{author}{\bibfnamefont{H.~R.} \bibnamefont{Philipp}},
  \bibinfo{year}{1965}, \bibinfo{journal}{Phys. Rev.}
  \textbf{\bibinfo{volume}{138}}, \bibinfo{pages}{A197}.

\bibitem[{\citenamefont{Takei and Kim}(2008)}]{Tak08}
\bibinfo{author}{\bibnamefont{Takei}, \bibfnamefont{S.}}, and
  \bibinfo{author}{\bibfnamefont{Y.~B.} \bibnamefont{Kim}},
  \bibinfo{year}{2008}, \bibinfo{journal}{Phys. Rev. B}
  \textbf{\bibinfo{volume}{78}}, \bibinfo{pages}{165401}.

\bibitem[{\citenamefont{Tan} \emph{et~al.}(2007)\citenamefont{Tan, Zhang,
  Bolotin, Zhao, Adam, Hwang, {Das Sarma}, Stormer, and Kim}}]{Tan:2007}
\bibinfo{author}{\bibnamefont{Tan}, \bibfnamefont{Y.~W.}},
  \bibinfo{author}{\bibfnamefont{Y.}~\bibnamefont{Zhang}},
  \bibinfo{author}{\bibfnamefont{K.}~\bibnamefont{Bolotin}},
  \bibinfo{author}{\bibfnamefont{Y.}~\bibnamefont{Zhao}},
  \bibinfo{author}{\bibfnamefont{S.}~\bibnamefont{Adam}},
  \bibinfo{author}{\bibfnamefont{E.}~\bibnamefont{Hwang}},
  \bibinfo{author}{\bibfnamefont{S.}~\bibnamefont{{Das Sarma}}},
  \bibinfo{author}{\bibfnamefont{H.}~\bibnamefont{Stormer}}, and
  \bibinfo{author}{\bibfnamefont{P.}~\bibnamefont{Kim}}, \bibinfo{year}{2007},
  \bibinfo{journal}{Phys. Rev. Lett.} \textbf{\bibinfo{volume}{99}},
  \bibinfo{pages}{246803}.

\bibitem[{\citenamefont{Tediosi} \emph{et~al.}(2007)\citenamefont{Tediosi,
  Armitage, Giannini, and van~der Marel}}]{Armitage}
\bibinfo{author}{\bibnamefont{Tediosi}, \bibfnamefont{R.}},
  \bibinfo{author}{\bibfnamefont{N.~P.} \bibnamefont{Armitage}},
  \bibinfo{author}{\bibfnamefont{E.}~\bibnamefont{Giannini}}, and
  \bibinfo{author}{\bibfnamefont{D.}~\bibnamefont{van~der Marel}},
  \bibinfo{year}{2007}, \bibinfo{journal}{Phys. Rev. Lett.}
  \textbf{\bibinfo{volume}{99}}, \bibinfo{pages}{016406}.

\bibitem[{\citenamefont{Terekhov} \emph{et~al.}(2008)\citenamefont{Terekhov,
  Milstein, Kotov, and Sushkov}}]{Terekhov:2008}
\bibinfo{author}{\bibnamefont{Terekhov}, \bibfnamefont{I.~S.}},
  \bibinfo{author}{\bibfnamefont{A.~I.} \bibnamefont{Milstein}},
  \bibinfo{author}{\bibfnamefont{V.~N.} \bibnamefont{Kotov}}, and
  \bibinfo{author}{\bibfnamefont{O.~P.} \bibnamefont{Sushkov}},
  \bibinfo{year}{2008}, \bibinfo{journal}{Phys. Rev. Lett.}
  \textbf{\bibinfo{volume}{100}}, \bibinfo{pages}{076803}.

\bibitem[{\citenamefont{Thomas}(1927)}]{Thomas:1927}
\bibinfo{author}{\bibnamefont{Thomas}, \bibfnamefont{L.~H.}},
  \bibinfo{year}{1927}, \bibinfo{journal}{Mat. Proc. Cambridge Phil. Soc.}
  \textbf{\bibinfo{volume}{23}}, \bibinfo{pages}{542}.

\bibitem[{\citenamefont{Tikhonenko}
  \emph{et~al.}(2009)\citenamefont{Tikhonenko, Kozikov, Savchenko, and
  Gorbachev}}]{TKSG10}
\bibinfo{author}{\bibnamefont{Tikhonenko}, \bibfnamefont{F.~V.}},
  \bibinfo{author}{\bibfnamefont{A.~A.} \bibnamefont{Kozikov}},
  \bibinfo{author}{\bibfnamefont{A.~K.} \bibnamefont{Savchenko}}, and
  \bibinfo{author}{\bibfnamefont{R.~V.} \bibnamefont{Gorbachev}},
  \bibinfo{year}{2009}, \bibinfo{journal}{Phys. Rev. Lett.}
  \textbf{\bibinfo{volume}{103}}, \bibinfo{pages}{226801}.

\bibitem[{\citenamefont{Tinkham}(1996)}]{tinkham:96}
\bibinfo{author}{\bibnamefont{Tinkham}, \bibfnamefont{M.}},
  \bibinfo{year}{1996}, \emph{\bibinfo{title}{Introduction to
  Superconductivity}} (\bibinfo{publisher}{McGraw-Hill}, \bibinfo{address}{New
  York}).

\bibitem[{\citenamefont{Titov and Beenakker}(2006)}]{Tit06}
\bibinfo{author}{\bibnamefont{Titov}, \bibfnamefont{M.}}, and
  \bibinfo{author}{\bibfnamefont{C.~W.~J.} \bibnamefont{Beenakker}},
  \bibinfo{year}{2006}, \bibinfo{journal}{Phys. Rev. B}
  \textbf{\bibinfo{volume}{74}}, \bibinfo{pages}{041401(R)}.

\bibitem[{\citenamefont{Titov} \emph{et~al.}(2007)\citenamefont{Titov, Ossipov,
  and Beenakker}}]{Tit07}
\bibinfo{author}{\bibnamefont{Titov}, \bibfnamefont{M.}},
  \bibinfo{author}{\bibfnamefont{A.}~\bibnamefont{Ossipov}}, and
  \bibinfo{author}{\bibfnamefont{C.~W.~J.} \bibnamefont{Beenakker}},
  \bibinfo{year}{2007}, \bibinfo{journal}{Phys. Rev. B}
  \textbf{\bibinfo{volume}{75}}, \bibinfo{pages}{0415417}.

\bibitem[{\citenamefont{Todd} \emph{et~al.}(2009)\citenamefont{Todd, Chou,
  Amasha, and Goldhaber-Gordon}}]{TCGA09}
\bibinfo{author}{\bibnamefont{Todd}, \bibfnamefont{K.}},
  \bibinfo{author}{\bibfnamefont{H.-T.} \bibnamefont{Chou}},
  \bibinfo{author}{\bibfnamefont{S.}~\bibnamefont{Amasha}}, and
  \bibinfo{author}{\bibfnamefont{D.}~\bibnamefont{Goldhaber-Gordon}},
  \bibinfo{year}{2009}, \bibinfo{journal}{Nano Lett.}
  \textbf{\bibinfo{volume}{9}}, \bibinfo{pages}{416}.

\bibitem[{\citenamefont{{T{\"o}ke} and Jain}(2007)}]{HCJ07}
\bibinfo{author}{\bibnamefont{{T{\"o}ke}}, \bibfnamefont{C.}}, and
  \bibinfo{author}{\bibfnamefont{J.~K.} \bibnamefont{Jain}},
  \bibinfo{year}{2007}, \bibinfo{journal}{Phys. Rev. B}
  \textbf{\bibinfo{volume}{75}}, \bibinfo{pages}{245440}.

\bibitem[{\citenamefont{{T{\"o}ke}}
  \emph{et~al.}(2006)\citenamefont{{T{\"o}ke}, Lammert, Crespi, and
  Jain}}]{TLCJ06}
\bibinfo{author}{\bibnamefont{{T{\"o}ke}}, \bibfnamefont{C.}},
  \bibinfo{author}{\bibfnamefont{P.~E.} \bibnamefont{Lammert}},
  \bibinfo{author}{\bibfnamefont{V.~H.} \bibnamefont{Crespi}}, and
  \bibinfo{author}{\bibfnamefont{J.~K.} \bibnamefont{Jain}},
  \bibinfo{year}{2006}, \bibinfo{journal}{Phys. Rev. B}
  \textbf{\bibinfo{volume}{74}}, \bibinfo{pages}{235417}.

\bibitem[{\citenamefont{Trevisanutto}
  \emph{et~al.}(2008)\citenamefont{Trevisanutto, Giorgetti, Reining, Ladisa,
  and Olevano}}]{Trevis08}
\bibinfo{author}{\bibnamefont{Trevisanutto}, \bibfnamefont{P.~E.}},
  \bibinfo{author}{\bibfnamefont{C.}~\bibnamefont{Giorgetti}},
  \bibinfo{author}{\bibfnamefont{L.}~\bibnamefont{Reining}},
  \bibinfo{author}{\bibfnamefont{M.}~\bibnamefont{Ladisa}}, and
  \bibinfo{author}{\bibfnamefont{V.}~\bibnamefont{Olevano}},
  \bibinfo{year}{2008}, \bibinfo{journal}{Phys. Rev. Lett.}
  \textbf{\bibinfo{volume}{101}}, \bibinfo{pages}{226405}.

\bibitem[{\citenamefont{Tyutyulkov}
  \emph{et~al.}(1998)\citenamefont{Tyutyulkov, Madjarova, Dietz, and
  Mullen}}]{TMDM98}
\bibinfo{author}{\bibnamefont{Tyutyulkov}, \bibfnamefont{N.}},
  \bibinfo{author}{\bibfnamefont{G.}~\bibnamefont{Madjarova}},
  \bibinfo{author}{\bibfnamefont{F.}~\bibnamefont{Dietz}}, and
  \bibinfo{author}{\bibfnamefont{K.}~\bibnamefont{Mullen}},
  \bibinfo{year}{1998}, \bibinfo{journal}{J. Phys. Chem. B}
  \textbf{\bibinfo{volume}{102}}, \bibinfo{pages}{10183}.

\bibitem[{\citenamefont{Uchoa} \emph{et~al.}(2005)\citenamefont{Uchoa, Cabrera,
  and {Castro Neto}}}]{Uch05}
\bibinfo{author}{\bibnamefont{Uchoa}, \bibfnamefont{B.}},
  \bibinfo{author}{\bibfnamefont{G.~G.} \bibnamefont{Cabrera}}, and
  \bibinfo{author}{\bibfnamefont{A.~H.} \bibnamefont{{Castro Neto}}},
  \bibinfo{year}{2005}, \bibinfo{journal}{Phys. Rev. B}
  \textbf{\bibinfo{volume}{71}}, \bibinfo{pages}{184509}.

\bibitem[{\citenamefont{Uchoa and {Castro Neto}}(2007)}]{Uch07}
\bibinfo{author}{\bibnamefont{Uchoa}, \bibfnamefont{B.}}, and
  \bibinfo{author}{\bibfnamefont{A.~H.} \bibnamefont{{Castro Neto}}},
  \bibinfo{year}{2007}, \bibinfo{journal}{Phys. Rev. Lett.}
  \textbf{\bibinfo{volume}{98}}, \bibinfo{pages}{146801}.

\bibitem[{\citenamefont{Uchoa and {Castro Neto}}(2009)}]{Uch09a}
\bibinfo{author}{\bibnamefont{Uchoa}, \bibfnamefont{B.}}, and
  \bibinfo{author}{\bibfnamefont{A.~H.} \bibnamefont{{Castro Neto}}},
  \bibinfo{year}{2009}, \bibinfo{journal}{Phys. Rev. Lett.}
  \textbf{\bibinfo{volume}{102}}, \bibinfo{pages}{109701}.

\bibitem[{\citenamefont{Uchoa}
  \emph{et~al.}(2008{\natexlab{a}})\citenamefont{Uchoa, Kotov, Peres, and
  {Castro Neto}}}]{Uch08b}
\bibinfo{author}{\bibnamefont{Uchoa}, \bibfnamefont{B.}},
  \bibinfo{author}{\bibfnamefont{V.~N.} \bibnamefont{Kotov}},
  \bibinfo{author}{\bibfnamefont{N.~M.~R.} \bibnamefont{Peres}}, and
  \bibinfo{author}{\bibfnamefont{A.~H.} \bibnamefont{{Castro Neto}}},
  \bibinfo{year}{2008}{\natexlab{a}}, \bibinfo{journal}{Phys. Rev. Lett.}
  \textbf{\bibinfo{volume}{101}}, \bibinfo{pages}{026805}.

\bibitem[{\citenamefont{Uchoa}
  \emph{et~al.}(2008{\natexlab{b}})\citenamefont{Uchoa, Lin, and {Castro
  Neto}}}]{Uch08a}
\bibinfo{author}{\bibnamefont{Uchoa}, \bibfnamefont{B.}},
  \bibinfo{author}{\bibfnamefont{C.-Y.} \bibnamefont{Lin}}, and
  \bibinfo{author}{\bibfnamefont{A.~H.} \bibnamefont{{Castro Neto}}},
  \bibinfo{year}{2008}{\natexlab{b}}, \bibinfo{journal}{Phys. Rev. B}
  \textbf{\bibinfo{volume}{77}}, \bibinfo{pages}{035420}.

\bibitem[{\citenamefont{Uchoa} \emph{et~al.}(2011)\citenamefont{Uchoa,
  Rappoport, and {Castro Neto}}}]{Uch10}
\bibinfo{author}{\bibnamefont{Uchoa}, \bibfnamefont{B.}},
  \bibinfo{author}{\bibfnamefont{T.~G.} \bibnamefont{Rappoport}}, and
  \bibinfo{author}{\bibfnamefont{A.~H.} \bibnamefont{{Castro Neto}}},
  \bibinfo{year}{2011}, \bibinfo{journal}{Phys. Rev. Lett.}
  \textbf{\bibinfo{volume}{106}}, \bibinfo{pages}{016801}.

\bibitem[{\citenamefont{Uchoa} \emph{et~al.}(2009)\citenamefont{Uchoa, Yang,
  Tsai, Peres, and {Castro Neto}}}]{Uch09b}
\bibinfo{author}{\bibnamefont{Uchoa}, \bibfnamefont{B.}},
  \bibinfo{author}{\bibfnamefont{L.}~\bibnamefont{Yang}},
  \bibinfo{author}{\bibfnamefont{S.~W.} \bibnamefont{Tsai}},
  \bibinfo{author}{\bibfnamefont{N.~M.~R.} \bibnamefont{Peres}}, and
  \bibinfo{author}{\bibfnamefont{A.~H.} \bibnamefont{{Castro Neto}}},
  \bibinfo{year}{2009}, \bibinfo{journal}{Phys. Rev. Lett.}
  \textbf{\bibinfo{volume}{103}}, \bibinfo{pages}{206804}.

\bibitem[{\citenamefont{Ugeda} \emph{et~al.}(2010)\citenamefont{Ugeda,
  Brihuega, Guinea, and G{\'o}mez-Rodr{\'i}guez}}]{UBGG10}
\bibinfo{author}{\bibnamefont{Ugeda}, \bibfnamefont{M.~M.}},
  \bibinfo{author}{\bibfnamefont{I.}~\bibnamefont{Brihuega}},
  \bibinfo{author}{\bibfnamefont{F.}~\bibnamefont{Guinea}}, and
  \bibinfo{author}{\bibfnamefont{J.~M.} \bibnamefont{G{\'o}mez-Rodr{\'i}guez}},
  \bibinfo{year}{2010}, \bibinfo{journal}{Phys. Rev. Lett.}
  \textbf{\bibinfo{volume}{104}}, \bibinfo{pages}{096804}.

\bibitem[{\citenamefont{Vafek}(2007)}]{Vaf07}
\bibinfo{author}{\bibnamefont{Vafek}, \bibfnamefont{O.}}, \bibinfo{year}{2007},
  \bibinfo{journal}{\prl} \textbf{\bibinfo{volume}{98}},
  \bibinfo{pages}{216401}.

\bibitem[{\citenamefont{Vafek}(2010)}]{Vafek10}
\bibinfo{author}{\bibnamefont{Vafek}, \bibfnamefont{O.}}, \bibinfo{year}{2010},
  \bibinfo{journal}{Phys. Rev. B} \textbf{\bibinfo{volume}{82}},
  \bibinfo{pages}{205106}.

\bibitem[{\citenamefont{Vafek and Case}(2008)}]{Vafek:2008}
\bibinfo{author}{\bibnamefont{Vafek}, \bibfnamefont{O.}}, and
  \bibinfo{author}{\bibfnamefont{M.~J.} \bibnamefont{Case}},
  \bibinfo{year}{2008}, \bibinfo{journal}{Phys. Rev. B}
  \textbf{\bibinfo{volume}{77}}, \bibinfo{pages}{033410}.

\bibitem[{\citenamefont{Vafek and Yang}(2010)}]{Vaf10}
\bibinfo{author}{\bibnamefont{Vafek}, \bibfnamefont{O.}}, and
  \bibinfo{author}{\bibfnamefont{K.}~\bibnamefont{Yang}}, \bibinfo{year}{2010},
  \bibinfo{journal}{Phys. Rev. B} \textbf{\bibinfo{volume}{81}},
  \bibinfo{pages}{041401(R)}.

\bibitem[{\citenamefont{Valenzuela and Vozmediano}(2008)}]{Val08}
\bibinfo{author}{\bibnamefont{Valenzuela}, \bibfnamefont{B.}}, and
  \bibinfo{author}{\bibfnamefont{M.~A.~H.} \bibnamefont{Vozmediano}},
  \bibinfo{year}{2008}, \bibinfo{journal}{New J. Phys.}
  \textbf{\bibinfo{volume}{10}}, \bibinfo{pages}{113009}.

\bibitem[{\citenamefont{Venezuela} \emph{et~al.}(2009)\citenamefont{Venezuela,
  Muniz, Costa, Edwards, Power, and Ferreira}}]{Ven09}
\bibinfo{author}{\bibnamefont{Venezuela}, \bibfnamefont{P.}},
  \bibinfo{author}{\bibfnamefont{R.~B.} \bibnamefont{Muniz}},
  \bibinfo{author}{\bibfnamefont{A.~T.} \bibnamefont{Costa}},
  \bibinfo{author}{\bibfnamefont{D.~M.} \bibnamefont{Edwards}},
  \bibinfo{author}{\bibfnamefont{S.~R.} \bibnamefont{Power}}, and
  \bibinfo{author}{\bibfnamefont{M.~S.} \bibnamefont{Ferreira}},
  \bibinfo{year}{2009}, \bibinfo{journal}{Phys. Rev. B.}
  \textbf{\bibinfo{volume}{80}}, \bibinfo{pages}{241413(R)}.

\bibitem[{\citenamefont{Visscher and Falikov}(1970)}]{Vis70}
\bibinfo{author}{\bibnamefont{Visscher}, \bibfnamefont{P.~B.}}, and
  \bibinfo{author}{\bibfnamefont{L.~M.} \bibnamefont{Falikov}},
  \bibinfo{year}{1970}, \bibinfo{journal}{Phys. Rev. B}
  \textbf{\bibinfo{volume}{3}}, \bibinfo{pages}{2541}.

\bibitem[{\citenamefont{Vojta}(2001)}]{Voj01a}
\bibinfo{author}{\bibnamefont{Vojta}, \bibfnamefont{M.}}, \bibinfo{year}{2001},
  \bibinfo{journal}{Phys. Rev. Lett.} \textbf{\bibinfo{volume}{87}},
  \bibinfo{pages}{097202}.

\bibitem[{\citenamefont{Vojta and Bulla}(2001)}]{Voj01b}
\bibinfo{author}{\bibnamefont{Vojta}, \bibfnamefont{M.}}, and
  \bibinfo{author}{\bibfnamefont{R.}~\bibnamefont{Bulla}},
  \bibinfo{year}{2001}, \bibinfo{journal}{Phys. Rev. B}
  \textbf{\bibinfo{volume}{65}}, \bibinfo{pages}{014511}.

\bibitem[{\citenamefont{Vojta and Fritz}(2004)}]{Voj04}
\bibinfo{author}{\bibnamefont{Vojta}, \bibfnamefont{M.}}, and
  \bibinfo{author}{\bibfnamefont{L.}~\bibnamefont{Fritz}},
  \bibinfo{year}{2004}, \bibinfo{journal}{Phys. Rev. B}
  \textbf{\bibinfo{volume}{70}}, \bibinfo{pages}{094502}.

\bibitem[{\citenamefont{Vojta} \emph{et~al.}(2010)\citenamefont{Vojta, Fritz,
  and Bulla}}]{Voj10}
\bibinfo{author}{\bibnamefont{Vojta}, \bibfnamefont{M.}},
  \bibinfo{author}{\bibfnamefont{L.}~\bibnamefont{Fritz}}, and
  \bibinfo{author}{\bibfnamefont{R.}~\bibnamefont{Bulla}},
  \bibinfo{year}{2010}, \bibinfo{journal}{Europhys. Lett.}
  \textbf{\bibinfo{volume}{90}}, \bibinfo{pages}{27006}.

\bibitem[{\citenamefont{Vozmediano}
  \emph{et~al.}(2005)\citenamefont{Vozmediano, L{\'o}pez-Sancho, Stauber, and
  Guinea}}]{VLSG05}
\bibinfo{author}{\bibnamefont{Vozmediano}, \bibfnamefont{M.~A.~H.}},
  \bibinfo{author}{\bibfnamefont{M.~P.} \bibnamefont{L{\'o}pez-Sancho}},
  \bibinfo{author}{\bibfnamefont{T.}~\bibnamefont{Stauber}}, and
  \bibinfo{author}{\bibfnamefont{F.}~\bibnamefont{Guinea}},
  \bibinfo{year}{2005}, \bibinfo{journal}{Phys. Rev. B}
  \textbf{\bibinfo{volume}{72}}, \bibinfo{pages}{155121}.

\bibitem[{\citenamefont{Wallace}(1947)}]{Wal47}
\bibinfo{author}{\bibnamefont{Wallace}, \bibfnamefont{P.~R.}},
  \bibinfo{year}{1947}, \bibinfo{journal}{Phys. Rev.}
  \textbf{\bibinfo{volume}{71}}, \bibinfo{pages}{622}.

\bibitem[{\citenamefont{Wang} \emph{et~al.}(2010)\citenamefont{Wang, Fertig,
  and Murthy}}]{Wang:2009}
\bibinfo{author}{\bibnamefont{Wang}, \bibfnamefont{J.}},
  \bibinfo{author}{\bibfnamefont{H.}~\bibnamefont{Fertig}}, and
  \bibinfo{author}{\bibfnamefont{G.}~\bibnamefont{Murthy}},
  \bibinfo{year}{2010}, \bibinfo{journal}{\prl} \textbf{\bibinfo{volume}{104}},
  \bibinfo{pages}{186401}.

\bibitem[{\citenamefont{Wang} \emph{et~al.}(2008)\citenamefont{Wang, Iyengar,
  Fertig, and Brey}}]{WIFB08}
\bibinfo{author}{\bibnamefont{Wang}, \bibfnamefont{J.}},
  \bibinfo{author}{\bibfnamefont{A.}~\bibnamefont{Iyengar}},
  \bibinfo{author}{\bibfnamefont{H.~A.} \bibnamefont{Fertig}}, and
  \bibinfo{author}{\bibfnamefont{L.}~\bibnamefont{Brey}}, \bibinfo{year}{2008},
  \bibinfo{journal}{Phys. Rev. B} \textbf{\bibinfo{volume}{78}},
  \bibinfo{pages}{165416}.

\bibitem[{\citenamefont{Wehling}
  \emph{et~al.}(2010{\natexlab{a}})\citenamefont{Wehling, Balatsky, Katsnelson,
  Lichtenstein, and Rosch}}]{Weh10b}
\bibinfo{author}{\bibnamefont{Wehling}, \bibfnamefont{T.~O.}},
  \bibinfo{author}{\bibfnamefont{A.~V.} \bibnamefont{Balatsky}},
  \bibinfo{author}{\bibfnamefont{M.~I.} \bibnamefont{Katsnelson}},
  \bibinfo{author}{\bibfnamefont{A.~I.} \bibnamefont{Lichtenstein}}, and
  \bibinfo{author}{\bibfnamefont{A.}~\bibnamefont{Rosch}},
  \bibinfo{year}{2010}{\natexlab{a}}, \bibinfo{journal}{Phys. Rev. B}
  \textbf{\bibinfo{volume}{81}}, \bibinfo{pages}{115427}.

\bibitem[{\citenamefont{Wehling} \emph{et~al.}(2007)\citenamefont{Wehling,
  Balatsky, Katsnelson, Lichtenstein, Scharnberg, and Wiesendanger}}]{Weh07}
\bibinfo{author}{\bibnamefont{Wehling}, \bibfnamefont{T.~O.}},
  \bibinfo{author}{\bibfnamefont{A.~V.} \bibnamefont{Balatsky}},
  \bibinfo{author}{\bibfnamefont{M.~I.} \bibnamefont{Katsnelson}},
  \bibinfo{author}{\bibfnamefont{A.~I.} \bibnamefont{Lichtenstein}},
  \bibinfo{author}{\bibfnamefont{K.}~\bibnamefont{Scharnberg}}, and
  \bibinfo{author}{\bibfnamefont{R.}~\bibnamefont{Wiesendanger}},
  \bibinfo{year}{2007}, \bibinfo{journal}{Phys. Rev. B}
  \textbf{\bibinfo{volume}{75}}, \bibinfo{pages}{125425}.

\bibitem[{\citenamefont{Wehling}
  \emph{et~al.}(2008{\natexlab{a}})\citenamefont{Wehling, Balatsky, Tsvelik,
  Katsnelson, and Lichtenstein}}]{Wetal08}
\bibinfo{author}{\bibnamefont{Wehling}, \bibfnamefont{T.~O.}},
  \bibinfo{author}{\bibfnamefont{A.~V.} \bibnamefont{Balatsky}},
  \bibinfo{author}{\bibfnamefont{A.~M.} \bibnamefont{Tsvelik}},
  \bibinfo{author}{\bibfnamefont{M.~I.} \bibnamefont{Katsnelson}}, and
  \bibinfo{author}{\bibfnamefont{A.~I.} \bibnamefont{Lichtenstein}},
  \bibinfo{year}{2008}{\natexlab{a}}, \bibinfo{journal}{Europhys. Lett.}
  \textbf{\bibinfo{volume}{84}}, \bibinfo{pages}{17003}.

\bibitem[{\citenamefont{Wehling}
  \emph{et~al.}(2010{\natexlab{b}})\citenamefont{Wehling, Dahal, Lichtenstein,
  Katsnelson, Manoharan, and Balatsky}}]{Weh10a}
\bibinfo{author}{\bibnamefont{Wehling}, \bibfnamefont{T.~O.}},
  \bibinfo{author}{\bibfnamefont{H.~P.} \bibnamefont{Dahal}},
  \bibinfo{author}{\bibfnamefont{A.~I.} \bibnamefont{Lichtenstein}},
  \bibinfo{author}{\bibfnamefont{M.~I.} \bibnamefont{Katsnelson}},
  \bibinfo{author}{\bibfnamefont{H.}~\bibnamefont{Manoharan}}, and
  \bibinfo{author}{\bibfnamefont{A.~V.} \bibnamefont{Balatsky}},
  \bibinfo{year}{2010}{\natexlab{b}}, \bibinfo{journal}{Phys. Rev. B}
  \textbf{\bibinfo{volume}{81}}, \bibinfo{pages}{085413}.

\bibitem[{\citenamefont{Wehling}
  \emph{et~al.}(2008{\natexlab{b}})\citenamefont{Wehling, Novoselov, Morozov,
  Vdovin, Katsnelson, Geim, and Lichtenstein}}]{Wetal08b}
\bibinfo{author}{\bibnamefont{Wehling}, \bibfnamefont{T.~O.}},
  \bibinfo{author}{\bibfnamefont{K.~S.} \bibnamefont{Novoselov}},
  \bibinfo{author}{\bibfnamefont{S.~V.} \bibnamefont{Morozov}},
  \bibinfo{author}{\bibfnamefont{E.~E.} \bibnamefont{Vdovin}},
  \bibinfo{author}{\bibfnamefont{M.~I.} \bibnamefont{Katsnelson}},
  \bibinfo{author}{\bibfnamefont{A.~K.} \bibnamefont{Geim}}, and
  \bibinfo{author}{\bibfnamefont{A.~I.} \bibnamefont{Lichtenstein}},
  \bibinfo{year}{2008}{\natexlab{b}}, \bibinfo{journal}{Nano Lett.}
  \textbf{\bibinfo{volume}{8}}, \bibinfo{pages}{173}.

\bibitem[{\citenamefont{Wehling}
  \emph{et~al.}(2010{\natexlab{c}})\citenamefont{Wehling, Yuan, Lichtenstein,
  Geim, and Katsnelson}}]{WehlingPRL}
\bibinfo{author}{\bibnamefont{Wehling}, \bibfnamefont{T.~O.}},
  \bibinfo{author}{\bibfnamefont{S.}~\bibnamefont{Yuan}},
  \bibinfo{author}{\bibfnamefont{A.~I.} \bibnamefont{Lichtenstein}},
  \bibinfo{author}{\bibfnamefont{A.~K.} \bibnamefont{Geim}}, and
  \bibinfo{author}{\bibfnamefont{M.~I.} \bibnamefont{Katsnelson}},
  \bibinfo{year}{2010}{\natexlab{c}}, \bibinfo{journal}{Phys. Rev. Lett.}
  \textbf{\bibinfo{volume}{105}}, \bibinfo{pages}{056802}.

\bibitem[{\citenamefont{Weisskopf}(1939)}]{Weiss}
\bibinfo{author}{\bibnamefont{Weisskopf}, \bibfnamefont{V.~F.}},
  \bibinfo{year}{1939}, \bibinfo{journal}{Phys. Rev.}
  \textbf{\bibinfo{volume}{56}}, \bibinfo{pages}{72}.

\bibitem[{\citenamefont{Weller} \emph{et~al.}(2005)\citenamefont{Weller,
  Ellerby, Saxena, Smith, and Skipper}}]{Wel05}
\bibinfo{author}{\bibnamefont{Weller}, \bibfnamefont{T.~E.}},
  \bibinfo{author}{\bibfnamefont{M.}~\bibnamefont{Ellerby}},
  \bibinfo{author}{\bibfnamefont{S.~S.} \bibnamefont{Saxena}},
  \bibinfo{author}{\bibfnamefont{R.~P.} \bibnamefont{Smith}}, and
  \bibinfo{author}{\bibfnamefont{N.~T.} \bibnamefont{Skipper}},
  \bibinfo{year}{2005}, \bibinfo{journal}{Nature Physics}
  \textbf{\bibinfo{volume}{1}}, \bibinfo{pages}{39}.

\bibitem[{\citenamefont{Williams} \emph{et~al.}(2007)\citenamefont{Williams,
  DiCarlo, and Marcus}}]{WDM07}
\bibinfo{author}{\bibnamefont{Williams}, \bibfnamefont{J.~R.}},
  \bibinfo{author}{\bibfnamefont{L.}~\bibnamefont{DiCarlo}}, and
  \bibinfo{author}{\bibfnamefont{C.~M.} \bibnamefont{Marcus}},
  \bibinfo{year}{2007}, \bibinfo{journal}{Science}
  \textbf{\bibinfo{volume}{317}}, \bibinfo{pages}{638}.

\bibitem[{\citenamefont{Wimmer} \emph{et~al.}(2010)\citenamefont{Wimmer,
  Akhmerov, and Guinea}}]{WAG10}
\bibinfo{author}{\bibnamefont{Wimmer}, \bibfnamefont{M.}},
  \bibinfo{author}{\bibfnamefont{A.}~\bibnamefont{Akhmerov}}, and
  \bibinfo{author}{\bibfnamefont{F.}~\bibnamefont{Guinea}},
  \bibinfo{year}{2010}, \bibinfo{journal}{Phys. Rev. B}
  \textbf{\bibinfo{volume}{82}}, \bibinfo{pages}{045409}.

\bibitem[{\citenamefont{Withoff and Fradkin}(1990)}]{Wit90}
\bibinfo{author}{\bibnamefont{Withoff}, \bibfnamefont{D.}}, and
  \bibinfo{author}{\bibfnamefont{E.}~\bibnamefont{Fradkin}},
  \bibinfo{year}{1990}, \bibinfo{journal}{Phys. Rev. Lett.}
  \textbf{\bibinfo{volume}{64}}, \bibinfo{pages}{1835}.

\bibitem[{\citenamefont{Wojs} \emph{et~al.}(2011)\citenamefont{Wojs, Moller,
  and Cooper}}]{WMC10}
\bibinfo{author}{\bibnamefont{Wojs}, \bibfnamefont{A.}},
  \bibinfo{author}{\bibfnamefont{G.}~\bibnamefont{Moller}}, and
  \bibinfo{author}{\bibfnamefont{N.~R.} \bibnamefont{Cooper}},
  \bibinfo{year}{2011}, \bibinfo{journal}{Acta Phys. Polon.}
  \textbf{\bibinfo{volume}{A119}}, \bibinfo{pages}{592}.

\bibitem[{\citenamefont{Wunsch}
  \emph{et~al.}(2008{\natexlab{a}})\citenamefont{Wunsch, Stauber, and
  Guinea}}]{WSG08}
\bibinfo{author}{\bibnamefont{Wunsch}, \bibfnamefont{B.}},
  \bibinfo{author}{\bibfnamefont{T.}~\bibnamefont{Stauber}}, and
  \bibinfo{author}{\bibfnamefont{F.}~\bibnamefont{Guinea}},
  \bibinfo{year}{2008}{\natexlab{a}}, \bibinfo{journal}{Phys. Rev. B}
  \textbf{\bibinfo{volume}{77}}, \bibinfo{pages}{035316}.

\bibitem[{\citenamefont{Wunsch} \emph{et~al.}(2007)\citenamefont{Wunsch,
  Stauber, Sols, and Guinea}}]{Wun07}
\bibinfo{author}{\bibnamefont{Wunsch}, \bibfnamefont{B.}},
  \bibinfo{author}{\bibfnamefont{T.}~\bibnamefont{Stauber}},
  \bibinfo{author}{\bibfnamefont{F.}~\bibnamefont{Sols}}, and
  \bibinfo{author}{\bibfnamefont{F.}~\bibnamefont{Guinea}},
  \bibinfo{year}{2007}, \bibinfo{journal}{New J. Phys.}
  \textbf{\bibinfo{volume}{8}}, \bibinfo{pages}{318}.

\bibitem[{\citenamefont{Wunsch}
  \emph{et~al.}(2008{\natexlab{b}})\citenamefont{Wunsch, Stauber, Sols, and
  Guinea}}]{WSSG08}
\bibinfo{author}{\bibnamefont{Wunsch}, \bibfnamefont{B.}},
  \bibinfo{author}{\bibfnamefont{T.}~\bibnamefont{Stauber}},
  \bibinfo{author}{\bibfnamefont{F.}~\bibnamefont{Sols}}, and
  \bibinfo{author}{\bibfnamefont{F.}~\bibnamefont{Guinea}},
  \bibinfo{year}{2008}{\natexlab{b}}, \bibinfo{journal}{Phys. Rev. Lett.}
  \textbf{\bibinfo{volume}{101}}, \bibinfo{pages}{036803}.

\bibitem[{\citenamefont{Yang} \emph{et~al.}(2006)\citenamefont{Yang, {Das
  Sarma}, and MacDonald}}]{YSM06}
\bibinfo{author}{\bibnamefont{Yang}, \bibfnamefont{K.}},
  \bibinfo{author}{\bibfnamefont{S.}~\bibnamefont{{Das Sarma}}}, and
  \bibinfo{author}{\bibfnamefont{A.~H.} \bibnamefont{MacDonald}},
  \bibinfo{year}{2006}, \bibinfo{journal}{Phys. Rev. B}
  \textbf{\bibinfo{volume}{74}}, \bibinfo{pages}{075423}.

\bibitem[{\citenamefont{Yao} \emph{et~al.}(2007)\citenamefont{Yao, Ye, Qi,
  Zhang, and Fang}}]{Yao07}
\bibinfo{author}{\bibnamefont{Yao}, \bibfnamefont{Y.}},
  \bibinfo{author}{\bibfnamefont{F.}~\bibnamefont{Ye}},
  \bibinfo{author}{\bibfnamefont{X.-L.} \bibnamefont{Qi}},
  \bibinfo{author}{\bibfnamefont{S.-C.} \bibnamefont{Zhang}}, and
  \bibinfo{author}{\bibfnamefont{Z.}~\bibnamefont{Fang}}, \bibinfo{year}{2007},
  \bibinfo{journal}{Phys. Rev. B} \textbf{\bibinfo{volume}{75}},
  \bibinfo{pages}{041401(R)}.

\bibitem[{\citenamefont{Yazyev}(2008)}]{Y08}
\bibinfo{author}{\bibnamefont{Yazyev}, \bibfnamefont{O.~V.}},
  \bibinfo{year}{2008}, \bibinfo{journal}{Phys. Rev. Lett.}
  \textbf{\bibinfo{volume}{101}}, \bibinfo{pages}{037203}.

\bibitem[{\citenamefont{Yazyev and Helm}(2007)}]{Yaz07}
\bibinfo{author}{\bibnamefont{Yazyev}, \bibfnamefont{O.~V.}}, and
  \bibinfo{author}{\bibfnamefont{L.}~\bibnamefont{Helm}}, \bibinfo{year}{2007},
  \bibinfo{journal}{Phys. Rev. B} \textbf{\bibinfo{volume}{75}},
  \bibinfo{pages}{125408}.

\bibitem[{\citenamefont{Ye and Sachdev}(1998)}]{Ye98}
\bibinfo{author}{\bibnamefont{Ye}, \bibfnamefont{J.}}, and
  \bibinfo{author}{\bibfnamefont{S.}~\bibnamefont{Sachdev}},
  \bibinfo{year}{1998}, \bibinfo{journal}{Phys. Rev. Lett.}
  \textbf{\bibinfo{volume}{80}}, \bibinfo{pages}{5409}.

\bibitem[{\citenamefont{Zeldovich and Popov}(1972)}]{Zeldovich:1972}
\bibinfo{author}{\bibnamefont{Zeldovich}, \bibfnamefont{Y.~B.}}, and
  \bibinfo{author}{\bibfnamefont{V.~S.} \bibnamefont{Popov}},
  \bibinfo{year}{1972}, \bibinfo{journal}{Sov. Phys. Uspekhi}
  \textbf{\bibinfo{volume}{14}}, \bibinfo{pages}{673}.

\bibitem[{\citenamefont{Zhang} \emph{et~al.}(2010)\citenamefont{Zhang, Min,
  Polini, and MacDonald}}]{Zha10}
\bibinfo{author}{\bibnamefont{Zhang}, \bibfnamefont{F.}},
  \bibinfo{author}{\bibfnamefont{H.}~\bibnamefont{Min}},
  \bibinfo{author}{\bibfnamefont{M.}~\bibnamefont{Polini}}, and
  \bibinfo{author}{\bibfnamefont{A.~H.} \bibnamefont{MacDonald}},
  \bibinfo{year}{2010}, \bibinfo{journal}{Phys. Rev. B}
  \textbf{\bibinfo{volume}{81}}, \bibinfo{pages}{041402}.

\bibitem[{\citenamefont{Zhang} \emph{et~al.}(2001)\citenamefont{Zhang, Hu, and
  Yu}}]{Zha01}
\bibinfo{author}{\bibnamefont{Zhang}, \bibfnamefont{G.-M.}},
  \bibinfo{author}{\bibfnamefont{H.}~\bibnamefont{Hu}}, and
  \bibinfo{author}{\bibfnamefont{L.}~\bibnamefont{Yu}}, \bibinfo{year}{2001},
  \bibinfo{journal}{Phys. Rev. Lett.} \textbf{\bibinfo{volume}{86}},
  \bibinfo{pages}{704}.

\bibitem[{\citenamefont{Zhang} \emph{et~al.}(2006)\citenamefont{Zhang, Jiang,
  Small, Purewal, Tan, Fazlollahi, Chudow, Stormer, Jaszczak, and
  Kim}}]{Zetal06}
\bibinfo{author}{\bibnamefont{Zhang}, \bibfnamefont{Y.}},
  \bibinfo{author}{\bibfnamefont{Z.}~\bibnamefont{Jiang}},
  \bibinfo{author}{\bibfnamefont{J.~P.} \bibnamefont{Small}},
  \bibinfo{author}{\bibfnamefont{M.~S.} \bibnamefont{Purewal}},
  \bibinfo{author}{\bibfnamefont{Y.-W.} \bibnamefont{Tan}},
  \bibinfo{author}{\bibfnamefont{M.}~\bibnamefont{Fazlollahi}},
  \bibinfo{author}{\bibfnamefont{J.~D.} \bibnamefont{Chudow}},
  \bibinfo{author}{\bibfnamefont{H.~L.} \bibnamefont{Stormer}},
  \bibinfo{author}{\bibfnamefont{J.~A.} \bibnamefont{Jaszczak}}, and
  \bibinfo{author}{\bibfnamefont{P.}~\bibnamefont{Kim}}, \bibinfo{year}{2006},
  \bibinfo{journal}{Phys. Rev. Lett.} \textbf{\bibinfo{volume}{96}},
  \bibinfo{pages}{136806}.

\bibitem[{\citenamefont{Zhang} \emph{et~al.}(2005)\citenamefont{Zhang, Tan,
  Stormer, and Kim}}]{Zhang:2005}
\bibinfo{author}{\bibnamefont{Zhang}, \bibfnamefont{Y.}},
  \bibinfo{author}{\bibfnamefont{Y.-W.} \bibnamefont{Tan}},
  \bibinfo{author}{\bibfnamefont{H.~L.} \bibnamefont{Stormer}}, and
  \bibinfo{author}{\bibfnamefont{P.}~\bibnamefont{Kim}}, \bibinfo{year}{2005},
  \bibinfo{journal}{Nature} \textbf{\bibinfo{volume}{438}},
  \bibinfo{pages}{201}.

\bibitem[{\citenamefont{Zhang} \emph{et~al.}(2009)\citenamefont{Zhang, Tang,
  Girit, Hao, Martin, Zettl, Crommie, Shen, and Wang}}]{Zha09}
\bibinfo{author}{\bibnamefont{Zhang}, \bibfnamefont{Y.}},
  \bibinfo{author}{\bibfnamefont{T.-T.} \bibnamefont{Tang}},
  \bibinfo{author}{\bibfnamefont{C.}~\bibnamefont{Girit}},
  \bibinfo{author}{\bibfnamefont{Z.}~\bibnamefont{Hao}},
  \bibinfo{author}{\bibfnamefont{M.}~\bibnamefont{Martin}},
  \bibinfo{author}{\bibfnamefont{A.}~\bibnamefont{Zettl}},
  \bibinfo{author}{\bibfnamefont{M.~F.} \bibnamefont{Crommie}},
  \bibinfo{author}{\bibfnamefont{Y.~R.} \bibnamefont{Shen}}, and
  \bibinfo{author}{\bibfnamefont{F.}~\bibnamefont{Wang}}, \bibinfo{year}{2009},
  \bibinfo{journal}{Nature} \textbf{\bibinfo{volume}{459}},
  \bibinfo{pages}{820}.

\bibitem[{\citenamefont{Zhao and Paramekanti}(2006)}]{Zha06}
\bibinfo{author}{\bibnamefont{Zhao}, \bibfnamefont{E.}}, and
  \bibinfo{author}{\bibfnamefont{A.}~\bibnamefont{Paramekanti}},
  \bibinfo{year}{2006}, \bibinfo{journal}{Phys. Rev. Lett.}
  \textbf{\bibinfo{volume}{97}}, \bibinfo{pages}{230404}.

\bibitem[{\citenamefont{Zhou} \emph{et~al.}(2009)\citenamefont{Zhou, Wang, Sun,
  Chen, Kawazoe, and Jena}}]{Zho09}
\bibinfo{author}{\bibnamefont{Zhou}, \bibfnamefont{J.}},
  \bibinfo{author}{\bibfnamefont{Q.}~\bibnamefont{Wang}},
  \bibinfo{author}{\bibfnamefont{Q.}~\bibnamefont{Sun}},
  \bibinfo{author}{\bibfnamefont{X.~S.} \bibnamefont{Chen}},
  \bibinfo{author}{\bibfnamefont{Y.}~\bibnamefont{Kawazoe}}, and
  \bibinfo{author}{\bibfnamefont{P.}~\bibnamefont{Jena}}, \bibinfo{year}{2009},
  \bibinfo{journal}{Nano Lett.} \textbf{\bibinfo{volume}{9}},
  \bibinfo{pages}{3867}.

\bibitem[{\citenamefont{Zhou} \emph{et~al.}(2007)\citenamefont{Zhou, Gweon,
  Fedorov, First, {de Heer}, Lee, Guinea, {Castro Neto}, and
  Lanzara}}]{Zhou:2007}
\bibinfo{author}{\bibnamefont{Zhou}, \bibfnamefont{S.~Y.}},
  \bibinfo{author}{\bibfnamefont{G.~H.} \bibnamefont{Gweon}},
  \bibinfo{author}{\bibfnamefont{A.~V.} \bibnamefont{Fedorov}},
  \bibinfo{author}{\bibfnamefont{P.~N.} \bibnamefont{First}},
  \bibinfo{author}{\bibfnamefont{W.~A.} \bibnamefont{{de Heer}}},
  \bibinfo{author}{\bibfnamefont{D.~H.} \bibnamefont{Lee}},
  \bibinfo{author}{\bibfnamefont{F.}~\bibnamefont{Guinea}},
  \bibinfo{author}{\bibfnamefont{A.~H.} \bibnamefont{{Castro Neto}}}, and
  \bibinfo{author}{\bibfnamefont{A.}~\bibnamefont{Lanzara}},
  \bibinfo{year}{2007}, \bibinfo{journal}{Nature Materials}
  \textbf{\bibinfo{volume}{6}}, \bibinfo{pages}{770}.

\bibitem[{\citenamefont{Zhou} \emph{et~al.}(2008)\citenamefont{Zhou, Siegel,
  Fedorov, and Lanzara}}]{Zhou08}
\bibinfo{author}{\bibnamefont{Zhou}, \bibfnamefont{S.~Y.}},
  \bibinfo{author}{\bibfnamefont{D.~A.} \bibnamefont{Siegel}},
  \bibinfo{author}{\bibfnamefont{A.~V.} \bibnamefont{Fedorov}}, and
  \bibinfo{author}{\bibfnamefont{A.}~\bibnamefont{Lanzara}},
  \bibinfo{year}{2008}, \bibinfo{journal}{Phys. Rev. B}
  \textbf{\bibinfo{volume}{78}}, \bibinfo{pages}{193404}.

\bibitem[{\citenamefont{Zhu and Ting}(2000)}]{Zhu00}
\bibinfo{author}{\bibnamefont{Zhu}, \bibfnamefont{J.-X.}}, and
  \bibinfo{author}{\bibfnamefont{C.~S.} \bibnamefont{Ting}},
  \bibinfo{year}{2000}, \bibinfo{journal}{Phys. Rev. B}
  \textbf{\bibinfo{volume}{63}}, \bibinfo{pages}{020506}.

\bibitem[{\citenamefont{Zhu} \emph{et~al.}(2009)\citenamefont{Zhu, Wang, Shi,
  Szeto, Chen, and Hou}}]{Zhu:2009}
\bibinfo{author}{\bibnamefont{Zhu}, \bibfnamefont{W.}},
  \bibinfo{author}{\bibfnamefont{Z.}~\bibnamefont{Wang}},
  \bibinfo{author}{\bibfnamefont{Q.}~\bibnamefont{Shi}},
  \bibinfo{author}{\bibfnamefont{K.~Y.} \bibnamefont{Szeto}},
  \bibinfo{author}{\bibfnamefont{J.}~\bibnamefont{Chen}}, and
  \bibinfo{author}{\bibfnamefont{J.~G.} \bibnamefont{Hou}},
  \bibinfo{year}{2009}, \bibinfo{journal}{Phys. Rev. B}
  \textbf{\bibinfo{volume}{79}}, \bibinfo{pages}{155430}.

\bibitem[{\citenamefont{Zhu} \emph{et~al.}(2010)\citenamefont{Zhu, Ding, and
  Berakdar}}]{Zhe10}
\bibinfo{author}{\bibnamefont{Zhu}, \bibfnamefont{Z.-G.}},
  \bibinfo{author}{\bibfnamefont{K.-H.} \bibnamefont{Ding}}, and
  \bibinfo{author}{\bibfnamefont{J.}~\bibnamefont{Berakdar}},
  \bibinfo{year}{2010}, \bibinfo{journal}{Europhys. Lett.}
  \textbf{\bibinfo{volume}{90}}, \bibinfo{pages}{67001}.

\bibitem[{\citenamefont{Zhuang} \emph{et~al.}(2009)\citenamefont{Zhuang, Sun,
  and Xie}}]{Zhu09}
\bibinfo{author}{\bibnamefont{Zhuang}, \bibfnamefont{H.-B.}},
  \bibinfo{author}{\bibfnamefont{Q.-F.} \bibnamefont{Sun}}, and
  \bibinfo{author}{\bibfnamefont{X.~C.} \bibnamefont{Xie}},
  \bibinfo{year}{2009}, \bibinfo{journal}{Europhys. Lett.}
  \textbf{\bibinfo{volume}{86}}, \bibinfo{pages}{58004}.

\end{thebibliography}

\end{document}